\documentclass[a4paper,11pt]{article}
\usepackage[margin=0.5in]{geometry}
\usepackage{jheppub}  
\usepackage{caption}
\usepackage{bbm,multicol}
\usepackage{esvect}
\usepackage{amstext}
\usepackage{subcaption}
\usepackage{hyperref}
\usepackage{filecontents}
\usepackage{amsfonts}
\usepackage{color,soul}
\usepackage{chapterbib}
\usepackage{float}
\usepackage{url}
\usepackage{slashed}
\usepackage{accents}
\usepackage{lipsum}
\usepackage{mathtools}
\usepackage{physics}
\usepackage[normalem]{ulem}

\begin{document}
\begin{titlepage}
    \begin{center}
        \vspace*{1cm}
        
        {\Large\textbf{ASYMMETRIC DARK MATTER FROM SCATTERING}}

        \vspace{0.5cm}

        \vspace{1.5cm}
        
        \textbf{THESIS SUBMITTED FOR THE DEGREE OF}\\
         \textbf{DOCTOR OF PHILOSOPHY (SCIENCE)}\\
         \vspace{0.2cm}
         \textbf{IN}\\ \vspace{0.2cm}\textbf{PHYSICS}\\\vspace{0.2cm} \textbf{BY}\\ \vspace{0.2cm}\textbf{DEEP GHOSH}\\ \vspace{0.2cm} \textbf{REGISTRATION NUMBER : 2020 03 06 01 02 035}

        \vspace{2.0 cm}
        \includegraphics[scale=0.6]{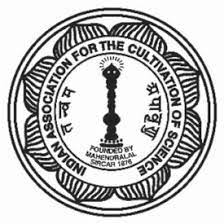}
        \vspace{2cm}
       
        \vfill
        
       \textbf{SCHOOL OF PHYSICAL SCIENCES}\\
       \vspace{0.4cm}
\textbf{INDIAN ASSOCIATION FOR THE CULTIVATION OF SCIENCE}\\\vspace{0.2cm}
\textbf{({\small DEEMED TO BE UNIVERSITY UNDER SECTION 3 OF UGC ACT, 1956})} \\
       
        \vspace{0.8cm}
         \Large{\textbf{2023}}        
        \vspace{1.2 cm}

    \end{center}
\end{titlepage}
\begin{center}
{\bf{\Large{Certificate}}}
\end{center}
\vspace{0.4 in}
I hereby certify that the matter described in the thesis titled,``\textbf{Asymmetric Dark Matter From Scattering}", has been carried out by \textbf{Mr. Deep Ghosh} at the \textbf{School of Physical Sciences (SPS)},Indian Association for the Cultivation of Science, Kolkata, India 
(a deemed to be University under the de novo category of UGC) under my supervision and that it has not been submitted elsewhere for the award of any degree or diploma.\\\\\\\\\\\\
Dr. Satyanarayan Mukhopadhyay\\
(Thesis Supervisor)
\vspace{2 in}
\newpage
\begin{center}
{\bf{\Large{Declaration}}}
\end{center}
I declare that the matter described in the thesis titled, ``\textbf{Asymmetric Dark Matter From Scattering}" is the result of investigation carried out by me at the \textbf{School of Physical Sciences (SPS)}, Indian Association for the Cultivation of Science, Kolkata, India (a deemed to be University under the de novo category of UGC) under the supervision of \textbf {Dr. Satyanarayan Mukhopadhyay} and that it has not been submitted elsewhere for theaward of any degree or diploma. I declare that this written document represents my ideas, in my own words and I have adhered to all principles of academic honesty and integrity and have not misrepresented or fabricated or falsified any idea/data/fact/source in my submission. In keeping with the general practice in reporting scientific observations, due acknowledgement and citation has been made whenever the work described based on the findings of other investigators. Any omission that might have occurred by oversight or error of judgement is regretted. \\\\\\
Date:\\\\
Place: Jadavpur, Kolkata 700032\\
\begin{flushright}
 Deep Ghosh\\
{\normalsize Reg. No. 2020 03 06 01 02 035 }
\end{flushright}
\newpage
\begin{center}
{\bf{\Large{Dedication}}}
\end{center}
\topskip0pt
\vspace*{\fill}
\begin{center}
{\itshape \LARGE{\textbf{
Sriramakrishnaarpanamastu}}}\\ \vspace{0.5cm}

{\large \textbf{(at the lotus feet of Sri Ramakrishna Paramhamsa})}
\end{center}
\vspace{0.2cm}
\vspace*{\fill}
\newpage
\begin{center}
{\bf{\Large{Acknowledgment}}}
\end{center}
This is the hardest part to write because whatever we achieve or arrive at in the present, the seed was planted in the past. We ourselves can not figure what triggered the present situation. Consequently, the acknowledgement part can never be complete. Since childhood many people have definitely influenced me, even beyond my perception. Therefore, what I would do here is - I would try to mention individuals who are related to this thesis and this course of time. 

The content of this thesis started with a discussion with my PhD. supervisor Dr. Satyanarayan Mukhopadhyay, who was interested to somehow categorize different dark matter (DM) production mechanisms in terms of different topologies. Eventually, we found that that is too difficult because of many-folded degeneracy in the DM study. Very recently, we have found a better template for explaining different production mechanisms starting from a general viewpoint and i.e. in terms of the phase-space distribution of DM. It took almost four years to reach up to this point. Long story short, I have experienced the development in understanding of subject and I strongly believe that this is not the end. I had similar experience with him while I was a teaching assistant of many of his courses. There too, I noticed how the nitty-gritty details are taken care of. Often new insights sprouted. I have seen good teachers but never have seen how they prepare themselves. This time, I saw the process and the end product. I believe that will be helpful in ensuing years.

In the context of this thesis, the next important person is Dr. Avirup Ghosh, with whom I have written three papers and it was a pleasant experience. Though during the collaborations we mostly communicated remotely, we had good understanding of our situations. Of late, we are going through a similar phase of our life and that too is a support for my professional career.        

In last four years, whenever I got stuck and needed a third person view I approached Rohan Pramanick, who is a PhD. student at IIT Kharagpur. The uniqueness about him is that he every time takes my problem as his own and is open to discussion for hours together. In my experience, this is one of the rarest qualities in these days in any field. I wish him all success in his career.  

In IACS, my go-to person was Dr. Sougata Ganguly, who is very rigorous in all aspects of life. We had many discussions over this period. During my first work, I everyday used to tell him about my project. In reply, he threatened me to write the paper on his own. 

During this period, I have learned many life lessons from a very practical man- Sourav Gope, unfortunately who is my immediate junior. In hostel, the duo of Sourav and Sougata used to take a `\textit{different}' class of which I was a quick learner.  

In IACS, we have a thriving group in particle physics. Prof. Sourov Roy is a benevolent person to discuss and ask for any help. I did a particle physics course offered by Prof. Dilip Kumar Ghosh, who is very persistent so that we learn the basics of particle physics. My interactions with Prof. Utpal Chattopadhyay were very pleasant. We mostly communicated in the context of setting up the cluster facilities for our department. Without him, the maintenance of the cluster would be impossible. In addition, his mild behaviors make everybody feel very comfortable. I am also thankful to Prof. Biswarup Mukhopadhyaya of IISER Kolkata for being my collaborator of late. The project with him taught me many lessons about the academia. Besides, I am also thankful Prof. Jayanta Kumar Bhattacharjee for listening to my doubts and answering those during my PhD. career. I also thank Dr. Sumanta Chakraborty for various discussion with him.

I did not face any serious administrative problems because of efficient people at our departmental office. I am thankful to Mr. Subrata Balti, Mr. Suresh Mandal and Mr. Bikash Darjee for their effort for smooth passage of official works. We had more than a professional relationship for their amicable behavior.

Now, it's time for my friends and colleagues, who have endured my many mood swings and a daily criticism. In particular, I thank Anirban Das for facing most of mischievous criticism in our daily adda sessions. The most calm and composed character in our friend's circle is Indrajit Sau, with whom I also share a musical interest. In this regard, the most proficient one is my junior Somsubhra Ghosh, who is the lead singer of our soiree. Our cricket sessions are the best bonding moments in terms of intensity and enthusiasm. I am thankful to my colleague Isha Ali, Aritra Pal, Ananya Tapadar ; my seniors Subhadip Sau, Dr. Disha Bhatia, Dr. Heerak Banerjee, Dr. Sourav Nandy, Dr. Abhijit Kumar Saha, Dr. Mainak Chakraborty, Dr. Purushottam Ghosh, Dr. Tapoja Jha, Dr. Arnab Paul, Dr. Rajeev and Dr. Nimmala Narendra for various interactions. My special thanks to Dr. Sankha Banerjee for helping me in several projects. I am also indebted to my juniors for listening to my free advice time-to-time despite warnings from Sougatada. Many many thanks to Sk. Jeesun, Pratick Sarkar, Tanmoy Kumar, Nandini Das, Sagnik Chaudhuri, Koustav Mukherjee, Vikramaditya Mondal, Kaustav Das, Ribhu Paul, Shouvik Biswas. I have a pleasant experience with Dr. Madhumita Sarkar, Dr. Roopayan Ghosh, Dr. Samadrita Mukherjee, Dr. Bhaswati Mondal, Tista Banerjee, Mainak Pal, Somsubhra Ghosh in sharing a common office. With Roopayanda, I spent many evenings discussing about Indian cricket. In the post-corona period, I had many personal discussions with Madhumitadi including the trio of Somsubhra, Mainak and Tista. I am also thankful to my IIT Kharagpur friends - Suman Das, Sayan Kumar Das, Sudipta Sikder, Sabyasachi Moulik, Sarthak Duary for sharing their experiences throughout this period. I am also grateful to my good friend, Dr. Shashwata Ganguly for listening to my thoughts and ideas. 

Finally, for whom I did nothing, I am pretty sure, I will not be able to do anything in future as well ; but they will do everything as long as they live - pranaams to my parents.       
\newpage
\begin{center}
{\bf{\Large{Abstract}}}
\end{center}
In dark matter (DM) cosmology, the central question is how the present-day density of DM is generated from some initial conditions in the early universe. There are several possible answers in the literature without any conclusive observational proof. In this thesis, one of the generic possibilities has been studied in details. In particular, the production mechanisms of DM are important in devising observational and experimental methods of DM for its detection in current and future experiments. In this context, thermal dark matter is historically most-studied scenario, in which DM is thermalized with the visible sector in the early universe. A system in thermodynamic equilibrium is described by its temperature and chemical potential, thereby initial conditions become irrelevant for subsequent dynamics. The chemical potential of DM is related to the difference (asymmetry) between particle-antiparticle number densities in the dark sector. The presence of chemical potential of DM modifies the \textit{`freeze-out'} process significantly by which the present-day density of DM is generated. In this work, we have conducted a detailed study on how the chemical potential of DM can be generated in the early universe and, subsequently, how the final density and composition of DM transpire. This type of thermal DM scenario is called `Asymmetric dark matter'(ADM).

The most-studied scenario in generating particle-antiparticle asymmetry of the DM, is either a decay or scattering of mother sector particles. In this thesis, we have considered scenarios in which DM asymmetry is generated from DM scatterings only. The mother sector and the daughter sector are the same, thereby these situations are kind of economic in terms of particle content of a theory. In particular, ADM from the semi-annihilation shows that a single process can produce an asymmetry in the DM sector as well as saturate the relic density of DM. In this case, subsequent pair annihilation is not needed for removing the symmetric part of the DM density, like most baryogenesis and leptogenesis scenarios.

Continuing discussion about ADM, we find a novel interplay between DM self-scatterings and annihilation in deciding the DM density and composition. This is a consequence of $S$-matrix unitarity that relates the CP-violation in self-scatterings and annihilation which eventually decide the DM cosmology. We are able to realize this feature in a model a complex scalar DM stabilized by a reflection symmetry but violate a global $U(1)$ symmetry (related to the DM number) can accommodate the expected correlation between CP-violating self-scatterings and annihilation. This simple model can explain both symmetric and asymmetric relic of DM depending on the rate of the CP-conserving pair annihilation of DM.  

We further investigate the role of CP-conserving processes in generating particle-antiparticle asymmetry in baryogenesis and leptogenesis scenarios. When the rate of CP-conserving processes become comparable with that of the CP-violating one, the number density of the mother particle is controlled by the former process. Thus, the efficiency of asymmetry generation in the daughter sector becomes suppressed due to diminishing number density of the mother particle. For ADM scenarios, the CP-conserving processes play a dual role, namely suppress the asymmetry generation at the early epoch, subsequently enhance the final asymmetry removing the symmetric part of DM.   
  
\newpage
\begin{center}
{\bf{\Large{List of Publications and Preprints}}}
\end{center}
\vspace{0.2cm}
\textbf{List of publications included in this thesis }
\begin{enumerate}
\item \textbf{Asymmetric dark matter from semi-annihilation}, Avirup Ghosh, Deep Ghosh, Satyanarayan Mukhopadhyay,
\href{https://link.springer.com/article/10.1007/JHEP08(2020)149}{\textcolor{blue} {JHEP 08 (2020) 149}}

\item \textbf{Cosmology of complex scalar dark matter: Interplay of self-scattering and annihilation}, Avirup Ghosh, Deep Ghosh, Satyanarayan Mukhopadhyay,
\href{https://journals.aps.org/prd/abstract/10.1103/PhysRevD.104.123543}{\textcolor{blue} {Phys.Rev.D 104 (2021) 12, 12}}

\item \textbf{Revisiting the role of CP-conserving processes in cosmological particle–antiparticle asymmetries}, Avirup Ghosh, Deep Ghosh, Satyanarayan Mukhopadhyay,
\href{https://link.springer.com/article/10.1140/epjc/s10052-021-09848-5}{\textcolor{blue} { Eur.Phys.J.C 81 (2021) 11, 1038}}
\end{enumerate}
\textbf{Preprints not included in this thesis}
\begin{enumerate}
\item \textbf{Revisiting big bang nucleosynthesis with a new particle species : effect of co-annihilation with neutrons},
Deep Ghosh, 
e-Print:\href{https://inspirehep.net/literature/2120653}{\textcolor{blue} {2207.10499}}
\end{enumerate}
\newpage
\tableofcontents
\newpage

\section{\Large{Introduction}}
\label{chap:chap1}
\hspace{0.5cm}
\subsection{Dark matter as a component of our universe}
\label{sec:sec11}
A plethora of astrophysical and cosmological evidence has made a strong case for the existence of a non-luminous component of our universe, which is called dark matter (DM). In particular, it contributes approximately a quarter of the total energy budget of our universe. This piece of information has gained strong support only after the incredibly precise measurement of the cosmic microwave background (CMB) radiation via different satellites (COBE, WMAP, Planck) with ever-increasing precision in the recent past. The analysis of the CMB anaisotropies provides us with measurements of the baryon density and the total matter density of the universe, which in turn indicate that the matter density mostly consists of non-baryonic components. The total matter density has also been inferred by studying large-scale universe structures (galaxy clusters and super-clusters).\footnote{This is mainly done by first observing the distribution of galaxies at different red-shifts, then fitting cosmological models to match the observation using N-body simulations. For details see Refs.\cite{VIRGOConsortium:1997rhx,2dFGRS:2001csf,2DFGRS:2001zay,Doroshkevich:2002mz}.} The baryonic part of the matter density has been estimated from the primordial abundances of light nuclei, e.g., hydrogen, deuterium, helium, etc., found mostly in the metal-poor galaxies. Most of the heavy metals are formed in the cores of stars via the fusion of light elements. Consequently, primordial light nuclei are used up in a galaxy with a large number of stars. Therefore, galaxies with a lower number of stars are good candidates for measuring the primordial abundances of light elements. From a theoretical perspective, the light matter abundance is exactly calculable within the framework of the Standard Model (SM) of particle physics and hot big bang cosmology \cite{Wagoner:1966pv,Bernstein:1988ad,Mukhanov:2003xs}. Given an initial baryon density, the light matter abundance depends on the rates of various nuclear processes and the expansion rate of the universe at the radiation-dominated epoch. The genesis of light nuclei in the early universe is called Big Bang Nucleosynthesis (BBN), which predicts the total baryon density. All these together indicate that SM baryons cannot saturate the total density of the universe, thereby introducing a non-baryonic component into the context of structure formation in the universe.   

Though the existence of DM is well-placed from the observation of the universe at large scales, historically the existence of DM sprouted from some enigmatic astrophysical observations. One of the most celebrated and earliest works in this field  was done by Fritz Zwicky, who was engaged in the mass measurement of nebulae \footnote{Nebulae are the luminescent part of galaxy where stars are usually created out of gas and cosmic dust. } of the Coma cluster. In particular, the mass estimate was performed in two different ways; the first was done by measuring the luminosity of nebulae, subsequently inferring the mass via a standard calibration between the luminosity and the mass of the astrophysical object. The second method was based on the measurement of the gravitational mass by observing internal rotations of nebulae. The discrepancy found in these two measurements led him to indicate that the cluster might have some non-luminous objects contributing to the total mass of nebulae \cite{Zwicky:1937zza}. 

The next note-worthy work was done by Vera C. Rubin and her colleagues who calculated the rotational velocities of the Andromeda nebula assuming the circular motions only, from a spectroscopic survey of different emission regions of the galaxy \cite{Rubin:1970zza}. They found so-called `flat' rotation curves in the outer portions of the spiral galaxy, apparently devoid of any visible matter. Such feature can be explained with the assumption of an embedding `dark' halo within the galaxy. To note, in proposing a DM component of the concerned galaxy Newtonian dynamics is assumed to be true.\footnote{As an alternative idea to DM at galactic scales, Modified Newtonian Dynamics (MOND) was proposed in the 1980's. There still exists a bone of contention between the DM hypothesis and the MOND proposition. In particular, MOND theory is argued mostly in explaining galaxy rotation curves whereas in the context of cosmology, it is hard to accommodate features like CMB anaisotropy structure formation of the universe. For a comprehensive study on the issue please see Refs.\cite{Milgrom:1983ca,Famaey:2011kh,Milgrom:2019cle,Skordis:2020eui} and references therein.}     

The above discussion points towards the gravitational imprints of DM both at small scales ( $\mathcal{O}(1-10)\,\rm Kpc$) and at large scales ($\gtrsim 1\,\rm Mpc$) of the universe. Barring the gravitational information we do not have any conclusive picture of the micro-physics of DM, unlike the visible sector of the universe for which we have substantial knowledge about its constituent particles and their fundamental properties. The difficulty in determining the nature of DM can be elucidated from the fact that the mass of DM can range from approximately $10^{-22}$ eV up to a thousand of solar mass ($\sim 10^{66}$ eV), i.e. roughly a spread of $90$ orders of magnitude. The lower bound of the mass is calculated with the consideration that DM must be confined in dwarf galaxies, the smallest DM-rich structures known with size around $1\,\rm Kpc$. Now, DM particles must be confined within the galaxy. The velocity of DM is measured roughly $10^{-4} c$, where $c$ is velocity of light in vacuum. Considering DM particles as bosons we get a generic lower bound on the DM mass to be roughly $10^{-22}$ eV \cite{Hu:2000ke}, exploiting Heisenberg's uncertainty principle. To note, this bound is modified significantly for fermionic  DM particles. Then, due to Pauli's exclusion principle the available phase space for DM particles becomes substantially suppressed, thereby the lower bound of the DM mass is roughly $750\,\rm eV$ for dwarf galaxies \cite{Madsen:1990pe}. The upper bound comes from the stability of the stellar cluster in the galaxy, which can be perturbed due to tidal forces generated by nearby DM particles \cite{Rubakov:2017xzr}. It is evident that such a huge degeneracy in the mass spectrum of DM makes our life difficult in constructing both theoretical and observational tests of DM, other than its gravitational interaction. Therefore, we need strong ansatzes in order to probe the micro-physics of DM in great details. In this thesis, we would treat DM as a fundamental particle. Historically, the particle DM paradigm is the most-studied scenario regarding the nature of DM. Here, we wish to discuss a specific class of particle DM, hence a holistic view of this ever-evolving field is beyond the scope of the present study. To a get a panorama of different particle DM scenarios see Refs.\cite{Bertone:2004pz,Jungman:1995df,Profumo:2019ujg} and references therein. The main message conveyed via these vast literature on this topic is that we need one or more beyond SM particles to accommodate necessary features of DM in general, namely

\begin{enumerate}
\item[\bf{I.}] \textbf{Stability}: DM is either absolutely stable or atleast its lifetime is greater than the age of the universe ($\tau_U \sim 13.8 ~{\rm Gyr}$), otherwise the structure would not form as we see today. In fact, the lifetime of DM can be constrained model-independently from the CMB data, considering the DM decaying to radiation and massive particles. The lower bound on the lifetime of DM is, $\tau_{\rm DM} \geq 160 ~{\rm Gyr}$ at $95\%$ confidence level   \cite{Audren:2014bca}, apparently ten times larger than the age of the universe. 
\item[\bf{II.}] \textbf{Charge-neutrality}: DM is electromagnetically neutral or it might have tiny charge, several orders of magnitude smaller than the charge of an electron. In fact, the bound on the electric charge of DM is given from the power spectrum analysis of CMB anisotropy \cite{Boddy:2018wzy,Dubovsky:2003yn} as well as from the  $21$-cm cosmology \cite{Liu:2019knx,Barkana:2018lgd}. This type of scenario is often termed as ``Milli-charged DM" in the literature.

\item[\bf{III.}] \textbf{Velocity dispersion}: DM particles should be non-relativistic at the matter-radiation equality epoch to initiate the structure formation during later epochs. In particular, observations \cite{Simon:2019nxf,Bullock:2000qf,SDSS:2004aee} showed that at high red-shifts there exists galaxies which implies the structure formation starts with smaller structures leading to larger structures like galaxy-clusters, super-clusters. Therefore, DM with relativistic velocity dispersion is ruled out as this predicts large structures to form first, contrary to cosmological observations. Consequently, the velocity dispersion should be well below the relativistic scale at the relevant epoch which is often associated with the `coldness' of DM.
\end{enumerate}    
The above-mentioned three characteristics rule out all known SM particles as long as we are assuming DM as a fundamental particle. However, DM particles have most commonality with SM neutrinos, but its tiny mass would make it relativistic at the time of matter-radiation equality epoch, consequently the structure formation is expected to be \textit{top-down}, which is contrary to cosmological observations. This is why, heavy neutrinos were first thought of as suitable DM candidates within some well-motivated theoretical framework \cite{Lee:1977ua,Feng:2000gh}.  

In postulating any well-motivated theoretical framework for the particle DM, the all-important observation to be satisfied is the present-day density of DM, which is reported as the relic density of DM, given below.
\begin{align}
\Omega_{DM} = \frac{\rho_{DM}}{\rho_c}, \,\hspace{1cm} \rho_c \approx 10^{-46}\, h^2\, {\rm GeV}^4
\label{eq:relicdef}
\end{align}  
where, $\rho_{DM}$ is the energy density of DM in the current universe and $\rho_c$ is the critical density of the universe. The observed relic density from recent data from Planck satellite is quoted as, $\Omega_{DM} h^2 = 0.120 \pm 0.001$ \cite{Planck:2018vyg}, where $h$ is the scaling factor of the Hubble expansion rate. This brings us to a pertinent question regarding the unknown DM sector - \textit{how is the present-day DM density created} ? This question is very engaging as different production mechanisms of DM predict different mass regimes, apparently trying to alleviate huge degeneracy in the mass spectrum. There are two broad classes of production mechanisms, namely, \textit{thermal} and \textit{non-thermal}. The classification is based on the thermodynamics details of the DM sector with or without any connection to the  SM thermal bath. Then the immediate question would be - how do we know that there was a SM thermal bath at all in the early universe ? The observation of CMB radiation (i.e. a black-body radiation with temperature, $ T_\gamma \sim 2.73\, \rm K$) and the expansion of the universe have confirmed that the early universe can be treated as a thermodynamic system with very high temperature compared to the present temperature of the universe (technically the temperature of the relic photons). This not only opens a pandora's box but also greatly simplifies our discussion in classifying particle DM in subsequent sections. 
 
\subsection{Particle Dark matter : Thermodynamic classification}
\label{sec:PDM}
\subsubsection{Preliminaries}
In an expanding universe, as we go back in time the density of matter or radiation are much higher than the today's universe. For example, in today's universe the number density of relic photons is $ n_\gamma \sim 410 \, \rm cm^{-3}$, whereas at the time of BBN, $n_\gamma \sim 10^{31} \, \rm cm^{-3}$. Consequently, there might be sufficient interaction between different components of the universe to create a thermodynamic system\footnote{Once there is lots of photons with sufficient energy it can produce electron-positron, quark-antiquark pairs from the vacuum. All these components can remain in thermal equilibrium from some epochs in the early universe depending on the relevant interaction strength.}. This opens a window of possibilities to study thermal properties of particle DM and its possible production mechanisms.Before, going to the thermodynamic scenarios of DM, let's take a detour to some basics of statistical physics, which would be instrumental to derive various aspects of DM physics in ensuing chapters. 

Any system of $N$ particles and volume $V$ at time $t$  is fully described by its phase-space distribution function, $f(p,q,t)$, where $p$, $q$ are generalised momentum and co-ordinates. In general, the determination of $f(p,q,t)$ is notoriously difficult to calculate as it requires the knowledge of exact initial conditions as well as all possible interactions of particles within the system. The situation is greatly simplified when we assume the \textit{ergodicity} of the system, i.e. the system is described independent of its pre-histories. In particular, the instantaneous value of a physical quantity $A$ is given by its the time-averaged value. This happens when the different parts of system interact long enough to create a chaotic motion in the total phase-space volume. Then, the time-averaged value of $A$ is same as the ensemble average of $A$, i.e.
\begin{align}
\expval{A} &= \lim_{\tau\to\infty}\frac{1}{\tau} \int^\tau_0  A(p(t),q(t)) \,dt \\
&=\frac{\int A(p,q)\, f(p,q) \,d^{3N}q\, d^{3N}p}{\int\, f(p,q) \,d^{3N}q\, d^{3N}p}
\label{eq:enaverage}
\end{align}      
The ergodic principle allows the existence of an time-independent phase-space distribution, $f(p,q)$, which led to the definition of thermal equilibrium via the local conservation law of phase-space points for a closed system as the following,
\begin{align}
\frac{df}{dt} =\frac{\partial f}{\partial t} + [f,H_m] = 0 ,
\label{eq:louville}
\end{align} 
where $H_m$ is the Hamiltonian of the system. The equilibrium condition, $\frac{\partial f}{\partial t} = 0$ implies $ [f,H_m] = 0 $. Hence, $f(p,q)$ is in general a combination of integrals of motion. For a closed static system, the total energy ($E_0$) of the system is an integral of motion. Then, the phase-space distribution function is given by \footnote{In principle, for a closed system the phase-space distribution must include all seven mechanical invariants (energy, three components of momentum and three components of angular momentum). The system is assumed to be at rest as a whole which allows to set all other invariants to zero except the total energy.}
\begin{align}
f(E)= \text{constant} \times \,\delta(E-E_0).
\label{eq:micro}
\end{align}    
This is so-called phase-space distribution function of a \textit{micro-canonical} ensemble. With this knowledge now we can calculate various thermodynamic quantities of our interest. The definition of entropy of the system in thermal equilibrium is given by 
\begin{align}
S= \log \tilde{\Omega} (E,V,N), \,\,\, k_B=1 ,
\end{align}  
where $\tilde{\Omega}$ being the total number of micro-states for the system defined earlier is calculated using the phase-space distribution function defined in Eq.\ref{eq:micro}. Now, we can define several thermodynamic macroscopic quantities (not limited to micro-canonical systems) as,
\begin{align}
\frac{1}{T} = \frac{\partial S}{\partial E}\,\,\, \text{(temperature)}; P= T\frac{\partial S}{\partial V}\,\,\, \text{(pressure)};\,\,\,\mu=-T\frac{\partial S}{\partial N}\,\,\, \text{(chemical potential)}  .
\label{eq:defintions}
\end{align}
Now, $S$ is an extensive variable, i.e. $S(\lambda E,\lambda V,\lambda N)=\lambda\, S(E,V,N)$. Choosing $\lambda=1/V$ we end up defining scale independent six `intensive' ($T,\mu,\rho \,(=E/V),n\, (=N/V),s \,(=S/V),P$) quantities, out of which any two quantities can be chosen independently to describe a thermodynamic system. Though the above discussion is based on micro-canonical system, i.e. there is no energy exchange or change in the number of particles, the main definitions are applicable in other types of ensembles too. In our case, the system is essentially a grand canonical one, where the number of particles does change. For example, the number of photons is not conserved. Given such a situation, we now want to discuss the issue of chemical potential when the system is in thermal equilibrium. For grand canonical systems, we have to look for conserved charges rather than conserved particle species. Therefore, we assign independent chemical potentials to conserved charges, which are of-course related to the chemical potential associated to individual particle species. Let's say, there are $k$ types of particles and $L(\leq k)$ conserved charges. Then the $l^{th}$ conserved charge can be written as
\begin{align}
\tilde{N}_{(l)}= \sum^k_{i=1} C^i_{(l)} \, N_i~ ,
\end{align}  
where $i$ is the index for particle species and $l$ is for a particular charge. $C^i_{(l)}$'s determine specific combination of particle species that conserves a particular charge. For example, in electron-positron pair annihilation the combination that conserves electric charge is $(N_{e^-}-N_{e^+})$, the number difference between electron and positron. To note, if there is no distinction between particle and antiparticle (e.g. for real scalar, Majorana particles), there is no conserved charge in the system. We now assign chemical potential, $\tilde{\mu}_{l}$ to the $l^{th}$ conserved charge and which relates to the chemical potential ($\mu_i$) of $i^{th}$ particle species as the following \cite{Ashoke}.
\begin{align}
\sum^L_{l=1} \tilde{\mu}_{(l)} \tilde{N}_{(l)} &= \sum^L_{l=1} \tilde{\mu}_{(l)} \sum^k_{i=1} C^i_{(l)} \, N_i=\sum^k_{i=1} \mu_i N_i ~. \nonumber\\
\mu_i &= \sum^L_{l=1} C^i_{(l)} \tilde{\mu}_{(l)}  ~.
\label{eq:chem}
\end{align}
Now, for a generic particle-antiparticle pair annihilation process, associated charge is conserved thereby the chemical potential of particle opposite to that of antiparticle in thermal equilibrium. It is apparent from the fact that in an annihilation process particle and antiparticle modifies the concerned charge in equal and opposite quantity. Therefore, 
\begin{align}
\tilde{N}_{(l)}=\sum^{k/2}_{i=1} C^i_{(l)} N_i + &\sum^{k/2}_{\bar{i}=1} C^{\bar{i}}_{(l)} N_{\bar{i}}=\sum^{k/2}_{i=1} C^i_{(l)} (N_i-N_{\bar{i}})~~.\nonumber\\
C^{i}_{(l)}&=-C^{\bar{i}}_{(l)}\,\,\, \text{for each $l$}~~,
\label{eq:charge}
\end{align}   
where $\bar{i}$ denotes the antiparticle species. Using Eq.\ref{eq:chem} and Eq.\ref{eq:charge}, we find the relation between particle  and antiparticle chemical potential as, $\mu_i = -\mu_{\bar{i}}$. In particular, for $\tilde{N}_{(l)}=0$, the chemical potential is zero for a particle species. This is what happens for photons in thermal equilibrium. There is no conserved charge associated to photons, prompting us to set $\mu_\gamma =0$.

\subsubsection{Classification w.r.t temperature}

With this brief interlude of thermodynamic definitions, now we are all set to venture various possibilities of the unknown DM sector by choosing temperature ($T_{DM}$) and chemical potential ($\mu_{DM}$) as the two independent thermodynamic quantities. In terms of DM temperature we can have following two scenarios, namely,
\begin{itemize}
\item[\bf{I.}] The DM sector equilibrates with the SM bath having temperature, $T_{SM} = T_{DM}$ in the early universe. 
\item[\bf{II.}] The DM sector equilibriates with itself with temperature, $T_{DM} \neq T_{SM}$ in the early universe. 
\end{itemize}
The generic assumption for the above categories is that in the early universe DM particles are thermalised due to some interaction either with SM bath particles or within itself. For case {\bf I}, elastic scatterings between DM particles and SM particles irrespective of any other interaction within the DM sector would be sufficient to set up a common temperature between these two sectors as long as the expansion rate of the universe is smaller than the scattering rate per DM particle at the relevant epoch. We can even estimate the thermal averaged elastic scattering cross-section ($\expval{\sigma v}_e$, defined as in Eq.\ref{eq:enaverage}) for which so-called \textit{kinetic equilibrium} is established by comparing the scattering rate per DM particle ($\Gamma$) and the expansion of the universe, denoted by the Hubble parameter ($H$). The necessary condition of thermalisation is that in one Hubble time atleast one collision between DM and SM particles should happen in a co-moving volume, which can be expressed as the following.
\begin{align}
\frac{\Gamma}{H} \gtrsim 1, \,\, \Gamma = n_{SM} \expval{\sigma v}_e, \,\, H \approx \sqrt{\frac{g_*}{10}}\, \frac{T^2_{SM}}{M_P},~ M_P=2.4\times 10^{18} \rm GeV ~~,
\label{eq:condition}
\end{align}
where the number density of SM particles is given by, $n_{SM} \sim g_* T^3_{SM}$ and $g_*$ is the relativistic degrees of freedom present at some temperature. $v$ is the relative velocity of colliding particles in their centre of momentum frame. The above condition gives a rough lower bound on the thermal averaged cross-section of the scattering, i.e. 
\begin{align}
\expval{\sigma v}_e \gtrsim 5\times 10^{-16} \,{\rm GeV^{-2}}\, \sqrt{\frac{1}{10 \,g_*}} \left(\frac{1 \,{\rm MeV}}{T_{SM}}\right) ~~.
\label{eq:crossbound}
\end{align}
However, the transition from a pre-thermal phase to a thermal phase of a species is not a very straightforward thing although it might seem from the above condition. Nevertheless, it provides the ballpark value of the interaction rate in this simplistic approach. To note, the universe is considered to be radiation-dominated at $T_{SM}=1\,\rm MeV$ for successful BBN prediction. Therefore, at this temperature, DM can thermalize with the existing SM bath. For $T_{SM}\gtrsim 1\,\rm MeV$, the universe can be radiation-dominated and the thermalization process can happen at relatively higher epoch, consequently the bound on the elastic scattering cross-section would get weaker than the previous one. In fact, the onset of the radiation-dominated universe after a plausible inflationary epoch is still unknown. In this thesis, we would assume the standard scenario in which the radiation domination starts right after the inflationary epoch ends. In this context, the lower  bound on the oft-quoted \textit{Reheating} temperature is $\mathcal{O}(\rm MeV)$ \cite{deSalas:2015glj}.

For case {\bf II} the interaction between the DM sector and the SM sector is too feeble \footnote{This is sometimes called `Feebly Interacting Massive particle' (FIMP) DM scenario. For a comprehensive study see Refs.\cite{Hall:2009bx,Yaguna:2011qn,Bernal:2017kxu}.} to equilibrate to a common temperature but the self-interaction of DM particles is sufficient enough to thermalize within itself. To contrast, with the former case, the transition from a pre-thermal phase to thermal phase is necessarily dependent on the self-interaction of DM particles. Now, the question is how much self-interaction is needed for internal thermalization. Can we derive similar bound for self-interaction as in Eq.\ref{eq:crossbound} ? The self-interaction rate per DM particle depends on the ambient number density of DM at some epoch, thereby in this case we must know the pre-thermal number density of DM unlike the former case. Therefore, the cosmological estimation of self-interaction would be dependent on the specific scenarios of the pre-thermal phase. For example, in Refs.\cite{Ghosh:2022hen,Moroi:2020has,Ala-Mattinen:2022nuj,Lebedev:2021tas,Garcia:2018wtq} the pre-thermal number density of DM is determined by the decay of inflationary scalar field. However, the self-interaction of DM has been estimated from astrophysical event like merging of clusters. In addition, various small scale issues related to galaxy formation and the distribution of DM in the galaxy have pointed out similar estimates of the DM self-interaction, i.e. $\sigma_{DM}/m_{DM} \sim 1 {\rm cm^2/g} \sim 10^3 \,{\rm GeV^{-2}}\left(1\, GeV/m_{DM} \right) $ \cite{Harvey:2015hha,Tulin:2017ara}, where $\sigma_{DM}$ is the self-interaction cross-section and $m_{DM}$ is the mass of the DM particle.  

The number density of a thermalized DM sector for both the cases discussed is completely determined by $T_{DM}$ and $\mu_{DM}$ in thermal equilibrium. For case {\bf I}, the DM temperature is same as the SM bath temperature, which is known from the CMB observation. On the other hand, for case {\bf II} the situation is grim as there is no direct observation of the DM temperature. Hence, the temperature of the DM sector would be very much model-dependent, thus the detailed discussion is beyond the scope of this thesis. However, the observation of  primordial light nuclei abundance puts constraints on the relativistic species present during BBN, in turn it restricts the DM temperature as the following, assuming an internally thermalized single species, fermion DM sector \cite{Ghosh:2022frt}.
\begin{align}
\left(T_{DM}/T_{SM}\right) \lesssim 0.6 \hspace{1cm} \text{at 68\% C.L. \cite{Fields:2019pfx}}~~.
\end{align}

\subsubsection{Classification w.r.t chemical potential}

Now, the last bit in this discussion is the chemical potential of DM. If the DM particles are described by real scalar, Majorana fermion or vector boson field for which there is no conserved charges, then $\mu_{DM}$ is identically zero. For all other cases, like a Dirac fermion or a complex scalar field, $\mu_{DM}$ can take non-zero value in general. Though DM is believed to be a singlet under SM gauge group, we can define a conserved charge related to DM particle-antiparticle count, just like the baryon number in the SM. For example, in a pair annihilation process, $DM+\bar{DM}\rightarrow SM+SM$, the particle number of the DM sector does not conserve but the difference in particle and anti-particle number is conserved. Therefore, we can assign a global $U(1)_{DM}$ to the DM sector under which DM particle takes $+1$ charge and DM anti-particle takes $-1$ charge. As discussed earlier, in such scenario the chemical potentials associated to particle and anti-particle become equal and opposite to each other. This brings further classification of the DM sector in terms of chemical potential as,  
\begin{itemize}
\item[\bf{A.}] The DM sector is with vanishing chemical potential, thereby there is no difference between the number densities of particle and anti-particle.   
\item[\bf{B.}] The DM sector is with non-vanishing chemical potential, thereby there is finite difference between the number densities of particle and antiparticle. 
\end{itemize}
\subsubsection{Thermal dark matter with $\mu_{DM}=0$}
\label{sub:sym DM}
Now, we can straightway write down the thermal distribution function of DM, assuming $T_{DM}=T_{SM}$ and $\mu_{DM}=0$ (case {\bf IA}) as 
\begin{align}
f(E)=\frac{1}{e^{ E/T_{SM}}\pm 1}\, \hspace{0.8cm} (\text{$+1$ for fermions and $-1$ for bosons} )
\label{eq:distfn}
\end{align}
For further calculation, let's assume the DM sector consists of single particle species described by a complex scalar field, $\chi$ with mass $m_\chi$. The number density of DM particle (antiparticle) $\chi$ ($\chi^{\dagger}$) is calculated using the above distribution function as,
\begin{align}
n_{\chi(\chi^\dagger)}(T_{SM}) = 
    \begin{cases}
     \displaystyle{\frac{\zeta(3)}{\pi^2}} T_{SM}^3 & \text{for $m_\chi<<T_{SM}$ (Relativistic case)}\\
      \left(\frac{m_\chi T_{SM}}{2\pi}\right)^{3/2}e^{-m_\chi/T_{SM}} & \text{for $m_\chi>>T_{SM}$ (Non-relativistic case)}
    \end{cases}
   \label{eq:dmnumber}    
\end{align}    
The another quantity of our interest is the energy density of $\chi$ which is given by 
\begin{align}
\rho_{\chi(\chi^\dagger)}(T_{SM}) = 
    \begin{cases}
     \displaystyle{\frac{\pi^2}{30}} T_{SM}^4 & \text{for $m_\chi<<T_{SM}$ (Relativistic case)}\\
     m_\chi\, \left(\frac{m_\chi T_{SM}}{2\pi}\right)^{3/2}e^{-m_\chi/T_{SM}} & \text{for $m_\chi>>T_{SM}$ (Non-relativistic case)}
    \end{cases}
   \label{eq:dmdensity}    
\end{align} 
Now, if the DM sector is thermalized with the SM thermal bath (more accurately with the SM photons having  $T_{SM}=2.72 \rm K $ ) till the current epoch then we can calculate the present density of DM adding both the contributions from particle and antiparticle. As discussed earlier that for structure formation the DM sector must be non-relativistic at the time of matter-radiation equality epoch ($T^{eq}_{SM}\sim 0.75 \,\rm eV$), then at the current epoch DM is undoubtedly non-relativistic. We take a conservative value of the DM mass, $m_\chi= 1\, \rm eV$ so that it becomes non-relativistic at $T_{SM} = 0.75\, \rm eV$. Hence, the relic density of DM is given by Eq.\ref{eq:relicdef}, 
\begin{align}
\Omega_{DM} &=\frac{\rho_\chi+\rho_{\chi^{\dagger}}}{\rho_c}\nonumber\\
\Omega_{DM} h^2 &=\frac{2 \rho_\chi}{10^{-46}}\approx 4.5\times 10^3\, e^{-4265} = 0 
\end{align}
This clearly indicates that DM particles can not be in thermal equilibrium  till date in order to satisfy the observed relic density. The DM sector must be decoupled from the visible sector at some point of time in the cosmic history. Often this type of scenario is called `\textit{Thermal} DM'. However, the question fosters how long DM can be in thermal equilibrium. This is again related to the growth of the matter perturbation at the matter-radiation equality epoch. If DM is thermalized  with the SM bath then the in-falling of matter in the gravitational well would be hindered due to the radiation pressure of DM-photon fluid. Therefore, the DM sector should be decoupled from the cosmic soup at the decoupling temperature, $T_d \geq T^{eq}_{SM}$. In particular, for standard WIMP (weakly interacting massive particle) DM scenario, the decoupling temperature is generally around a GeV for $m_\chi=100\,\rm GeV$. Now, we can estimate the DM mass from the \textit{free streaming length} constrained by the Ly-$\alpha$ forest observation \cite{SDSS:2004aee}. Here, we present a very crude estimate of the free streaming length as the following. Details can be found in refs.\cite{Kolb:1990vq,Boyarsky:2008xj}. The free streaming length is a characteristic scale upto which the gravitation clustering of matter is suppressed for a collisionless fluid. The physical free streaming length ($\lambda_{FS}$) of DM particles at $T^{eq}_{SM}$ is given by \cite{Boyarsky:2008xj},
\begin{align}
\lambda_{FS} = 2 \pi \expval{v} H^{-1}_{eq}~~,
\label{eq:jeans}
\end{align}    
where $H_{eq}$ is the Hubble parameter and $\expval{v}$ is the velocity dispersion of DM particles at $T=T^{eq}_{SM}$. Now, the free streaming length scale in Eq.\ref{eq:jeans} corresponds to the scale at the present epoch as $\lambda_0 = (1+z_{eq})\lambda_{FS}$, where $z_{eq} = 3.2\times 10^3$, the redshift corresponding to the matter-radiation equality epoch and $2 \pi (1+z_{eq}) H^{-1}_{eq} = 640 \,\rm Mpc$. Now, the smallest DM structure is dwarf galaxies which are formed from the gravitational collapse of spatial size around $0.1 \, \rm Mpc$, therefore $\lambda_0 \leq 0.1 \,\rm Mpc$. Putting all these we get the velocity dispersion of DM to be
\begin{align}
\expval{v} \leq 10^{-4} \,c ~~,
\end{align} 
which signifies `\textit{coldness}' of DM at the matter-radiation equality epoch. Now, if DM decouples non-relativistically at $T_d=T^{eq}_{SM}$, then $\expval{v}= \sqrt{3\, T^{eq}_{SM}/m_\chi}$ assuming Maxwell-Boltzmann velocity distribution. This gives a lower bound on the DM mass as the following. 
\begin{align}
m_\chi \geq 225\,\rm MeV~~.
\end{align} 
However, the most stringent bound in this regard comes from the measurement of the extra relativistic freedom at the time of BBN and also from the CMB data. The mass bound for a particle species which elastically interacts with the SM thermal bath calculated from the observation convoluted into the effective number of extra neutrino species, defined as the following. \footnote{The definition is valid until electron-positron freeze-out temperature, $T_\gamma \sim 0.5 ~ {\rm MeV}$, as thereafter the temperature of neutrinos becomes different from the photon temperature.}    
\begin{align}
\Delta N_{eff} =\frac{8}{7}\frac{\rho_\chi}{\rho_\gamma}~~,
\label{eq:Neff}
\end{align}   
where $\rho_\gamma$ is the energy density of the SM photons. Eventually, the mass of a thermally coupled particle species roughly becomes, $m_\chi \gtrsim \mathcal{O}(MeV)$. For detailed discussion see Refs.\cite{Kolb:1986nf,Boehm:2013jpa,An:2022sva} and references therein.
     
The other alternative to the above paradigm is that DM decouples from the SM bath relativisically, but becomes non-relativistic at $T^{eq}_{SM}$. This implies $T_d$ is sufficiently earlier than $T^{eq}_{SM}$. In this case, the temperature of the dark sector at some cosmic time $t$, after the decoupling epoch becomes
\begin{align}
T_{DM}(t) = \left(\frac{g_*(T)}{g_*(T_d)}\right)^{1/3} T_{SM}(t)~~.
\end{align}
This is derived from the fact that the entropy of the DM sector and the visible sector are conserved separately after the decoupling of DM from the cosmic soup. To note, the decoupling of DM in this scenario must be before the BBN epoch, otherwise BBN observations would be messed up due to extra degrees of freedom contributing to the expansion of universe. Then the velocity dispersion at $T^{eq}_{SM}$ becomes $\expval{v}= \left(\frac{g_*(T_{eq})}{g_*(T_d)}\right)^{1/3} T^{eq}_{SM}/m_\chi$. The lower bound on the DM mass becomes,
\begin{align}
m_\chi \geq 2.5 \,{\rm KeV} \left(\frac{100}{g_{*}(T_d)}\right)^{1/3}~~~.
\end{align}
This scenario is called as `\textit{warm}' DM \cite{Gorbunov:2011zzc,Dodelson:1993je}, which provides a lower bound on the mass of thermal DM in a model-independent way. In passing we note that this formalism is also valid for purely non-thermal scenarios, in which $\expval{v}$ is calculated from the non-thermal phase-space distribution \cite{Ghosh:2022hen, Haque:2021mab}.   

Let's discuss about some details of thermal decoupling of DM in the present context. As we have discussed earlier that the elastic scatterings between DM and SM particles are enough to set thermal equilibrium. But, we need additional number changing processes which can reduce the DM number density to satisfy the observed relic density. These type of \textit{inelastic} processes set a chemical equilibrium between the DM sector and the visible sector and eventually go out-of-equilibrium due to the expansion of the universe. Eventually, the processes completely seize and the co-moving number density ($Y_\chi = n_\chi/s$, $s$ is the entropy density of the universe) of DM becomes `\textit{frozen}' at a particular value. This is called `chemical decoupling' of DM. The canonical example in this context is WIMP annihilation in which a pair of DM particles annihilates to lighter SM final states via the interaction with strength comparable to that of the SM weak interaction. To elaborate, let's write the relic density of DM in terms of co-moving number density as,
\begin{align}
\Omega_{DM} &= \frac{s_0}{\rho_c}\, m_\chi (Y_\chi (T_d) + Y_{\chi^{\dagger}}(T_d))\nonumber\\
\Omega_{DM} h^2 &= 4\times 10^8 \, m_\chi Y_{\chi} (T_d)~~.
\label{eq:symDM}
\end{align}     
$s_0$ is the present entropy density of the universe and $Y_\chi (T_d)$ is the co-moving number density of DM at the time of chemical decoupling. In the absence of chemical potential, $Y_\chi = Y_{\chi^{\dagger}}$ as before. Now, $Y_\chi (T_d)$ can be estimated by similar analysis as in Eq.\ref{eq:condition}. Assuming the chemical decoupling happening in the radiation-dominated universe we get,
\begin{align}
Y_{\chi} (T_d) = \frac{3}{4}\frac{1}{ M_P\,  T_d \expval{ \sigma v}_{ann} g_* (T_d)^{1/2}}~~,
\end{align}
where $\expval{\sigma v}_{ann}$ is the thermally averaged annihilation cross-section. For the non-relativistic `\textit{freeze-out}' of DM, the relic density depends weakly on the mass of DM as the scaled freeze-out temperature, $m_\chi/T_d$ is logarithmically dependent on the DM mass. We notice from Eq.\ref{eq:wimp}, the correct relic abundance is set only by the pair annihilation cross-section. 
\begin{align}
\frac{\Omega_{DM} h^2}{0.12} = \left(\frac{3 \times 10^{-9}\, \rm GeV^{-2}}{\expval{\sigma v}}\right)\, \left(\frac{m_\chi/T_d}{20}\right) \left(\frac{100}{g_* (T_d)}\right)^{1/2} ~~.
\label{eq:wimp}
\end{align}
Now, we can estimate typical mass of DM assuming some theoretical framework for calculating $\expval{\sigma v}$. Here we show two examples, one is contact interaction between scalar particles and another is four fermi-type interaction with Dirac fermions, as it was done historically by Lee-Weinberg \cite{Lee:1977ua}.
\begin{align}
\expval{\sigma v}_{ann} &=\frac{g^2}{32\pi m^2_\chi}\hspace{1cm} (\text{contact interaction})~~.\\
\expval{\sigma v}_{ann} &=\frac{G^2_F m^2_\chi}{2\pi }\hspace{1cm} (\text{four-fermi interaction})~~.
\end{align}
For weak interactions, $G_F= \frac{\sqrt{2}\, g^2}{8\, m^2_W} = 1.16 \times 10^{-5}\,{\rm GeV^{-2}}$, where $m_W$ is the mass of $W$ boson. For contact interaction we find $m_\chi \sim 1\, {\rm TeV} $ and for four-fermi interaction, $m_\chi \sim 10 \,{\rm GeV} $. Here, the decoupling temperature becomes approximately $m_\chi/20$, noticeably before the matter-radiation equality epoch.
\subsubsection{Thermal dark matter $\mu_{DM} \neq 0$ }

Now, we move to case {\bf IB} i.e. $T_{DM}=T_{SM}$ and $\mu_{DM}\neq 0$, for which the distribution functions for particle and antiparticle are different due to $\mu_\chi=-\mu_{\chi^{\dagger}}$, $\mu_\chi$ is the chemical potential assigned to $\chi$ particle. 
\begin{align}
f(E)=\frac{1}{e^{ (E\mp\mu_\chi)/T_{SM}} + 1}\, \hspace{0.8cm} (\text{$-\mu_\chi$ for particle and $+\mu_{\chi}$ for antiparticle} )~~.
\label{eq:distfn1}
\end{align}
The total number density of the DM sector in the relativistic regime (considering particle and antiparticle as two different species) is given by,
\begin{align}
n_\chi+n_{\chi^{\dagger}} = \frac{T^3_{SM}}{2\pi^2}\int^\infty_0 x^2 \, dx\left( \frac{1}{e^{x-\mu_\chi/T_{SM}}-1}+\frac{1}{e^{x+\mu_\chi/T_{SM}}-1} \right)~~,
 \label{eq:dmnumber1}    
\end{align}    
where $ E \approx p $ ; $p$ is the momentum of the particle and $x= p/T_{SM}$. The condition, $\mu_\chi << T_{SM}$, reproduces the case {\bf IA}. In the non-relativistic regime, the total energy density becomes
\begin{align}
\rho_{DM}=\rho_\chi+\rho_{\chi^{\dagger}} = 2 m_\chi\, \left(\frac{m_\chi T_{SM}}{2\pi}\right)^{3/2}e^{-m_\chi/T_{SM}} \cosh \left(\frac{\mu_\chi}{T_{SM}}\right)~~.   
\end{align}
One of the obvious observations here is that the energy density of DM with chemical potential is larger than the thermal DM with $\mu_\chi=0$ at any temperature as $\cosh y \geq 1$ for any real $y$. To note, in the previous case, the thermal number density of DM is ridiculously small due to Boltzmann-suppression, whereas here we can avoid such suppression due to an undetermined chemical potential which can be taken as $\mu_\chi \sim \mathcal{O}(m_\chi)$. Thus, we can match observed relic abundance of DM with the fine-tuning of $m_\chi$ and $\mu_\chi$ values. The velocity dispersion only depends on the temperature and the mass of DM, thereby the DM sector with non-zero chemical potential must also go out-of-equilibrium as argued for the case with $\mu_\chi=0$.

Now, in a pair annihilation of DM particle the difference in DM particle and antiparticle number densities would be conserved and that would affect the freeze-out dynamics as discussed in \ref{sub:sym DM}. In equilibrium the `\textit{asymmetry}' in number densities of the DM sector is given by, 
\begin{align}
n_\chi - n_{\chi^{\dagger}} &= \frac{T_{SM}^3}{\pi^2} \sinh{\left(\frac{\mu_\chi}{T_{SM}}\right)} \int^\infty_0 \frac{x^2\, dx}{e^{2x}-2 e^{x} \cosh (\mu_\chi/T_{SM})+1}\nonumber\\
&=\frac{T_{SM}^3}{3} \left(\frac{\mu_\chi}{T_{SM}}\right)\,\hspace{1cm} \text{for $\mu_\chi << T_{SM}$}
\label{netnumber}
\end{align}
The main difference between the present scenario and the former one is in the initial conditions. In this case, both particle and antiparticle number densities follow equilibrium distribution with an additional constraint, written in terms of co-moving number densities.
\begin{align}
Y_\chi - Y_{\chi^\dagger} = C ~~,
\label{eq:constraint}
\end{align} 
where $C$ is set at some earlier epoch and remains constant thereafter. For standard WIMP scenario, $C=0$. To get the freeze-out number density of DM particles via $2\rightarrow 2$ annihilation process, we need to solve \textit{Boltzmann Transport} equations (detailed in the next chapter) for $Y_\chi$ and $Y_{\chi^\dagger}$ in general. Using Eq.\ref{eq:constraint}, we can eliminate one of them and subsequently solve only the evolution equation of $Y_\chi$ again in the radiation-dominated universe with scaled temperature, $x=m_\chi/T_{SM} $ as
\begin{align}
\frac{d Y_\chi}{d x} = - \frac{\lambda \expval{\sigma v}_{ann}}{x^2}\left(Y_\chi (Y_\chi- C)-Y^{eq^2}_{\chi}\right)~~,
\label{eq:asywimp}
\end{align}  
where $\lambda=1.32 \,M_{\rm P}\,m_{\chi} g^{1/2}_*$ and $Y^{eq}_{\chi}$ is the co-moving equilibrium number density calculated using Eq \ref{eq:dmnumber}. Solving the above equation in asymptotic limit, i.e. $x \rightarrow \infty$, we get the solutions as the following
\begin{align}
Y_\chi(x\rightarrow \infty) =\frac{C}{1-e^{-\lambda \expval{\sigma v}_{ann}C x^{-1}_d}}~~,\nonumber\\
Y_{\chi^{\dagger}}(x\rightarrow \infty) =\frac{C}{e^{\lambda \expval{\sigma v}_{ann}C x^{-1}_d}-1}~~.
\label{eq:solution}
\end{align} 
where, $x_d=m_\chi/T_d$, $T_d$ being the decoupling temperature of pair annihilation of DM. The relic density of DM is given by \cite{Graesser:2011wi, Iminniyaz:2011yp},
\begin{align}
\Omega_{\rm DM} = \frac{s_0}{\rho_c}m_{\chi} C \coth \left(\frac{C\lambda \langle \sigma v \rangle_{ann}} {2 x_d} \right)~~. 
\label{eq:relicsemi}
\end{align} 
It is apparent that for $C \rightarrow 0$, we restore `\textit{symmetric}' WIMP scenario as in Eq.\ref{eq:symDM} recognizing $Y_\chi(T_d) = \frac{2 x_d}{\lambda\, \expval{\sigma v}_{ann}}$. The stark difference from the standard WIMP scenario is that the relic density has non-trivial dependence on the pair annihilation cross-section of DM. Now, we can define a measure of the final asymmetry stored in the DM sector assuming mostly DM particles survive.
\begin{align}
\eta = \frac{Y_{\chi} (x\rightarrow\infty)+Y_{\chi^\dagger}(x\rightarrow\infty)}{Y_{\chi}(x\rightarrow\infty)}= 1+ e^{-\frac{C\lambda \langle \sigma v \rangle_{ann}} {x_d}}~~.
\end{align}
It is evident that for large pair annihilation antiparticles (in this parametrization) get washed away to give \textit{completely asymmetric} DM for which $\eta \rightarrow 1$ whereas for WIMP scenario, $\eta\rightarrow 2$. This is called `\textit{symmetric}' DM scenario. For partially asymmetric DM, i.e. the unequal and non-zero proportion of both particle and antiparticle densities, $1 < \eta < 2$. For completely asymmetric DM scenario, the relic density is determined by the initial asymmetry, reminiscent of the baryon asymmetry in the visible sector. We find the mass of DM is controlled by the asymmetry in the DM sector as the following.
\begin{align}
m_\chi \simeq 5\, {\rm GeV} \left(\frac{0.12}{\Omega_{DM} h^2} \right)\left(\frac{C}{0.86\times 10^{-10}}\right)~~.
\label{eq:share}
\end{align}   
If we assume the baryon asymmetry (more accurately, baryon-to-entropy ratio) and the DM asymmetry is of the same order then $m_\chi \sim 5\,m_p$, $m_p$ being the mass of the proton. From observations we know $\Omega_{DM} = \Omega_b$, $\Omega_b$ being the relic density for baryons. This implies the final number density of DM and baryons might be the same in some specific scenarios. Nevertheless, we do not have any prior knowledge about the amount of asymmetry present in the DM sector, if at all. Therefore, this is very special scenario in the context of asymmetric dark matter (ADM). 

\subsection{Generation of dark matter asymmetry}
\label{sec:gdm}
In this section, we discuss how the non-zero chemical potential (related to $C$) for DM is generated in the early universe. To generate a particle-antiparticle asymmetry from a symmetric initial condition three Sakharov conditions \cite{Sakharov:1967dj} must be satisfied. These were proposed in the context of baryogenesis, namely
\begin{enumerate}
\item[ {\bf I.}] Particle number must be violated.
\item[{\bf II.}] $C$ and $CP$ violation are necessary.
\item[{\bf III.}] The concerned particle species must go \textit{out-of-equilibrium} from the thermal bath.
\end{enumerate} 
As discussed earlier, particle number is associated with some $U(1)$ global symmetry under which particle and antiparticle are oppositely charged. Therefore, to create a difference between particle and antiparticle number densities primarily we need violation of the global $U(1)$ symmetry. For example, for baryon (B) asymmetry $B$-number violation in necessary. To understand, the role of $CP$-violation we consider a current of $\chi$ as $J^\mu_\chi(\vec{x},t)$, which transforms under $CP$ as $J^{\mu}_\chi(\vec{x},t)\stackrel{CP}{\implies} -J^{\mu}(-\vec{x},t)$. Now, the average value of the corresponding charge is given by using Eq.\ref{eq:enaverage}
\begin{align}
\expval{Q(t)}=\int J^0(x,t) f(x,p) d^3p\, d^3 x~~.
\end{align}   
If $CP$ symmetry is preserved then $[O_{CP},H_m]=0$, where $O_{CP}$ is the generator of $CP$- transformation. For thermal equilibrium, $f=f(H_m)$, hence $[O_{CP},f]=0$, but the particle current is odd under $CP$-transformation, thereby
\begin{align}
O_{CP}\expval{Q(t)}O^{-1}_{CP} &= \int O_{CP}\, J^0_\chi(x,t)f(H_m(p))\,O^{-1}_{CP} \, d^3p\, d^3x \nonumber\\
&= \int O_{CP}\, J^0_\chi(x,t) \,O^{-1}_{CP} \, f(H_m(p)) d^3p\, d^3x \nonumber \\
&= -\expval{Q(t)}~~.
\end{align}
Consequently, even if there is particle number violation the net charge would  be zero due to $CP$-invariance of the underlying interaction. As long as the above derivation is concerned, we have already assumed the thermal distribution for the phase-space distribution function, which eventually indicates towards the third Sakharov condition, i.e. the out-of-equilibrium condition. However, this condition can be an intriguing one for different scenarios which are elucidated at length in Chapter \ref{chap:chap5} of this thesis. Despite the $CP$-violation, the ambient asymmetry in a particle sector is washed out in thermal equilibrium.    

There are many ways to generate an asymmetry in the DM sector.  We would like to discuss the issue in terms of two broad categories. The classification is based on the interpretation of Eq.\ref{eq:share}, namely
\begin{enumerate}
\item[{\bf A.}] The DM asymmetry is of the same order that of the baryon asymmetry.
\item[{\bf B.}] The DM asymmetry is independent of the baryon asymmetry.
\end{enumerate} 
The case {\bf A} is well-studied and spans over a wide range of models in the literature. Here we just sketch out the generic theme of those scenarios following Refs \cite{Petraki:2013wwa,Zurek:2013wia}. We first assign a \textit{dark baryon number}, $B_\chi$ to the DM sector particles and assume two linearly independent combinations of baryon number in the visible sector ($B_V$) and the dark baryon number, out of which one is broken and another one is conserved in an appropriate theory at high temperatures. Without loss of any generality, we can take the conserved and broken baryon charges as
\begin{align}
B_{c}= B_V-B_\chi~,\nonumber\\
B_{b}=B_V+B_\chi~~.
\label{eq:assign}
\end{align}
This kind of choice is a reminiscent of the baryogenesis via leptogenesis scenario in which an asymmetry in the lepton sector can be transferred to the baryon sector via the instanton-like electroweak process or the \textit{sphaleron} process which preserves ($B-L$) charge but violates ($B+L$) charge, where $L$ denotes the lepton number. At the some temperature, $T=M$ (mass of the right handed Majorana neutrinos) a non-zero ($B-L$) charge is generated from the zero charge configuration via generating excess leptons over antileptons. Later on, the sphaleron process coverts the net lepton number to the net baryon number. However, in the current scenario, we apriori do not know similar conversion techniques in the context of DM physics. Here, we concern ourselves with sharing of the asymmetry in both sectors. Therefore, we would impose the constraint, i.e. $C(B_c)=0$, where $C$ signifies the net charge (number) density per entropy density of the universe, while $C(B_b)=\eta$. This constraint implies similar number density of the baryon and the DM sector as discussed earlier. Therefore, $C (B_V)=C(B_\chi)=C (B_b)/2$. This scenario necessarily needs particular connection between the SM and the DM sector so that one of the two linear combinations of charges is conserved by a full renormalizable theory or an effective operator of the form,
\begin{align}
\mathcal{L}\supset\mathcal{O}_{B_V}~\mathcal{O}_{B_\chi}~~,
\end{align} 
where $\mathcal{O}_{B_V}$ consists visible sector field configuration which carries non-zero $B_V$ charge and $\mathcal{O}_{B_\chi}$ is the DM counterpart. The operator is invariant under $U(1)_{B_V-B_\chi}$ global symmetry. The stability of both baryon and DM is ensured by separate conservation laws of $B_V$ and $B_\chi$ number at low temperatures, which suggests a residual $U(1)_{B_V}$ symmetry in the visible sector and $U(1)_{B_\chi}$ symmetry in the DM sector. In SM, $B_V$ is identified as $B-L$ charge as the associated current is non-anomalous, i.e. conserved even at quantum level unlike the charge directly related to the baryon number.  

The case {\bf B} is more open-ended scenario, in which the DM asymmetry can be generated completely independent of the baryon asymmetry. This thesis deals with few examples of this kind. To contrast with the former case, we do not need operators that conserve some linear combination of charges of the visible sector and DM sector. In particular, the DM sector can be completely secluded from the SM sector and asymmetric. From a perspective of symmetry structure of a QFT, only requirement here is the violation of $U(1)_{B_\chi}$ symmetry corresponding to the DM number. In this type of DM models the stability of DM particles is ensured by  assigning some extra charge associated to a discrete symmetry. In chapters \ref{chap:chap3} and \ref{chap:chap4} we illustrate the generation of ADM with concrete models in which DM is stabilized by $\mathcal{Z}_3$  and $\mathcal{Z}_2$ symmetry respectively, while violating the global  $U(1)_{B_\chi}$.        
\newpage
\section{\Large{Boltzmann Transport Equations}}
\label{chap:chap2}
\hspace{0.5cm}
\subsection{Preliminaries}
As discussed in the previous chapter, the generic feature in producing the DM density in correct abundance, the out-of-equilibrium phenomena is absolutely necessary for both symmetric and asymmetric DM scenarios. In this chapter, we demonstrate the key tool that handles the thermodynmaic evolution of particles of different species throughout the cosmic history of our universe. The foundation of the out-of-equilibrium dynamics of a particle species has already been indicated in Eq.\ref{eq:louville}. We need to the study the equation in an expanding homogeneous isotropic background including all relevant collision terms, which are symbolically represented by the collision operator, $C[f]$ in the covariant form of the local conservation law of phase-space points given below.
\begin{equation}
\displaystyle{\frac{d f}{d \lambda}=C[f]}~~.
\end{equation}
 $\lambda$ here is the affine parameter related to \textit{Friedmann–Lemaître–Robertson–Walker} (FLRW) metric, which represents a homogeneous isotropic expanding universe at large scales. We use the following form of the metric with vanishing spatial curvature.
 \begin{equation}
 \displaystyle{ds^2=d\lambda^2=dt^2-a^2(t)\bigg[dx^2+dy^2+dz^2\bigg]}~~,
 \label{eq:rellu}
\end{equation}  
where, $a(t)$ accounts for the scale factor of the expansion and $x,y,z$ are co-moving co-ordinates. We can define canonical momentum corresponding to space-time co-ordinates as 
\begin{align}
\displaystyle{\frac{dt}{d\lambda}=P^0=\frac{E}{m}}, \hspace{0.5cm}\displaystyle{\frac{dx^i}{d\lambda}=P^i}~,
\end{align}
which are related to the magnitude of physical momentum $p$ as $p^2= a^2(t)\, \delta_{ij}P_i P_j$. Now for time-like geodesic $(ds/d\lambda)^2=1$ in our metric convention, which reproduces the energy-momentum relation for an on-shell massive particle, i.e. $m^2=E^2-p^2$. In general, the phase-space distribution function can be cast as a function of the co-moving co-ordinates, the magnitude of physical momenta or the energy, three directions of momenta and time. With the assumptions of homogeneity and isotropy, we can take the distribution function dependent only on the magnitude of the physical momenta and time, then the L.H.S of the Eq.\ref{eq:rellu} becomes
\begin{align}
 \frac{df}{d\lambda}&= \frac{dt}{d\lambda}\left[\frac{\partial f}{\partial t}+ \frac{\partial f}{\partial p} \frac{dp}{dt} \right]\nonumber\\
 &= \frac{E}{m}\left[\frac{\partial f}{\partial t}-Hp \frac{\partial f}{\partial p} \right]~,
\label{bl}
\end{align} 
where $dp/dt= H p$ and the Hubble parameter, $H=\frac{d \log a}{dt}$.
Now, the collision operator represents the rate of change of particles in unit phase-space volume. The co-variant form of $C[f]$ can be written as the following.
\begin{align}
C[f]=\frac{d f}{d \lambda}\bigg|_{coll^n}=\frac{E}{m}\frac{d f}{d t}\bigg|_{coll^n}= \frac{E}{m}\,(2\pi)^3\frac{dN}{g\, d^3 p\, d^4x}~,
\label{eq:colldef}
\end{align}
where $g$ is degrees of freedom the particle. For explicit calculation of $C[f]$, we need to assume some process. Here, for demonstration we take a generic $2\rightarrow 2$ scattering process with four different particles, with momentum assignments as, $p_1+p_2\rightarrow p_3+p_4$. Now, we want to write the collision term for particle 1 with momentum $p_1$. Evidently, in the forward scattering process, we lose a particle 1 and in the backward process we gain one. Therefore, the collision terms for $f_1$, i.e. distribution function of particle 1 with momentum $p_1$ becomes
\begin{align}
C[f_1]=\frac{E_1}{m_1}(2\pi)^3\frac{dN^{gain}-dN^{loss}}{g_1 d^3 p_1 d^4x}~~.
\end{align} 
The scattering is seen in the center of mass frame of initial the particles so that the volume of scattering is defined as the cylinder having a length of $v_{rel}dt=|v_1-v_2|dt$ and an area of $A$. The loss term originating from the forward scattering process and the gain term coming from the backward scattering are given by
\begin{align}
dN^{loss}=\displaystyle{\bigg[g_1f_1(p_1,t)d^3 p_1 d^{3}x\bigg]\bigg[g_2 f_2(p_2,t)d^3p_2\bigg]\bigg[Av_{rel}d t\bigg]\Gamma_{12\rightarrow 34}\,dt}~~,\nonumber\\
dN^{gain}=\displaystyle{\bigg[g_3f_3(p_3,t)d^3 p_3 d^{3} x\bigg]\bigg[g_4 f_4(p_4,t)d^3p_4\bigg]\bigg[A\bar{v}_{rel}d \bar{t}\bigg]\Gamma_{34\rightarrow 12}d\bar{t}}~~.
\end{align}   
where $\Gamma_{i\rightarrow j}$ is the transition probability from $i^{th}$ initial state to $j^{th}$ final state. $\Gamma_{i\rightarrow j}$ is calculated from a particular particle physics model. The generic form of the transition probabilities of the forward and backward processes are given by,
\begin{align}
\displaystyle{d\Gamma_{12\rightarrow 34}=\frac{1}{Av_{rel}dt}|M|^2_{12\rightarrow34}(2\pi)^4 \delta^4(p_1+p_2-p_3-p_4)\frac{1}{2E_1 2E_2}\frac{d^3p_3}{(2\pi)^3 2E_3}\frac{d^3p_4}{(2\pi)^3 2E_4}}~,\nonumber\\
\displaystyle{d\Gamma_{34\rightarrow 12}=\frac{1}{A\bar{v}_{rel}d\bar{t}}|M|^2_{34\rightarrow12}(2\pi)^4 \delta^4(p_1+p_2-p_3-p_4)\frac{1}{2E_3 2E_4}\frac{d^3p_1}{(2\pi)^3 2E_1}\frac{d^3p_2}{(2\pi)^3 2E_2}}~,
\end{align}
where $|M|^2_{i\rightarrow j}$ is the probability amplitude of the process, $i\rightarrow j$. The total collision term for particle 1 with momentum $p_1$ becomes after integrating all other undetermined momenta, summing over all  final spins and averaged over initial spins,
\begin{align}
C[f_1]= \frac{E_1}{m_1}\int (2\pi)^4 \delta^4(p_1+p_2-p_3-p_4)\prod^{4}_{i=2}\frac{d^3 p_{i}}{(2\pi)^3 2E_{i}}\frac{1}{2E_1}\bigg[g_2 f_{1}(p_1,t)f_2(p_2,t)\frac{1}{g_1 g_2}\sum_{\text{all spins}}|M|^2_{12\rightarrow 34}\nonumber\\ \displaystyle{- \frac{g_3g_4}{g_1} f_3(p_3,t)f_{4}(p_4,t)\frac{1}{g_3 g_4} \sum_{\text{all spins}}|M|^2_{34\rightarrow 12}\bigg]}~.
\label{eq:twocoll}
\end{align} 
The Eq.\ref{eq:twocoll} simplifies a bit when the charge conjugation parity (CP) symmetry is respected in the concerned process, thereby $|M|^2_{12\rightarrow 34} = |M|^2_{34\rightarrow 12}$. We note, with this simplification we can define a single cross-section for both forward and backward scattering processes by the standard particle physics definition as the following.
\begin{align}
\sigma=\frac{1}{g_1 g_2}\frac{1}{2E_1 2E_2 v_{rel}}\int\sum_{\text{all spins}}|M|^2_{12\rightarrow34}(2\pi)^4 \delta^4(p_1+p_2-p_3-p_4)\frac{d^3p_3}{(2\pi)^3 2E_3}\frac{d^3p_4}{(2\pi)^3 2E_4}~.
\end{align}
Now, we combine Eq.\ref{bl} and Eq.\ref{eq:twocoll} to write the dynamical equation for the phase-space distribution function of particle $1$ with momentum $p_1$ as,
\begin{align}
\displaystyle{\frac{\partial f_1(p_1,t)}{\partial t}- H p_1\frac{\partial f_1(p_1,t)}{\partial p_1}=-g_2\int \frac{d^{3}p_{2}}{(2\pi)^3}    \sigma v_{rel}\bigg[f_{1}(p_1,t)f_2(p_2,t)-f_{3}(p_1,t)f_4(p_2,t)\bigg]}~.
\label{boltz}
\end{align}
In deriving the collision terms we implicitly have assumed particles of distinguishable whereas in defining the cross-section quantum nature of particles is embedded by the construction of underlying quantum field theory. Therefore, the above expression demands some modifications depending on the spin of the colliding particles. Let's say, we are considering forward scattering process in which particle 3 is produced with momentum $p_3$ and some spin state from the initial state. Now, there is some finite probability that particular state already occupied as we are dealing with distribution of particles. If the particle is a fermion then due to Pauli's exclusion principle, then the filling of that state would be blocked. To take this effect we modify the forward scattering term by multiplying $(1-f_3(p_3))$ for particle 3. Similar terms should be multiplied accordingly for all other particles involved in the collision process. If the particles are bosons then there would be an enhancement factor, $(1+f_i)$ for $i^{th}$ particle multiplied similarly. As one can see that these arguments are rather heuristic, for a formal derivation of the quantum Boltzmann equation from the first principle of the thermal quantum field theory (QFT) can be found in Ref.\cite{Dighera:2016urx}. Anyway, including the quantum nature of colliding particles, the Eq.\ref{boltz} becomes \footnote{Here we can not perform $p_3$ and $p_4$  integral separately as in Eq.\ref{boltz}.}
\begin{align}
\displaystyle{\frac{\partial f_1(p_1,t)}{\partial t}-H p_1\frac{\partial f_1(p_1,t)}{\partial p_1}=-g_2\int d^{3}p_{2} \,   \sigma v_{rel}\bigg[f_{1}(p_1,t)f_2(p_2,t)(1\pm f_3(p_3,t))(1\pm f_4(p_4,t))}\nonumber\\
-f_{3}(p_3,t)f_4(p_4,t)(1\pm f_1(p_1,t))(1\pm f_2(p_2,t))\bigg]~.
\label{eq:genboltz}
\end{align}  
At this level already we can demonstrate the chemical equilibrium of the system. The chemical equilibrium is established in a static universe when the forward and backward reaction rates are equal, which automatically implies that $\frac{\partial f_1}{\partial t} =0$ and this marks the thermal equilibrium of the system. In the expanding universe, there is an additional factor coming from the expansion term (second term in the L.H.S of Eq.\ref{eq:genboltz}) which is responsible of red-shift of the momentum distribution function of particles. Nevertheless, in the chemical equilibrium, the phase-space distribution function remains in its equilibrium shape with scaled thermodynamic parameters. Now, assuming all particle are bosons, we get the following relation in the chemical equilibrium, which is sometimes termed as the \textit{detailed balance} of a reaction \cite{JKB}.
\begin{align}
&\frac{f_1(p_1,t)}{1+f_1(p_1,t)} ~\frac{f_2(p_2,t)}{1+f_2(p_2,t)}=\frac{f_3(p_3,t)}{1+f_3(p_3,t)}~\frac{f_4(p_4,t)}{1+f_4(p_4,t)~.}\nonumber\\
&\log\left(\frac{f_1(p_1,t)}{1+f_1(p_1,t)}\right)+ \log\left(\frac{f_2(p_2,t)}{1+f_2(p_2,t)}\right)=\text{collision invariant}~.
\end{align}
It is apparent from the above expression that the collision invariant is additive in nature. The collision invariant in $2\rightarrow 2$ scattering is total three momenta and total energy. Now, in the centre-of-mass frame total three momenta of particle 1 and particle 2 is zero which leaves the total energy ($E=E_1+E_2$) as the only collision invariant. Hence, we parametrize the equilibrium form of $f_1(E_1,t)$ as 
\begin{align}
\log\left(\frac{f^{eq}_1(E_1,t)}{1+f^{eq}_1(E_1,t)}\right) = \alpha - \beta E_1~,
\end{align}  
where $\alpha$ and $\beta$ are independent of energy and particle species. The above expression with appropriate normalization reproduces Bose-Einstein distribution function as in Eq.\ref{eq:distfn1}. We can identify $\beta$ as $1/T$, $T$ being the common temperature, which is essentially a function of time in an expanding universe as explained before. So, at every instant we have thermal equilibrium for particular particle species with different temperature at different epochs. In passing we notice that the principle of chemical equilibrium is simplified with the assumption of the $CP$ symmetry. We shall return to the issue of the chemical equilibrium when the $CP$ symmetry is necessarily violated for generating particle-antiparticle asymmetry in the next section.     

When there is no chemical equilibrium, then the general solution of $f_1(p_1)$ can be found out after completing the system of equations for the rest of particles involved in the scattering. Eventually there would be a set of coupled \textit{integro-differential} equations, very challenging to solve even numerically. For standard WIMP scenario, we fortunately need to solve just one equation due to some simplified physical conditions at hand. In particular, we consider the process, $\chi(p_1)+\chi^{\dagger}(p_2)\rightarrow \psi(p_3)+\bar{\psi}(p_4)$, where $\psi$ represents some unspecified SM particles which are in thermal equilibrium throughout the pair annihilation process. With dilute gas approximation we completely ignore the above-mentioned quantum effect and the phase-space distribution function for SM states becomes, $f^{eq}_\psi (E) = e^{E/T_{SM}}$ with zero chemical potential. The counter part of DM particles is parametrized assuming kinetic equilibrium during the relevant epoch as, $f_\chi(E,t)= f^{eq}_\chi(E) \phi (t)=e^{E/T_{SM}}\phi(t)$, where $\phi(t)$, an explicit function of time captures the amount deviation from the equilibrium form of the phase-space distribution, in turn the thermal number density of DM.
\begin{align}
\phi(t)=\frac{f_\chi(E,t)}{f^{eq}_\chi (E)}=\frac{n_\chi(t)}{n^{eq}_\chi}~.
\end{align}
To note, $\phi(t)$ evolves differently for DM particle and antiparticle in general. Anyhow, the subsequent analysis is absolutely valid irrespective of the form of $\phi(t)$. Now, we integrate over $p_1$ integral multiplying both sides of Eq.\ref{eq:genboltz} with $\displaystyle{\frac{g_1 d^3p_1}{(2\pi)^3}}$ to get the dynamical equation for $n_\chi(t)$ as,
\begin{align}
\frac{dn_\chi}{dt}+3Hn_\chi = -g_1 g_2\int \frac{d^{3}p_{1}}{(2\pi)^3}\frac{d^{3}p_2}{(2\pi)^3} f^{eq}_\chi(p_1)f^{eq}_\chi(p_2)\, \sigma v_{rel} \left[\frac{n_\chi n_{\chi^\dagger}}{n^{eq^2}_\chi}-1\right]~.
\end{align}
In deriving the above equation we have used the energy-momentum conservation to relate the equilibrium distribution function of SM states with that of DM particles as,
 \begin{equation}
f^{eq}_\chi(p_1)f^{eq}_{\chi^\dagger}(p_2) = e^{-(E_1+E_2)/T_{SM}}=e^{-(E_3+E_4)/T_{SM}}=f^{eq}_\psi(p_3) f^{eq}_\psi(p_4)~.
\label{as}
\end{equation}
As the scattering process is happening in a thermal background, therefore we define the thermal averaged cross-section for WIMP annihilation as,
\begin{align}
\expval{\sigma v_{rel}}=\frac{1}{n^{eq^2}_\chi}\int \frac{d^{3}p_{1}}{(2\pi)^3}\frac{d^{3}p_2}{(2\pi)^3} \,f^{eq}_\chi(p_1)f^{eq}_\chi(p_2)\, \sigma v_{rel}~.
\label{av}
\end{align} 
In averaging appropriate symmetry factor must be included for identical particles. To match the notation of thermal averaged annihilation cross-section as in Eq.\ref{eq:asywimp} we define $\expval{\sigma v}_{ann} = \expval{\sigma v_{rel}}$ and the Boltzmann equation for WIMP becomes,
\begin{align}
\frac{d n_\chi}{dt} + 3Hn_\chi = - \expval{\sigma v}_{ann}\left(n_\chi n_{\chi^\dagger} - n^{eq^2}_\chi \right)~.
\end{align} 
We can recast the above equation in terms of dimensionless variables, i.e. $Y_\chi = n_\chi/s$ and $x=m_\chi/T_{SM}$. Using the entropy conservation, i.e. $sa^3=const$, we can relate time and temperature and in turn we can relate scaled temperature, $x$ with time $t$ as $\frac{dx}{dt} = H x$. Then we get, 
\begin{align}
\frac{dY_\chi}{dx} = - \frac{ s \expval{ \sigma v}_{ann}}{H x}\left(Y_\chi Y_{\chi^\dagger} - Y^{eq^2}_\chi \right)~.
\label{eq:wimpboltz}
\end{align}
The Boltzmann equation for $\chi^{\dagger}$ is achieved by replacing $\chi$ by $\chi^{\dagger}$ in the above equation where $Y^{eq}_\chi=Y^{eq}_{\chi^\dagger}$. As its consequence, for WIMP annihilation we end up with Eq.\ref{eq:constraint} as follows.
\begin{align}
\frac{dY_\chi}{dx}-\frac{dY_{\chi^{\dagger}}}{dx} = 0 \nonumber\\
\implies Y_\chi - Y_{\chi^{\dagger}} = C~~.
\label{eq:diff}
\end{align}
For the WIMP scenario, $C$ is assumed to be zero and hence the Eq.\ref{eq:wimpboltz} is sufficient to solve for the co-moving number densities of DM particles and antiparticles.
\subsection{Boltzmann Equation with CP-violation and S-matrix Unitarity}
We are interested in those scenarios where $C$ is non-zero and generated via dynamically from the initial condition, i.e. $C=0$. As explained in the previous chapter, in such scenarios the $CP$ symmetry must be violated in the process that generates particle-antiparticle asymmetry. In addition, The process must necessarily go out-of-equilibrium of that process to have non-zero asymmetry in a particle sector. Therefore, the above formalism should be done more carefully to incorporate the effect of the $CP$-violation. In fact,  in the presence of the $CP$-violation, the thermal equilibrium is achieved only when we take all possible final states given an arbitrary initial state, which is one of the implication $S$-matrix unitarity in the context of thermodynamics \cite{Weinberg:1995mt}. Hence we very briefly revisit the implications of $S$-matrix unitary and the $CPT$ to illustrate the idea following Ref.\cite{Bhattacharya:2011sy}.

The probability conservation in a QFT is encoded in the unitarity of $S$-matrix as the following.
\begin{align}
S^{\dagger}S = \mathbb{I} =S S^\dagger~~.
\label{eq:Smatrix} 
\end{align}
Now, for an arbitrary initial state $i$ and final state $f$ we can decompose $S$ in two parts, one of which contains the process, $i=f$ and another process is $i\neq f$, a distinct final state. Then $S$ matrix is written as,
\begin{align}
S= \mathbb{I}+iT~~.
\end{align}   
We define the transition amplitude to a state $f$ from a state $i$ is given by $|T_{fi}|^2$, where
\begin{align}
T_{fi}=\langle f | T | i \rangle~~.
\end{align}
Now, from Eq.\ref{eq:Smatrix} we get an important relation which is often termed as the \textit{generalised optical theorem}.
\begin{align}
-i \,\langle f \, | \left(T-T^{\dagger}\right)|\, i \rangle = \sum_n T_{fn} T^*_{in} = \sum_n T^*_{nf} T_{ni}~~.
\label{eq:Tmatrix}
\end{align} 
In the above expression, the sum is over all momenta and all other discreet indices (like, helicities, colors, particle species ) of an arbitrary intermediate state $n$. For $i=f$, we can relate total forward transition amplitude with the total backward amplitude as,
\begin{align}
\sum_n |T_{in}|^2 = \sum_n |T_{ni} |^2 \hspace{1cm} \text{for a given i}~~.
\label{eq:fwdbwd}
\end{align}
In addition, $CPT$ theorem \footnote{Example of $CPT$ transformation: $e^+(p_1,h_1)+e^{-}(p_2,h_2)\rightarrow \mu^{+}(p_3,h_3)+\mu^{-}(p_4,h_4)\stackrel{CP}{\implies}e^-(-p_1,-h_1)+e^{+}(-p_2,-h_2)\rightarrow \mu^{-}(-p_3,-h_3)+\mu^{+}(-p_4,-h_4)\stackrel{T}{\implies}\mu^{-}(p_3,h_3)+\mu^{+}(p_4,h_4)\rightarrow e^-(p_1,h_1)+ e^{+}(p_2,h_2)$ ; where $p_i$ and $h_i$ are momentum and helicity indices respectively.} ensures the probability of transition from an initial $i$ to a final state $f$ is identical to that of the transition of corresponding $CP$ conjugate states $\bar{f}$ to $\bar{i}$, i.e. $T_{fi}=T_{\bar{i}\bar{f}}$.
Now, we define, $|T|^2_{ni}= (2\pi)^4 \delta^{(4)}(p_i-p_n)\,|M|^2_{i\rightarrow n}$ to write the implication of unitarity along with $CPT$ as the following.  
\begin{align}
\sum_n \int dPS_n |M|^2_{i\rightarrow n} = \sum_n \int dPS_n |M|^2_{n\rightarrow i} \stackrel{CPT}{=} \sum_n \int dPS_n |M|^2_{\bar{i}\rightarrow \bar{n}} ~~,
\label{eq:implication}
\end{align} 
where, $dPS_n$ is the Lorentz-invariant phase space integral including the four dimensional delta function and now the sum indicates only the discreet indices. It is apparent from Eq.\ref{eq:implication} that not only the total forward transition amplitude is identical with the total backward transition amplitude, but also identical with the CP-conjugate counterpart of the forward processes. The takeaway of this is that in the presence of $CP$-violation the transition amplitude of an individual process is not same as its backward or $CP$-conjugate process. Hence, we can not make the simplification as done in Eq.\ref{boltz}, rather we have to include all possible processes given the initial state.

Another important observation of Eq.\ref{eq:implication} is that the processes which violate the $CP$-symmetry, the amount of the $CP$-violation is related by the unitarity relation for a given initial state. For a simple demonstration we consider two CP-violating processes, namely $\chi(p_1)+\chi(p_2)\rightarrow \psi(p_3)+\psi(p_4)$ and $\chi(p_1)+\chi(p_2)\rightarrow \phi(p_3)+\phi(p_4)$. We further simply situation by assuming all particles as complex scalars so that the helicity transformations become trivial. Now, for these two processes Eq.\ref{eq:implication} becomes
\begin{align}
\int dPS_2 \,|M|^2_{\chi\chi\rightarrow \psi\psi}+ \int dPS_2 \,|M|^2_{\chi\chi\rightarrow \phi\phi} = \int dPS_2 \,|M|^2_{\chi^{\dagger}\chi^{\dagger}\rightarrow \psi^\dagger\psi^\dagger}+ \int dPS_2 \,|M|^2_{\chi^{\dagger}\chi^{\dagger}\rightarrow \phi^\dagger\phi^\dagger}\nonumber\\
\int dPS_2 \,\left(|M|^2_{\chi\chi\rightarrow \psi\psi}-|M|^2_{\chi^{\dagger}\chi^{\dagger}\rightarrow \psi^\dagger\psi^\dagger}\right)=- \int dPS_2 \,\left(|M|^2_{\chi\chi\rightarrow \phi\phi}-|M|^2_{\chi^{\dagger}\chi^{\dagger}\rightarrow \phi^\dagger\phi^\dagger}\right)~~.
\end{align} 
We define the measure of the $CP$-violation for each final state $f$, which will be used abundantly in ensuing chapters, as
\begin{align}
\epsilon_f = \frac{|M|^2_{i\rightarrow f}-|M|^2_{i^\dagger\rightarrow f^\dagger}}{|M|^2_{i\rightarrow f}+|M|^2_{i^{\dagger}\rightarrow f^\dagger}}=\frac{|M|^2_{i\rightarrow f}-|M|^2_{i^\dagger\rightarrow f^\dagger}}{|M|^2_f}
\label{eq:defasy}
\end{align} 
Using this definition for the above two process we arrive at
\begin{align}
\int dPS_2\, \epsilon_\psi \,|M|^2_\psi + \int dPS_2\, \epsilon_\phi \,|M|^2_\phi = 0~~.
\end{align}
Therefore, the amount of CP-violation in the first process is exactly opposite that of the second process. In addition, the simultaneous presence of two processes is inevitable to respect $S$-matrix unitarity. The implication of this principle is exploited to a great extent in chapter \ref{chap:chap4}. For the present scenario, as all the particles have same spins, the integrand of the above expression becomes identically zero as $|M|^2_f$'s are necessarily a real positive quantity and $0\leq |\epsilon_f|\leq 1$.
\begin{align}
\epsilon_\psi |M|^2_\psi +\epsilon_\phi |M|^2_\phi = 0~~.
\label{eq:epr}
\end{align}
Before returning to the Boltzmann equation let's discuss how one can calculate the amount of $CP$-violation, i.e. basically the numerator of Eq.\ref{eq:defasy}. For this we rewrite Eq.\ref{eq:Tmatrix} as
\begin{align}
T_{fi} = i\sum_n T_{fn} T^*_{in} + T^*_{\bar{f}\,\bar{i}}~~.
\end{align}
The difference in amplitudes of two $CP$-conjugate processes is calculated to be
\begin{align}
\displaystyle{|T_{fi}|^2- |T_{\bar{f}\,\bar{i}}|^2 = 2 \, Im \left(\sum_n T_{fn}T^*_{in} T_{if}\right) + \bigg|\sum_n T^*_{fn}T_{in}\bigg|^2}~~.
\end{align}
This shows the amount of $CP$-violation is calculated from the interference of tree-level and loop-level amplitudes of a particular process, where the non-zero imaginary part stems from the on-shell intermediate states. Now, we return to the discussion of Boltzmann equations. We rewrite Eq.\ref{boltz} including example processes ignoring the quantum effect, as
\begin{align}
\frac{\partial f_\chi(p_1,t)}{\partial t}- H p_1\frac{\partial f_\chi(p_1,t)}{\partial p_1}=-\frac{1}{2E_1}\int \prod^4_{i=2}\frac{d^{3}p_{i}}{(2\pi)^3 2E_i}    \bigg[f_{\chi}(p_1,t)f_\chi(p_2,t)\left(|M|^2_{\chi\chi\rightarrow \psi\psi}+|M|^2_{\chi\chi\rightarrow \phi\phi}\right)\nonumber\\
\hspace{2.5cm}-f_{\psi}(p_3,t)f_\psi(p_4,t) |M|^2_{\psi\psi\rightarrow \chi\chi}-f_{\phi}(p_3,t)f_\phi(p_4,t) |M|^2_{\phi\phi\rightarrow \chi\chi}\bigg]~~.
\label{eq:boltzcp}
\end{align}
Noticeably, the establishment of thermal equilibrium distribution for interacting particle is not straightforward as before. In particular, we now need all possible processes involving $\psi$ and $\phi$ particles and write similar equations to retrieve the detailed balance equation. For that, we have to write an expression for entropy and by the virtue of the second law of thermodynamics the thermal equilibrium is achieved. This is often termed as \textit{Boltzmann H-theorem} and the $H$-function is given by
\begin{align}
\tilde{H}=\sum_\alpha f_\alpha \log f_\alpha \,\,; \,\hspace{0.5cm} \frac{d\tilde{H}}{dt}\leq 0~~, 
\end{align}
where $\alpha$ runs for all particle species and all possible momenta of a particle. For thermodynamic equilibrium, $d\tilde{H}/dt=0$. For a comprehensive study see Refs.\cite{Weinberg:1995mt,aharony,Kolb:1979qa} in the context of particle phenomenology. Nevertheless, we work out the explicit proof with some simplified assumptions in the appendix \ref{app:A}. As of now, we would assume the detailed balance equation conventional way and see that due to the unitarity relation the collision term vanishes even the $CP$-symmetry is violated in a particular channel. In equilibrium, $f_\chi(p_1)f_\chi(p_2)=f_\psi(p_3)f_\psi(p_4)=f_\phi(p_3)f_\phi(p_4)$ and the collision term of Eq.\ref{eq:boltzcp} becomes 
\begin{align}
C[f_1]=-\frac{1}{2E_1}\int\prod^4_{i=2}\frac{d^{3}p_{i}}{(2\pi)^3 2E_i} f_{\chi}(p_1,t)f_\chi(p_2,t)\bigg[|M|^2_{\chi\chi\rightarrow \psi\psi}+|M|^2_{\chi\chi\rightarrow \phi\phi}-|M|^2_{\psi\psi\rightarrow \chi\chi}-|M|^2_{\phi\phi\rightarrow \chi\chi}\bigg].
\end{align}  
Now, we can do $p_3$ and $p_4$ integrals to get $C[f_1]=0$, enforcing Eq.\ref{eq:fwdbwd} as the following.   
\begin{align}
\int\frac{d^3 p_3}{(2\pi)^3 2E_3}\frac{d^3 p_4}{(2\pi)^3 2E_4}\bigg[|M|^2_{\chi\chi\rightarrow \psi\psi}+|M|^2_{\chi\chi\rightarrow \phi\phi}-|M|^2_{\psi\psi\rightarrow \chi\chi}-|M|^2_{\phi\phi\rightarrow \chi\chi}\bigg]=0,
\end{align}
where $|M|^2_{\chi\chi\rightarrow f}\neq |M|^2_{f\rightarrow \chi\chi} $ for each $f$. Now, we integrate $p_1$ momentum in Eq.\ref{eq:boltzcp} putting the equilibrium form for $f_\psi$ and $f_\phi$ as for the previous scenario without $CP$-violation, defining $|M|^2_T=|M|^2_{\chi\chi\rightarrow \psi\psi}+|M|^2_{\chi\chi\rightarrow \phi\phi}$ and corresponding the thermal average of cross-section as $\expval{\sigma v}_T$ using Eq.\ref{av}. We write  the Boltzmann equation in terms of dimension-less variable as before.
\begin{align}
\frac{dn_\chi}{dx} =-\frac{s \expval{\sigma v}_T}{Hx} \left(Y^2_\chi-Y^{eq^2}_\chi\right).
\end{align}
We can write similar equation for $\chi^{\dagger}$ using $S$-matrix uniatrity and $CPT$ theorem appropriately and get the evolution equation of $Y_\chi-Y_{\chi^\dagger}$ as, 
\begin{align}
\frac{d(Y_\chi-Y_{\chi^\dagger})}{dx}= -\frac{s \expval{\sigma v}_T}{Hx}\left(Y^2_\chi-Y^2_{\chi^\dagger}\right).
\label{eq:diffcp}
\end{align}
To contrast with Eq.\ref{eq:diff}, the difference in the co-moving number density of DM particle and antiparticle becomes dynamic although we can not generate a non-zero difference from a symmetric condition due to correlated $CP$-violation in the concerned processes. This becomes vivid once we recast Eq.\ref{eq:diffcp} by defining new variables, $Y_{\Delta \chi}=Y_\chi-Y_{\chi^\dagger}$ and $Y_S=Y_\chi+Y_{\chi^\dagger}$. 
\begin{align}
\frac{dY_{\Delta\chi}}{dx} = -\frac{s \expval{\sigma v}_T}{Hx}\, Y_{\Delta\chi} Y_S~.
\label{eq:rediff}
\end{align} 
If the initial condition, $Y_{\Delta \chi}=0$, then the situation is equivalent to the WIMP scenario. For non-zero initial condition, i.e. $Y_{\Delta \chi}=C$, the dynamics becomes more involved as both the difference and the total number density of DM change in a correlated fashion, evident from Eq.\ref{eq:rediff} and Eq.\ref{eq:resum}.
\begin{align}
\frac{dY_S}{dx}=-\frac{s \expval{\sigma v}_T}{2Hx}\,\left(Y^2_S+Y^2_{\Delta\chi}-4Y^{eq^2}_\chi\right).
\label{eq:resum}
\end{align}
This is an illustration of $CP$-violating DM annihilation in which the final abundance would depend both on the asymmetry ($C$) of the DM and the symmetric component ($Y_S$) in general. In the ensuing chapters we would work with examples in which we can generate non-zero asymmetry from the symmetric initial condition. Consequently, we would see explicitly how the out-of-equilibrium condition is enforced to have ADM.

\newpage
%%%%%%%%%%%%%%%%%%%%%%%%%%%%%%%%%%%%%%%

%%%%%%%%%%%%%%%%%%%%%%%%%%%%%%%%%%%%%%%%%%%%%%%%%%%%%%%%%%%%%%%%%%%%%
%%%%%%%%%%%%%%%%%%%%%%%%%%%%%%%%%%%%%%%%%%%%%%%%%%%%%%%%%%%%%%%%%%%%%%%
\section{\Large{Asymmetric Dark matter from semi-annihilation}}
\label{chap:chap3}
\subsection{Introduction}
The  production mechanism for dark matter (DM) particles in the early Universe span a broad range of possibilities, ranging from processes in the thermal bath, to non-thermal mechanisms as explained in Chapter \ref{chap:chap1}. If the DM states were in local kinetic and chemical equilibrium in the cosmic plasma at some epoch, its number-changing reactions would determine its final abundance observed today. Such number changing interactions can take place either entirely within the dark sector, or may involve the standard model (SM) particles as well. Here, we assume the existence of some conserved discrete or continuous global symmetry that can distinguish between the two sectors. 

The DM states can in general be either self-conjugate or have a distinct anti-particle. In the latter case, the number densities of DM particles and anti-particles can be different, if there is a conserved charge carried by the DM states which has a non-zero density in the Universe~\cite{Weinberg:2008zzc}. The generation of such an asymmetry requires DM number violating interactions, processes that violate charge conjugation ($C$) and charge conjugation-parity ($CP$), and departures from thermal equilibrium in the early Universe. Such Sakharov conditions~\cite{ Sakharov:1967dj} are known to be realized in different ways in baryogenesis mechanisms to produce matter-antimatter asymmetry in the SM sector~\cite{Yoshimura:1978ex, Ignatiev:1978uf, Kuzmin:1985mm, Affleck:1984fy, Weinberg:1979bt, Yoshimura:1979gy, Fukugita:1986hr}. In general, the asymmetries in the dark sector and visible sector may or may not be related, and in the latter case the asymmetry generation in the dark sector can be independently studied. A large number of mechanisms have been proposed for generating asymmetric DM, many of which connecting the asymmetries in the visible and dark sectors~\cite{Nussinov:1985xr, Petraki:2013wwa, Zurek:2013wia, Kaplan:2009ag, Kaplan:1991ah, Buckley:2010ui, Shelton:2010ta, Ibe:2011hq, Falkowski:2011xh, MarchRussell:2011fi, Bhattacherjee:2013jca, Fukuda:2014xqa}. 

\begin{figure}
\centering
\includegraphics[scale=0.75]{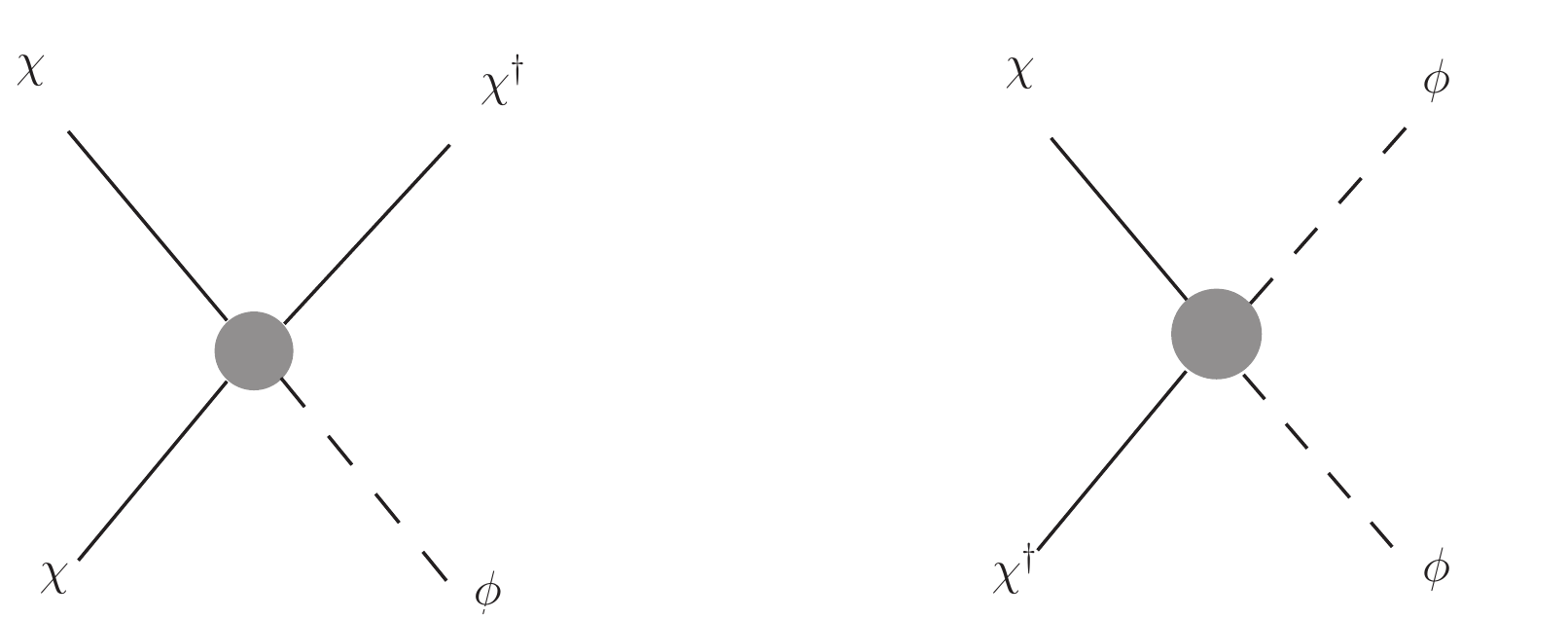}
\caption{\small{\em Effective interactions for the semi-annihilation (left) and pair-annihilation (right) processes. The former violates DM number by three units (and hence can generate a DM anti-DM asymmetry), while the latter conserves DM number.}}
\label{fig:eff_int}
\end{figure}

Among the DM number changing topologies, the simplest topologies with two DM, or two anti-DM, or one DM and one anti-DM particles in the initial state can involve either zero or one (anti-)DM particle in the final state, if there is a conserved stabilizing symmetry. The former final state corresponds to the standard pair-annihilation employed in the weakly interacting massive particle (WIMP) scenario, while the latter is the so-called semi-annihilation process~\cite{DEramo:2010keq}. If we assign a DM number of $n_\chi=1$ to the DM particle ($\chi$) and $n_\chi=-1$ to the anti-DM state ($\chi^\dagger$), then the annihilation of a $\chi \chi^\dagger$ pair does not change DM number $\Delta n_\chi = n_\chi^{\rm final} - n_\chi^{\rm initial} = 0$. On the other hand, a semi-annihilation process, for example, $\chi + \chi \rightarrow \chi^\dagger + \phi$, where $\phi$ is an unstable state not in the dark sector that can mix with or decay to SM states, can in general violate DM number (in the above reaction $\Delta n_\chi = -3$).  Thus, in the presence of semi-annihilations, the first Sakharov condition of DM number violation may easily be satisfied. We illustrate these effective interactions in Fig.~\ref{fig:eff_int}.

$CP-$violation in DM annihilation processes requires both the presence of residual complex phases in the Lagrangian (that cannot be removed by field re-definitions), as well as the interference between tree and loop level graphs, where the loop amplitudes develop a non-zero imaginary piece with intermediate states going on-shell. As we shall see in the subsequent discussion, the most minimal scenario with a complex scalar field dark matter with cubic self-interactions can satisfy both these requirements. This is one of the primary results of this chapter. We compute the $CP-$violation that can be generated using this minimal setup, including the relevant loop-level amplitudes. 

The final Sakharov condition of out-of-equilibrium reactions can easily be realized in an expanding Universe, since the reaction time scales may become larger than the inverse Hubble scale at a given temperature, thereby leading to a departure from local thermal equilibrium. In our scenario, we achieve the out-of-equilibrium condition through the semi-annihilation process. As this process freezes out, a net difference in DM and anti-DM number densities is generated, starting from a symmetric initial condition. We formulate the set of coupled Boltzmann equations for the DM and anti-DM states, and study the evolution of their number densities as a function of the temperature scale to determine the resulting asymmetry, as well as the present net DM number density. 

As we shall see in the following, it is sufficient to have only the semi-annihilation process to generate a nearly maximal asymmetry in the DM sector with the required abundance, in which either only the DM or only the anti-DM survives in the present epoch. This is realized when the CP-violation in the process is large. For smaller CP-violation, the generated asymmetry is a partial one, with an unequal mixture of both DM and anti-DM states surviving. Thus in a scenario in which only the semi-annihilation process changes DM number in the thermal bath, or changes it sufficiently fast to achieve chemical equilibrium, this process entirely determines all the properties of asymmetric DM.

However, even in simple scenarios that realize the semi-annihilation process, including CP-violation through the interference of one-loop graphs with tree level ones, additional fast DM number-changing processes may also be present. In this class of models, there will be an interplay of semi-annihilation with these other processes in chemical equilibrium, such as the pair-annihilation process. In particular, if the semi-annihilation freezes out before the pair-annihilation, then the resulting ratio between DM and anti-DM co-moving number densities may be further enhanced. This results in the possibility that even with a tiny CP-violation in the DM sector, a maximal asymmetry may be achieved. Thus in this latter scenario one generically requires lower CP-violation for any amount of asymmetry, compared to the scenario in which only semi-annihilation is present.

Although studies on generating particle anti-particle asymmetries in both the matter sector and the dark matter sector have largely focussed on generating the asymmetries through CP-violating out-of-equilibrium decay of a particle (or multiple particles), asymmetry generation through CP-violating $2 \rightarrow 2$ annihilations has also been explored. This includes studies in baryogenesis and leptogenesis~\cite{Bento:2001rc,Nardi:2007jp,Gu:2009yx} and baryogenesis through WIMP annihilations~\cite{Cui:2011ab, Bernal:2012gv, Bernal:2013bga, Kumar:2013uca}, where the DM sector remains symmetric. In most previous studies on asymmetric DM, the primordial DM asymmetry is taken to be an input parameter, which is then evolved through the pair-annihilation process, using a set of coupled Boltzmann equations~\cite{Scherrer:1985zt, Griest:1986yu, Graesser:2011wi, Iminniyaz:2011yp, Lin:2011gj}.

The general possibility of generating particle anti-particle asymmetry in the dark sector from annihilations was studied in Refs.~\cite{Baldes:2014gca, Baldes:2015lka}. In particular, in Ref.~\cite{Baldes:2014gca} the general considerations of CPT and unitarity were imposed on a toy model involving two Dirac fermion fields in the dark sector pair-annihilating to the SM sector. In our study, however, we show that a minimal scenario with one complex scalar in the DM sector can lead to asymmetry generation through the semi-annihilation process. Furthermore, in Ref.~\cite{Baldes:2014gca}, the symmetric component of the DM was large at the end of asymmetry production, and it was necessary to introduce large particle antiparticle pair-annihilation cross-sections to remove this component. As discussed above, in our scenario, the pair-annihilation is not necessary to generate a DM asymmetry with the required abundance, but may be present in addition.

We now summarize the contents and the primary results of the subsequent sections. In Sec.~\ref{sec:mi}, we describe a model independent setup that encapsulates the role of the semi-annihilation process in generating a DM and anti-DM asymmetry in the present universe. We formulate a coupled set of Boltzmann equations involving the thermally averaged semi-annihilation rate, and a thermal average of the semi-annihilation rate times a suitably defined CP-violation parameter. {\em We find that for a large CP-violation, semi-annihilation alone gives rise to nearly complete asymmetry in the DM sector, with no symmetric component surviving at its decoupling}. For a given DM mass, larger the CP-violation, a correspondingly larger value of the semi-annihilation rate is required to satisfy the observed DM relic density. {\em Using S-matrix unitarity to bound the  semi-annihilation rate from above, we obtain an upper bound of $15$ GeV on the DM mass in this scenario, for maximal CP-violation and asymmetry.}

In Sec.~\ref{sec:interplay} we then introduce an additional number changing mechanism in the DM sector, namely the pair-annihilation process, and obtain the modified set of Boltzmann equations for this scenario to study the interplay of the two annihilation processes. We then go on to find a simple estimate of the present relic abundance in terms of the CP-violation, the annihilation rates and the dark matter mass. We obtain these estimates first in the case in which the symmetric component is completely annihilated away, and then compare it with results in which part of the symmetric component survives in the present Universe. {\em We find that in the presence of subsequent pair-annihilations, the required CP-violation to generate a complete DM asymmetry is much smaller, compared to the first scenario above with only semi-annihilation.} The required values of the pair-annihilation rates are also generically higher than in the standard WIMP scenario. {\em Imposing S-matrix unitarity to bound the pair-annihilation rate from above, we obtain an upper bound of around $25$ TeV on the DM mass, for a completely asymmetric scenario, which is to be contrasted with the result for only semi-annihilation above.} We show that a simple phase-diagram in the plane of the two annihilation rates summarizes the occurrence of symmetric and asymmetric DM, depending upon the values of these two rates. 

Finally, in Sec.~\ref{model} we describe a minimal example DM scenario that can lead to asymmetric DM production through the semi-annihilation mechanism, involving a complex scalar DM particle with a cubic self-interaction. The interplay of the semi- and pair-annihilation processes is realized in this scenario. We compute the CP-violation parameter explicitly in this model at one-loop level, and compare its values, and the correlation of the CP-violation parameter with the DM annihilation rates, with the ones obtained in the model-independent setup. We find that the required values of the physical parameters that can satisfy the observed DM abundance can be reproduced in this minimal scenario.

\subsection{Asymmetric dark matter from semi-annihilation}
\label{sec:mi}
To illustrate the main idea, we shall first consider the model independent parametrization of an example scenario involving only the semi-annihilation process, in which asymmetric dark matter through DM annihilations can be realized. The minimal number of DM degrees of freedom with which this can be implemented involves a complex scalar field ($\chi$). As mentioned in the Introduction, in the semi-annihilation process, two dark matter particles annihilate to produce an anti-dark matter particle and a neutral unstable state $\phi$: $\chi + \chi \rightarrow \chi^\dagger + \phi$. Here the state $\phi$ is not in the dark sector and can mix with or decay to standard model states. For production of on-shell $\phi$ particles from non-relativistic DM annihilation, we require $m_{\phi} < m_{\chi}$. We shall parametrize the next-to-leading-order cross-section for this process by $\sigma_S$, evaluated including the tree-level and one-loop diagrams. The corresponding CP-conjugate process is $\chi^\dagger + \chi^\dagger \rightarrow \chi +\phi$, with cross-section $\sigma_{\overline{S}}$, also evaluated at next-to-leading order. In general, since CP can be violated in the semi-annihilation process from the interference of the tree-level and one-loop graphs, $\sigma_{\overline{S}} \neq \sigma_S$. 

For temperatures $T>T_S$, where $T_S$ is the freeze-out temperature of the semi-annihilation process, using the conditions of detailed balance for the reactions $\chi + \chi \rightarrow \chi^\dagger + \phi$ and $\chi^\dagger + \chi^\dagger \rightarrow \chi +\phi$, we obtain the relation between the chemical potentials $\mu_\chi = \mu_{\chi^\dagger}=\mu_\phi$. For the cases when $\mu_\phi=0$, this implies that $\mu_\chi = \mu_{\chi^\dagger}=0$. During the freeze-out of the semi-annihilation, the third Sakharov condition of out-of-equilibrium is satisfied, and a DM anti-DM asymmetry may be generated. Since in this scenario for $T<T_S$, the DM particles are not in chemical equilibrium through any reactions, we do not assign it a chemical potential for these temperatures, but a pseudo-chemical potential may be defined as shown below in Eq.~\ref{eq:pseudo}. Furthermore, in this case, since no other number-changing processes are active for $T<T_S$, the  present particle anti-particle number density ratio ($n^0_\chi/n^0_{\chi^\dagger}$) is entirely determined by the semi-annihilation process. 

In addition to the cross-section $\sigma_S$, the other relevant parameters that determine the DM abundance are the mass of $\chi$ ($m_{\chi}$) and a CP-violation parameter $\epsilon$. Here, the CP-violation parameter is defined as:
\begin{equation}
\epsilon =\frac{ |M|^2_{\chi\chi\rightarrow \chi^{\dagger} \phi}-|M|^2_{\chi^{\dagger}\chi^{\dagger}\rightarrow \chi \phi}}{|M|^2_{\chi\chi\rightarrow \chi^{\dagger} \phi}+|M|^2_{\chi^{\dagger}\chi^{\dagger}\rightarrow \chi \phi}},
\label{eq:epsilon}
\end{equation}
where $|M|^2$ denotes the matrix element for the process. As for the cross-section difference between the CP-conjugate processes, the interference of the tree and one-loop amplitudes for the semi-annihilation process determines the value of $\epsilon$. 

The Boltzmann equation for the evolution of the DM number density $n_\chi$ can be expressed in terms of the squared matrix elements of the above processes as follows:
\begin{align}
\label{eq:boltz_semi}
\dfrac{dn_\chi}{dt}+3Hn_{\chi} &= -\int \prod^{4}_{i=1} \frac{d^3 p_i}{(2\pi)^3 2 E_{p_i}}g^2_{\chi}   (2 \pi)^4 \delta^{(4)}(p_1+p_2-p_3-p_4)
\bigg[2 f_{\chi}(p_1)f_{\chi}(p_2)\overline{|M|^2}_{\chi\chi\rightarrow \chi^{\dagger}\phi} \nonumber \\ 
&- 2 f_{\chi^{\dagger}}(p_3)f_{\phi}(p_4)\overline{|M|^2}_{\chi^{\dagger}\phi\rightarrow \chi\chi}
- f_{\chi^{\dagger}}(p_1)f_{\chi^{\dagger}}(p_2)\overline{|M|^2}_{\chi^{\dagger}\chi^{\dagger}\rightarrow \chi\phi}  \nonumber  \\ 
&+ f_{\chi}(p_3)f_{\phi}(p_4)\overline{|M|^2}_{\chi\phi\rightarrow \chi^{\dagger}\chi^{\dagger}} 
\bigg],
\end{align}
where $g_\chi$ denotes the number of internal degrees of freedom of $\chi$, and $\overline{|M|^2}$ is the squared matrix element for the given process, summed over final spins, and averaged over initial spins, {\em with appropriate factors for identical initial or final state particles included}. We can also write a similar Boltzmann equation for the evolution of the anti-particle number density $n_\chi^\dagger$, by replacing the symbol $\chi$ with the symbol $\chi^\dagger$ everywhere in Eqn.~\ref{eq:boltz_semi}. The distribution functions $f_i(p)$ in the above equation take the standard form 
\begin{equation}
f_{i}(p,t)=e^{-\frac{E_{i}}{T}}e^{\frac{\mu_{i}(t)}{T}},
\label{eq:pseudo}
\end{equation}
where we have set the Boltzmann constant $k_B=1$. The pseudo-chemical potential $\mu_{i}(t)$ parametrizes the small departure from the equilibrium distribution for the particle species $i$, and it approaches the chemical potential of the particle in chemical equilibrium~\cite{Dodelson:2003ft}. We note that CPT conservation can be used to relate the matrix elements for different processes above. For example, we have $|M|^2_{ \chi^{\dagger}\phi\rightarrow \chi\chi} =|M|^2_{ \chi^{\dagger}\chi^{\dagger}\rightarrow \chi\phi}$, where, since we are dealing with scalar particles only, the helicities of the states do not appear. 

Using energy conservation for the initial and final state particles,  and defining dimensionless variables (namely, $Y_i=n_i/s$ and  $x=m_\chi/T$, where $s$ is the entropy density per comoving volume), the coupled set of Boltzmann equations for the dark matter particle and anti-particle number densities take the following form:
\begin{eqnarray}
\dfrac{d Y_{\chi}}{d x} &=& -\dfrac{s}{H x}\left[A_S\left(Y^2_{\chi}+\dfrac{Y_0 Y_{\chi}}{2}\right)-B_S\left(\dfrac{Y^2_{\chi^{\dagger}}} {2}+Y_0 Y_{\chi^{\dagger}}\right)\right]  \nonumber  \\ 
\dfrac{d Y_{\chi^{\dagger}}}{d x} &=& -\dfrac{s}{H x}\left[B_S\left(Y^2_{\chi^{\dagger}}+\dfrac{Y_0 Y_{\chi^{\dagger}}}{2}\right)-A_S\left(\dfrac{Y^2_{\chi}} {2}+Y_0 Y_{\chi}\right)\right].
\label{boltz1_semi}
\end{eqnarray}  
Here, $H$ is the Hubble constant. We have also defined $A_S = \expval{\sigma v}_S+\expval{\epsilon \sigma v}_{S}$ and $B_S = \expval{\sigma v}_S-\expval{\epsilon \sigma v}_{S}$, with $\expval{\sigma v}_S$ and $ \expval{\epsilon \sigma v}_{S}$ being the thermally averaged cross-sections for the semi-annihilation process, without and with the asymmetry factor $\epsilon(p_i)$ included, respectively. In particular, 
\begin{equation}
\expval{\epsilon \sigma v}_{s} = \dfrac{\int \prod^{4}_{i=1} \frac{d^3 p_i}{(2\pi)^3 2 E_{p_i}}  (2 \pi)^4 \delta^{(4)}(p_1+p_2-p_3-p_4) \epsilon(p_i) \overline{|M_0|^2}f_0(p_1)f_0(p_2)}{\int \dfrac{d^3 p_1}{(2\pi)^3} \dfrac{d^3 p_2}{(2\pi)^3} f_0(p_1)f_0(p_2)} \hspace{0.5cm}
\end{equation}
with $f_0(p)=e^{-\frac{E}{T}}$ being the equilibrium distribution function when the chemical potential vanishes, and 
\begin{equation}
|M_0|^2 = |M|^2_{\chi\chi\rightarrow \chi^{\dagger} \phi}+|M|^2_{\chi^{\dagger}\chi^{\dagger}\rightarrow \chi \phi}.
\end{equation}
Finally, $Y_0$ is defined as $Y_0 =  \frac{1}{s} \int  \frac{d^3 p_i} {(2\pi)^3 } g_{\chi} f_0(p)$. We have assumed that throughout the evolution of the $\chi$ and $\chi^\dagger$ particles until the freeze-out of the semi-annihilation processes, the $\phi$ particle is in thermal equilibrium with the SM plasma with a vanishing chemical potential. We note that the equilibrium distribution with zero chemical potential $Y_0$ is not a solution of the coupled Eqs.~\ref{boltz1_semi}. This is because only the CP-violating process $\chi \chi \rightarrow \chi^\dagger \phi$ and its conjugate have been included while writing the collision term here. In other words, Eqs.~\ref{boltz1_semi} are valid when all the other processes in the thermal bath involving the $\chi$ and $\chi^\dagger$ particles have decoupled, by which time $Y_0$ is no longer a solution to the Boltzmann equations by the Boltzmann H-theorem~\cite{Weinberg:1995mt}. At even higher temperatures there must be other such processes with the same initial states, in order for the T-matrix element sum rules to be consistent with the requirements of CPT and S-matrix unitarity as shown in previous chapter.

\subsubsection{Results}
In order to determine the DM relic abundance in a model-independent setup, we consider the thermally averaged cross-section for the semi-annihilation process ($\langle \sigma v \rangle_S$) as a free parameter. In addition, we define an effective CP-violation parameter $\epsilon_{\rm eff}=\langle \epsilon \sigma v \rangle_S/\langle \sigma v \rangle_S$. Therefore, there are three parameters appearing in the Boltzmann equations determining the DM and anti-DM number densities, as shown in Eq.~\ref{boltz1_semi}, namely, $m_\chi$, $\langle \sigma v \rangle_S$ and $\epsilon_{\rm eff}$. We see from Eq.~\ref{eq:epsilon} that $0<\epsilon<1$, whereby $\epsilon=0$ corresponds to no CP-violation in the semi-annihilation process, and $\epsilon=1$ to maximal CP-violation. We note that in general since $\epsilon$ is a function of the four-momenta of the particles, $\epsilon$ and $\epsilon_{\rm eff}$ are different. However, when the annihilation rates are dominated by the s-wave contributions, they become equal, and independent of the temperature. We shall work in this approximation in the model-independent analyses in Sec.~\ref{sec:mi} and Sec.~\ref{sec:interplay}.

\begin{figure}[t!]
\centering
\includegraphics[scale=0.55]{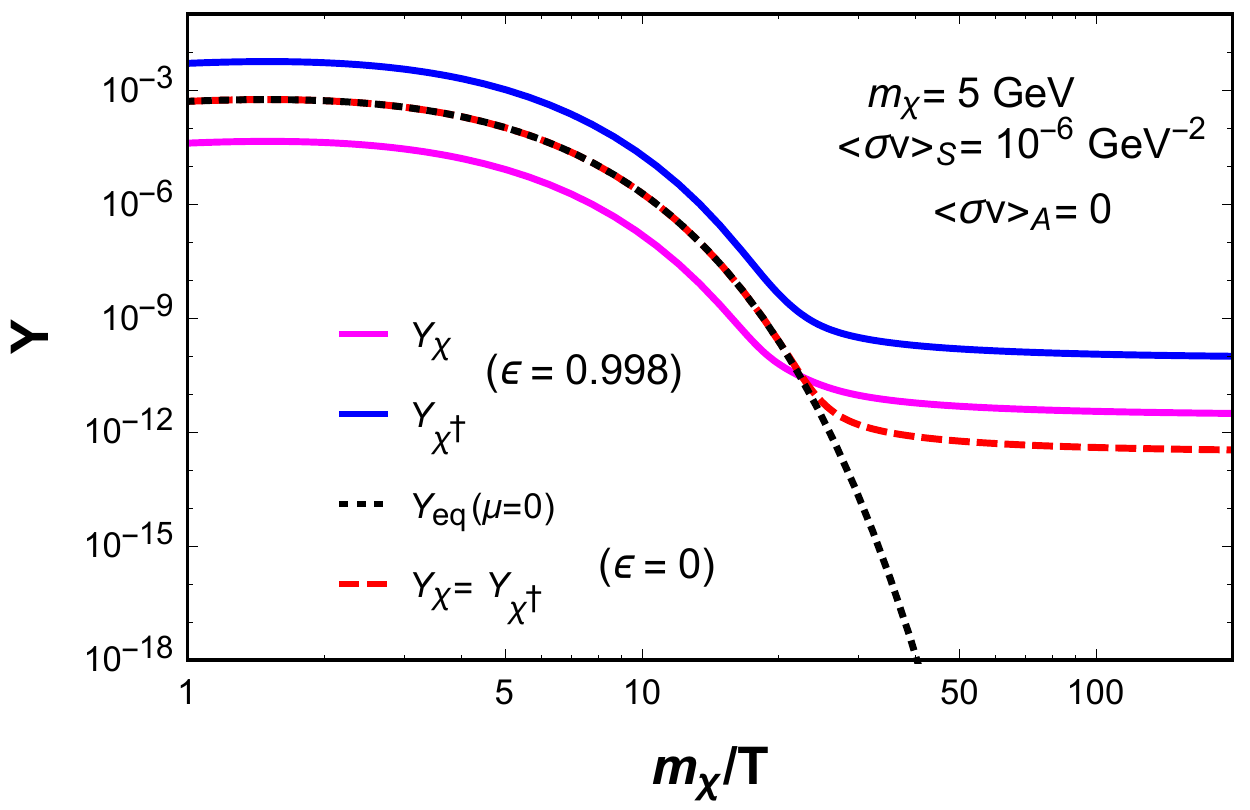} \hspace{0.9cm}
\includegraphics[scale=0.55]{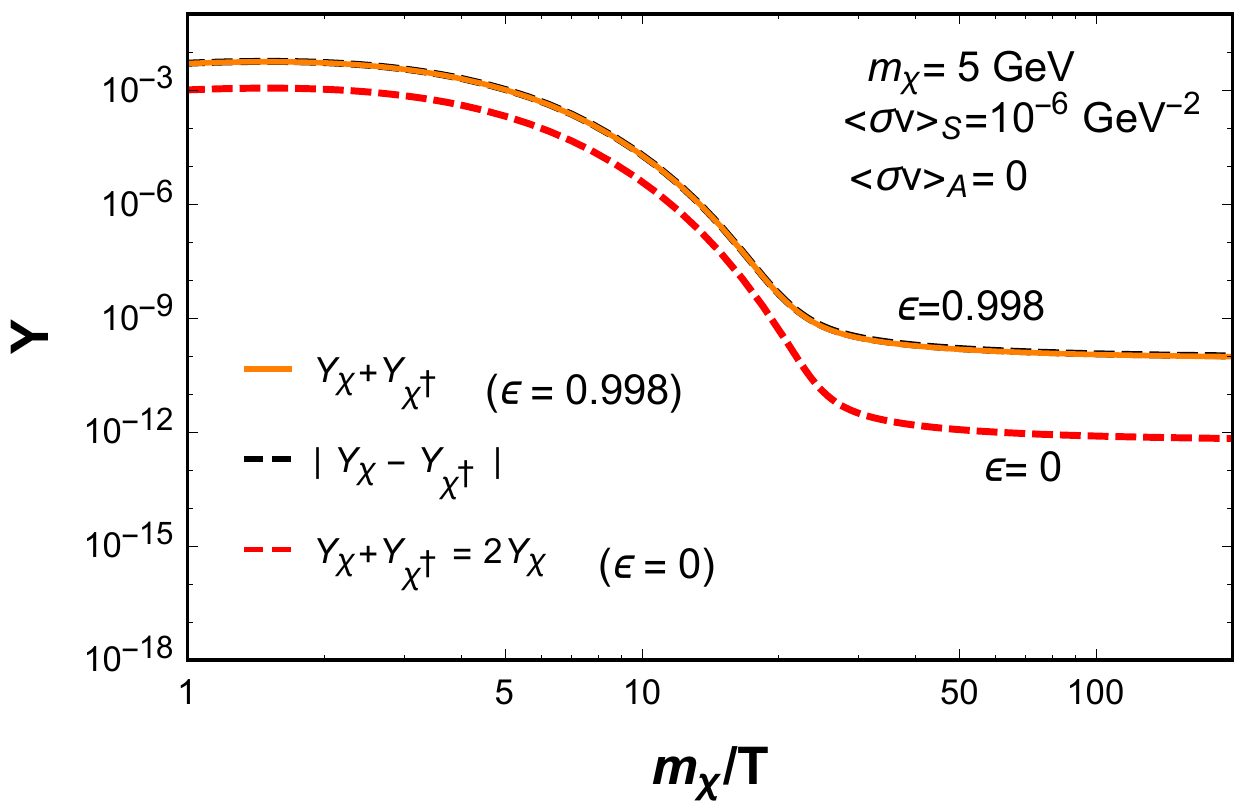}
\caption{\small{\em Dark matter ($Y_{\chi}$) and anti-dark matter yields ($Y_{\chi^\dagger}$) as a function of $x=m_\chi/T$, for the scenarios with no CP-violation ($\epsilon=0$) and nearly maximal CP-violation ($\epsilon=0.998$) without any pair annihilation, thereby $\expval{\sigma v}_A =0$. The individual yields are shown in the left figure, while their sum and difference are shown in the right.  See text for details.}}
\label{fig:semi_results0}
\end{figure}

We numerically solve the coupled Boltzmann equations in Eq.~\ref{boltz1_semi} to understand the parameter space in which the observed relic density of DM can be obtained. In Fig.~\ref{fig:semi_results0} (left panel), we show the dark matter ($Y_{\chi}$) and anti-dark matter yields ($Y_{\chi^\dagger}$) as a function of $x = m_\chi/T$, for the scenarios with no CP-violation ($\epsilon=0$, red dashed line) and nearly maximal CP-violation ($\epsilon=0.998$, pink and blue solid lines for DM and anti-DM respectively). Here, the semi-annihilation rate has been fixed to be $\langle \sigma v \rangle_S = 10^{-6} ~{\rm GeV}^{-2}$, with the DM mass $m_\chi=5$ GeV, to reproduce the observed central value of the DM relic density  $\Omega h^2 = 0.12$~\cite{Planck:2018vyg}. To contrast the results of this section with the ones in the next, in which we shall introduce DM pair-annihilation as a possible additional number-changing reaction, we have explicitly noted in this figure that the pair-annihilation rate vanishes in this scenario, i.e., $ \langle \sigma v \rangle_A=0$. In the left panel, we also show the corresponding equilibrium abundance $Y_{\rm eq}$ for the zero chemical potential scenario ($\mu=0$, black dashed line). 

In the right panel, we see that for $\epsilon=0.998$, the $\left(Y_{\chi}+Y_{\chi^\dagger}\right)$ (orange solid) and $|Y_{\chi}-Y_{\chi^\dagger}|$ (black dashed) lines almost identically trace each other. This demonstrates that for large $\mathcal{O}(1)$ CP-violation, the generated asymmetry is nearly maximal, and therefore the  (anti-)dark matter dominates the net yield. We also see from this figure that the total DM and anti-DM yield in the large CP-violation scenario ($\epsilon=0.998$) is larger than the yield for the CP-conserving case ($\epsilon=0$, red dashed line), for the same semi-annihilation rate $\langle \sigma v \rangle_S$. This is because of the large non-zero DM chemical potential in the CP-violating case. Therefore, in order to reproduce the observed relic abundance, we 
require correspondingly larger values of the semi-annihilation rate in the scenario with CP-violation.

\begin{figure}[t!]
\centering
\includegraphics[scale=0.5]{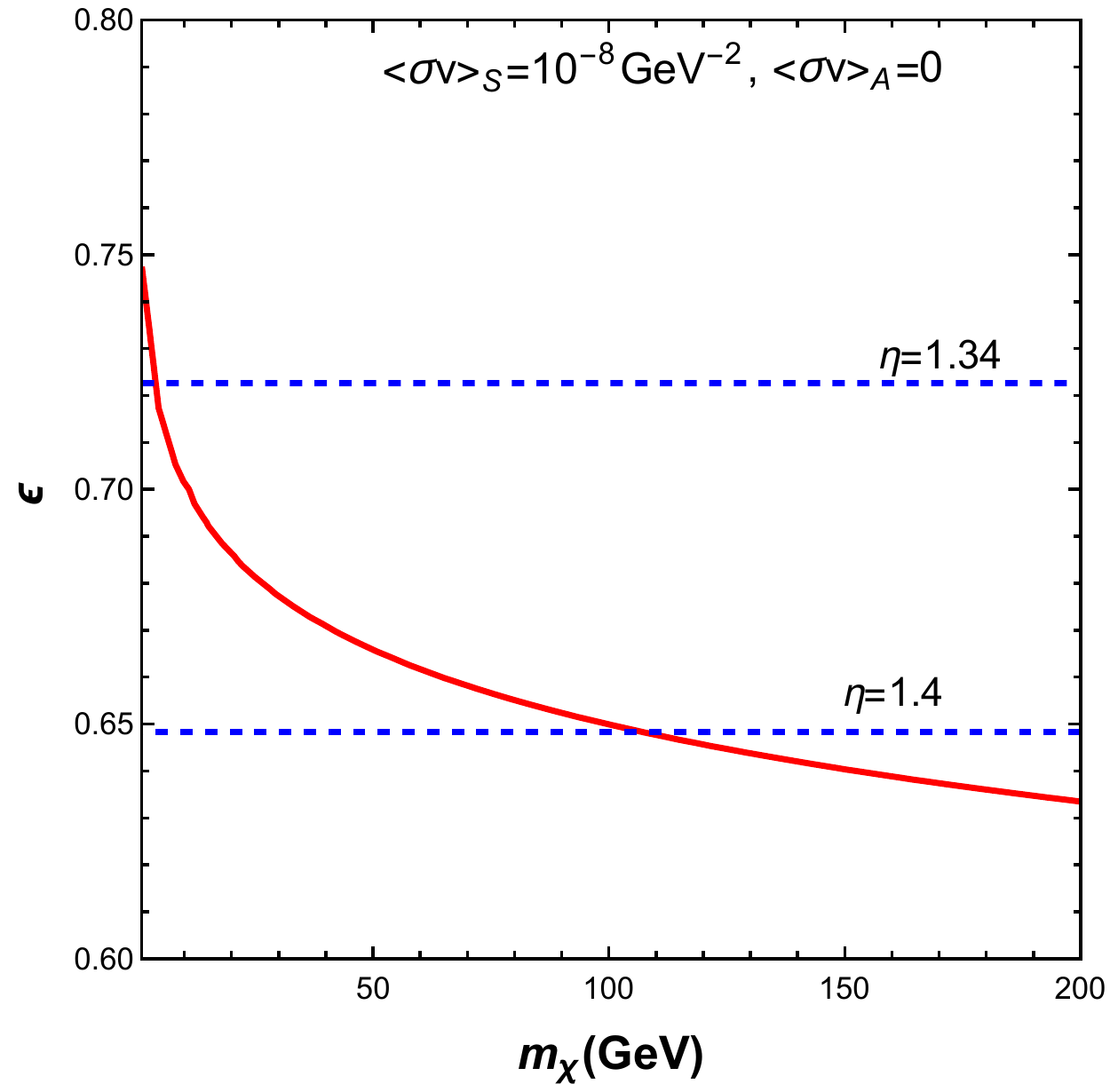}
\includegraphics[scale=0.48]{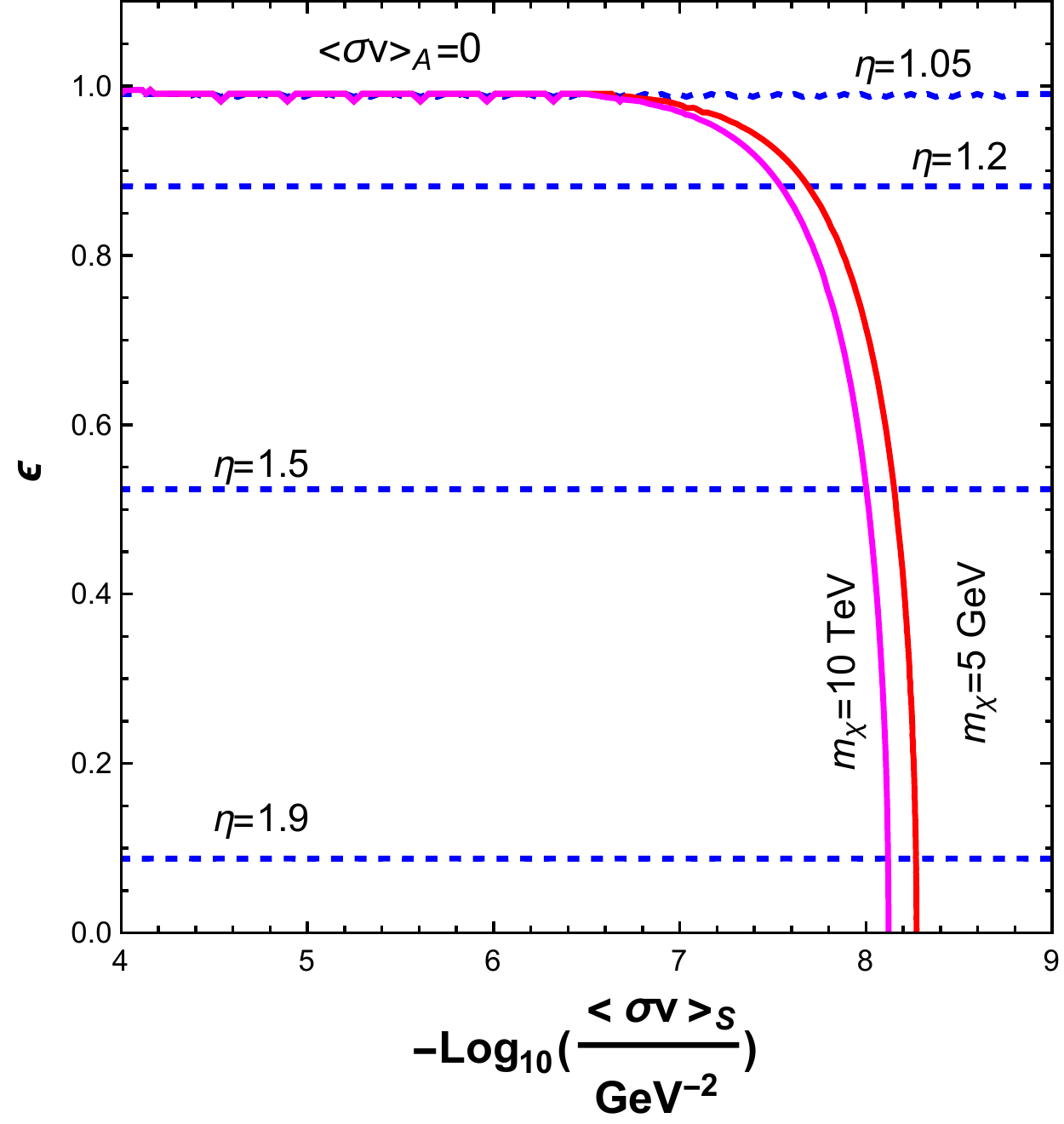}
\caption{\small{\em Contour in the $m_\chi - \epsilon$ plane (left panel, red solid line) in which the central value of the DM relic density  $\Omega h^2 = 0.12$ is reproduced, for a fixed value of the semi-annihilation rate. Same contours in the $\langle \sigma v \rangle_S - \epsilon$ plane for two different values of $m_\chi$ (right panel, red and pink solid lines). Also shown in both panels are contours of constant values of the relative abundance of DM and anti-DM, parametrized as $\eta = (Y_{\chi} (\infty)+Y_{\chi^\dagger}(\infty))/Y_{\chi^\dagger}(\infty)$, with $\eta \rightarrow 1$ being completely asymmetric DM. Only the semi-annihilation process has a non-zero rate in both the figures. See text for details.}}
\label{fig:semi_results1}
\end{figure}

In Fig.~\ref{fig:semi_results1} (left), we show the contour in the $m_\chi - \epsilon$ plane (red solid line) in which the central value of the DM relic density  $\Omega h^2 = 0.12$ is reproduced. For this figure, we have fixed the value of the semi-annihilation rate to be $\langle \sigma v \rangle_S = 10^{-8} ~{\rm GeV}^{-2}$. As before, for both the figures $ \langle \sigma v \rangle_A=0$. We also show contours in the $m_\chi - \epsilon$ parameter space for constant values of the relative abundance of DM and anti-DM, parametrized as
\begin{equation}
\eta = \frac{Y_{\chi} (\infty)+Y_{\chi^\dagger}(\infty)}{Y_{\chi^\dagger}(\infty)}, 
\label{eq:eta}
\end{equation}
where, the yield $Y_{\chi}(x)$ is evaluated at the present epoch with $x \rightarrow \infty$. Since for $\epsilon>0$, only  the $\chi^\dagger$ states survive for a scenario in which the symmetric component is completely annihilated away, in this limit, $\eta \rightarrow 1$. In scenarios in which the symmetric component partially survives, $1 < \eta <  2$. As we can see from this figure, for a fixed value of 
$\langle \sigma v \rangle_S$, higher values of $\epsilon$ imply a lower DM mass $m_\chi$ in which the relic density is reproduced. This is because, higher the CP-violation $\epsilon$, the higher is the difference in the number densities of the DM and anti-DM particles, which in turn implies a large pseudo-chemical potential. For a fixed value of the semi-annihilation rate, this also implies that the resulting frozen out number densities are higher, thus requiring a lower DM mass to saturate the same DM abundance. As is also clear, higher $\epsilon$ implies values of the relative abundance parameter $\eta$ closer to $1$. 

For a fixed DM mass, if we in turn keep increasing the CP violation $\epsilon$, the reaction rate $\langle \sigma v \rangle_S$ also needs to be correspondingly higher, for the same reason as described above. This is shown in Fig.~\ref{fig:semi_results1} (right), where for two fixed values of $m_\chi$ ($5$ GeV and $10$ TeV), we show the contours in the $\langle \sigma v \rangle_S - \epsilon$ plane (red and pink solid lines respectively) in which the central value of the DM relic density  $\Omega h^2 = 0.12$ is reproduced. The approach to $\epsilon=1$ in this figure is asymptotic, where the small numerical differences are not clear from the plot shown (which, however, we have checked numerically). For $\epsilon \rightarrow 1$, we see from this figure that $\eta \rightarrow 1$, with the surviving DM state being almost entirely the anti-DM.

\begin{figure}[t!]
\centering
\includegraphics[scale=0.5]{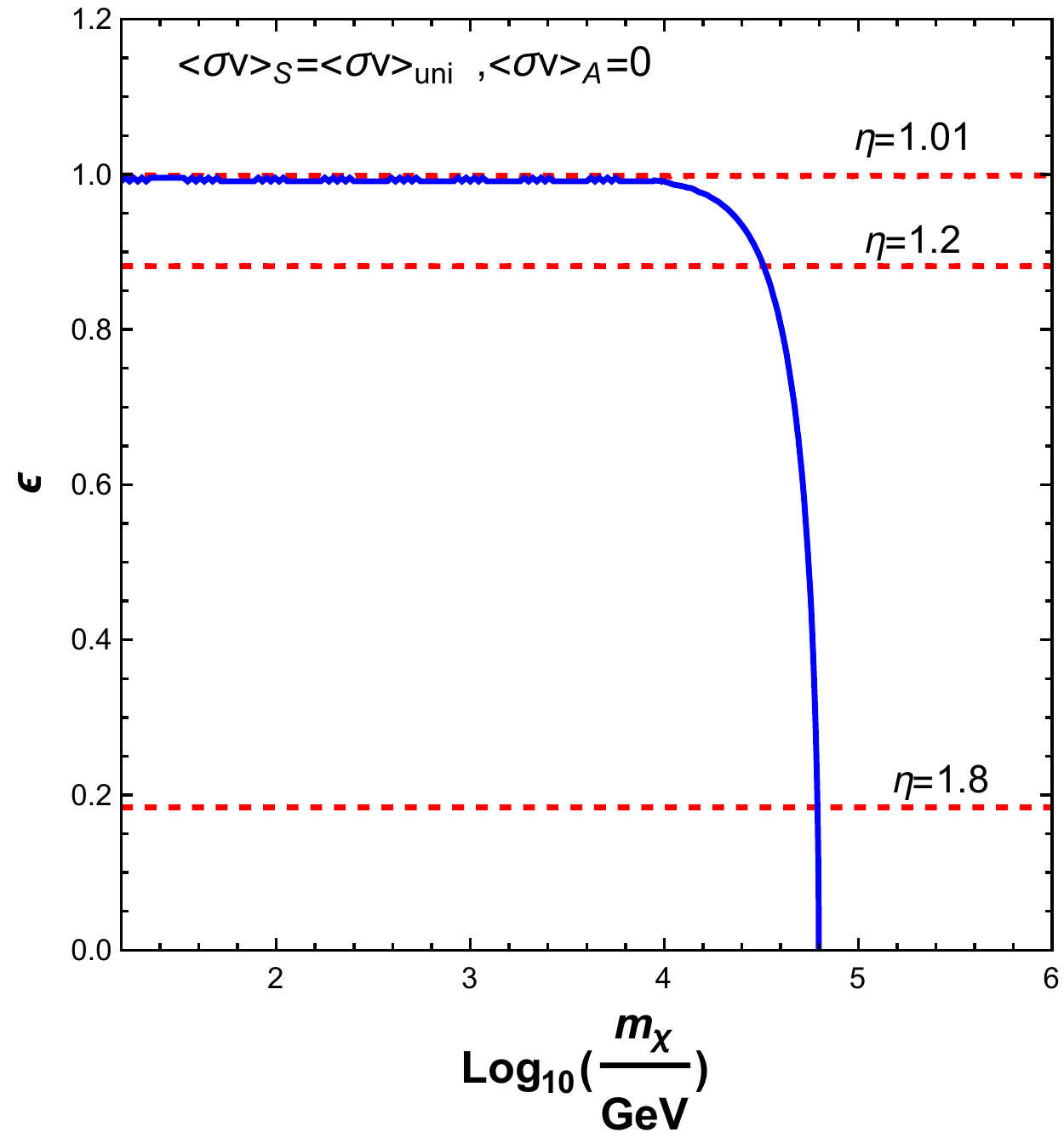}
\includegraphics[scale=0.55]{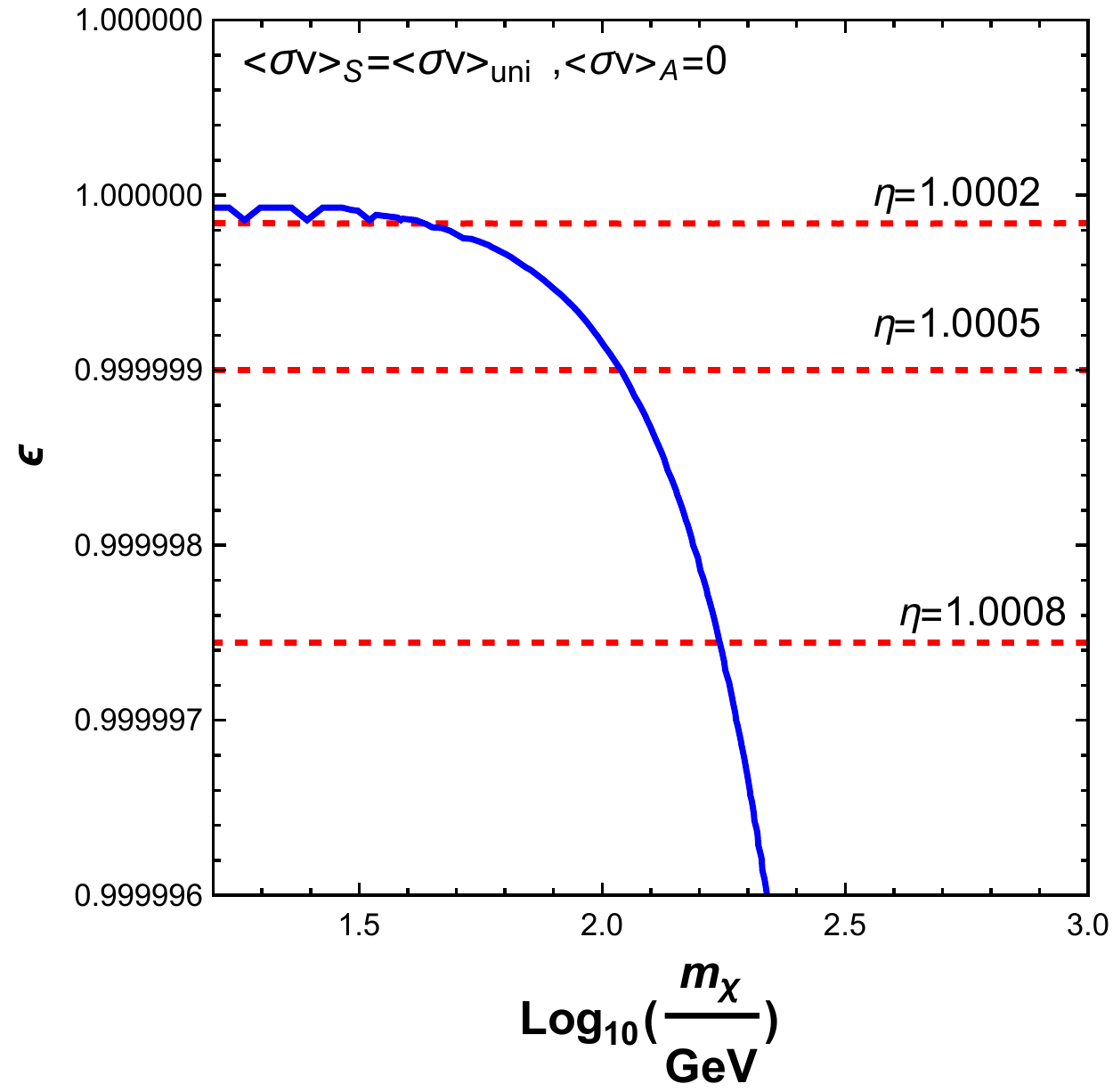}
\caption{\small{\em Contours in the $m_\chi - \epsilon$ plane in which the central value of the DM relic density  $\Omega h^2 = 0.12$ is reproduced, with the semi-annihilation rate fixed at its s-wave upper bound implied by S-matrix unitarity, $\langle \sigma v \rangle_S = \langle \sigma v \rangle_{\rm uni} = (4 \pi/ m_\chi^2)(x_F/ \pi)^{1/2}$. The $\eta$ parameter is the same as defined for Fig.~\ref{fig:semi_results1}, with $\eta \rightarrow 1$ being completely asymmetric DM. The right panel is the zoomed in version of the left panel focussing on a narrower region in the CP-violation parameter $\epsilon$. See text for further details.}}
\label{fig:semi_results2}
\end{figure}

How high can we go in the rate $\langle \sigma v \rangle_S$? We can use partial-wave S-matrix unitarity to bound the semi-annihilation cross-section from above. This in turn will also translate into an upper bound for possible values of the dark matter mass. The maximum allowed value of the cross-section determines the lowest possible number density of dark  matter today, which in turn determines the highest possible mass, if this single dark matter component saturates the observed abundance. In Fig.~\ref{fig:semi_results2} we impose the unitarity bound on $\langle \sigma v \rangle_S = \langle \sigma v \rangle_{\rm uni}$, where, for s-wave annihilation the unitarity upper bound is given by~\cite{Griest:1989wd, Hui:2001wy}:
\begin{equation}
 \langle \sigma v \rangle_{\rm uni} = (4 \pi/ m_\chi^2)(x_F/ \pi)^{1/2}.
 \label{eq:uni}
 \end{equation}
Here, $x_F=m_\chi/T_F$, with $T_F$ being the freeze-out temperature of the corresponding process. For both the plots in Fig.~\ref{fig:semi_results2}, $x_F=20$ is set as a benchmark value. With the semi-annihilation cross-section set at the s-wave unitarity upper bound, we show the contour in the $m_\chi - \epsilon$ plane (blue solid line) for which in  $\Omega h^2 = 0.12$ is reproduced in the left plot of Fig.~\ref{fig:semi_results2}. This figure shows the maximum possible DM mass allowed for a particular value of $\epsilon$, and as discussed earlier, higher values of $\epsilon$ imply that the upper bound on the DM mass is stronger. In order to understand the approach towards $\epsilon \rightarrow 1$ better, we show in the right panel of Fig.~\ref{fig:semi_results2} a narrower region along the $\epsilon$ axis. From this figure we observe a number of important results:
\begin{enumerate}
\item {\em With the semi-annihilation process alone, one can obtain a scenario giving rise to a nearly complete asymmetry in the DM sector, in which only the (anti-)DM state survives today}. This is obtained for a large value of the CP violation parameter $\epsilon$. Smaller values of $\epsilon$ correspond to scenarios with a mixed present abundance of DM, with both the particle and anti-particle states present.

\item As mentioned above, here we explicitly observe that the approach to $\epsilon \rightarrow 1$ is asymptotic, and correspondingly to $\eta \rightarrow 1$.

\item For $\epsilon \rightarrow 0$, the upper bound on the DM mass is obtained to be $80$ TeV, which is the bound for purely symmetric semi-annihilation scenario, with no CP-violation. 

\item {\em For $\epsilon \rightarrow 1$, the upper bound on the DM mass is obtained to be around $15$ GeV, which is the bound for purely asymmetric semi-annihilation scenario, with maximal CP-violation.} We note that this is much stronger than the unitarity bounds obtained for asymmetric DM scenarios where strong subsequent pair-annihilations are necessarily present, which we consider in the next section~\cite{Baldes:2017gzw}. The above upper bound of $15$ GeV is obtained  for $\eta \simeq 1.0002$, which represents a scenario with a nearly complete present asymmetry in the DM sector (2 particles in 10,000 anti-particles). We have checked that if we reduce $\eta$ further closer to $1$, the consequent change in this upper bound on the DM mass is very small. 

\item We see that being entirely within the limits of maximal possible semi-annihilation rate and the maximal possible value of CP-violation, we can indeed obtain a completely asymmetric DM scenario, with no requirement of subsequent pair-annihilations to remove the symmetric component. This is one of the primary important observations of this chapter.
\end{enumerate}

\subsection{The interplay of semi-annihilation and pair-annihilation}
\label{sec:interplay}
We now consider the second scenario, in which both the semi-annihilation and pair-annihilation processes are active, and their interplay determines the resulting DM properties. In the latter process, a dark matter particle annihilates with an anti-dark matter particle, creating a pair of unstable states $\phi$, $\chi + \chi^\dagger \rightarrow \phi + \phi$, where as earlier $\phi$ can mix with or decay to the SM states. We shall parametrize the leading-order cross-section for this process by $\sigma_A$, which is an additional parameter in this scenario. 

\begin{figure}[htb!]
\label{fig:temp}
\centering
\includegraphics[scale=0.3]{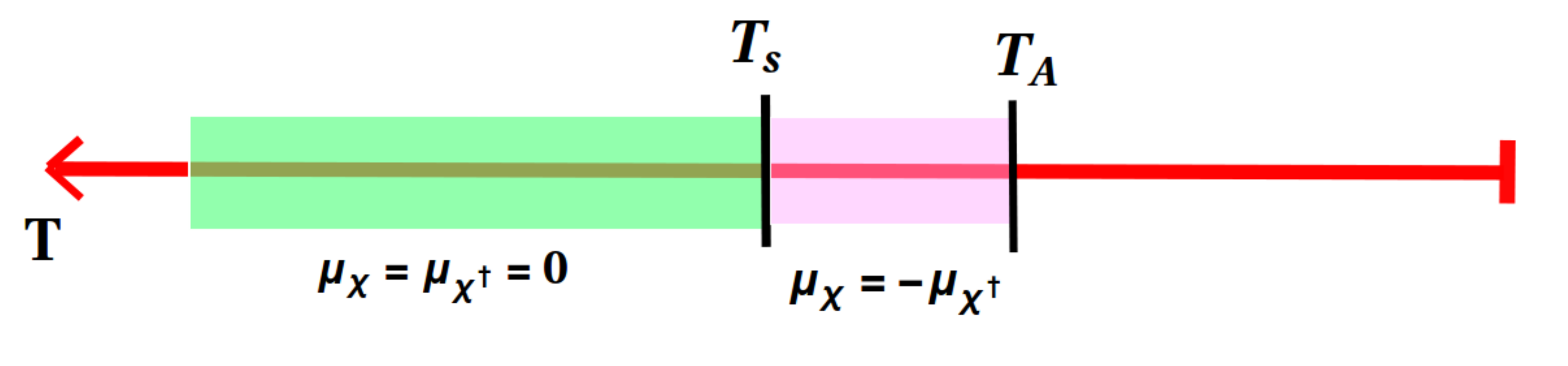}
\caption{\small{\em One of the two possible hierarchies of the freeze-out temperature for the semi-annihilation ($T_S$) and pair-annihilation ($T_A$) processes, and the chemical potentials of the DM ($\mu_\chi$) and anti-DM ($\mu_{\chi^\dagger}$) states at different temperature intervals.}}
\label{fig:temperature}
\end{figure}

We assume that initially at high enough temperatures, both the semi-annihilation and the pair annihilation processes are in chemical equilibrium, with their freeze-out temperatures being $T_S$ and $T_A$ respectively. If the freeze-out temperatures have the hierarchy $T_S > T_A$, the semi-annihilation process freezes out earlier, as schematically shown in Fig.~\ref{fig:temperature}. For temperatures 
$T>T_S > T_A$, using the conditions of detailed balance for the reactions $\chi + \chi \rightarrow \chi^\dagger + \phi$, $\chi + \chi^\dagger \rightarrow \phi+ \phi$ and $\chi^\dagger + \chi^\dagger \rightarrow \chi +\phi$, we obtain the 
relation between the chemical potentials $\mu_\chi = \mu_{\chi^\dagger}=\mu_\phi$. For the cases when $\mu_\phi=0$, this implies that $\mu_\chi = \mu_{\chi^\dagger}=0$.

For $T_A<T < T_S$, the semi-annihilation process freezes out, keeping only the pair annihilation in chemical equilibrium. This would imply that $\mu_\chi + \mu_{\chi^\dagger}=2\mu_\phi$, and if $\mu_\phi=0$ we obtain $\mu_\chi = - \mu_{\chi^\dagger}$. Hence, in this temperature regime, the $\chi$ particle can have a non-zero chemical potential, and therefore, a particle anti-particle asymmetry in the $\chi$ sector is generically possible. Such an asymmetry is generated by the freeze-out of the semi-annihilation process once all the Sakharov conditions are satisfied. In this case, since the pair annihilation process is active for $T<T_S$, the  final particle anti-particle number density ratio ($n^0_\chi/n^0_{\chi^\dagger}$) is determined by both the reaction rates. 

For the opposite hierarchy $T_S < T_A$, there cannot be any chemical potential for the $\chi$ particle for temperatures $T>T_S$, with $\mu_\phi=0$. After the freeze-out of the semi-annihilation, asymmetry may again be generated, as discussed in Sec.~\ref{sec:mi} for the scenario with only semi-annihilation. In particular, since the pair annihilation process is no longer active for $T<T_S$, the ratio ($n^0_\chi/n^0_{\chi^\dagger}$) is entirely determined by the semi-annihilation process. Thus this scenario is identical to the scenario considered in Sec.~\ref{sec:mi} as far as the present DM properties are concerned.

With the pair-annihilation process included in addition to the two CP-conjugate semi-annihilation channels, there are now three relevant processes in the thermal bath that can change the number of DM particles $\chi$. Consequently, the Boltzmann equations~\ref{eq:boltz_semi} are now modified to include an additional collision term as follows:
\begin{align}
\label{eq:boltz_both}
\dfrac{dn_\chi}{dt}+3Hn_{\chi} &= C_{\rm semi}-\int \prod^{4}_{i=1} \frac{d^3 p_i}{(2\pi)^3 2 E_{p_i}}g^2_{\chi}   (2 \pi)^4 \delta^{(4)}(p_1+p_2-p_3-p_4) \nonumber \\ 
& \left[ \overline{|M|^2}_{\chi\chi^{\dagger}\rightarrow \phi\phi}\left[f_{\chi}(p_1)f_{\chi^{\dagger}}(p_2)-f_{\phi}(p_3)f_{\phi}(p_4)\right]
\right],
\end{align}
where $C_{\rm semi}$ is the collision term from the semi-annihilation processes given in the RHS of Eq.~\ref{eq:boltz_semi}, all other notations being the same as in Eq.~\ref{eq:boltz_semi}.

Following the same procedure as in the previous section, the coupled set of Boltzmann equations for the dark matter particle and anti-particle co-moving number densities are now modified to take the following form:
\begin{eqnarray}
\dfrac{d Y_{\chi}}{d x} &=& -\dfrac{s}{H x}\left[A_S\left(Y^2_{\chi}+\dfrac{Y_0 Y_{\chi}}{2}\right)-B_S\left(\dfrac{Y^2_{\chi^{\dagger}}} {2}+Y_0 Y_{\chi^{\dagger}}\right)+\expval{\sigma v}_A \left(Y_{\chi}Y_{\chi^{\dagger}}-Y^2_0\right)\right] \nonumber \\ 
\dfrac{d Y_{\chi^{\dagger}}}{d x} &=& -\dfrac{s}{H x}\left[B_S\left(Y^2_{\chi^{\dagger}}+\dfrac{Y_0 Y_{\chi^{\dagger}}}{2}\right)-A_S\left(\dfrac{Y^2_{\chi}} {2}+Y_0 Y_{\chi}\right)+\expval{\sigma v}_A \left(Y_{\chi}Y_{\chi^{\dagger}}-Y^2_0\right)\right]
\label{boltz1_both}
\end{eqnarray}  
Here, $\expval{\sigma v}_A$ is the thermally averaged pair-annihilation cross-section. As before, we have assumed that throughout the evolution of the $\chi$ and $\chi^\dagger$ particles until the freeze-out of the semi-annihilation and the pair-annihilation processes, the $\phi$ particle is in thermal equilibrium with the SM plasma with a vanishing chemical potential.

\subsubsection{Estimate of relic abundance}
\label{sec:estimate}
Before proceeding to the discussion of the numerical solutions for the coupled Boltzmann equations, we first provide a rough estimate of the relation between the DM relic density ($\Omega_\chi$), its mass ($m_\chi$), and the CP-violation parameter $\epsilon$.  For this estimate, we shall assume that there is complete asymmetry between the dark matter and anti-matter states in the current Universe, i.e., either only the particle or the anti-particle states survive. It then follows that the present DM relic density $\Omega_{\rm DM} = m_\chi s_0 \left(Y_\chi^\infty +Y_{\chi^\dagger}^\infty \right)/\rho_c =  m_\chi s_0 Y_\chi^\infty /\rho_c$, in the scenario when only the $\chi$ particles survive today, where $s_0$ and $\rho_c$ are the present entropy density and the critical density respectively. 

After the freeze-out of the semi-annihilation process, in the absence of subsequent pair-annihilations, both the $\chi$ and $\chi^\dagger$ co-moving number densities ($Y_\chi $ and $Y_{\chi^\dagger}$) remain constants. However, in the presence of subsequent pair annihilations, namely, the process $\chi \chi^\dagger \rightarrow \phi \phi$, at temperatures below $m_\chi$ (when the backward process is not active), each reaction reduces both $\chi$ and $\chi^\dagger$ numbers by one unit. Therefore, in this latter case, only $Y_\chi - Y_{\chi^\dagger}$ remains a constant, which we can, therefore, equate to $Y_\chi^\infty$, assuming the symmetric part is completely annihilated, and only $\chi$ particles survive today. We now define the net co-moving charge density in the dark matter sector at the temperature $T$ to be $\Delta B (T) = Q \left(n_\chi (T)-n_{\chi^\dagger} (T) \right)$, where $Q$ is  the charge assigned to one DM particle. We can then express the present relic abundance of DM as 
\begin{equation}
\Omega_{\rm DM} = \frac{m_\chi s_0 \Delta B (T_S)}{\rho_c \,s(T_S) Q}, 
\end{equation}
where $T_S$ is the freeze-out temperature for the semi-annihilation process.

In the semi-annihilation reaction, $\chi \chi \rightarrow \chi^{\dagger} \phi$, the net change in $\chi$ charge per reaction is negative ($\Delta Q=-3Q$), while in the CP-conjugate process $\chi^{\dagger} \chi^{\dagger}\rightarrow \chi \phi$, the net change in $\chi$ charge per reaction is positive ($\Delta Q=3Q$). Hence, the probability of having a positive change is $P_{+} = \sigma_{\bar{S}} / \left(\sigma_S + \sigma_{\bar{S}}\right)$, while the probability for a negative change is $P_{-} =\sigma_S / \left(\sigma_S + \sigma_{\bar{S}}\right)$. Therefore, $\Delta Q$ produced per semi-annihilation and its CP-conjugate reaction is $(3QP_{+} -3QP_{-})=-3Q\epsilon$, where $\epsilon$ is defined as in Eq.~\ref{eq:epsilon}. Here, we have used the fact that the final state phase space elements are the same for the two CP-conjugate processes. Finally, the net DM charge density produced is $\Delta B = - 3 \epsilon Q n_{\chi}^{\rm eq} (T_S)$, 
assuming that the near-equilibrium distribution with zero chemical potential, $n_{\chi}^{\rm eq} (T_S)$, is being maintained by fast pair-annihilation reactions, and therefore, $n_{\chi}^{\rm eq} (T_S) \simeq n_{\chi^\dagger}^{\rm eq} (T_S)$.
Since this assumption is invalid for the scenario discussed in Sec.~\ref{sec:mi}, our estimate of the relic abundance in this section does not apply for that scenario. In particular, with only the CP-violating semi-annihilation reaction active in the thermal bath, for large CP-violation (which is necessary to get a complete asymmetry with only semi-annihilation) the DM and the anti-DM particles have large and different pseudo-chemical potentials, and therefore do not follow the equilibrium distribution.

Plugging in the expression for $\Delta B$ as obtained above, we can now write the relic abundance of DM particles today as follows
\begin{align}
\Omega_{\rm DM} &= \frac{3~m_\chi s_0 |\epsilon| ~n_\chi^{\rm eq} (T_S)}{\rho_c ~ s(T_S) } ~~~{\rm (completely ~asymmetric ~scenario)} .
\label{eq:estimate}
\end{align}
Since we assumed the DM state $\chi$ to survive in the present Universe, $\epsilon<0$ in this case, while if the anti-DM state $\chi^\dagger$ survives, $\epsilon>0$, as can be seen from Eq.~\ref{eq:epsilon}. We can re-write the above expression in terms of a set of particular choices of the parameters as
\begin{equation}
\Omega_{\rm DM}h^2 = 0.12 \times ~ \bigg(\frac{m_\chi}{100 ~\rm GeV}\bigg) \bigg(\frac{100}{g_{*S}(x_F)}\bigg) ~ \bigg( \frac{|\epsilon| ~x_S^{3/2}~ e^{-x_S}}{10^{-9}} \bigg).
\label{eq:estimate2}
\end{equation}
This shows that apart from the implicit dependence of $\Omega_{\rm DM}$ on $m_\chi$ and $\epsilon$ through the value of $x_S(=m_\chi /T_S)$, there is an explicit linear proportionality with both these parameters expected. This is to be contrasted with the simple scenarios of asymmetry generation through the out-of-equilibrium decay of a heavy particle, where the resulting particle density today is proportional to $\epsilon$ only, and not to the mass of the decaying heavy particle~\cite{Kolb:1990vq}. Furthermore, in the decay scenario, the asymmetry parameter $\epsilon$ is independent of the particle momenta, unlike in the annihilation scenario~\cite{Kolb:1990vq}. For a typical value of $x_S = 20$, we see that $|\epsilon| \simeq 5.4 \times 10^{-3}$ can reproduce the present DM abundance, for $m_\chi = 100$ GeV.  {\em In contrast to the scenario with only semi-annihilation discussed in Sec.~\ref{sec:mi}, we see that the CP-violation required to generate complete asymmetry here is very small.}

Unlike in the previous case, for pair-annihilation cross-sections that are not sufficient to completely remove the symmetric component, there is an explicit dependence of the DM relic density on the pair-annihilation rate $\langle \sigma v \rangle_A$. In this case, the coupled Boltzmann equations can be integrated piecewise in different temperature regimes, firstly near the freeze-out of the semi-annihilation process, in which the pair-annihilation rate is not relevant, and then near the freeze-out of the pair-annihilation process, but now with an initial asymmetry in the DM sector generated by the earlier freeze-out of the semi-annihilation. The resulting relic abundance already calculated in Eq.\ref{eq:relicsemi} as 
\begin{equation}
\Omega_{\rm DM} = \frac{s_0}{\rho_c}m_{\chi} C \coth \left(\frac{C\lambda \langle \sigma v \rangle_A} {2 x_A} \right) ~~~~{\rm (partially ~asymmetric ~scenario),}
\end{equation}
where, $x_A=m_{\chi}/T_A$, with $T_A$ being the freeze-out temperature of the pair-annihilation process,  $C=Y_\chi (T)- Y_{\chi^\dagger}(T)$, for all $T<T_S$, and $\lambda=1.32 M_{\rm Pl}m_{\chi} g^{1/2}_*$. In the limit $C \rightarrow 0$, the above expression reduces to the well-known result for symmetric WIMP scenario
\begin{equation}
\Omega_{\rm DM}= \frac{2 s_0 m_{\chi} x_A}{\rho_c \lambda \langle \sigma v \rangle_A} ~~~~~~{\rm (symmetric ~WIMP ~scenario)}.
\end{equation}

\subsubsection{Numerical results}
We shall now solve the coupled Boltzmann equations~\ref{boltz1_both} numerically, with four free parameters. The three parameters $m_\chi$, $\langle \sigma v \rangle_S$ and $\epsilon_{\rm eff}$ are the same as in Sec.~\ref{sec:mi}, with the additional parameter being the pair-annihilation rate $\langle \sigma v \rangle_A$. Since we have already discussed the role of the first three parameters in determining the DM properties in the previous section, the primary aim of this section is to understand the impact of pair-annihilation, in particular its interplay with the semi-annihilation process. Following our general discussion above, therefore, the relevant temperature hierarchy is $T_S > T_A$, in which the semi-annihilation freezes out earlier. The opposite hierarchy, $T_S < T_A$, is exactly equivalent to the scenario in Sec.~\ref{sec:mi}, as far as the DM asymmetry and relic density today are concerned.

\begin{figure}[htb!]
\begin{center}
\includegraphics[scale=0.6]{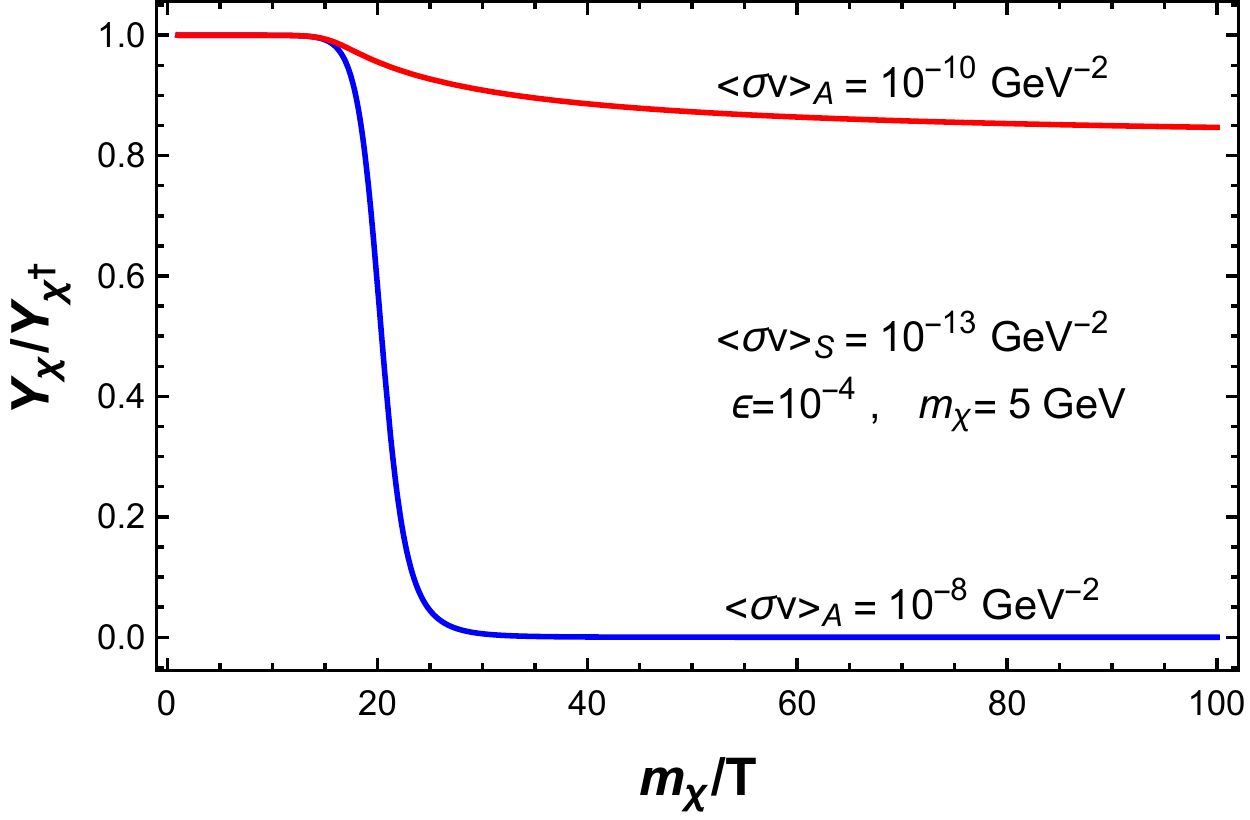} 
\caption{\small{\em Variation of the ratio of dark matter and anti-dark matter yields, $Y_\chi / Y_{\chi^\dagger}$, as a function of $m_\chi/T$, for different values of the pair-annihilation rate, $\langle \sigma v \rangle_A=10^{-10} {~\rm GeV}^{-2}$ (red solid line) and $\langle \sigma v \rangle_A=10^{-8} {~\rm GeV}^{-2}$ (blue solid line), with fixed values of $m_\chi$, $\langle \sigma v \rangle_S$ and $\epsilon$.}}
\label{fig:Yield_pair}
\end{center}
\end{figure}
For  $T_S > T_A$, the essential role of the pair-annihilation process is to remove the symmetric component of dark matter, as illustrated in Fig.~\ref{fig:Yield_pair}. As we can see from this figure, with fixed values of $m_\chi$, $\langle \sigma v \rangle_S$ and $\epsilon$, if we increase the value of the pair-annihilation rate from $\langle \sigma v \rangle_A=10^{-10} {~\rm GeV}^{-2}$ (red solid line) to $\langle \sigma v \rangle_A=10^{-8} {~\rm GeV}^{-2}$ (blue solid line), the ratio of dark matter and anti-dark matter yields $Y_\chi / Y_{\chi^\dagger}$ decreases rapidly.

\begin{figure}
\includegraphics[scale=0.52]{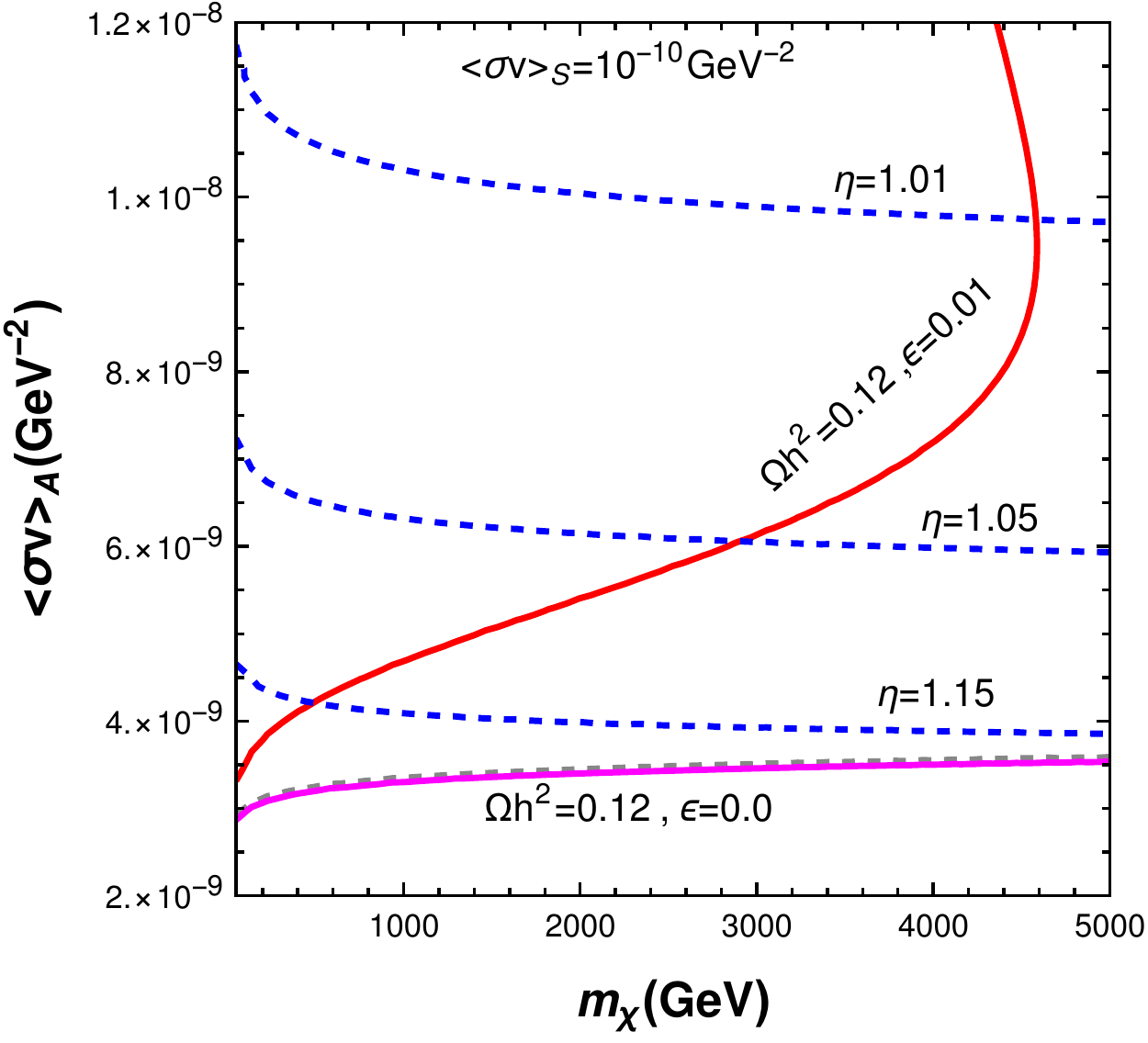} \hspace*{1.5cm}
\includegraphics[scale=0.52]{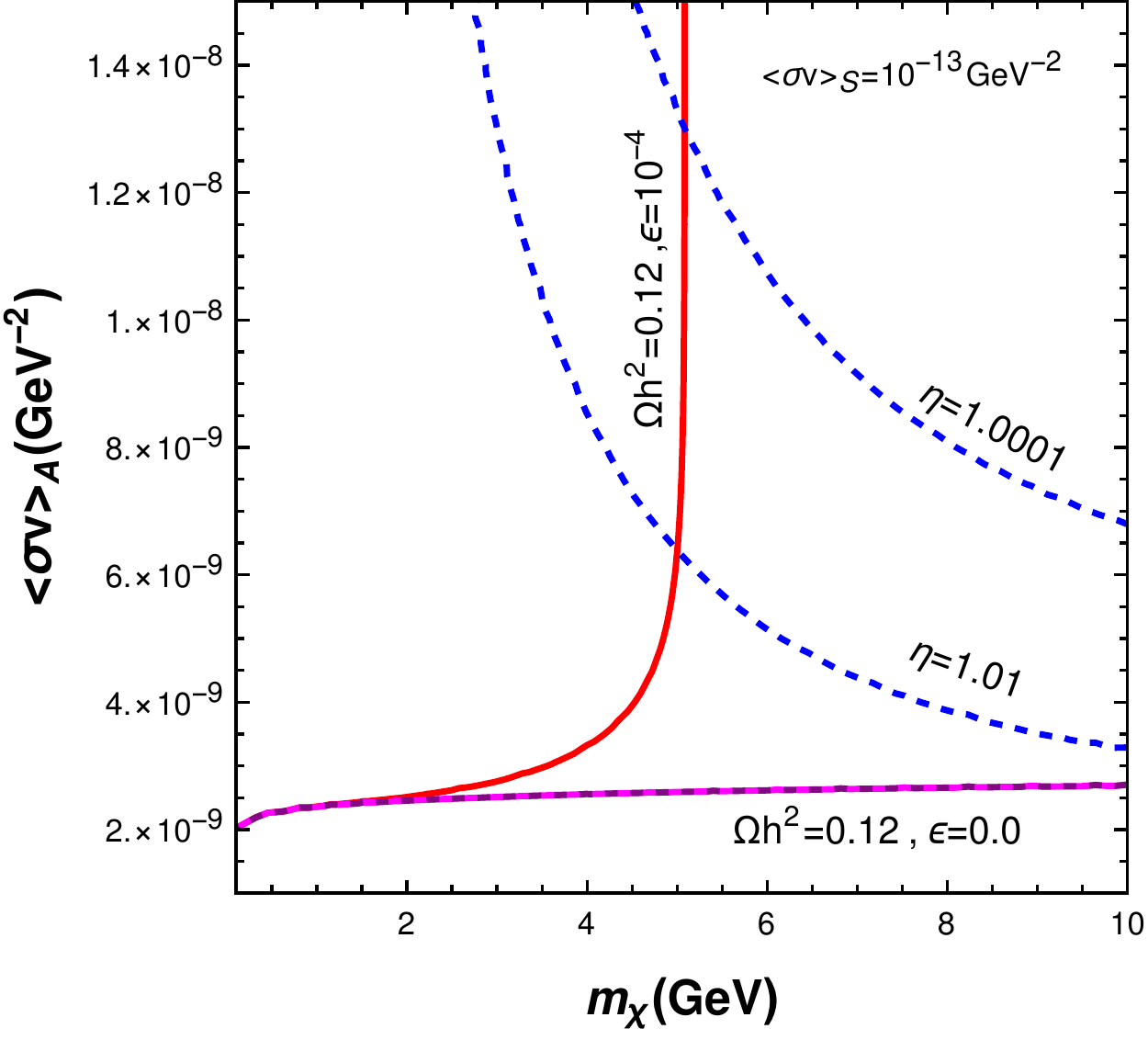}
\caption{\small{\em Required values of the pair-annihilation cross-section $\langle \sigma v \rangle_A$ as a function of the DM mass $m_\chi$ that can reproduce the observed DM relic abundance (red solid line). The results are shown for different values of the CP-violation parameter $\epsilon$ and the semi-annihilation rate $\langle \sigma v \rangle_S$ (left and right plots). Corresponding results for the symmetric case ($\epsilon=0$) are also shown for comparison, with semi-annihilation (black dashed line) and without it (pink solid line), the latter case being the pure WIMP scenario. Also shown are the contours for the present DM relative abundance parameter $\eta$. See text for details.}}
\label{Fig:param1}
\end{figure}

In order to understand the typical values of the cross-sections required to reproduce the observed relic abundance today, we show in Fig.~\ref{Fig:param1} the regions in the $\langle \sigma v \rangle_A$ and $m_\chi$ parameter space in which the central value of the DM relic density  $\Omega h^2 = 0.12$ is reproduced. For both the plots in this figure (left and right), the values of $\epsilon$ and $\langle \sigma v \rangle_S$ have been kept fixed. We show the results for $\epsilon=0.01$ and $\langle \sigma v \rangle_S = 10^{-10} {~\rm GeV}^{-2}$ in the left figure, and for $\epsilon=10^{-4}$ and $\langle \sigma v \rangle_S = 10^{-13} {~\rm GeV}^{-2}$ in the right figure. We also show contours in the $m_\chi - \langle \sigma v \rangle_A$ parameter space for constant values of the present relative abundance of DM and anti-DM, parametrized by $\eta$, as defined in Eq.~\ref{eq:eta}. 

As expected from our discussion in Sec.~\ref{sec:estimate}, in particular Eq.~\ref{eq:estimate2}, as we increase $\epsilon$, the value of $\langle \sigma v \rangle_S$ required to reproduce the relic density is also correspondingly increased. This is primarily due to the exponential suppression of $\Omega h^2$ from $x_S$, which is increased for larger $\langle \sigma v \rangle_S$, thereby requiring larger $\epsilon$. This estimate is applicable only in the case when the symmetric component today is negligible, i.e., $\eta \simeq 1$. As we see in Fig.~\ref{Fig:param1}, in the entire parameter space under consideration, $\eta$ is close to $1$. We also show the contour for $\Omega h^2 = 0.12$, in the $\epsilon=0$ limit (black dashed line), which is found to overlap with the corresponding contour (pink solid line) in the case in which only the pair-annihilation is active (i.e., $\langle \sigma v \rangle_S=0$ as well). This is not surprising, since for such small values of $\langle \sigma v \rangle_S$, which is at least an order of magnitude below the values of $\langle \sigma v \rangle_A$, semi-annihilation is essentially not relevant in determining the present DM abundance as long as $\epsilon=0$. 

The scenario, however, changes dramatically with the introduction of a small CP-violation with a non-zero $\epsilon$, when semi-annihilation becomes the key process in determining the present density. The role of $\langle \sigma v \rangle_A$ for non-zero $\epsilon$ is then to eliminate the symmetric component of DM that is left over at the freeze-out of the semi-annihilation process. As we have already seen in Sec.~\ref{sec:mi}, for large $\mathcal{O}(1)$ values of $\epsilon$, no other number changing process plays any role in determining the relic abundance. This is because such a scenario leads to a large violation of CP in the DM sector, thereby producing an almost completely asymmetric DM already at the freeze-out of the semi-annihilation process at temperature $T_S$. Since almost no symmetric component is left out in this case at $T=T_S$, the pair-annihilation process is not relevant. 

In the limit $\eta \rightarrow 1$, we see from Eq.~\ref{eq:estimate2} that for a fixed value of $x_S$ (which in turn is obtained for a fixed value of $\langle \sigma v \rangle_S$ in this limit) and $\epsilon$, the dark matter mass is also fixed. In particular, as we see from Fig.~\ref{Fig:param1}, with $\epsilon=0.01$ and $\langle \sigma v \rangle_S = 10^{-10} {~\rm GeV}^{-2}$, we obtain $m_\chi \sim 4600$ GeV, while for $\epsilon=10^{-4}$ and $\langle \sigma v \rangle_S = 10^{-13} {~\rm GeV}^{-2}$ , $m_\chi \sim 5$ GeV. Away from the region in the parameter space for which $\eta \rightarrow 1$, we find it non-trivial to obtain a semi-analytic solution to the Boltzmann equations. However, it is clear from Fig.~\ref{Fig:param1} that the DM mass is no longer uniquely fixed for such a case, but varies with $\langle \sigma v \rangle_A$. This is essentially because the symmetric component is not completely removed in such scenarios.

We note in passing that the parameter values $\epsilon=10^{-4}$ and $\langle \sigma v \rangle_S = 10^{-13} {~\rm GeV}^{-2}$ predict a DM mass of around $5$ GeV in the completely asymmetric DM limit. Since this value of the DM mass is around five times the proton mass, we expect the current number densities of the surviving DM particle and protons to be similar in this scenario. As is well-known, such a DM mass is also expected in scenarios which dynamically relate the DM and baryon number densities in the current Universe~\cite{Petraki:2013wwa, Zurek:2013wia}. Such a mechanism to relate the two asymmetries might be possible through semi-annihilation.

In the pure WIMP scenario, with $\epsilon=0$ and $\langle \sigma v \rangle_S=0$, in the freeze-out approximation, the dependence of $\Omega h^2$ on the DM mass is logarithmic, while it is inversely proportional to $\langle \sigma v \rangle_A$. Therefore, we see in Fig.~\ref{Fig:param1} that the value of $\langle \sigma v \rangle_A$ required (around $3.5 \times 10^{-9} {~\rm GeV}^{-2}$) to reproduce $\Omega h^2=0.12$ is largely independent of $m_\chi$ (pink solid line in both figures). As discussed above, this value remains unchanged with the introduction of a small $\langle \sigma v \rangle_S$, when the CP-violation is zero ($\epsilon=0$). In the $\eta \rightarrow 1$ limit, for non-zero $\epsilon$, the requirement of $\langle \sigma v \rangle_A$ is larger, and it increases with increasing $\epsilon$. 

\begin{figure}[htb!]
\centering
\includegraphics[scale=0.52]{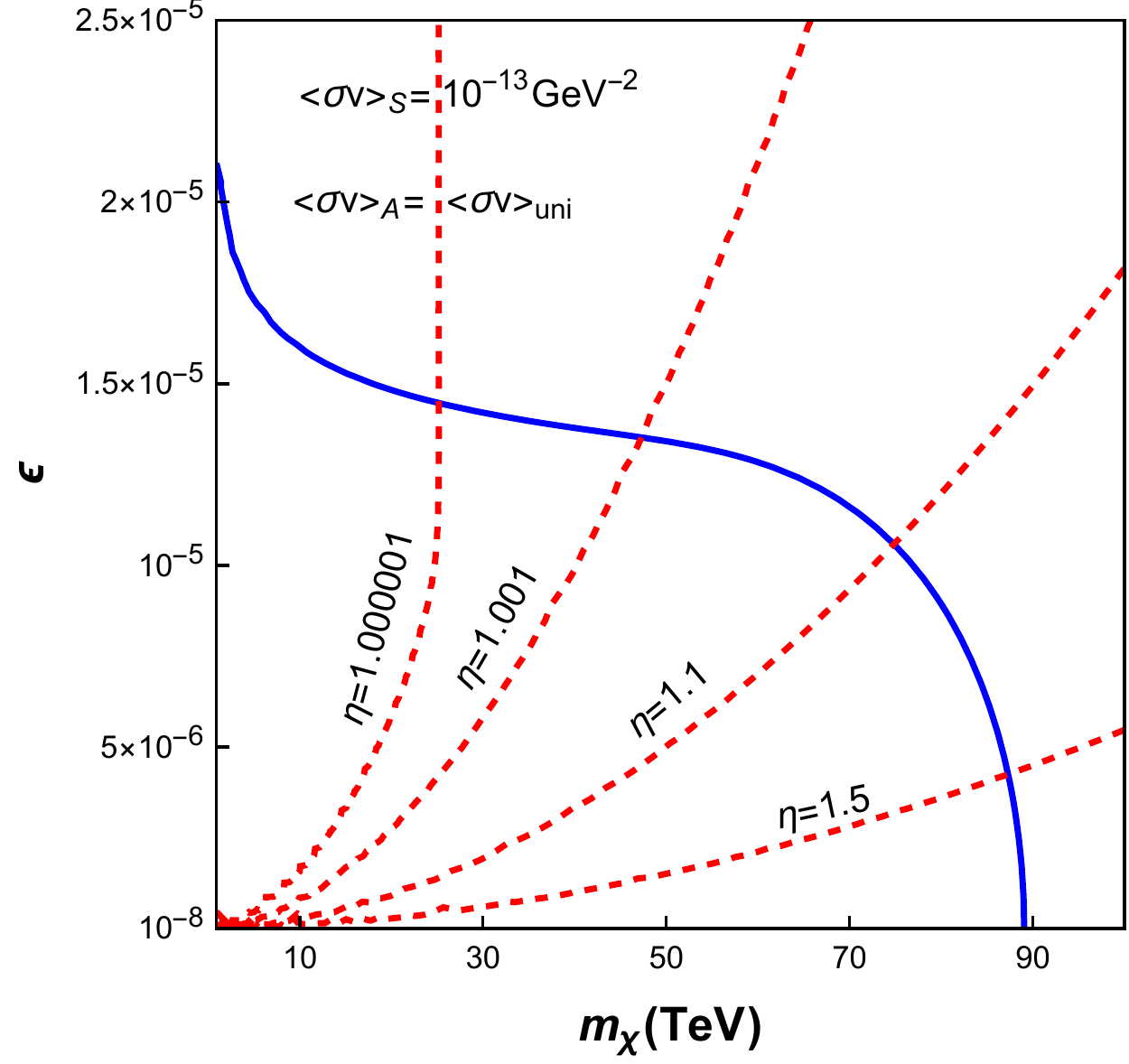}
\caption{\small{\em Contours in the $m_\chi - \epsilon$ plane in which the central value of the DM relic density  $\Omega h^2 = 0.12$ is reproduced, with the pair-annihilation rate fixed at its s-wave upper bound implied by S-matrix unitarity, $\langle \sigma v \rangle_A = \langle \sigma v \rangle_{\rm uni} = (4 \pi/ m_\chi^2)(x_F/ \pi)^{1/2}$. The semi-annihilation cross-section is fixed to ensure $ \langle \sigma v \rangle_{S} <  \langle \sigma v \rangle_A$, for all values of $m_\chi$ considered in this figure, such that the freeze-out temperature hierarchy $T_S > T_A$ is satisfied.}}
\label{fig:uni}
\end{figure}

As in Sec.~\ref{sec:mi}, we can obtain an upper bound for possible values of the dark matter mass by using partial-wave unitarity to bound the annihilation (or semi-annihilation) cross-sections from above. For the scenario in which $T_S>T_A$, the annihilation cross-section must be larger than the semi-annihilation cross-section, and therefore, we impose the unitarity bound on $\langle \sigma v \rangle_A = \langle \sigma v \rangle_{\rm uni}$, where, $\langle \sigma v \rangle_{\rm uni}$ is as given in Eq.~\ref{eq:uni}. In this case, we show the resulting upper bound on the dark matter mass as a function of the CP-violation parameter $\epsilon$ in Fig.~\ref{fig:uni}. We have fixed the value of the semi-annihilation cross-section to be $ \langle \sigma v \rangle_S = 10^{-13} {~\rm GeV}^{-2}$, which is chosen to ensure that  $ \langle \sigma v \rangle_{S} <  \langle \sigma v \rangle_A =  \langle \sigma v \rangle_{\rm uni}$, for all values of $m_\chi$ considered in this figure.

In Fig.~\ref{fig:uni}, the observed relic abundance $\Omega h^2 = 0.12$ is satisfied along the solid blue line. As in Fig.~\ref{fig:semi_results2}, we see that as the CP-violation parameter $\epsilon$ decreases, the resulting mass bound becomes stronger. Furthermore, higher values of $\epsilon$ lead to larger present asymmetry in the dark matter sector, and therefore a value of $\eta$ closer to $1$. The general result obtained in Sec.~\ref{sec:mi}  that the bound on $m_\chi$ for asymmetric DM is stronger compared to the symmetric DM scenario, continues to hold in this scenario as well. In the complete asymmetric limit, i.e., $\eta \rightarrow 1$, the upper bound on the DM mass is found to be around $25$ TeV (numerically for $\eta=1.000001$), while for $\eta \rightarrow 2$ it's around $90$ TeV, assuming s-wave annihilation. 
For the opposite hierarchy of the freeze-out temperatures, i.e., $T_S<T_A$, the semi-annihilation cross-section must be larger than the pair-annihilation cross-section, and therefore, the unitarity bound must be imposed on $\langle \sigma v \rangle_S$, which has already been discussed in Sec.~\ref{sec:mi}, in particular in Fig.~\ref{fig:semi_results2}.

\begin{figure}[htb!]
\begin{center}
\includegraphics[scale=0.52]{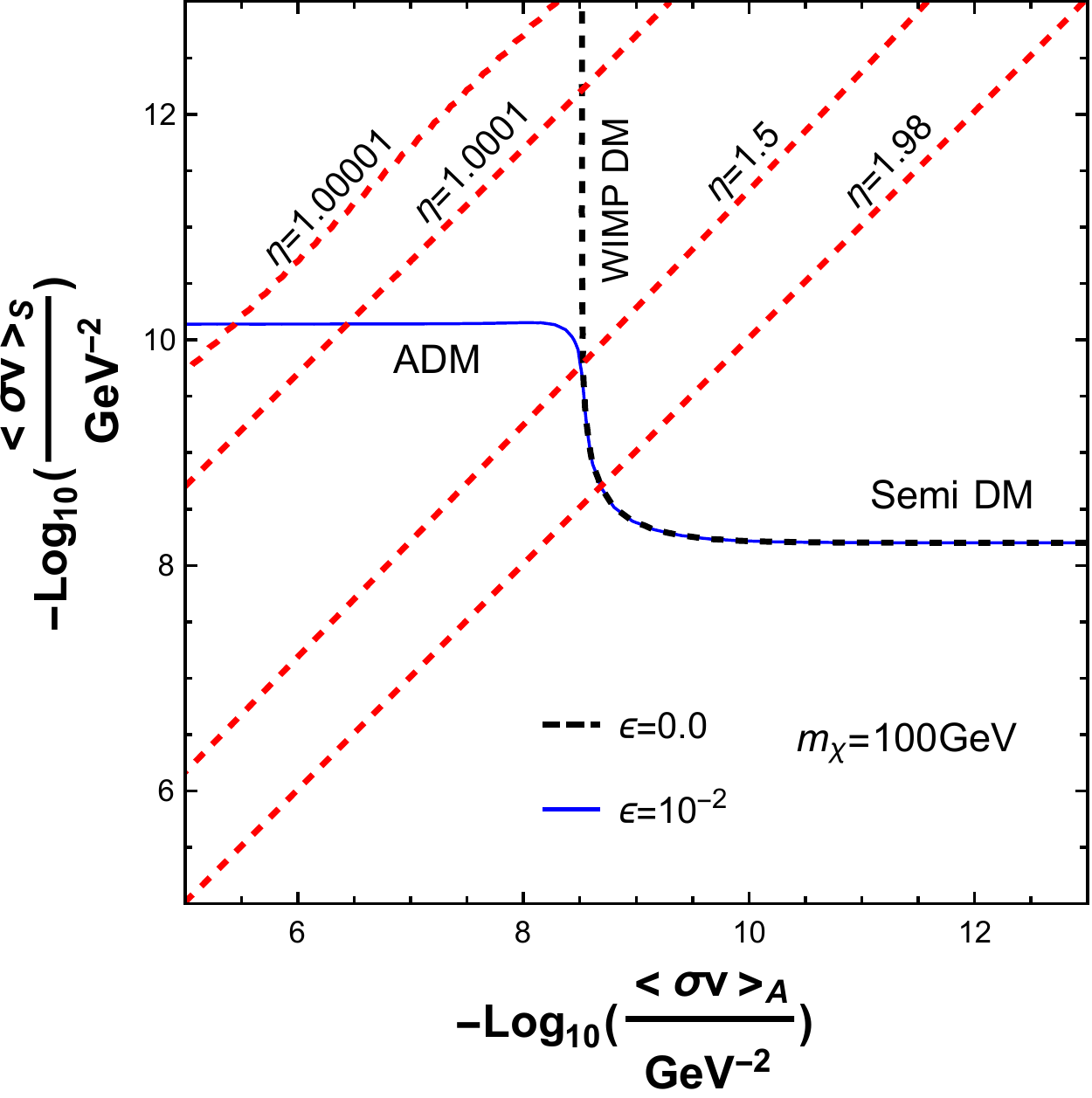} 
\caption{\small{\em Phase diagram showing the interplay between the semi-annihilation and pair-annihilation rates in determining the asymmetric DM properties. The observed relic abundance $\Omega h^2 = 0.12$ is satisfied, with and without CP-violation, along the blue solid and black dashed lines, respectively. The relative abundance parameter $\eta \rightarrow 2$ represents the symmetric phase, while $\eta \rightarrow 1$ represents the asymmetric phase. Here, ADM, Semi DM and WIMP DM denote the dominantly asymmetric dark matter, symmetric semi-annihilating DM and symmetric pair-annihilating DM phases, respectively. See text for details.}}
\label{Fig:param2}
\end{center}
\end{figure}
We can summarize our discussion of the interplay between the semi-annihilation and pair-annihilation rates in determining the asymmetric DM properties using an instructive phase diagram, as shown in Fig.~\ref{Fig:param2}. In this figure, we study the values of $\langle \sigma v \rangle_S$ and $\langle \sigma v \rangle_A$ for which the observed relic abundance $\Omega h^2 = 0.12$ is satisfied, with and without CP-violation. When the CP-violation vanishes, i.e., with $\epsilon=0$, the relic abundance is satisfied along the black dashed contour~\cite{DEramo:2010keq}. Since for $\epsilon=0$, both semi-annihilation and pair-annihilation can reproduce the observed relic abundance, with either or both of them contributing, we obtain an approximate upper bound of $10^{-8} {~\rm GeV}^{-2}$ for both the rates, for a fixed DM mass of $m_\chi=100$ GeV. In contrast, when CP-violation is turned on, i.e., for $\epsilon=10^{-2}$ in Fig.~\ref{Fig:param2}, a symmetric phase and  an asymmetric phase appear in which the relic density is satisfied, as seen in the solid blue line. The two phases can be distinguished by constant values of the DM relative abundance parameter $\eta$. The symmetric phase, with $\eta \rightarrow 2$ is identical to the $\epsilon=0$ scenario, and hence the blue solid line and the black dashed lines overlap. In this phase, the pair-annihilation rate is not large enough to remove the symmetric component efficiently. On the other hand, the asymmetric phase appears when $\langle \sigma v \rangle_A$ is larger than the previously obtained upper bound of around $10^{-8} {~\rm GeV}^{-2}$, for $m_\chi=100$ GeV. In contrast, $\langle \sigma v \rangle_S$ is much smaller in this phase. Thus, to summarize, there are two ways to produce asymmetric DM in the absence of any wash-out processes, namely, 
\begin{enumerate}
\item have a large CP-violation $\epsilon$ as in Sec.~\ref{sec:mi}, in which case semi-annihilation is sufficient to create a complete DM asymmetry, and no subsequent number changing process is necessary, or, 

\item produce a small asymmetry through a small CP-violation $\epsilon$, and then have a sufficiently large pair-annihilation rate to remove the symmetric component, as shown in this section, and as is clear from Fig.~\ref{Fig:param2}.
\end{enumerate}

\subsection{Complex scalar DM with cubic self-interaction}
\label{model}
We now discuss a simple toy model in which the generic scenario described in Sec.~\ref{sec:interplay} with both the semi- and pair-annihilation processes can be realized. The minimal new field content that can lead to a particle-antiparticle asymmetry through the semi-annihilation process include a complex scalar $\chi$ which is charged under a $Z_3$ symmetry (we assign the charge $\omega$ to $\chi$, where $\omega^3=1$) and a real scalar $\phi$, which is a singlet under this symmetry, as well as the SM gauge interactions. The SM fields are also singlets under the discrete $Z_3$ symmetry. The $Z_3$ symmetry ensures the stability of $\chi$, making it the DM candidate. For earlier studies involving different aspects of $Z_3$ symmetric DM, see, for example, Refs.~\cite{Belanger:2012vp, Belanger:2012zr, Hochberg:2014dra, Bernal:2015bla, Hektor:2019ote, Choi:2015bya, Choi:2016tkj}. The effective low-energy interaction Lagrangian involving the $\chi$ and the $\phi$ particles is given by
\begin{equation}
\mathcal{L} \supset \frac{1}{3!} \left(\mu \chi^3 + {\rm h.c.} \right) + \frac{1}{3!} \left(\lambda \chi^3 \phi + {\rm h.c.} \right) + \frac{\lambda_1}{4} \left(\chi^\dagger \chi \right)^2 +   \frac{\lambda_2}{2} \phi^2 \chi^\dagger \chi + \mu_1 \phi  \chi^\dagger \chi +  \frac{\mu_2}{3!} \phi^3 +  \frac{\lambda_3}{4!} \phi^4.
\label{eq:lag1}
\end{equation}
Here, the couplings $\mu$ and $\lambda$ can be complex in general. However, one of the phases can be rotated away by an appropriate re-definition of the field $\chi$. Therefore, in this general effective low-energy theory, there is one residual complex phase, which is necessary to generate a CP-asymmetry in the DM sector. We take $\mu$ to be real, and $\lambda$ to have a non-zero imaginary part, with a phase $\theta$. The parameters in the scalar potential in Eq.~\ref{eq:lag1} can be suitably chosen to ensure that the $\chi$ field does not develop a VEV, thereby ensuring that the $Z_3$ symmetry is not spontaneously broken.

In addition to the interaction terms involving the $\chi$ and the $\phi$ fields in Eq.~\ref{eq:lag1}, there can be two dimension-four and one dimension-three couplings to the SM Higgs doublet $H$ as well, namely, $\lambda_{\chi H} ~(\chi^\dagger \chi  |H|^2) + \lambda_{\phi H} ~(\phi^2 |H|^2) + \mu_{\phi H} ~(\phi |H|^2)$. For $m_{\chi}> m_H$, the $\lambda_{H\chi}$ term contributes in exactly the same way as the $\lambda_2$ term in Eq.~\ref{eq:lag1}, and therefore we do not consider it separately here. Furthermore, the $\lambda_{H \phi}$ and $\mu_{H \phi}$ terms lead to interactions of the $\phi$ field with the $H$ field, which will thermalize the $\phi$ field with the SM plasma. Since we assume the $\phi$ particles to be in equilibrium with the SM bath with zero chemical potential, the effect of these terms are also included. To note, $\phi$ mixes with the Higgs field via $\mu_{H_\phi}$ coupling, in turn it introduces interaction between $\chi$ and physical Higgs ($h$), e.g.  $\lambda \sin\beta ~h \chi^3  $, $\mu_1\sin\beta ~ h \chi^\dagger\chi$, where $\beta$ is the mixing angle between $\phi$ and $H$ after the electroweak symmetry breaking. Consequently, Higgs can decay to a DM pair via $\mu_1$ coupling for DM masses, $m_\chi \leq m_h/2$, where $m_h$ is the mass of the physical Higgs. However, we find that the decay is well suppressed for the benchmark point, $m_\chi= 5 ~\rm GeV$ used in the right panel of Fig.\ref{fig:epsilon_model}. Detailed analysison this can be found in Appendix \ref{App.B}.

The above interaction Lagrangian in Eq.~\ref{eq:lag1} leads to several class of $2 \rightarrow 2$, $2 \rightarrow 3$ and $3 \rightarrow 2$ processes. We find that in different regions of the multi-dimensional parameter space, different class of diagrams (or combinations thereof) may dominate. Since in this section we are presenting a toy model that realises the general features of the model-independent setup discussed in the previous section, we shall focus on a restricted region of the parameter space in which a subset of the $2 \rightarrow 2$ diagrams dominate. In particular, we shall consider the values of the dimensionful parameters to be small compared to the DM mass scale, i.e., $\mu / m_\chi << 1$ and $\mu_1 / m_\chi << 1$. We shall also take the cubic and quartic self-couplings of the $\phi$ field to be small, which does not alter the qualitative features of the scenario.

\begin{figure}[htb!]
\centering
\includegraphics[scale=0.55]{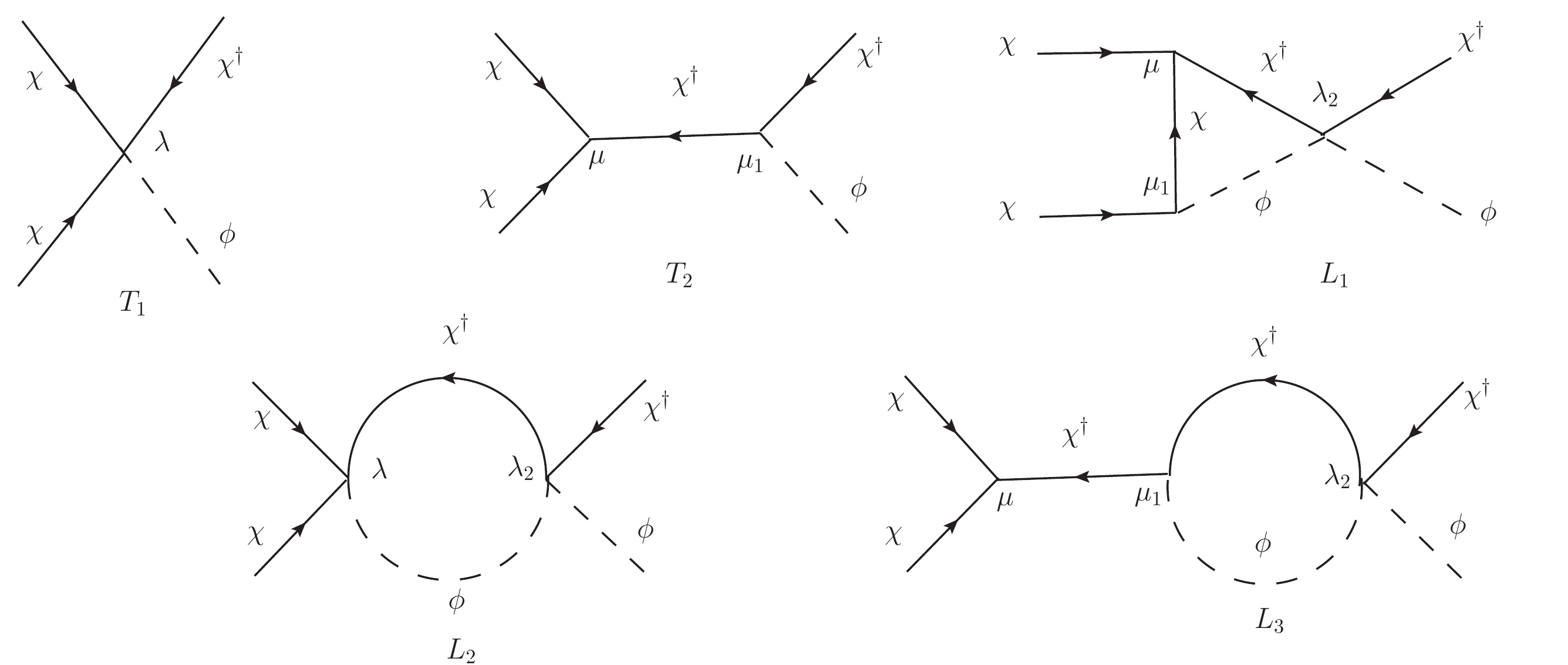}
\caption{\small{\em  Relevant $2 \rightarrow 2$ tree-level and one-loop Feynman diagrams for the semi-annihilation process $\chi \chi \rightarrow \chi^\dagger \phi$.}}
\label{Fig:diagrams}
\end{figure}
%%%%%%%%%%%%%%%%%%%%%%%%%%
The relevant tree-level and one-loop Feynman diagrams for the semi-annihilation process $\chi \chi \rightarrow \chi^\dagger \phi$ are shown in Fig.~\ref{Fig:diagrams}. At tree-level there are two Feynman diagrams contributing to this process: one involving a contact interaction (diagram $T_1$), and the other with an intermediate $\chi$ propagator (diagram $T_2$). The second diagram gives a contribution to the matrix element proportional to $\left (\mu \mu_1/ m_\chi^2 \right)$, in the non-relativistic limit for the $\chi \chi$ initial state, with the centre of mass energy squared $s \simeq 4 m_\chi^2$. Therefore, for  $\mu / m_\chi << 1$ and $\mu_1 / m_\chi << 1$, the contact interaction dominates.

In order to determine the CP-asymmetry generated by the semi-annihilation process, we compute the interference between the tree-level and loop-level diagrams shown in Fig.~\ref{Fig:diagrams}. In general the CP-asymmetry is proportional to  $\Im \left({M}_{\rm tree} (g_i)^* {M}_{\rm loop} (g_j)\right)$, which in turn is proportional to $\Im \left( \prod_{i,j} g_i^* g_j \right) \times \Im ( I)$, where $I$ is the loop factor which acquires an imaginary part when the particles in the loop go on-shell. The latter requirement is ensured by the condition $m_{\phi} < m_{\chi}$. We find that diagram $T_2$ gives a non-zero contribution to the CP-asymmetry, resulting from its interference with the loop diagram $L_2$, while diagram $T_1$ leads to a non-zero contribution from its interference with $L_1$ and $L_3$. Furthermore, the contributions from the interference of $T_2$ and $L_2$, and that from $T_1$ and $L_3$ cancel identically. Therefore, the only relevant contribution is from the interference of $T_1$ and $L_1$. The resulting difference in matrix elements squared that contribute to $\epsilon$ as defined in Eq.~\ref{eq:epsilon}, is given as:
\begin{equation}
|{M}|^2_{\chi\chi\rightarrow \chi^{\dagger} \phi}-|{M}|^2_{\chi^{\dagger}\chi^{\dagger}\rightarrow \chi \phi} = \frac{4 |\lambda| \mu \mu_1 \lambda_2 \sin{\theta}}{16 \pi \sqrt{s(s-4m^2_{\chi})}} \log \left[ \frac{m_\chi^2+m^2_{\phi}-s+\beta_1}{m_\chi^2+m^2_{\phi}-s-\beta_1}\right] 
\end{equation}
where, $s$ is the centre of mass energy squared, and 
\begin{equation}
\beta_1 =\sqrt{\frac{(s-4 m_\chi^2)(m_\chi^4+(m^2_{\phi}-s)^2-2 m_\chi^2(s+m^2_{\phi}))}{s}}.  
\end{equation}

\begin{figure}
\includegraphics[scale=0.55]{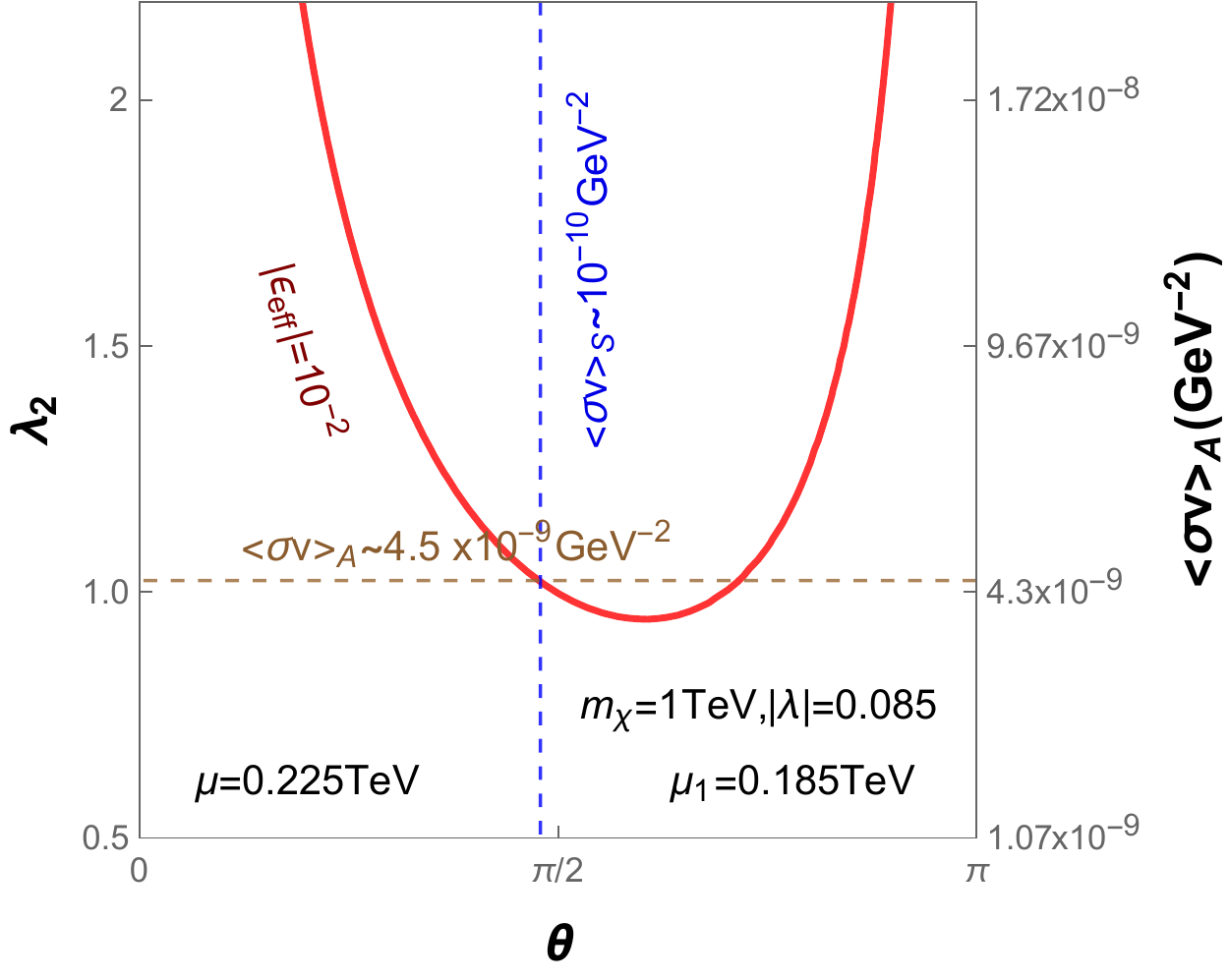} \hspace*{1.0cm}
\includegraphics[scale=0.55]{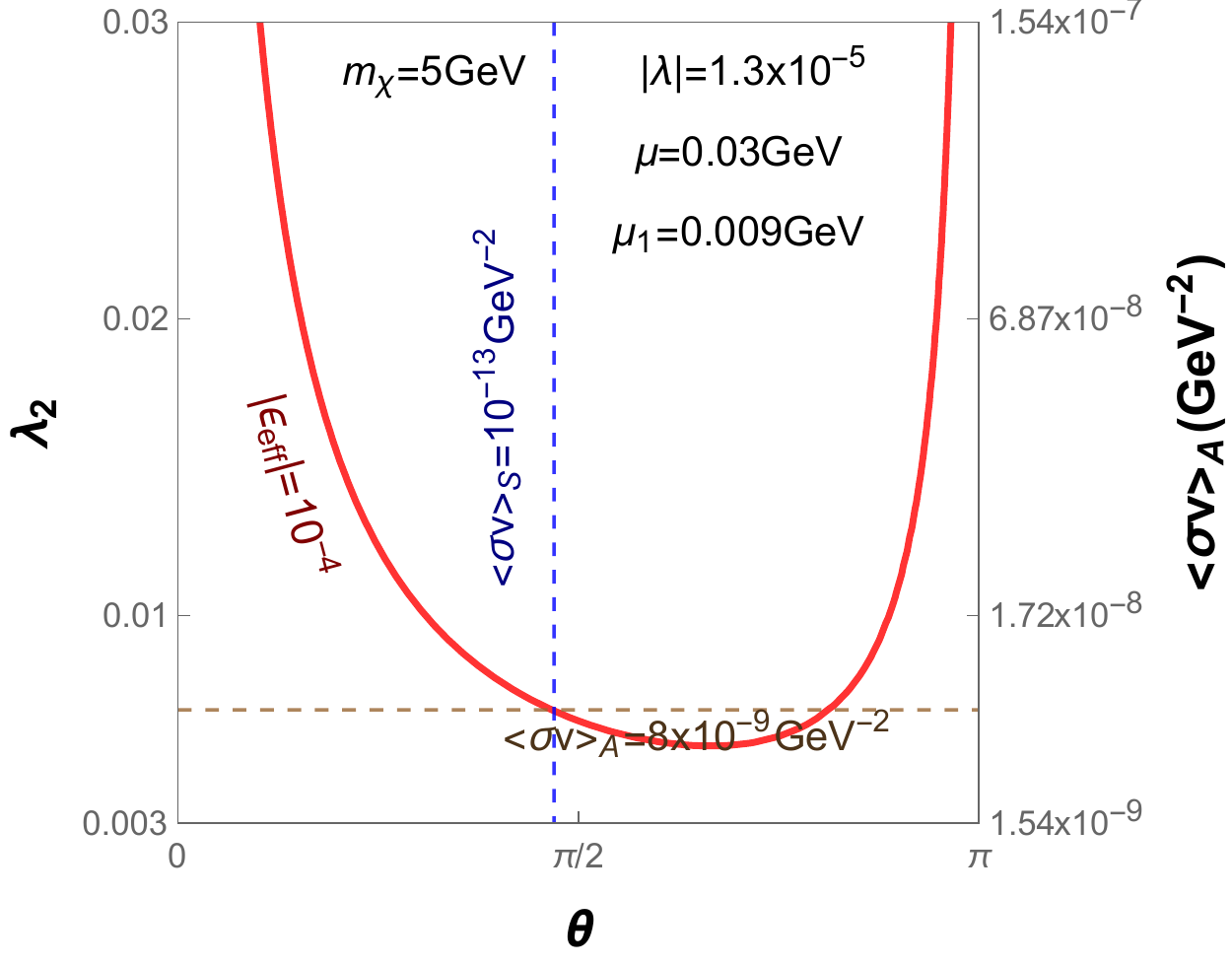}
\caption{\small{\em Contours of fixed effective CP-violation parameter $|\epsilon_{\rm eff}|$ (red solid lines), as a function of the complex phase $\theta=\arg(\lambda)$ and the effective pair-annihilation coupling $\lambda_2$. The results are shown for two different DM mass values $m_\chi=1$ TeV (left panel) and $5$ GeV (right panel). The required values of the annihilation rates and $\epsilon_{\rm eff}$ are reproduced, as indicated. See text for details, and Fig.~\ref{Fig:param1} for comparison with the results obtained in the model-independent analysis.}}
\label{fig:epsilon_model}
\end{figure}

In the model-independent setup discussed in Sec.~\ref{sec:mi} and Sec.~\ref{sec:interplay}, the different annihilation rates and the effective CP-violation parameter were treated as independent free parameters. However, in a model in which such processes are realized, these parameters are often correlated, and are determined in terms of the common set of couplings and masses. Therefore, in order to understand whether the simple model described by Eq.~\ref{eq:lag1} can accommodate the required values of the relevant physical parameters found in the previous section, we study in Fig.~\ref{fig:epsilon_model} the correlation between the effective CP-violation $\epsilon_{\rm eff}$, and the annihilation rates, as a function of the CP-violating phase $\theta$, and the relevant couplings $\lambda_2$ and $|\lambda|$. 

In Fig.~\ref{fig:epsilon_model}, we show contours of fixed effective CP-violation parameter $|\epsilon_{\rm eff}|$ (red solid lines), as a function of the complex phase $\theta=\arg(\lambda)$ and the effective pair-annihilation coupling $\lambda_2$. We have also shown the corresponding values of the annihilation rate $\langle \sigma v \rangle_A$ in both panels. The results are shown for two different DM mass values $m_\chi=1$ TeV (left panel) and $5$ GeV (right panel).  As we can see from this figure, the values of the annihilation rates and $\epsilon$ required to satisfy the DM relic abundance can be obtained in this model, as indicated by the dashed horizontal and vertical lines. This can be observed by comparison with the $\Omega h^2=0.12$ contour in Fig.~\ref{Fig:param1}, where the results were obtained in the model-independent analysis.

A few comments are in order. First of all, as mentioned earlier, in Fig.~\ref{fig:epsilon_model} we ensure $\mu / m_\chi << 1$ and $\mu_1 / m_\chi << 1$, for which our restriction to the class of $2 \rightarrow 2$ diagrams in Fig.~\ref{Fig:diagrams} remains valid.    Since the loop amplitudes in this model depend upon the coupling $\lambda_2$, the pair-annihilation process is necessarily present whenever the CP-violation in the semi-annihilation process is sufficiently large. Thus the first scenario with only the semi-annihilation process discussed in Sec.~\ref{sec:mi} is not obtained in this model, while the second scenario in Sec.~\ref{sec:interplay} with both semi- and pair-annihilations can be easily realized. Additional structures are therefore necessary to have loop graphs with sufficiently large imaginary parts, which do not induce significant tree-level pair annihilation.

\begin{figure}[htb!]
\begin{center}
\includegraphics[scale=0.55]{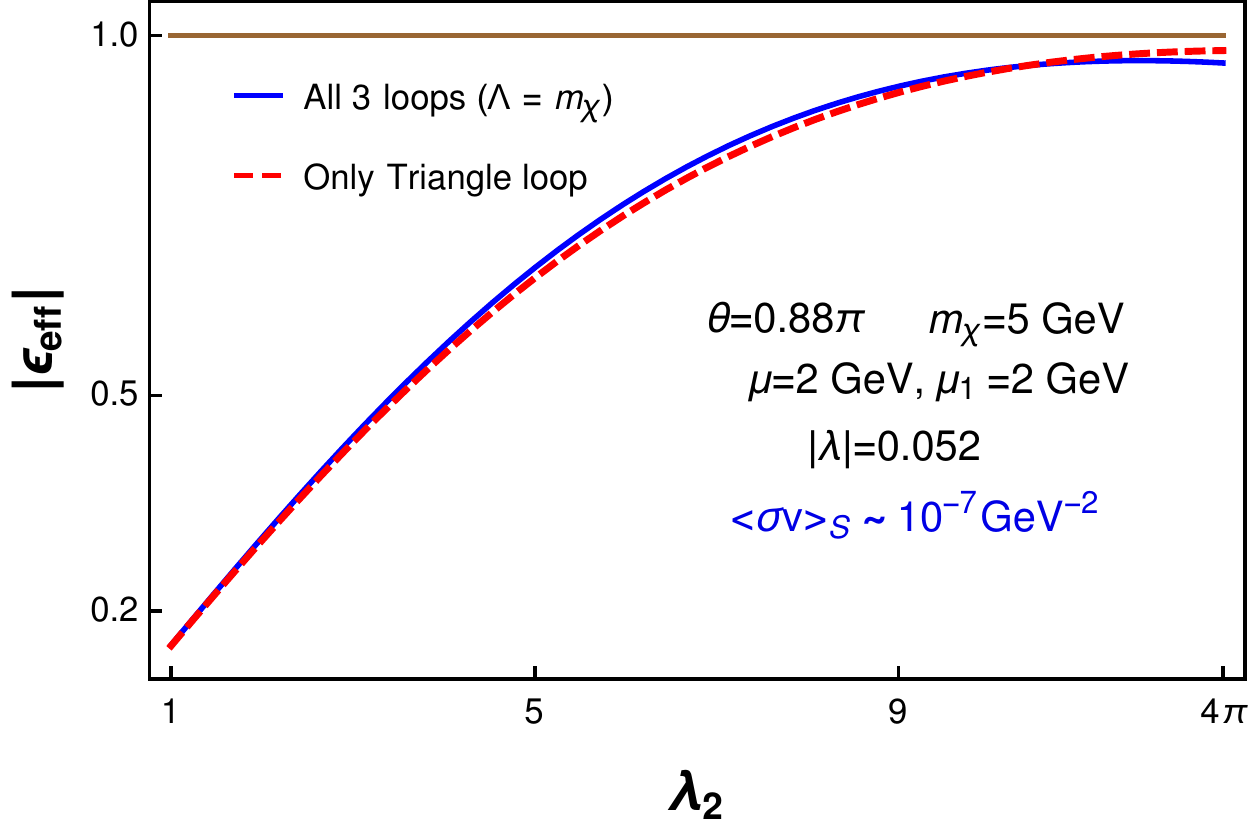} 
\caption{\small{\em The effective CP-violation parameter $|\epsilon_{\rm eff}|$, as a function of the pair-annihilation coupling $\lambda_2$. The results are shown for fixed values of all other parameters, including the DM mass value of $m_\chi=5$ GeV. The triangle loop gives the dominant contribution, as shown by comparison of $|\epsilon_{\rm eff}|$ computed only with the triangle loop (red dashed line), and all the three loops (blue solid line).}}
\label{Fig:eps_eff}
\end{center}
\end{figure}
We find that a scenario where the effective CP-violation parameter $|\epsilon_{\rm eff}|$ is close to unity, thereby leading to the present DM asymmetry $\eta \rightarrow 1$, can be realized in the model described by Eq.~\ref{eq:lag1}, for values of the model parameters within perturbative limits. We show in Fig.~\ref{Fig:eps_eff} the variation of $|\epsilon_{\rm eff}|$ as a function of the coupling $\lambda_2$, which appears in all the relevant loop graphs in Fig.~\ref{Fig:diagrams}. The results are shown for fixed values of all the other parameters, including the DM mass value of $m_\chi=5$ GeV. Even though here the dimensionful parameters $\mu$ and $\mu_1$ are taken to be only a factor of two smaller than $m_\chi$, we have checked that only the $2 \rightarrow 2$ processes in Fig.~\ref{Fig:diagrams} dominate. We find that the triangle loop gives the dominant contribution, as shown by comparison of $|\epsilon_{\rm eff}|$ computed only with the triangle loop (red dashed line), and all the three loops (blue solid line). While the triangle loop ($L_1$ in Fig.~\ref{Fig:diagrams}) does not lead to any ultraviolet (UV) divergence, the other two loops (namely, $L_2$ and $L_3$) are divergent in the UV, and therefore $|\epsilon_{\rm eff}|$ computed with all three loops has a renormalization scale dependence, which is found to be rather weak. For our computations, we have set the renormalization scale to be $\Lambda = m_\chi$. We emphasize that even though in this example, the value of the pair-annihilation coupling $\lambda_2$ is large, the choice of parameters for which $|\epsilon_{\rm eff}| \rightarrow 1$ essentially belongs to the first scenario we studied, in which the semi-annihilation process almost entirely determines the present DM properties, including its asymmetry. This is because, for $|\epsilon_{\rm eff}| \rightarrow 1$, at the decoupling of the semi-annihilation process, the symmetric component is already negligible, and hence the subsequent pair-annihilation is largely irrelevant. Therefore, as is clear from Fig.~\ref{Fig:eps_eff}, we can achieve $|\epsilon_{\rm eff}|$ close to $1$ in this scenario, being within perturbative limit of all the relevant parameters including $\lambda_2$. Furthermore, we can also essentially realize, albeit in the presence of the coupling $\lambda_2$, the first scenario in which the semi-annihilation process almost entirely determines the present DM properties.

\newpage
\section{\Large{Cosmology of complex scalar dark matter: Interplay of self-scattering and annihilation}}
\label{chap:chap4}

\subsection{Introduction}
\label{sec:sec41}
Perhaps the simplest studied scenario for thermal dark matter (DM) is that of a real scalar field ($S$), which is a singlet under the standard model (SM) gauge interactions of $SU(3) \times SU(2) \times U(1)$~\cite{zee, McDonald}. The new degree of freedom added to the SM field content here is at an absolute minimum of one. Such a DM particle can be stable due to the existence of an effective reflection symmetry ($Z_2$ symmetry), under which the DM field is odd ($S \rightarrow -S$) and the SM particles are even. A renormalizable interaction of the real scalar singlet with the SM Higgs doublet ($H$) exists, namely, $S^2 |H|^2$, which is allowed by the reflection symmetry~\cite{zee, McDonald, Burgess, Patt}. The cosmology of this scenario is also very simple -- depending upon the mass of the singlet $S$, it may go through pair-annihilations to different SM particles, achieve and stay in thermodynamic equilibrium with the SM sector in the early Universe, and eventually freeze-out with a thermal relic abundance, which is determined by the two unknown parameters of the singlet mass and coupling to the Higgs. Such a scenario also presents with different signals at direct, indirect and collider searches -- and is strongly constrained by them in a large region of the parameter space providing the required DM abundance~\cite{cline, Athron:2017kgt, Athron:2018ipf}. 

We have already noticed in the earlier chapter that the DM particle may or may not be self-conjugate and it may be charged under a global symmetry of DM number. What new phenomena do we expect in such a case in general? Here, there are two broad possibilities. The first is that of a restricted scenario in which we impose the strict conservation of this global DM number symmetry, while the second possibility is of a more general scenario in which such a symmetry can be broken by different interaction terms. In the first case, the phenomena are essentially same as that of the self-conjugate DM, albeit with twice the new degrees of freedom. However, there is no strong argument behind the conservation of such global continuous symmetries. We know, for example, that it is very likely that the global $U(1)_{\rm B}$ symmetry of baryon number must be broken in order to dynamically generate the baryon asymmetry of the Universe~\cite{Weinberg:1979sa}. Similarly, the SM gauge quantum numbers of neutrinos allow for a Majorana mass term -- whose existence is being searched for -- and such a term would violate the global $U(1)_{\rm L}$ symmetry of lepton number. Such global symmetry breaking effects are generically found in unified theories of the strong, weak and electromagnetic interactions~\cite{Pati:1974yy, Georgi:1974sy}. It is also generally argued that global continuous symmetries may not be conserved  due to quantum gravitational effects~\cite{Witten:2017hdv}. Discrete symmetries, on the other hand, may remain in effective low energy theories, in some cases as a remnant of larger continuous symmetries~\cite{Witten:2017hdv}, thereby make the DM candidate absolutely stable.

Thus the second general possibility of including interactions that do not conserve the global symmetry of DM number should be explored in detail. And we expect to encounter new phenomena in studying the cosmology of such a scenario. A complex scalar singlet DM constitutes the version of the most minimal thermal DM scenario, with now a non-self-conjugate DM particle instead of a self-conjugate one. In this chapter, we carry out a detailed study of this scenario, and find several interesting effects in its cosmological history, some of them novel. No complete study of such a general scenario for a complex scalar DM has been carried out so far, to our knowledge, and the previous studies of the complex scalar DM are restricted to symmetric weakly interacting massive particle (WIMP)-like behaviour~\cite{Barger:2008jx, Chiang:2017nmu}.  The different interaction terms present in such a scenario and their role in the various scattering processes are discussed in Sec.~\ref{sec:sec2} in the following.

The primary effect observed is that of the close interplay between the self-scattering and annihilation processes in determining the present DM density and composition. In particular, the fact that DM number may now be violated with all renormalizable interactions included, leads to the possibility of creation of a particle-antiparticle asymmetry in the DM sector. Such an asymmetry generation of course crucially depends on the CP-violation in the relevant scattering rates, which is non-zero only when all of the self-scattering and annihilation processes are simultaneously present. The absence of any one of them leads to the vanishing of the CP-violation, and hence the asymmetry -- by very general arguments of unitarity sum rules. This close interplay of the self-scatterings and annihilations is a very novel phenomenon, which we discuss in Sec.~\ref{sec:sec3}, after setting up the Boltzmann kinetic equations for the DM and anti-DM number densities. We then present the approximate analytic solutions to these equations (Sec.~\ref{sec:sec31}), which clearly demonstrate the role of different type of scattering processes in determining the net DM yields as well as any possible asymmetries, which are then compared with the general numerical solutions (Sec.~\ref{sec:sec32}). 

In Sec.~\ref{sec:sec4} we compute the CP-violation (at one-loop level) and the thermally averaged symmetric and asymmetric scattering rates in the complex singlet scenario, that are inputs to the kinetic equations, to understand the typical values of these quantities that one might expect in this model. We then go on to discuss the close interplay of the coupling parameters that determine the self-scattering and annihilation rates in fixing the DM relic density and asymmetry. We summarize our results in Sec.~\ref{sec:sec6}. The details of the CP-violation calculation for the sub-leading coversion process $\chi+\phi \rightarrow \chi^{\dagger}+\phi$ are provided in Appendix~\ref{App.A}.

%%%%%%%%%%%%%%%%%%%%%%%%%%%%%%%%%%%%%%%
\subsection{Gauge singlet complex scalar dark matter: general scenario}
\label{sec:sec2}
The most general low-energy effective Lagrangian of a SM gauge singlet complex scalar ($\chi$), that is odd under a $Z_2$ reflection symmetry (with $\chi \rightarrow -\chi$), is given as follows:
\begin{equation}
\mathcal{L} \supset (\partial_\mu \chi)^\dagger \partial^\mu \chi - m_{\chi}^2 \chi^\dagger \chi - \frac{1}{2} \left(\tilde{\mu}^2 \chi^2 + {\rm h.c.}\right)-\lambda_{\chi H} |\chi|^2 |H|^2,
\label{eq:lag0}
\end{equation}
where $H$ is the SM Higgs doublet. The interaction with the Higgs boson, however, is severely constrained by spin-independent direct detection probes for a large range of the DM mass, as well as by the search for invisible Higgs decay for a DM that is lighter than half of the Higgs mass. With such constrained values of the Higgs portal coupling, it is difficult to obtain the observed DM abundance in a large range of the DM mass. Therefore, we shall discuss the DM cosmology, assuming, for all practical purposes, that  the coupling $\lambda_{\chi H} \simeq 0$. As we shall see in the following, the effects of the Higgs coupling-like term in the DM cosmology will be captured by a new interaction that is then necessary for DM thermalization. The $\tilde{\mu}^2 \chi^2 $ term and its hermitian conjugate both break the larger global $U(1)_{\chi}$ \footnote{We replace previously used suffix $B_\chi$  by $\chi$ for simplicity.} symmetry of DM number under which the DM particle (denoted by $\chi$) has charge $+1$ and its antiparticle (denoted by $\chi^\dagger$) has a charge of $-1$.

We shall assume the $U(1)_{\chi}$ breaking $\tilde{\mu}^2$ term to be much smaller than the corresponding $U(1)_{\chi}$ conserving mass term $m_{\chi}^2$. In such a case, the DM mass eigenstates will essentially correspond to the particle states with $U(1)_{\chi}$ charge of $+1$ and $-1$, throughout the cosmological evolution of interest in this study. In late epochs, after the chemical decoupling of the DM species, particle-antiparticle oscillations can be caused by the presence of the $U(1)_{\chi}$ breaking mass term, whose effects are discussed in Refs.~\cite{Buckley:2011ye, Cirelli:2011ac, Tulin:2012re}. In particular, the oscillation probability is obtained to be $P(\chi\rightarrow\chi^{\dagger})= \sin^2\left({\mid\tilde{\mu}\mid^2 t}/{m_\chi}\right)$. In order for the oscillation dynamics to not affect the freeze-out process, the characteristic time scale of oscillations ($\tau=\left({\mid\tilde{\mu}\mid^2}/{m_\chi}\right)^{-1}$) should be larger than the freeze-out time scale ($t_F \sim H^{-1} (T=T_F)$), where $H(T=T_F)$ is the Hubble parameter at the freeze-out temperature. 
With $T_F \sim m_\chi$, requiring $\tau > t_F$ implies that as long as $|\tilde{\mu}|^2 \lesssim m_\chi^3/M_{\rm Pl}$, the mass and charge eigenstates can be taken to be essentially the same during the cosmological evolution time scale of the DM species. 

For example, with a typical order of the DM mass of $m_\chi = 10\, \rm TeV$, which is found to be relevant in the subsequent analyses, this upper bound on the possible size of $|\tilde{\mu}|$ is around $80$ keV. The upper bound on $|\tilde{\mu}|$ is modified appreciably in the presence of additional scattering processes which delay the onset of oscillations (see Ref.\cite{Cirelli:2011ac} for details). In our model, we find that for $|\tilde{\mu}|=10\, \rm GeV$ and $|\tilde{\mu}|=100 \,\rm GeV$, the oscillation starts at temperatures $T \sim1\, \rm GeV$ and $T \sim 10 \,\rm GeV$, respectively, for $m_\chi=10\,\rm TeV$, whereas the corresponding freeze-out dynamics takes place at  temperature $T \sim \mathcal{O}(500 \,\rm GeV)$. Therefore, we can safely carry out the subsequent analysis of the Boltzmann system in the charge basis, as the oscillation phenomenon is not relevant before or during the freeze-out process, as long as $|\tilde{\mu}|$ is around two orders of magnitude smaller than the DM mass.

With the condition that $\lambda_{\chi H} \simeq 0$, as discussed above, in order to thermalize the DM state with the SM sector, so that a viable thermal production mechanism for the DM density is obtained, we are forced to introduce a mediator degree of freedom in the scenario. A minimal new addition in such a case would be a real scalar field $\phi$, which is also a singlet under the SM gauge interactions, but is even under the $Z_2$ reflection symmetry. Since $\phi$ can couple to both the DM state as well as the SM sector, it can effectively bring the two sectors to thermal equilibrium. The additional interaction terms in the Lagrangian density are now as follows:
\begin{align}
-\mathcal{L_{\rm int}}  \supset  \mu  \chi^\dagger \chi  \phi+ \left(\frac{\mu_1}{2} \chi^2 \phi+ {\rm h.c.} \right)  + \frac{\lambda_1}{4} \left(\chi^\dagger \chi \right)^2 +   \left(\frac{\lambda_2}{4!}\chi^4 +{\rm h.c.} \right)+\left(\frac{\lambda_3}{4}\chi^2 \phi^2 +{\rm h.c.} \right)\nonumber\\
+\left(\frac{\lambda_4}{3!}\chi^3\chi^{\dagger} +{\rm h.c.} \right)
+ \frac{\lambda_5}{2} \phi^2 \chi^\dagger \chi  +  \frac{\mu_\phi}{3!} \phi^3 +  \frac{\lambda_\phi}{4!} \phi^4 + \frac{\lambda_{\phi H}}{2} \phi^2 |H|^2 + \mu_{\phi H} \phi |H|^2\,\,\,
\label{eq:lag2}
\end{align}
In this most general renormalizable interaction Lagrangian involving the $\chi$ and $\phi$ fields, the interaction terms with couplings $\mu_1, \lambda_2, \lambda_3$ and $\lambda_4$ break the global $U(1)_{\chi}$ symmetry, in addition to the $\tilde{\mu}^2$ term in Eq.~\ref{eq:lag0} above. All of these terms, in general, can also have complex coupling parameters, one of which can be chosen to be real by an appropriate redefinition of the $\chi$ field. 

The scalar mass and quartic terms are chosen appropriately to ensure that the $\chi$ and $\phi$ fields do not obtain any non-zero vacuum expectation values, where the former condition also implies that the $Z_2$ reflection symmetry remains unbroken, rendering the DM stable. With this condition, the couplings between the $\phi$ and the SM Higgs fields will eventually lead to $\phi$ mixing with the Higgs particle after electroweak symmetry breaking through the $\phi |H|^2$ term only. Therefore, the current Higgs coupling measurements will strongly constrain the $\mu_{\phi H}$ coupling. On the other hand, the $\lambda_{\phi H}$ coupling is not as strongly bounded, even if $\phi$ is much lighter than the Higgs. This is because, if the Higgs boson decays to two $\phi$ particles, each of the resulting $\phi$ will further decay to visible SM final states through its small mixing with the Higgs. If the decay width of $\phi$ is so small that the $\phi$ particles decay outside the LHC detectors, there will be an upper bound on $\lambda_{\phi H}$ from the search for invisible decays of the Higgs boson. Using the $95 \%$ C.L. upper limit on the Higgs invisible branching ratio obtained by a combination of the $7,8$ and $13$ TeV data from the CMS collaboration~\cite{CMS:2018yfx}, we find $\lambda_{\phi H} \lesssim 10^{-2}$. Such couplings are however sufficient for the $ \lambda_{\phi H} $ and $\mu_{\phi H}$ terms to keep the $\phi$ particles in thermal equilibrium with the SM sector. 

The small $\phi$ mixing with the Higgs however can induce signals in DM direct detection probes through the $\chi^\dagger \chi  \phi$ and $\chi^2 \phi$ terms, which will induce $\chi$ scattering with nucleons. Therefore, to a first approximation, we also set $\mu \simeq 0$ and $\mu_1 \simeq 0$. As we shall see in the following, the essential aspects of the cosmology of $\chi$ particles are effectively captured even in such a case. With these parameter choices, we have checked that the $U(1)_\chi$ symmetry breaking interaction terms do not generate a large $\tilde{\mu}^2$ term through loop corrections.\footnote{As is well-known from the example of scalar field theory with a $\phi^4$ interaction, at one-loop level, the quartic couplings do not generate any quantum corrections to the $\tilde{\mu}^2$ term. Furthermore, since we have also set the trilinear couplings to be vanishing to begin with, the leading quantum correction to the $\tilde{\mu}^2$ term will be generated only at the two loop level. Such two-loop corrections to $|\tilde{\mu}|$ are, however, expected to be at least two orders of magnitude smaller than $m_\chi$, which is the typical upper bound required for oscillation phenomenon to occur only much after the DM freeze-out.}. We can thus consistently treat the $U(1)_\chi$ charge eigenstates as the DM mass eigenstates for the cosmological evolution until the freeze-out time scale, which is the subject matter of this study. For the astrophysical dynamics in later epochs, the effect of these terms should be included.

Through their interactions with $\phi$, the DM particles and anti-particles are assumed to remain in kinetic equilibrium with the SM thermal bath throughout the evolution of the chemical processes that determine its density. The chemical reactions that affect the DM density and composition, following from the interactions in Eq.~\ref{eq:lag2} are as follows:
\begin{enumerate}
\item {\em Self-scattering reaction leading to particle-antiparticle conversion}: $\chi+\chi \rightarrow \chi^{\dagger} + \chi^{\dagger}$. Due to the presence of a complex coupling parameter, this reaction can be CP-violating and can create  a particle-antiparticle asymmetry in the DM sector. Moreover, it violates the DM number of $U(1)_{\chi}$ by four units, but cannot change the total DM and anti-DM number. 

\item {\em Self-scattering reaction leading to particle-antiparticle conversion}: $\chi+\chi \rightarrow \chi + \chi^{\dagger}$. Similar to the previous process, this can also lead to CP-violation and asymmetry in the DM sector. This process violates the DM number by two units, but does not modify the total DM number.

\item {\em DM number violating annihilation}: $\chi+\chi \rightarrow \phi+\phi$. This process can also violate CP-symmetry and create asymmetric DM, due to the complex coupling of $\lambda_3$. It violates DM number by two units, and also changes the total DM and anti-DM number by the same amount.

\item {\em DM number conserving annihilation}: $\chi+\chi^\dagger \rightarrow  \phi+\phi$. This process is CP-conserving, and cannot by itself create a DM-anti-DM asymmetry. However, as shown in Ref.~\cite{Ghosh:2021ivn}, it can indirectly affect the generated asymmetry by controlling the out-of-equilibrium number density of relevant particles in the thermal bath. This process conserves the $U(1)_{\chi}$ DM number symmetry, but changes the total DM and anti-DM number by two units. 
\end{enumerate}

In addition to the processes included in the DM and anti-DM number-density evolution equations, the process, $\chi+\phi \rightarrow \chi^{\dagger}+\phi$ can also change the DM number by $-2$ units, whereas its CP-conjugate process,  $\chi^{\dagger}+\phi\rightarrow \chi+\phi$ changes the DM number by $+2$ units. Therefore, if the CP-violation in this process is substantial, it can lead to particle-antiparticle conversions, thereby affecting the DM asymmetry. However, as discussed in Sec.~\ref{sec:sec4}, and detailed in Appendix~\ref{App.A}, we find the contribution of this process to the CP-violation to be negligible in the parameter region of interest in this study, and therefore, we have not included the effect of this scattering process in generating the DM asymmetry.

%%%%%%%%%%%%%%%%%%%%%%%%%%%%%%%%%%%%%%%
\subsection{Boltzmann equations and unitarity sum rules}
\label{sec:sec3}
We shall compute, in Sec.~\ref{sec:sec4}, the reaction rates for the  processes described in the previous section, and the possible CP-violation in them in terms of the coupling and mass parameters appearing in Eqs.~\ref{eq:lag0} and \ref{eq:lag2}. However, to understand the evolution dynamics  of the number densities of the $\chi$ and $\chi^\dagger$ particles in the thermal bath, it is useful to first study the kinetic equations with the thermally averaged reaction rates as inputs. Through this process, we shall arrive at an understanding of the relationships between these reaction rates and their consequences --- in the DM density, the CP-violation in the above reactions, and the resulting possible particle-antiparticle asymmetry in the DM sector. We take up this study in this section.

The Boltzmann kinetic equations~\cite{Kolb:1990vq} governing the number densities of DM particle and antiparticle are given as the following, appropriately accounting for distinct processes along with their conjugate ones.

\begin{align}
\frac{dn_\chi}{dt}+3Hn_\chi = -\int \prod^{4}_{i=1} \frac{d^3 p_i}{(2\pi)^3 2 E_{p_i}}g^2_{\chi}   (2 \pi)^4 \delta^{(4)}(p_1+p_2-p_3-p_4)\bigg[2\bigg(|M|^2_{\chi\chi \rightarrow \chi^{\dagger}\chi^{\dagger}}f_\chi (p_1)f_\chi(p_2)\nonumber\\
-|M|^2_{\chi^{\dagger}\chi^{\dagger} \rightarrow \chi\chi}f_{\chi^{\dagger}} (p_3)f_{\chi^{\dagger}}(p_4)\bigg)
+\bigg(|M|^2_{\chi\chi \rightarrow \chi^{\dagger}\chi}f_\chi (p_1)f_\chi(p_2)-|M|^2_{\chi^{\dagger}\chi \rightarrow \chi\chi}f_{\chi^{\dagger}} (p_3)f_\chi(p_4)\bigg)\nonumber\\
-\bigg(|M|^2_{\chi^{\dagger}\chi^{\dagger} \rightarrow \chi^{\dagger}\chi}f_{\chi^{\dagger}} (p_1)f_{\chi^{\dagger}}(p_2)-|M|^2_{\chi^{\dagger}\chi \rightarrow \chi^{\dagger}\chi^{\dagger}}f_{\chi^{\dagger}} (p_3)f_\chi(p_4)\bigg)\nonumber\\
+2\bigg(|M|^2_{\chi\chi \rightarrow \phi\phi}f_\chi (p_1)f_\chi(p_2)-|M|^2_{\phi\phi \rightarrow \chi\chi}f_\phi (p_3)f_\phi(p_4)\bigg)\nonumber\\
+\bigg(|M|^2_{\chi^{\dagger}\chi \rightarrow \phi\phi}f_{\chi^{\dagger}} (p_1)f_\chi(p_2)-|M|^2_{\phi\phi \rightarrow \chi^{\dagger}\chi}f_{\phi} (p_3)f_\phi(p_4)\bigg)\bigg].
\label{eq:boltz}
\end{align} 
Here, $H$ denotes the Hubble expansion parameter, $f_i(p_i)$ are the distribution functions of the particle type $i$, and $g_\chi$ is the number of internal degrees of freedom of the $\chi$ field in general, which is equal to $1$ in the present case. The corresponding equation for $n_{\chi^\dagger}$ can be obtained by replacing $\chi$ by $\chi^\dagger$ everywhere in the above equation, and vice versa. We have not explicitly shown the appropriate symmetry factors for identical particles in the initial and final states, which have, however, been incorporated in our computations in the following. 

In terms of the standard dimensionless variables $Y_\chi=n_\chi/s$ and $x=m_\chi/T$, where $s$ and $T$ are the entropy density and temperature of the radiation bath, respectively, the Boltzmann equations can be re-written as
\begin{align}
\frac{dY_\chi}{dx} &= -\frac{s}{2 H x}\bigg[\left(\expval{\sigma v}_1+\frac{\expval{\sigma v}_2}{2}+\expval{\sigma v}_3\right)\left(Y^2_\chi-Y^2_{\chi^{\dagger}}\right)+\expval{\sigma v}_3\left(Y^2_{\chi^{\dagger}}-Y^2_0\right)\nonumber\\
&+ \expval{\epsilon\sigma v}_1\left(Y^2_{\chi^{\dagger}}-Y^2_0\right)+\expval{\epsilon\sigma v}_2\left(Y_{\chi^{\dagger}}Y_{\chi}+\frac{Y^2_{\chi^{\dagger}}}{2}-\frac{Y^2_\chi}{2}-Y^2_0\right)+2\expval{\sigma v}_A \left(Y_{\chi^{\dagger}}Y_{\chi}-Y^2_0\right)\bigg]\nonumber\\
\frac{dY_{\chi^{\dagger}}}{dx} &= -\frac{s}{2 H x}\bigg[\left(\expval{\sigma v}_1+\frac{\expval{\sigma v}_2}{2}+\expval{\sigma v}_3\right)\left(Y^2_{\chi^{\dagger}}-Y^2_{\chi}\right)+\expval{\sigma v}_3\left(Y^2_{\chi}-Y^2_0\right)\nonumber\\
&- \expval{\epsilon\sigma v}_1\left(Y^2_{\chi}-Y^2_0\right)-\expval{\epsilon\sigma v}_2\left(Y_{\chi^{\dagger}}Y_{\chi}+\frac{Y^2_\chi}{2}-\frac{Y^2_{\chi^{\dagger}}}{2}-Y^2_0\right)+2\expval{\sigma v}_A \left(Y_{\chi^{\dagger}}Y_{\chi}-Y^2_0\right)\bigg].\nonumber\\
\label{eq:boltz_p}
\end{align}
Here,  $Y_0 =  \frac{1}{s} \int  \frac{d^3 p} {(2\pi)^3 } g_\chi f_0(p)$, with $f_0(p)=e^{-\frac{E(p)}{T}}$ being the equilibrium distribution function with zero chemical potential. We have also introduced the thermally averaged symmetric and asymmetric reaction rates $\expval{\sigma v}_f$ and $\expval{\epsilon \sigma v}_f$ for a particular final state $f$, respectively, where the latter is defined as follows:
\begin{equation}
\expval{\epsilon \sigma v}_{f} = \dfrac{\int \prod^{4}_{i=1} \frac{d^3 p_i}{(2\pi)^3 2 E_{p_i}}  (2 \pi)^4 \delta^{(4)}(p_1+p_2-p_3-p_4) \,\epsilon_f(p_i) |M_0|^2_f f_0(p_1)f_0(p_2)}{\int \dfrac{d^3 p_1}{(2\pi)^3} \dfrac{d^3 p_2}{(2\pi)^3} f_0(p_1)f_0(p_2)} \hspace{0.5cm},
\label{eq:cross}
\end{equation}
with   $|M_0|^2_f = |M|^2_{\chi\chi\rightarrow f}+|M|^2_{\chi^{\dagger}\chi^{\dagger}\rightarrow f^{\dagger}} $, and $\epsilon_f(p_i)$ is given by
\begin{align}
\epsilon_f (p_i) =\frac{ |M|^2_{\chi\chi\rightarrow f}-|M|^2_{\chi^{\dagger}\chi^{\dagger}\rightarrow f^{\dagger}}}{|M|^2_{\chi\chi\rightarrow f}+|M|^2_{\chi^{\dagger}\chi^{\dagger}\rightarrow f^{\dagger}}}.
\label{eq:def_eps}
\end{align} 
The corresponding thermally averaged symmetric reaction rates $\expval{\sigma v}_f$ can be computed using Eq.~\ref{eq:cross} by removing the $\epsilon_f(p_i)$ factors. Finally, we denote the reaction rates for $\chi+\chi \rightarrow \chi^{\dagger} + \chi^{\dagger}$ by $\expval{\sigma v}_1$ and  $\expval{\epsilon \sigma v}_1$, for $\chi+\chi \rightarrow \chi + \chi^{\dagger}$ by $\expval{\sigma v}_2$ and $\expval{\epsilon \sigma v}_2$ and for $\chi+\chi \rightarrow \phi+\phi$  by $\expval{\sigma v}_3$ and $\expval{\epsilon \sigma v}_3$. For the CP-conserving reaction $\chi+\chi^\dagger \rightarrow  \phi+\phi$ we denote the average reaction rate by $\expval{\sigma v}_A$.

We have parametrized the CP-violation resulting from the first three reactions discussed in Sec.~\ref{sec:sec2} in terms of $\expval{\epsilon \sigma v}_1$ and $\expval{\epsilon \sigma v}_2$ in Eqs.~\ref{eq:boltz_p}, and have eliminated $\expval{\epsilon \sigma v}_3$ from the Boltzmann equations using unitarity sum rules. This stems from the fact that the amplitudes for the CP-violating processes are related by CPT and S-matrix unitarity as follows~\cite{Kolb:1979qa, Baldes:2014gca, Baldes:2015lka}:
\begin{align}
\sum_f \int dPS_f |M|^2_{\chi\chi \rightarrow f }=\sum_f \int dPS_f  |M|^2_{f \rightarrow \chi\chi}= \sum_f \int dPS_f  |M|^2_{\chi^{\dagger}\chi^{\dagger}\rightarrow f^{\dagger}}, 
\label{eq:sum}
\end{align}
where we have used the S-matrix unitarity in the first equality, and CPT conservation in the second one. Here, the integral over $dPS_f$ sum over the momenta of the particles in the state $f$, as well as any discrete label that may be carried by $f$. There is a further sum over all possible final states $f$ that may be obtained starting from the initial state of $\chi \chi$, which has been explicitly indicated. This unitarity sum rule in Eq.~\ref{eq:sum}, together with the definition of the CP-violation parameter $\epsilon_f$ in Eq.~\ref{eq:def_eps}, imply the following sum rule relating the CP-violation in all the channels

\begin{align}
\sum_f \int dPS_f\,\, \epsilon_f\, |M_0|^2_f = 0, 
\label{eq:eps_unitarity}
\end{align}
where, the quantity $|M_0|^2_f $ has been defined above. Using this sum rule in Eq.~\ref{eq:eps_unitarity} for the three relevant CP-violating reactions with the initial state of $\chi \chi$ and final states in the set $f=\{\chi^{\dagger} + \chi^{\dagger}, \chi + \chi^{\dagger}, \phi+\phi\}$, we can express one of the asymmetric rates $\expval{\epsilon \sigma v}_f$ in terms of the other two, and therefore, can write the Boltzmann equations in terms of only two asymmetric reaction rates as in Eq.~\ref{eq:boltz_p}.

%%%%%%%%%%%%%%%%%%%%%%%%%%%%%%%%%%%%%%%
\subsubsection{Analytic solutions}
\label{sec:sec31}
We now discuss approximate analytic solutions to the Boltzmann equations~\ref{eq:boltz_p}. For this, it is convenient to define the symmetric and asymmetric yields, $Y_S = Y_\chi+Y_{\chi^\dagger}$ and $Y_{\Delta \chi}=Y_\chi-Y_{\chi^\dagger}$, respectively, and rewrite the Boltzmann equations in terms of these variables as
\begin{align}
\dfrac{dY_S}{dx}&=-\dfrac{s}{2Hx}\left[\expval{\sigma v}_A\left(Y^2_S-Y^2_{\Delta \chi}-4Y^2_0\right)+\expval{\sigma v}_3\bigg(\dfrac{Y^2_S+Y^2_{\Delta \chi}-4Y^2_0}{2}\bigg)-\expval{\epsilon\sigma v}_S Y_S Y_{\Delta \chi}\right]\nonumber\\
\dfrac{dY_{\Delta \chi}}{dx}&=-\dfrac{s}{2Hx}\left[\expval{\epsilon\sigma v}_S\left(\dfrac{Y^2_S-4Y^2_0}{2}\right)+\expval{\epsilon\sigma v}_D \dfrac{Y^2_{\Delta \chi}}{2}+\expval{\sigma v}_{all}\,Y_S Y_{\Delta \chi}\right].
\label{eq:asym}
\end{align} 
In order to write these equations in a compact form, we have further defined the following quantities: $\expval{\epsilon\sigma v}_S =\expval{\epsilon \sigma v}_1+\expval{\epsilon \sigma v}_2 $, $\expval{\epsilon\sigma v}_D=\expval{\epsilon \sigma v}_1-\expval{\epsilon \sigma v}_2$ and $\expval{\sigma v}_{all}=2\expval{\sigma v}_1+\expval{\sigma v}_2+\expval{\sigma v}_3$. Let us discuss the analytic solutions in two different regions of the scaled temperature variable $x$. In the first region, we consider values of $x$ upto its freeze-out value of $x_F=m_\chi/T_F$, i.e., $1\leq x \leq x_F$. Here, at the temperature $T_F$, all the relevant chemical reactions involving the DM particle freeze-out.  In the second region, we consider values $x>x_F$. 

For $1\leq x \leq x_F$, we can parametrize the number densities of the DM and anti-DM particles as 
\begin{equation}
Y_{\chi(\chi^\dagger)} =Y_0(1+\delta_{1(2)}),
\end{equation}
where, $\delta_{1}$ and $\delta_2$ parametrize the small deviations away from their common equilibrium value with zero chemical potential, $Y_0$. We expect that in this region, the deviations will satisfy the condition $|\delta_{1,2}|<< 1$. In the presence of CP-violating scatterings, in general, $|\delta_1| \neq |\delta_2|$. In terms of this parametrization, we have $Y_S = 2Y_0 (1+\delta)$ and 
$Y_{\Delta\chi}=Y_0 \bar{\delta}$, where $\delta=(\delta_1+\delta_2)/2$ and $\bar{\delta}=\delta_1-\delta_2$. 

Dropping terms quadratic in the small deviations, i.e., terms proportional to  $\delta^2$, $\bar{\delta}^2$ and $\delta \bar{\delta}$, we can solve Eqs.~\ref{eq:asym} algebraically. The total DM and anti-DM yield is obtained to be as follows:
\begin{equation}
Y_S(x) = 2Y_0\left[1+\dfrac{Hx}{ s Y_0}\dfrac{\expval{\sigma v}_{all}}{2\expval{\sigma v}_{ann}\expval{\sigma v}_{all}+\expval{\epsilon \sigma v}^2_S}\right]~~~~{\rm (for~} 1\leq x \leq x_F) ,
\label{eq:analytic1}
\end{equation}
whereas, the yield for the DM-anti-DM asymmetry is given by
\begin{equation}
|Y_{\Delta\chi}(x)|= \dfrac{2 H x}{s}\,\dfrac{\expval{\epsilon\sigma v}_S}{2\expval{\sigma v}_{ann}\expval{\sigma v}_{all}+\expval{\epsilon \sigma v}^2_S}~~~~{\rm (for~} 1\leq x \leq x_F),
\label{eq:analytic2}
\end{equation}
where, we have defined $\expval{\sigma v}_{ann} = \expval{\sigma v}_A + \expval{\sigma v}_3/2 $. In deriving these expressions, we have taken the non-relativistic form of $Y_0$, which is applicable near the freeze-out temperature. We can draw several important conclusions from these solutions:
\begin{enumerate}
\item The CP-conserving DM and anti-DM annihilation process, $\chi+\chi^\dagger \rightarrow  \phi+\phi$, can indirectly control the asymmetric yield. This can be seen in Eq.~\ref{eq:analytic2}, where we find that $|Y_{\Delta\chi}(x)|$ is inversely proportional to $\expval{\sigma v}_A $. This is because the out-of-equilibrium number density of the relevant species can be controlled by this CP-conserving reaction, which, in turn, controls the asymmetric yield. We have discussed the role of CP-conserving processes in producing cosmological particle-antiparticle asymmetries in detail in Ref.~\cite{Ghosh:2021ivn}.

\item In our parametrization of CP-violation in the Boltzmann equations~\ref{eq:boltz_p}, $\expval{\epsilon\sigma v}_S =\expval{\epsilon \sigma v}_1+\expval{\epsilon \sigma v}_2$ acts as a source term for the asymmetry, as seen in Eq.~\ref{eq:analytic2}. This is because, as explained earlier,  $\expval{\epsilon \sigma v}_3$ cannot be independently varied, and is determined in terms of $\expval{\epsilon \sigma v}_1$ and $\expval{\epsilon \sigma v}_2$ through the unitarity sum rules in Eq.~\ref{eq:eps_unitarity}. 

\item We see in Eq.~\ref{eq:analytic1}  that the the total DM yield is significantly affected by the pair annihilation process $\chi+\chi^\dagger \rightarrow  \phi+\phi$. This is because, we expect $\expval{\epsilon \sigma v}^2_S << \expval{\sigma v}_{ann}\expval{\sigma v}_{all}$, as the asymmetric scattering rates stem from the interference of tree and loop level amplitudes, and are therefore suppressed compared to the symmetric scattering rates. With this approximation, we have
\begin{equation}
Y_S \sim 2Y_0 \left(1+\dfrac{Hx}{sY_0 (2\expval{\sigma v}_A+ \expval{\sigma v}_3)} \right)
\label{eq:YS_approx_1}
\end{equation}
\end{enumerate}
In the second region with $x>x_F$, we can no longer ignore the terms quadratic in the deviations $\delta$ and $\bar{\delta}$, and the analytic solutions in closed form become cumbersome. However, the results are tractable under certain simplifying assumptions. We note that for $x > x_F$, the self-scattering reactions that can generate the asymmetry are mostly decoupled since the annihilation rate is proportional to the DM velocity, as detailed in the next section. Thus we see that in the Boltzmann equation for $Y_{\Delta \chi}$ in Eq.~\ref{eq:asym} the terms in the right hand side (RHS) proportional to $\expval{\epsilon\sigma v}_S$ and $\expval{\epsilon\sigma v}_D$ will be very small as well. Furthermore, since $Y_S \propto 1/\expval{\sigma v}_A$, the last term in the RHS will be proportional to the ratio $\expval{\sigma v}_{all}/\expval{\sigma v}_A$. Therefore, in the regime $\expval{\sigma v}_{all} << \expval{\sigma v}_A$, i.e., when the dominant scattering process is the CP-even annihilation, $Y_{\Delta \chi}$ essentially remains a constant, being frozen at its value around $x=x_F$. With this as input, we can solve the equation for $Y_S$ in Eq.~\ref{eq:asym}, and for dominantly s-wave contribution to $\expval{\sigma v}_A$, the solution is given by
\begin{align}
Y_S(x>x_F) &= |Y^F_{\Delta \chi}|\,\,\dfrac{1+r_F \exp[\lambda\, |Y^F_{\Delta \chi}|\,(x^{-1}-x^{-1}_F)]}{1-r_F \exp[\lambda\, |Y^F_{\Delta \chi}|\,(x^{-1}-x^{-1}_F)]},
\label{eq:YS_approx_2}
\end{align}
where,
\begin{equation}
r_F =\dfrac{Y^F_S-|Y^F_{\Delta \chi}|}{Y^F_S+|Y^F_{\Delta \chi}|}.
\end{equation}
Here, $\lambda = 1.32\, m_\chi M_{Pl}\,g^{1/2}_{*}\expval{\sigma v}_A$, and $Y^F_S$ and $Y^F_{\Delta \chi}$ represent the symmetric and asymmetric yields of DM obtained at $x=x_F$ using Eqs.~\ref{eq:analytic1} and~\ref{eq:analytic2}. 
We shall compare these approximate analytical results with the numerical solutions in the next sub-section.

%%%%%%%%%%%%%%%%%%%%%%%%%%%%%%%%%%%%%%%
\subsubsection{Numerical results}
\label{sec:sec32}
\begin{figure}[htb!]
\centering
\includegraphics[scale=0.5]{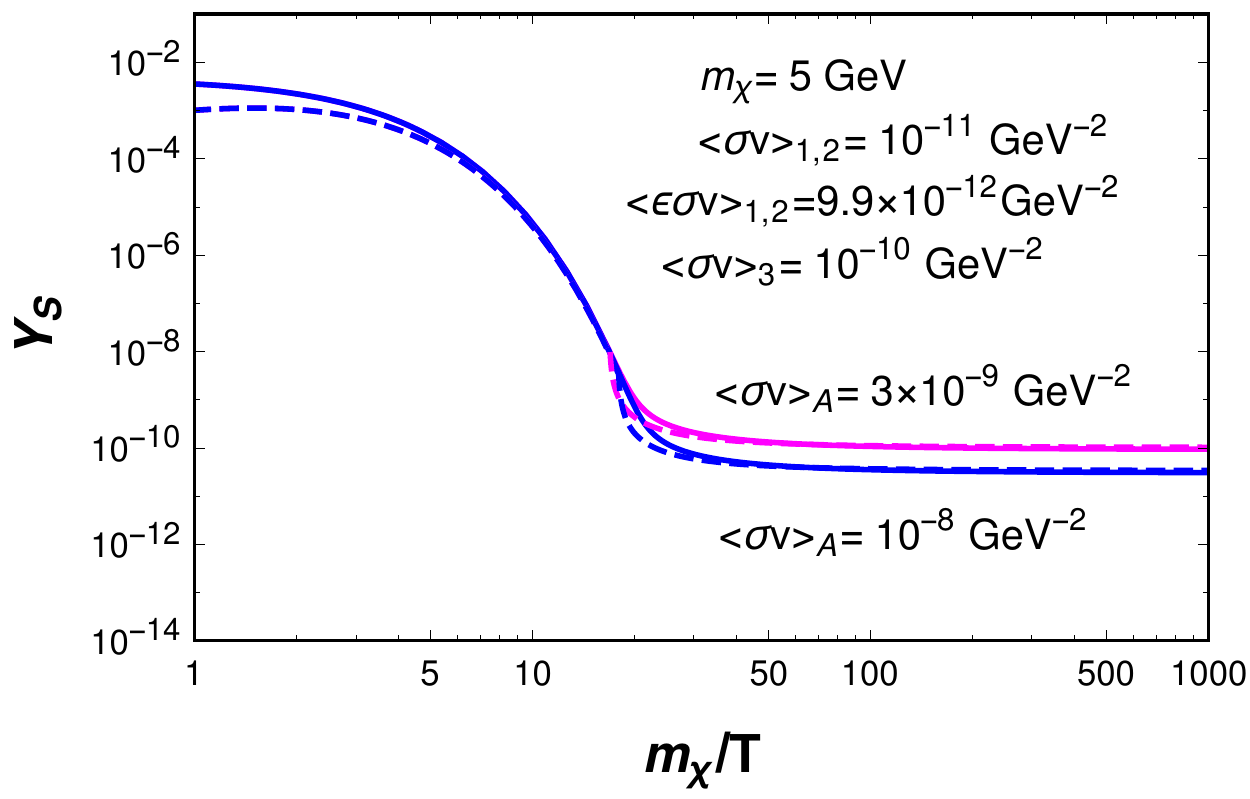}
\includegraphics[scale=0.5]{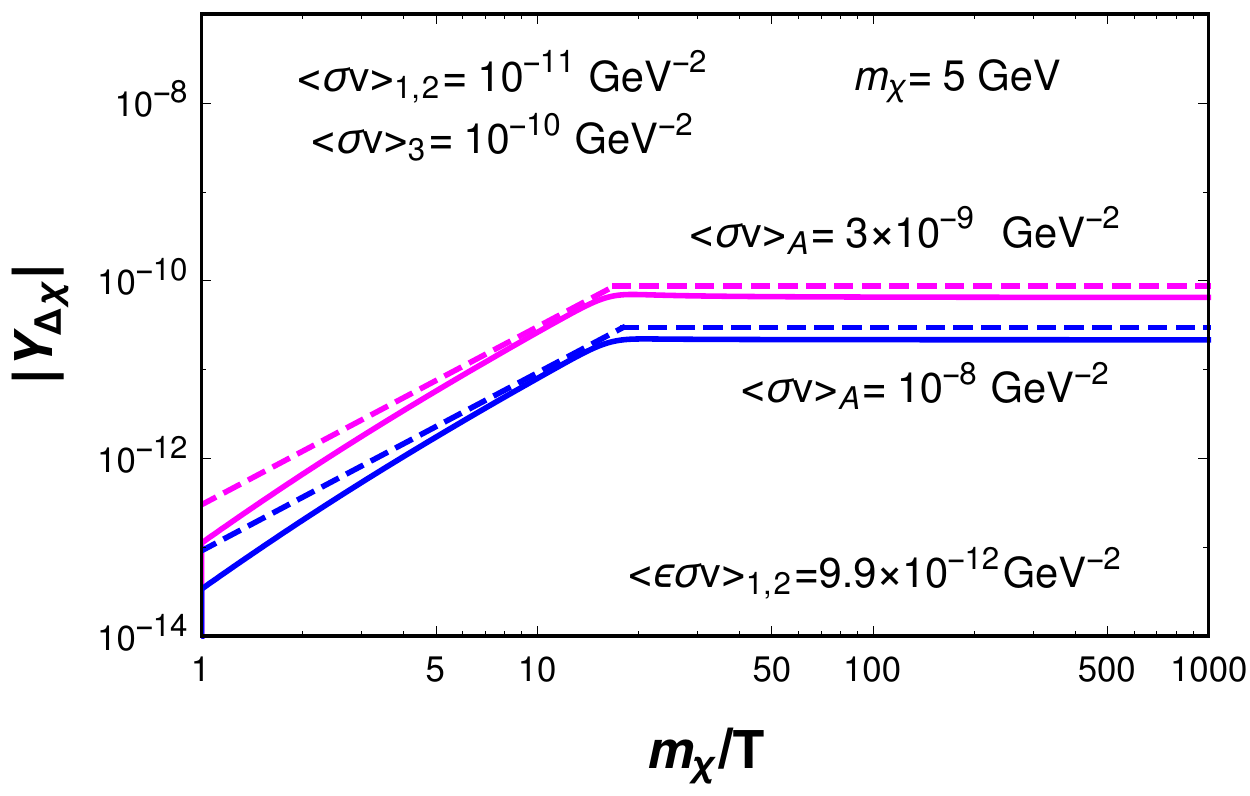}
\caption{\small{\em Evolution of the symmetric (left panel, with $Y_S=Y_\chi+Y_{\chi^\dagger}$) and the asymmetric yields (right panel, with $|Y_{\Delta \chi}|=|Y_\chi-Y_{\chi^\dagger}|$), as a function of the scaled inverse temperature variable $m_\chi/T$. The solid lines in both the figures represent the exact numerical solution of Eqs.~\ref{eq:boltz_p}. The dashed lines show the predictions based on the approximate analytic solutions, namely, Eqs.~\ref{eq:analytic1} and \ref{eq:analytic2}. The dark matter mass and CP-violating symmetric and asymmetric reaction rates have been kept fixed. We have shown the results for two different choices of the CP-conserving pair annihilation rate $\expval{\sigma v}_A $ for the process $\chi+\chi^{\dagger} \rightarrow \phi + \phi$ (by the blue and pink lines) to demonstrate its impact not only on the total yield, but also in indirectly controlling the particle anti-particle asymmetry as well.
}}
\label{fig:evolve}
\end{figure}

Having obtained the approximate analytic solutions to the sum and difference of DM and anti-DM yields for different ranges of the temperature in the thermal bath, we now go on to solve Eqs~\ref{eq:boltz_p} numerically to understand the exact relationship between the DM properties at freeze-out and the microscopic parameters. In Fig.~\ref{fig:evolve} we have shown the evolution of the symmetric (left panel, with $Y_S=Y_\chi+Y_{\chi^\dagger}$) and the asymmetric yields (right panel, with $|Y_{\Delta \chi}|=|Y_\chi-Y_{\chi^\dagger}|$), as a function of the scaled inverse temperature variable $m_\chi/T$. The solid lines in both the figures represent the exact numerical solution of Eqs.~\ref{eq:boltz_p}. The dashed lines show the predictions based on the approximate analytic solutions, namely, Eqs.~\ref{eq:analytic1} and \ref{eq:analytic2}. The dark matter mass and CP-violating symmetric and asymmetric reaction rates have been kept fixed. We have shown the results for two different choices of the CP-conserving pair annihilation rate $\expval{\sigma v}_A $ for the process $\chi+\chi^{\dagger} \rightarrow \phi + \phi$, by the blue and pink lines, to demonstrate its impact not only on the total yield, but also in the particle anti-particle asymmetry as well. As we can see, by increasing $\expval{\sigma v}_A $, one obtains a lower value of $|Y_{\Delta \chi}|$ throughout the evolution. We also see from Fig.~\ref{fig:evolve} (right panel) that a small out-of-equilibrium asymmetry $|Y_{\Delta \chi}|$ is generated from around $x \sim 1$, which eventually grows and freezes at around $x \sim x_F$. 

The analytic approximation is found to be in good agreement with the exact numerical solutions, the maximum difference between the two being of the order of $10\%$. The larger differences are found for smaller values of $x$. This is because in order to obtain an approximate closed form analytic solution to Eqs.~\ref{eq:boltz_p}, we have taken $dY_0/dx \sim -Y_0$, and have dropped an additional term proportional to $1/x$. This approximation is justified for $x > 1$, which is the main region of interest for determining the relic abundance. However, it is no longer a valid approximation for $x \sim 1$, where the additional $x-$dependent term also contributes, and an accurate (better than the $10\%$ accuracy obtained here) closed form solution cannot be obtained for $x \sim 1$ while retaining such a term. 

We note that the patching of the analytic solutions in the regions $1 \leq x \leq x_F$ and $x>x_F$ requires an input of the value of $x_F$ itself, and is therefore sensitive to how precisely we can estimate $x_F$. However, the asymptotic values of the yields $Y_S$ and $|Y_{\Delta \chi}|$ are found to be well approximated by the analytic solutions. Here, we have estimated $x_F$ as follows. Since we have $Y_S = 2Y_0 (1+\delta)$, and in the first patch the net deviation from equilibrium $\delta < 1$, we can use this condition in the expression for $Y_S$ in Eq.~\ref{eq:YS_approx_1}, and obtain a relation which determines the boundary of the two regions $x_F$ as
\begin{equation}
\frac{H(x_F) x_F}{s(x_F)Y_0(x_F) (2\expval{\sigma v}_A+\expval{\sigma v}_3)} \sim 1.
\end{equation} 
An iterative solution of this equation gives a good estimate of the value of the scaled freeze-out temperature $x_F$.

%%%%%%%%%%%%%%%%%%%%%%%%%%%%%%%%%%%%%%%
\subsection{CP-violation and scattering rates in the complex singlet model}
\label{sec:sec4}
We now compute the relevant CP-violation parameters and the thermally averaged symmetric and asymmetric scattering rates in the complex scalar singlet model, as a function of the model parameters. We then go on to study the dependence of the DM relic abundance and the possible particle-antiparticle asymmetry in the DM sector on such parameters, with a particular emphasis on the interplay between the DM self-scattering and annihilation in controlling the DM density and composition. 

We first  take up the three CP-violating processes one by one, and discuss the corresponding results at next-to-leading order in perturbation theory, following which we show the leading order results for the CP-conserving process.
\subsubsection{Process 1: Self-scattering $\chi+\chi \rightarrow \chi^{\dagger} + \chi^{\dagger}$}

\begin{figure}[htb!]
\centering
\includegraphics[scale=0.39]{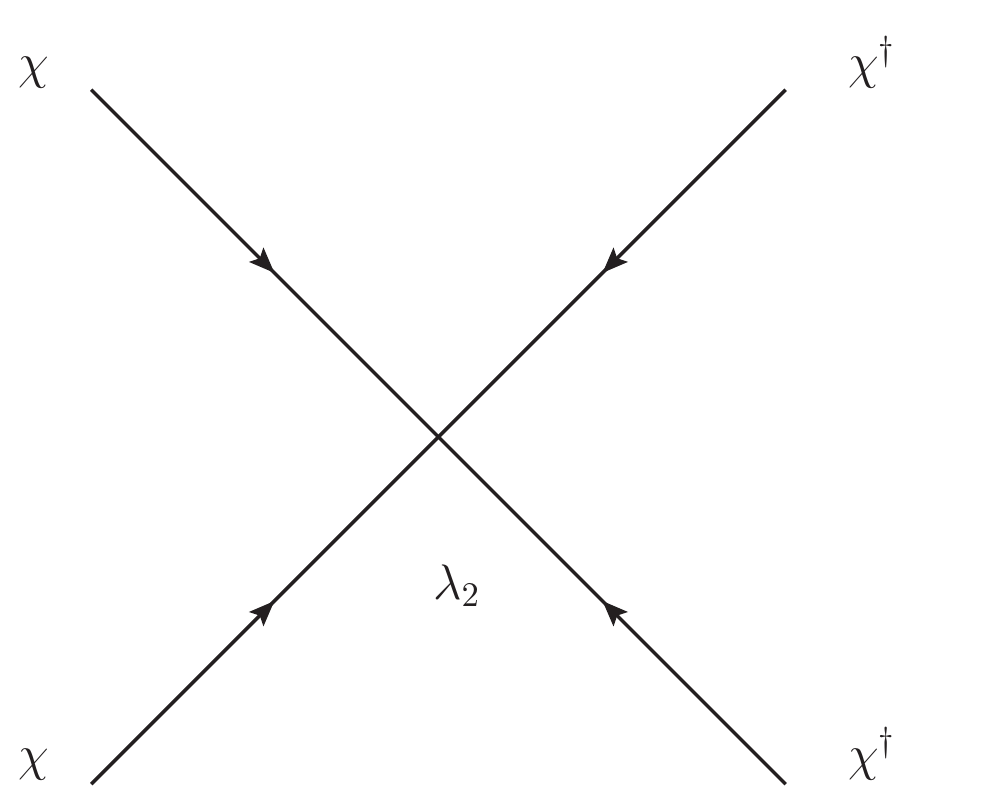}
\includegraphics[scale=0.44]{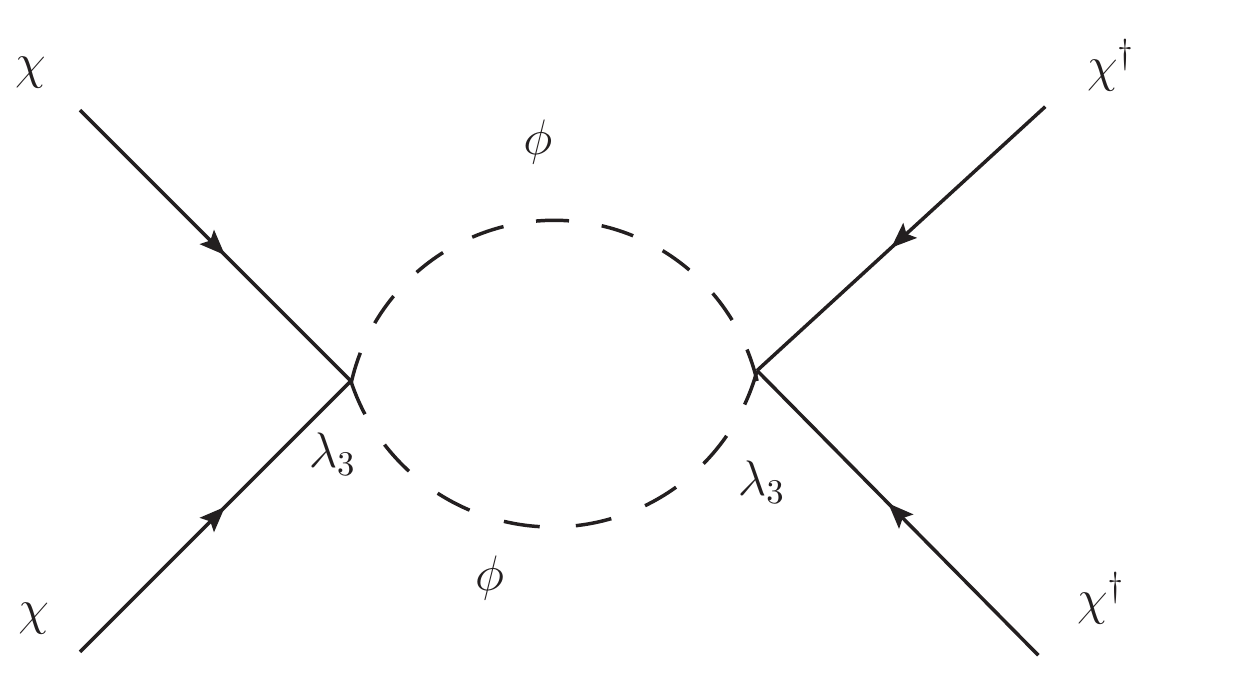}
\includegraphics[scale=0.44]{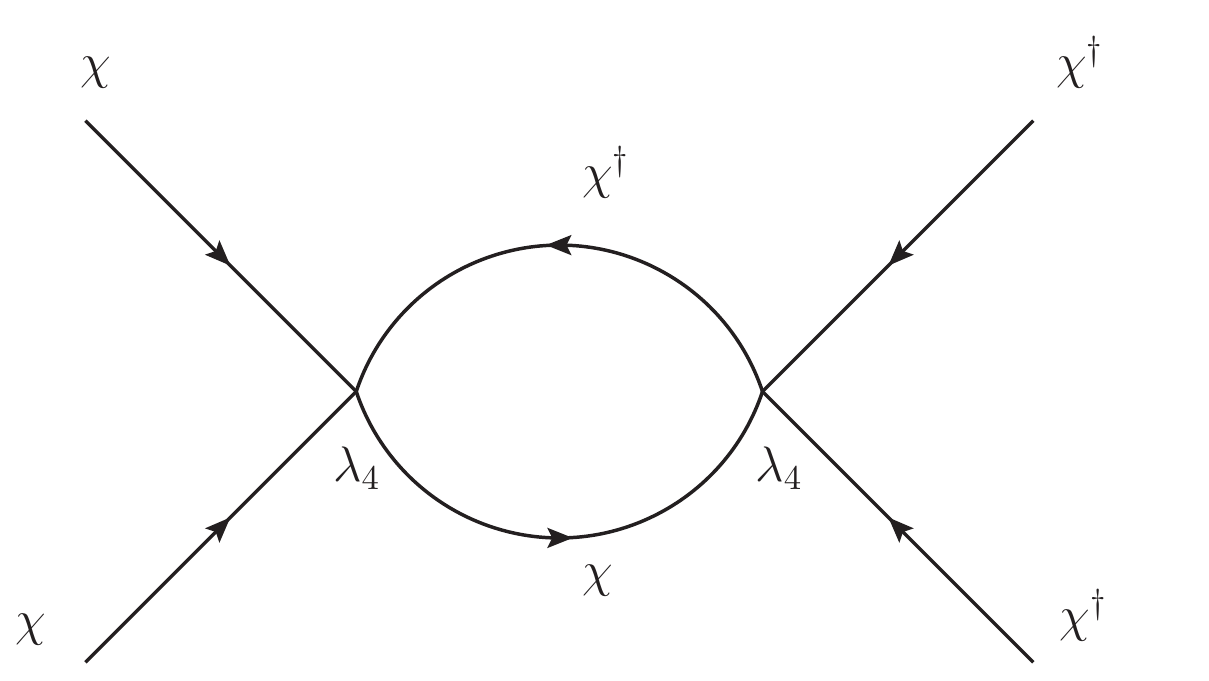}
\caption{\small{\em Relevant tree and one-loop level Feynman diagrams for the CP-violating self-scattering process $\chi+\chi \rightarrow \chi^{\dagger} + \chi^{\dagger}$.}}
\label{fig:diag1}
\end{figure}

The relevant tree and one-loop level Feynman diagrams for the self-scattering process $\chi+\chi \rightarrow \chi^{\dagger} + \chi^{\dagger}$ are shown in Fig.~\ref{fig:diag1}. 
The CP-violation in this process arises from the interference of the tree level and one-loop graphs shown in Fig.~\ref{fig:diag1}. 
There are additional diagrams for this process at the one-loop level at the same order in the couplings. However, those diagrams do not give rise to any CP-violating imaginary parts. Therefore, we have restricted ourselves to the diagrams shown in Fig.~\ref{fig:diag1}. The resulting difference in the matrix elements squared of the two CP-conjugate processes is given by
\begin{align}
|M|^2_{\chi\chi\rightarrow \chi^{\dagger}\chi^{\dagger}}-|M|^2_{\chi^{\dagger}\chi^{\dagger}\rightarrow \chi\chi}=4\,\left[\rm Im(\lambda^*_2\lambda^2_3)\,Im\, I_{\phi\phi}+\,\rm Im(\lambda^*_2\lambda^2_4)\,Im\, I_{\chi^{\dagger}\chi}\right].
\label{eq:ME1}
\end{align}
The imaginary part of the loop amplitudes are non-zero when internal lines in the loop go onshell, and using the Cutkosky rules~\cite{Peskin:1995ev} we obtain them to be
\begin{align}
\rm Im\, I_{\phi\phi}=\dfrac{\beta_\phi}{32 \pi},\hspace{0.5cm} Im\, I_{\chi^{\dagger}\chi}=\dfrac{\beta_\chi}{16 \pi}.
\label{eq:imaginary}
\end{align}
Here, the factor of $\beta_f = (1-4m^2_f/s)^{1/2}$ comes from the two-body phase-space integral, with the usual Mandelstam variable $s=(p_1+p_2)^2$, where $p_1$ and $p_2$ are the four-momenta of the two initial state particles. Since we take $m_\phi << m_\chi$, we can approximate $\beta_\phi \sim 1$. 

With the amplitudes in hand, we can now compute the thermally averaged symmetric rate of this self-scattering process as follows:
\begin{align}
\expval{\sigma v}_1 = \dfrac{|\lambda_2|^2}{16\pi m^2_\chi}\dfrac{1}{\sqrt{\pi x}},
\end{align}
where we have used the partial wave expansion of the annihilation rate, keeping only the leading term in the non-relativistic limit.
As mentioned in the previous sections, and as we can see from this expression, the leading term in the self-scattering rate is proportional to the DM velocity in the thermal bath. Therefore, in the non-relativistic regime, this rate may be suppressed. It is sufficient to consider only the tree-level amplitude for computing the leading contribution to the symmetric annihilation rate. In obtaining the thermal average, we have included the final state symmetry factors here, while the initial state symmetry factors have already been taken into account in the Boltzmann equations.

We can, similarly, obtain the thermal average of the asymmetric reaction rate, which stems from the interference of the tree level and one-loop level amplitudes. The resulting expression is given as
\begin{align}
\expval{\epsilon \sigma v}_1 &= \dfrac{1}{(16\pi m_\chi)^2}\left[\dfrac{1}{\sqrt{\pi x}} \text{Im}(\lambda^*_2\lambda^2_3)+\dfrac{3}{2 x} \text{Im}(\lambda^*_2\lambda^2_4)\right].
\label{eq:eps_sigmav_1}
\end{align}
Here again, the leading term is proportional to the DM velocity, whereas the sub-leading term is proportional to the square of the DM velocity. With this, the effective CP-violation parameter can be obtained as follows: 
\begin{equation}
\epsilon^{1}_{eff} = \expval{\epsilon \sigma v}_1/\expval{\sigma v}_1.
\end{equation}
For $\mathcal{O}(1)$ values of the absolute values of $\lambda_i$, and the phase of the complex coupling combination $\lambda_2^* \lambda_3^2$ (denoted by $\theta_{23}$) set to be $\pi/2$, we obtain for $x>>1$:
\begin{align}
\epsilon^{1}_{eff} \simeq  \dfrac{1}{16\pi} , \,\,\,\, \text{for}\,\, x >> 1.
\label{epsestimate}
\end{align}
Thus for $\mathcal{O}(1)$ couplings, $\epsilon^{1}_{eff} $ is expected to be small, around $0.02$, but can of course be made larger with larger couplings.

\subsubsection{Process 2: Self-scattering $\chi+\chi \rightarrow \chi^{\dagger} + \chi$}

\begin{figure}[htb!]
\centering
\includegraphics[scale=0.39]{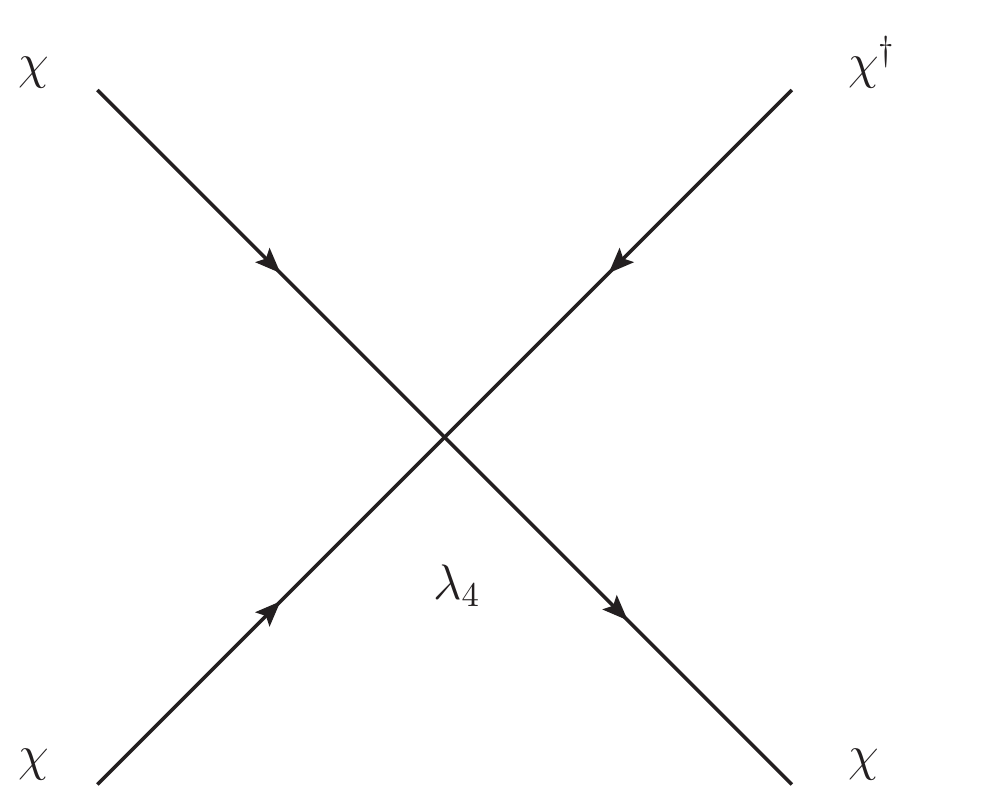}
\includegraphics[scale=0.44]{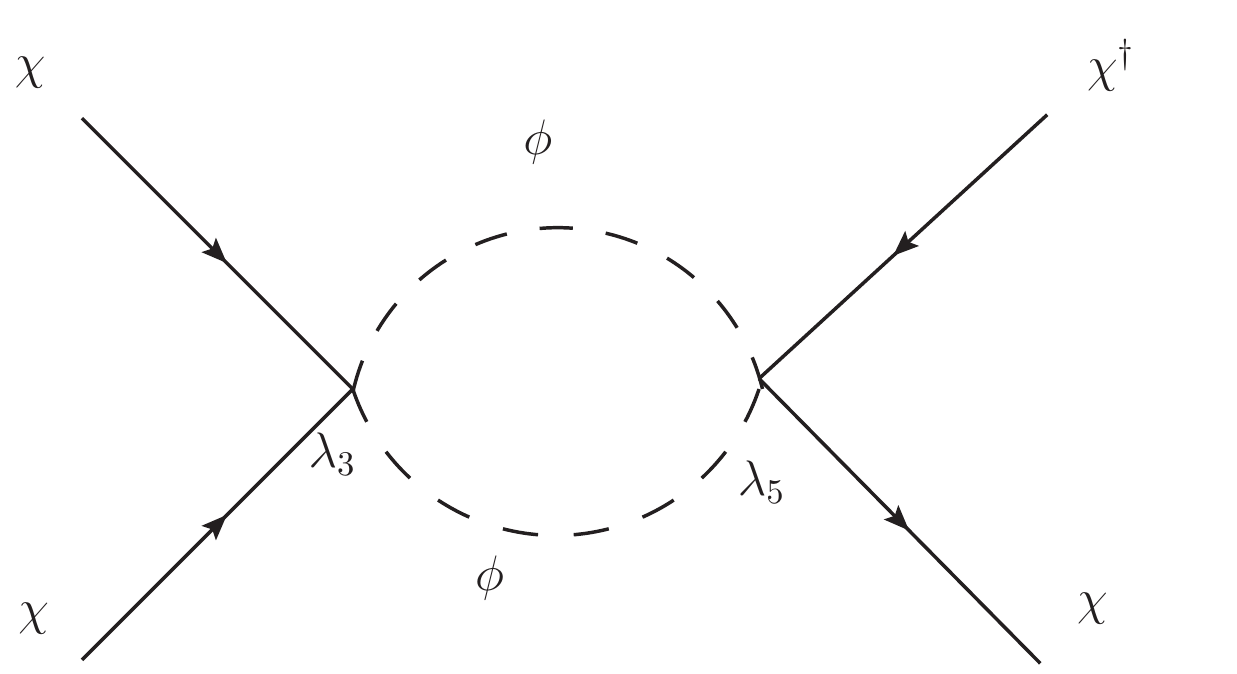}
\includegraphics[scale=0.44]{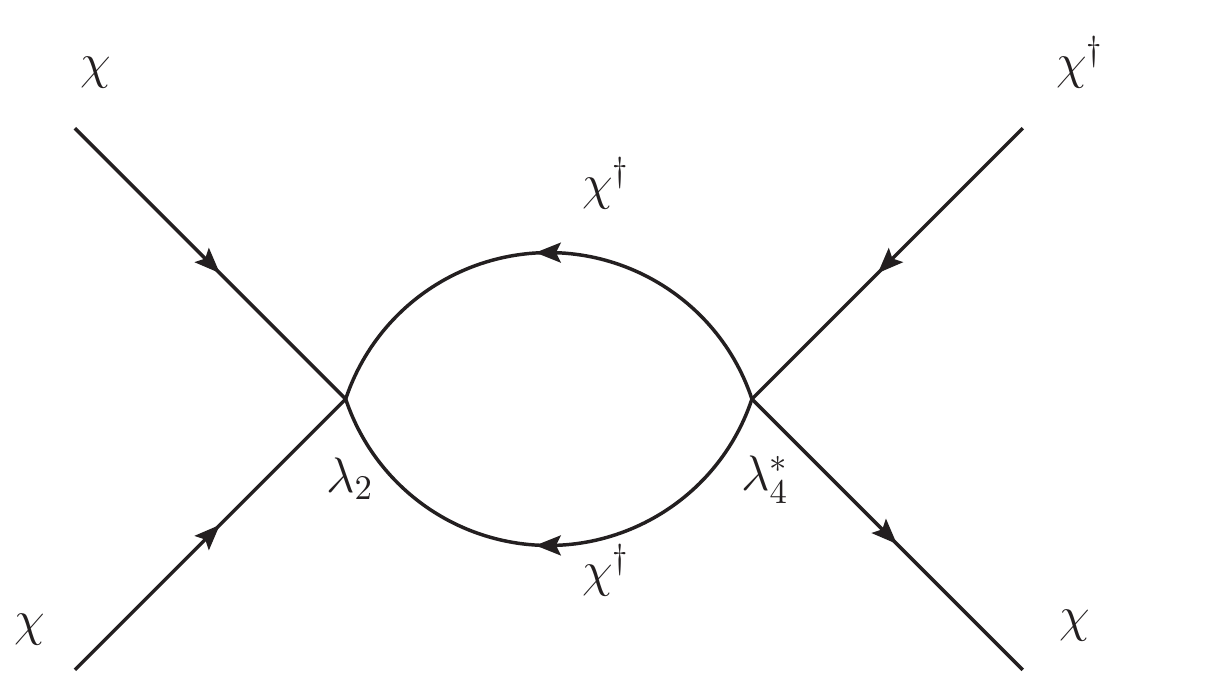}
\caption{\small{\em Relevant tree and one-loop level Feynman diagrams for the CP-violating self-scattering process $\chi+\chi \rightarrow \chi^{\dagger} + \chi$.}}
\label{fig:diag2}
\end{figure}

The second CP-violating self-scattering process, $\chi+\chi \rightarrow \chi^{\dagger} + \chi$ shows very similar behaviour as far as the reaction rates are concerned, except the fact that different coupling combinations appear. In particular, the coupling $\lambda_5$, which controls the rate of CP-conserving $\chi+\chi^{\dagger} \rightarrow \phi + \phi$ annihilation, now appears in the dominant loop contribution to the $\chi+\chi \rightarrow \chi^{\dagger} + \chi$ process, as can be seen from the Feynman diagrams in Fig.~\ref{fig:diag2}. This makes the rate of CP-violation in this process and the rate of CP-conserving annihilations related. 
We can now similarly compute the relevant difference between the squared matrix elements from the interference of the graphs in Fig.~\ref{fig:diag2}
\begin{align}
|M|^2_{\chi\chi\rightarrow \chi^{\dagger}\chi}-|M|^2_{\chi^{\dagger}\chi^{\dagger}\rightarrow\chi^{\dagger}\chi}=4\,\left[\rm Im(\lambda_2\lambda^{*2}_4)\,Im\, I_{\chi^{\dagger}\chi^{\dagger}}+\,\rm Im(\lambda^*_4\lambda_3)\lambda_5\,Im\, I_{\phi\phi}\right],
\label{eq:ME2}
\end{align}
with the following imaginary part of the new loop amplitude, in addition to the ones shown in Eq.~\ref{eq:imaginary}:
\begin{align}
\rm Im\, I_{\chi^{\dagger}\chi^{\dagger}}=\dfrac{\beta_\chi}{32 \pi}.
\end{align}
The thermal averaged symmetric annihilation rate shows the same velocity dependence as before, and is given by
\begin{align}
\expval{\sigma v}_2 = \dfrac{|\lambda_4|^2}{8\pi m^2_\chi}\dfrac{1}{\sqrt{\pi x}}\hspace{1cm}.
\end{align}
Similarly the $x-$dependence of the asymmetric annihilation rate is also similar, but now with the explicit appearance of the $\lambda_5$ coupling
\begin{align}
\expval{\epsilon \sigma v}_2 &= \dfrac{1}{(16\pi m_\chi)^2}\left[\dfrac{3}{2 x} \text{Im}(\lambda_2\lambda^{*2}_4)+\dfrac{2}{\sqrt{\pi x}} \text{Im}(\lambda^*_4\lambda_3)\lambda_5\right].
\label{eq:eps_sigmav_2}
\end{align}
Thus, the CP-violation will proportionally increase if we increase the pair-annihilation rate $\expval{\sigma v}_A$. We can also write the effective CP-violation parameter as 
\begin{align}
\epsilon^{2}_{eff} \simeq \dfrac{\lambda_5}{16\pi} , \,\,\,\, \text{for}\,\, x >> 1.
\end{align}
Here, as before, we have set $|\lambda_i|=\mathcal{O}(1)$, except for $\lambda_5$ which is kept variable, and all the effective phases to be $\pi/2$. We shall see in the following that to obtain a large asymmetry in the DM sector, one generally requires a large rate of $\expval{\sigma v}_A$, and hence a large $\lambda_5$. In such cases, we may have $\epsilon^{2}_{eff} \simeq 0.25$, for $\lambda_5 \sim \mathcal{O}{(4\pi)}$.

\subsubsection{Process 3: CP-violating Annihilation $\chi+\chi \rightarrow \phi + \phi$}
\begin{figure}[htb!]
\centering
\includegraphics[scale=0.39]{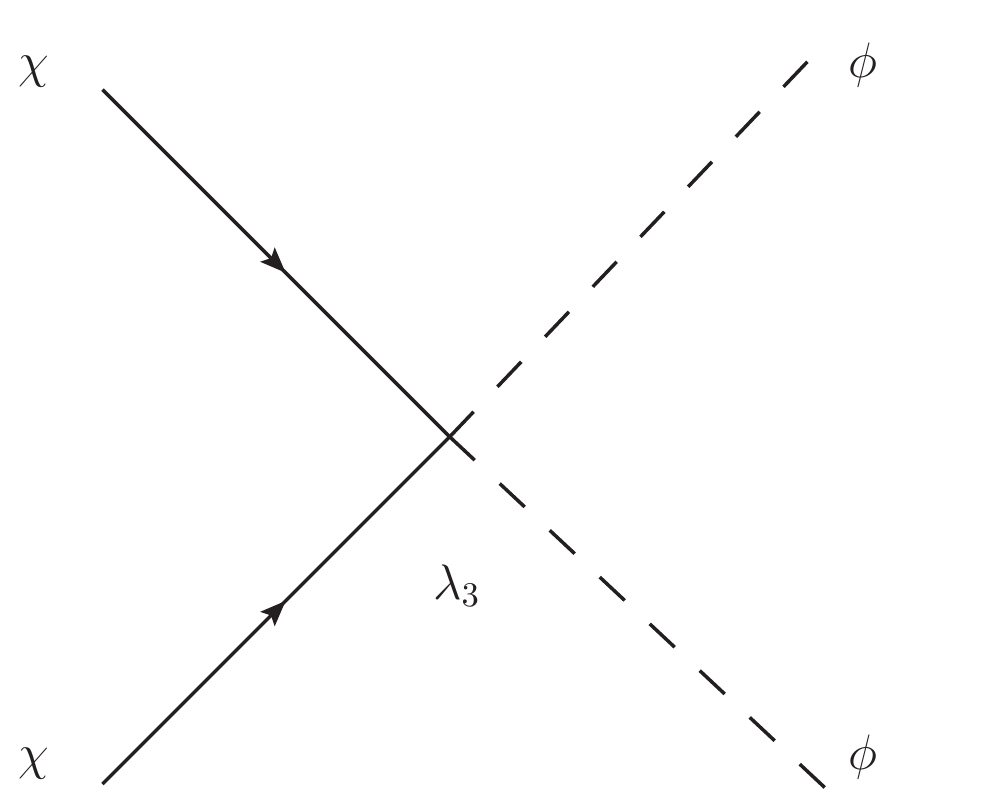}
\includegraphics[scale=0.43]{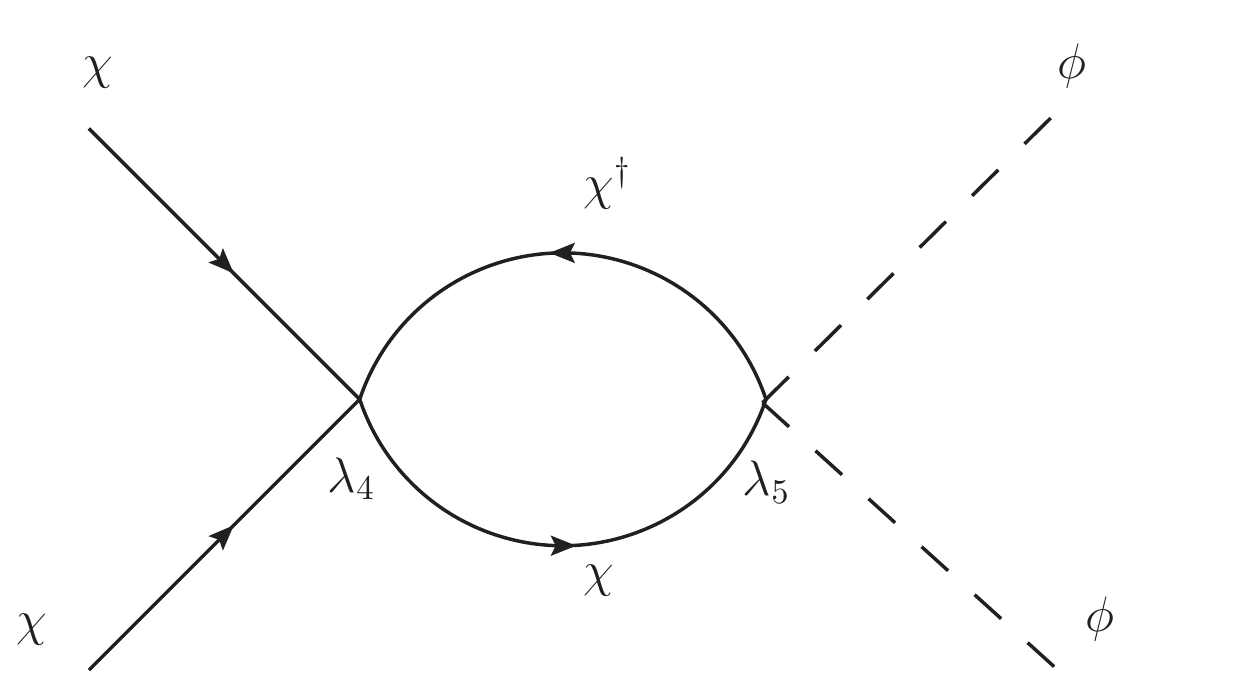}
\includegraphics[scale=0.43]{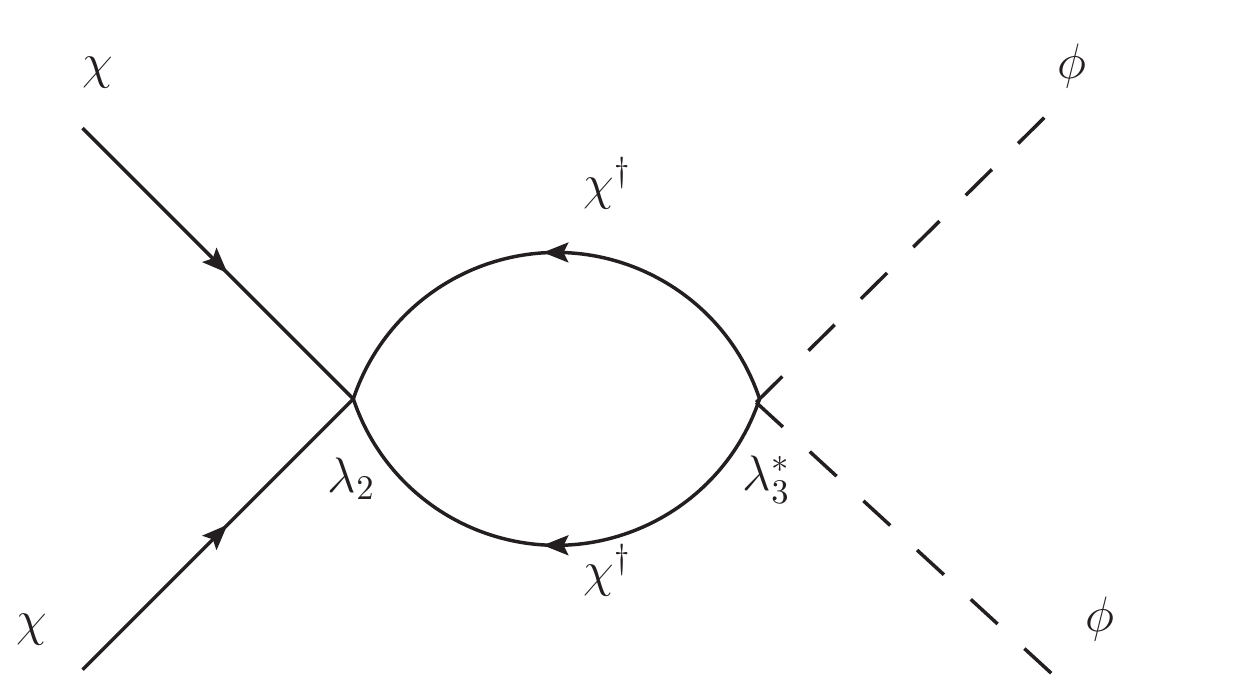}
\caption{\small{\em Relevant tree and one-loop level Feynman diagrams for the CP-violating annihilation process $\chi+\chi \rightarrow \phi + \phi$.}}
\label{fig:diag3}
\end{figure}
Finally, we come to the third CP-violating process, which is an annihilation that reduces the total DM and anti-DM number as well, unlike the previous two self-scattering processes. The relevant Feynman diagrams are shown in Fig.~\ref{fig:diag3}. One of the loop diagrams here is again proportional to the CP-conserving $\lambda_5$ coupling. The difference between the matrix elements squared of the CP-conjugate processes is
\begin{align}
|M|^2_{\chi\chi\rightarrow \phi\phi}-|M|^2_{\chi^{\dagger}\chi^{\dagger}\rightarrow\phi\phi}=4\,\left[\rm Im(\lambda^*_3\lambda_4)\lambda_5\,Im\, I_{\chi^{\dagger}\chi}+\,\rm Im(\lambda_2\lambda^{*2}_3)\,Im\, I_{\chi^{\dagger}\chi^{\dagger}}\right]
\label{eq:ME3}
\end{align}
Unlike in the self-scatterings, the thermally averaged symmetric scattering rate is no longer velocity suppressed in the leading term, and is given as
\begin{align}
\expval{\sigma v}_3 = \dfrac{|\lambda_3|^2}{32\pi m^2_\chi}\left(1+\dfrac{3}{4x}\right).
\end{align}
The asymmetric scattering rate also shows a somewhat different behaviour, with both terms in the thermal average now suppressed by a single power of the DM velocity:
\begin{align}
\expval{\epsilon \sigma v}_3 &= \dfrac{1}{(16\pi m_\chi)^2}\dfrac{1}{\sqrt{\pi x}}\bigg[\text{Im}(\lambda_2\lambda^{*2}_3)+ 2\, \text{Im}(\lambda^*_3\lambda_4)\lambda_5 \bigg].
\label{eq:eps_sigmav_3}
\end{align}
Thus, the effective CP-violation parameter, for $|\lambda_i| \sim \mathcal{O}(1)$ and the phases in the relevant coupling combinations set to $\pi/2$, now carries an explicit $x-$dependence even in the leading term:
\begin{align}
 \epsilon^{3}_{eff} \simeq -\dfrac{1}{8\pi}\dfrac{1+2\lambda_5}{\sqrt{\pi x}} , \,\,\,\, \text{for}\,\, x >> 1.
\end{align}

With all our computations in place, we can check the unitarity sum rules explicitly upto the next-to-leading order in perturbation theory. First of all, using Eqs.~\ref{eq:ME1},~\ref{eq:ME2} and~\ref{eq:ME3}, we see that the sum rule in Eq.~\ref{eq:eps_unitarity} holds, namely
\begin{equation}
\int dPS_2 \,\epsilon_1 \,|M_0|^2_1 + \int dPS_2 \,\epsilon_2 \,|M_0|^2_2 + \int dPS_2 \,\epsilon_3 \,|M_0|^2_3 = 0.
\end{equation}
The same unitarity sum rule can also be written in terms of the thermally averaged asymmetric reaction rates:
\begin{align}
\expval{\epsilon \sigma v}_1 + \expval{\epsilon \sigma v}_2 + \expval{\epsilon \sigma v}_3 =0,
\label{eq:unitarity_sum_epsilon}
\end{align}
which we can again easily check at this order in perturbation theory using the expressions in Eqs.~\ref{eq:eps_sigmav_1}, \ref{eq:eps_sigmav_2} and \ref{eq:eps_sigmav_3}.

We now point out a key observation of our study. If the CP-violating annihilation process is absent, i.e., if we set $\lambda_3=0$, then we can see from Eq.~\ref{eq:unitarity_sum_epsilon} that $\expval{\epsilon \sigma v}_1 + \expval{\epsilon \sigma v}_2=0$. Equivalently, this can be verified also by setting $\lambda_3=0$ in Eqs.~\ref{eq:eps_sigmav_1} and \ref{eq:eps_sigmav_2}. And therefore, by Eq.~\ref{eq:asym}, the asymmetric yield will not be generated, starting from a symmetric initial condition. This is because $\expval{\epsilon \sigma v}_S(=\expval{\epsilon \sigma v}_1 + \expval{\epsilon \sigma v}_2)$ acts as the source term for the asymmetry in our parametrization of the Boltzmann equations, as seen from both Eq.~\ref{eq:asym}, and the approximate analytic solution in Eq.~\ref{eq:analytic2}. Thus we see that the presence of only the CP-violating self-scatterings cannot generate the asymmetry. Similarly, only the presence of the CP-violating annihilation cannot generate an asymmetry either, by the unitarity sum rule. It is the simultaneous presence of both the self-scatterings and the annihilation that can lead to a generation of particle-antiparticle asymmetries.

A remark about the number of independent phases in the coupling combinations that appear in the $\expval{\epsilon \sigma v}_i$ is in order. Defining each coupling as $\lambda_i =|\lambda_i| e^{i\theta_i}$, we can write the three rates in terms of two relative phase angles, namely $\theta_{23} = 2\theta_3-\theta_2$ and $\theta_{24} = 2\theta_4-\theta_2$. The third relevant angle is $\theta_4-\theta_3=(\theta_{24}-\theta_{23})/2$, and hence is not independent. In what follows, we shall set $\theta_{23}= - {\theta}_{24}=\pi/2$, to maximize the CP-violating effects.

\subsubsection{Process 4: CP-conserving Annihilation $\chi+\chi^\dagger \rightarrow \phi + \phi$}
We have already mentioned the role of the CP-conserving annihilation in indirectly controlling the particle-antiparticle asymmetry in Sec.~\ref{sec:sec3}. In addition to the asymmetry, such a pair annihilation process, if present, also strongly affects the final relic abundance of the DM and anti-DM particles. The thermally averaged annihilation rate for this process in the complex singlet scenario, keeping the first two terms in the partial-wave expansion, is given as follows:
\begin{align}
 \expval{\sigma v}_A = \dfrac{|\lambda_5|^2}{64\pi m^2_\chi}\left(1+\dfrac{3}{4x}\right).
\end{align}

\subsubsection{The $\chi+\phi \rightarrow \chi^{\dagger}+\phi$ process}
As mentioned in Sec.~\ref{sec:sec2}, we can also have the CP-violating DM anti-DM conversion process $\chi+\phi \rightarrow \chi^{\dagger}+\phi$. As shown by the explicit computation described in Appendix~\ref{App.A}, the net contribution to CP-violation from all the Feynman diagrams contributing to this process is 
\begin{align}
\int dPS_2\left(|M|^2_{\chi\phi\rightarrow \chi^\dagger\phi}-|M|^2_{\chi^{\dagger}\phi \rightarrow \chi\phi}\right) \propto{\rm Im(\lambda^*_3\mu_1)} \times {\rm ~Terms ~of}~\mathcal{O}({\hat\mu^3/m_\chi^4}),
\end{align}
where, $\hat{\mu}$ is either $\mu$, $\mu_1$ or $\mu_\phi$. This is in contrast to the CP-violation stemming from the processes discussed in Sec.~\ref{sec:sec2}, which is independent of the dimensionful couplings. Since we have assumed $\mu_1 \simeq 0$ in order to avoid the stringent constraints from spin-independent DM direct detection probes, clearly, the contribution to CP-violation from the $\chi\phi\rightarrow \chi^\dagger\phi$ process is much smaller compared to the ones we have included in the Boltzmann equations for the (anti-)DM number densities.

\subsubsection{Dark matter density and asymmetry: the interplay of different processes}
\begin{figure}[htb!]
\centering
\includegraphics[scale=0.55]{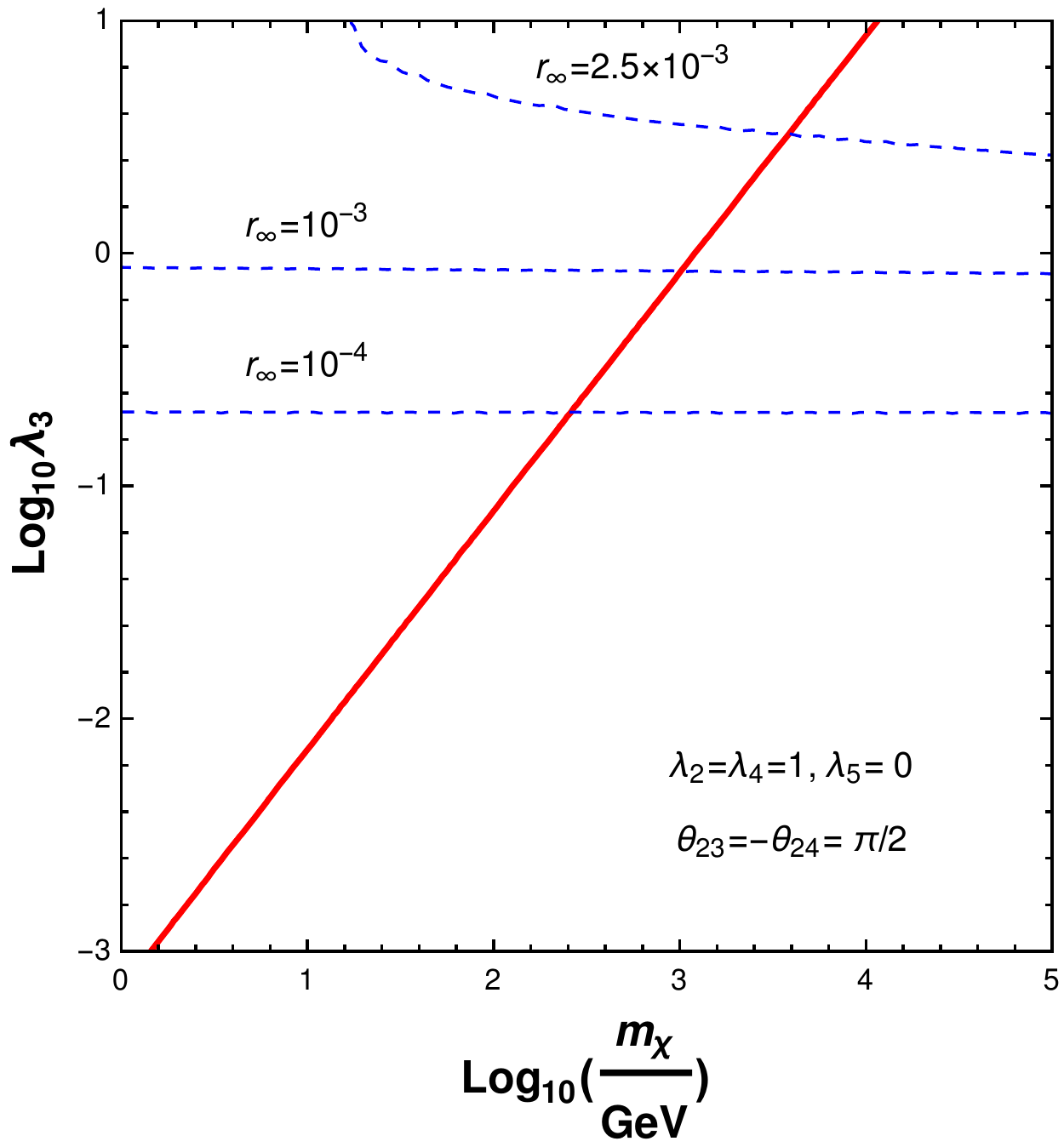}
\caption{\small{\em Contour in the $\lambda_3 - m_\chi$ plane for which the DM and anti-DM densities saturate the DM density of the Universe, with $\Omega h^2 \simeq 0.12$ (red solid line). Also shown are the contours of the constant final asymmetry parameter $r_\infty=\lvert Y_{\Delta\chi} \rvert / Y_S$ (blue dashed lines). Here, the coupling $\lambda_3$ determines the symmetric annihilation rate for the $\chi + \chi \rightarrow \phi + \phi$ process, and also the CP-violation through loop effects. For this figure, we have set the CP-conserving pair-annihilation coupling $\lambda_5=0$.}}
\label{fig:mass_lambda3}
\end{figure}  

Having determined the CP-violation and the symmetric and asymmetric annihilation rates in the complex scalar singlet model, we can now compute the relic abundance of the DM and anti-DM system. Due to the possible CP-violation, there can also be a resulting particle-antiparticle asymmetry in the DM sector. We have defined the measure of the final particle-antiparticle asymmetry parameter as $r_\infty=\lvert Y_{\Delta\chi} \rvert / Y_S$, where the asymptotic values of the yields are considered. Clearly, we have $0 \leq r_\infty \leq 1$. Here, $r_\infty=0$ corresponds to the completely symmetric limit, in which the asymptotic yields of the DM and anti-DM are the same. On the other hand, $r_\infty=1$ corresponds to the completely asymmetric limit, in which only either the DM or the anti-DM species survives. As we shall see in the following, depending upon the dominance of either the CP-conserving or the CP-violating pair-annihilation couplings, we obtain two different regimes for the DM $-$ a nearly symmetric regime, and an asymmetric regime.

\begin{figure}[htb!]
\centering
\includegraphics[scale=0.56]{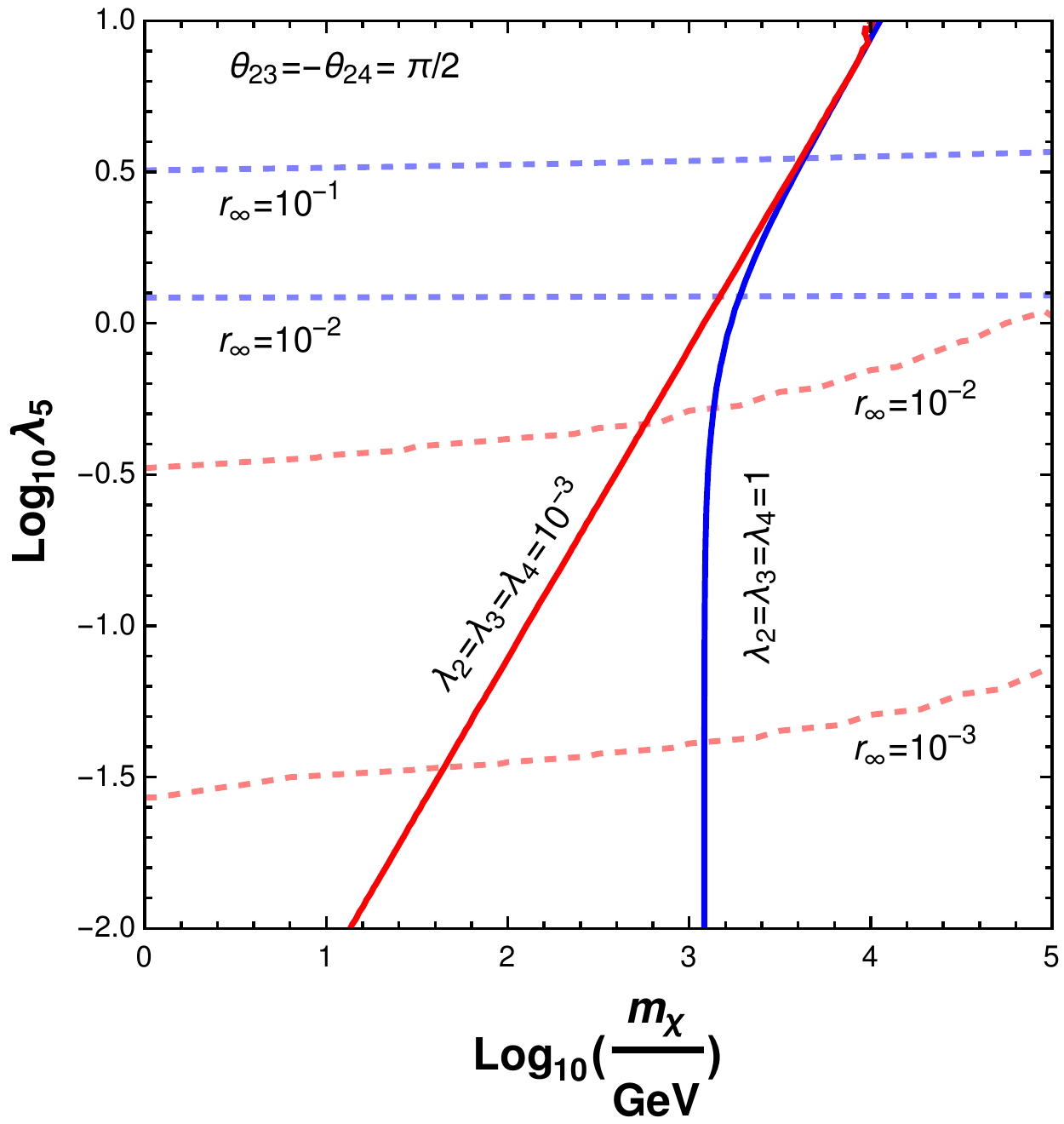}
\caption{\small{\em Contours in the $\lambda_5 - m_\chi$ plane for which the DM and anti-DM densities saturate the DM density of the Universe, with $\Omega h^2 \simeq 0.12$ (blue and red solid lines), for two different choices of $\lambda_2,\lambda_3$ and $\lambda_4$, as indicated against the lines. Also shown are the colour-coded contours of the constant final asymmetry parameter $r_\infty=\lvert Y_{\Delta\chi} \rvert / Y_S$ (with the blue and red  dashed lines corresponding to the scenarios with $\lambda_2=\lambda_3=\lambda_4=1$ and $\lambda_2=\lambda_3=\lambda_4=10^{-3}$, respectively). Here, the coupling $\lambda_5$ determines the CP-conserving pair annihilation rate for the $\chi + \chi^\dagger \rightarrow \phi + \phi$ process, and also the CP-violation through loop effects.}}
\label{fig:mass_lambda_5}
\end{figure}

To begin with, let us consider the range of DM mass that is allowed in this scenario. We find that the DM relic abundance required for saturating the observed DM density of $\Omega h^2 \simeq 0.12$~\cite{Planck:2018vyg} can be obtained for $m_\chi$ in the range from $\mathcal{O}({\rm GeV})$ to $\mathcal{O}(10 {~\rm  TeV})$, with the relevant couplings kept within their perturbative limits. Since both $\lambda_3$ and $\lambda_5$ can play a dominant role in determining the DM relic density, as seen in Sec.~\ref{sec:sec3} and in particular in Eqs.~\ref{eq:YS_approx_1} and \ref{eq:YS_approx_2}, we show the allowed DM mass range as a function of these two couplings separately in Figs.~\ref{fig:mass_lambda3} and ~\ref{fig:mass_lambda_5}. 

\begin{figure}[htb!]
\centering
\includegraphics[scale=0.6]{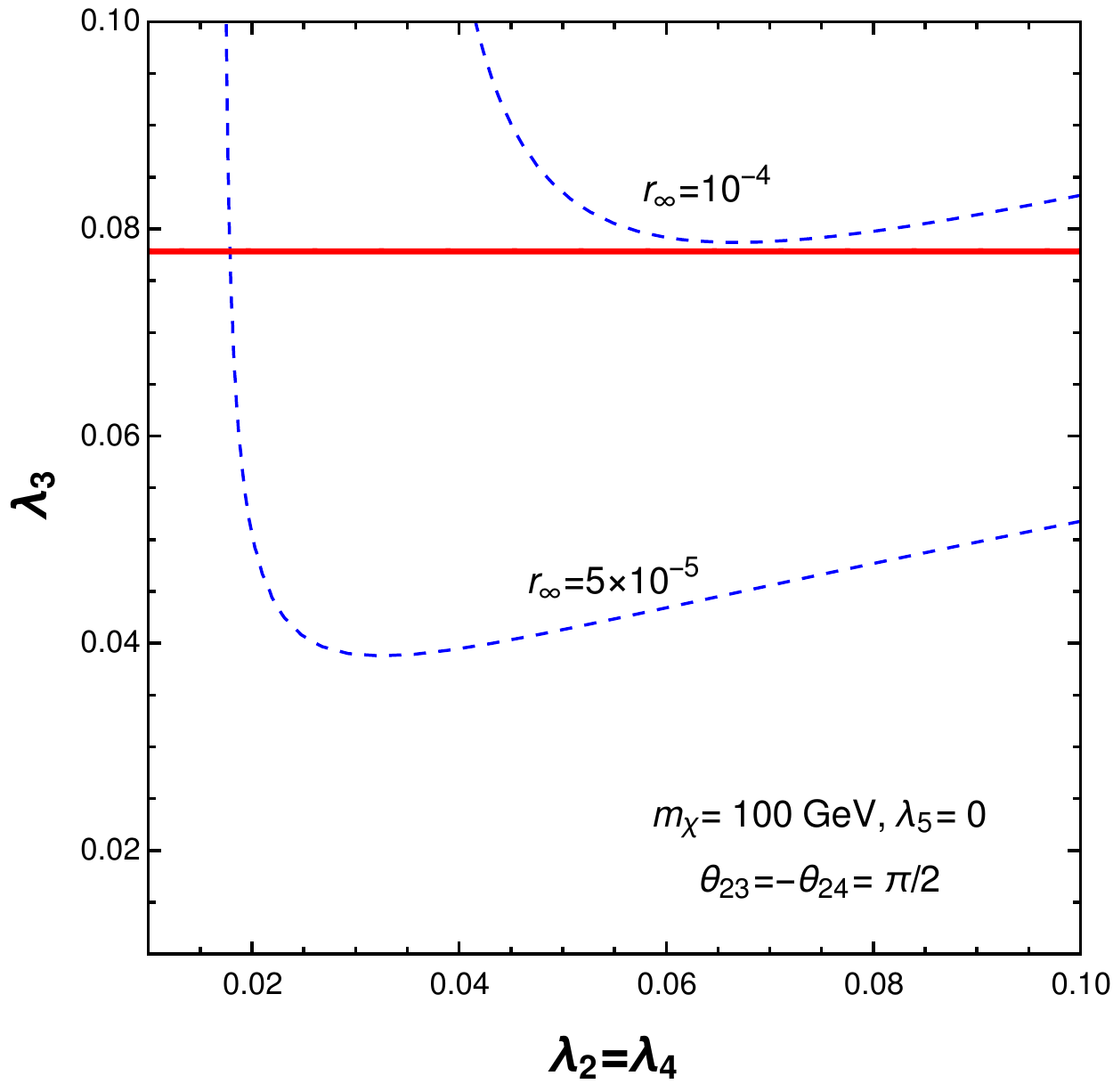}
\caption{\small{\em Contour in the $\lambda_3 - \lambda_{2,4}$ plane  for which the DM and anti-DM densities saturate the DM density of the Universe, with $\Omega h^2 \simeq 0.12$ (red solid line). Also shown are the contours of the constant final asymmetry parameter $r_\infty=\lvert Y_{\Delta\chi} \rvert / Y_S$ (blue dashed lines). We have set the two coupling parameters determining the DM self-scattering rates to be equal ($\lambda_2 = \lambda_4$) for simplicity. The CP-conserving coupling has been fixed at $\lambda_5=0$, see text for details.}}
\label{fig:results_wthout_ann}
\end{figure}

In Fig.~\ref{fig:mass_lambda3}, we show the contour in the $\lambda_3 - m_\chi$ plane for which the DM and anti-DM densities furnish $\Omega h^2 \simeq 0.12$ (red solid line). We also show the contours of the constant final asymmetry parameter $r_\infty$ defined above (blue dashed lines). For this figure, we have fixed the DM self-couplings as $\lambda_2=\lambda_4=1$, and have
set the CP-conserving pair-annihilation coupling $\lambda_5=0$. The reason for setting $\lambda_5=0$ is to demonstrate the role of $\lambda_3$ in determining the density. The effective CP-violating phases are set to $\pi/2$. As we can see from this figure, increasing $\lambda_3$ implies a corresponding increase in the mass $m_\chi$ that saturates the DM abundance. The final asymmetry parameter $r_\infty$ is found to be rather small, of the order of $10^{-4}$ to $10^{-3}$, indicating a nearly symmetric DM. This is primarily because with these choices of parameters the CP-violation is not large, and with the CP-conserving pair-annihilation switched off, the symmetric part is not subsequently removed as well. In addition to that, in this scenario, the asymmetric reaction rate for the CP-violating processes are also reduced, as seen in, for example, Eq.~\ref{eq:eps_sigmav_2}.

How do the DM density and the asymmetry parameter vary as a function of the pair-annihilation ($\lambda_3$) and self-scattering couplings ($\lambda_2, \lambda_4$)? We show this correlation in Fig.~\ref{fig:results_wthout_ann}. For this figure, we have kept the DM mass fixed at $m_\chi=100$ GeV, and continue to consider $\lambda_5=0$. As we can see, the DM asymmetry continues to be small, and in the parameter range of interest can vary from $10^{-5}$ to $10^{-4}$. Furthermore, with such a small asymmetry, the DM self-scatterings play no significant role in deciding the DM density, which is fixed by $\lambda_3$. However, the DM composition varies with the variation of the self-interaction couplings.

We now move to the scenario in which the CP-conserving pair-annihilation coupling is non-zero, and potentially large. We find that in such a case, a larger asymmetry is obtained. Furthermore, we find a novel and interesting effect that the self-scattering couplings now play an equally dominant role in deciding both the DM  density and asymmetry, as do the annihilation couplings. We first show the contours in the $\lambda_5 - m_\chi$ plane for which the DM and anti-DM densities saturate the DM density of the Universe in Fig.~\ref{fig:mass_lambda_5}. The mass range obtained is again similar to the one obtained in Fig.~\ref{fig:mass_lambda3}. For the blue solid line, we have fixed $\lambda_2=\lambda_3=\lambda_4=1$, while for the red solid line we have set $\lambda_2=\lambda_3=\lambda_4=10^{-3}$, for showing two illustrative cases. We see that when $\lambda_3=1$, there is no significant variation in the DM density for $\lambda_5 < 1$, since $\lambda_3$ also plays a dominant role in determining it. However, for $\lambda_5>\lambda_3$, the density changes rapidly with varying  $\lambda_5$, as seen from the $\lambda_5 > 1$ region along the blue line, and the entire region along the red line. 

\begin{figure}[htb!]
\centering
\includegraphics[scale=0.59]{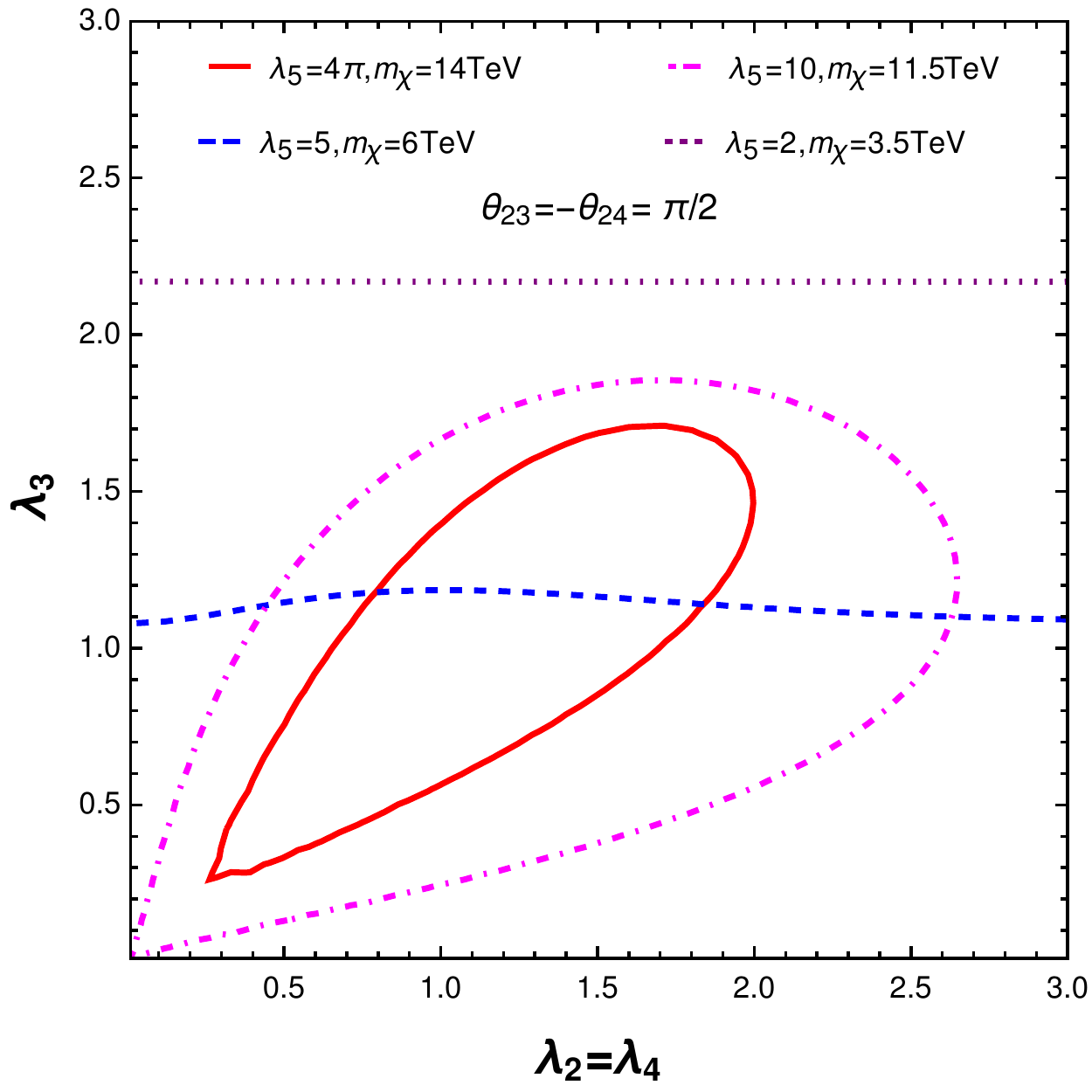}
\includegraphics[scale=0.59]{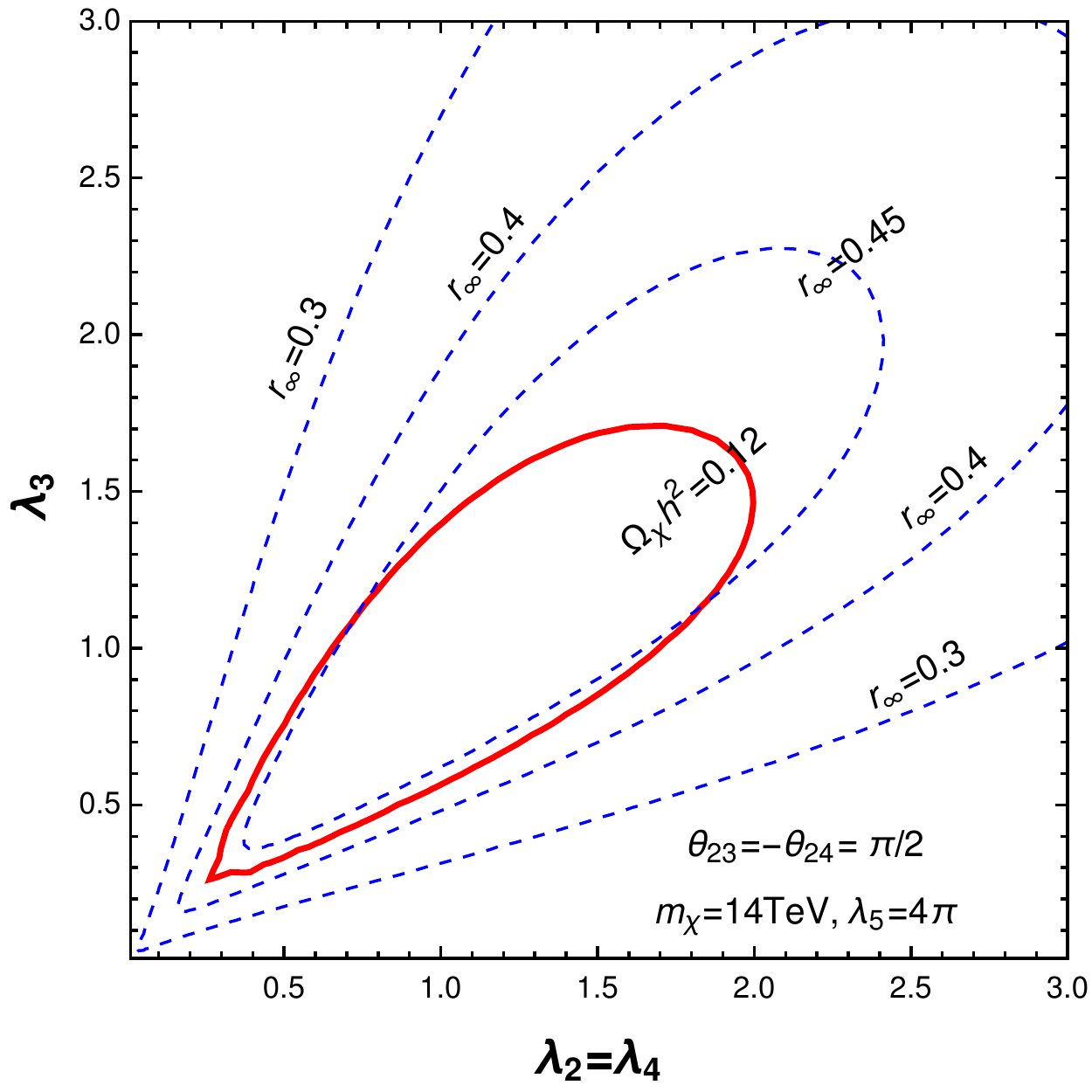}
\caption{\small{\em (Left Panel:) Contours in the $\lambda_3 - \lambda_{2,4}$ plane  for which the DM and anti-DM densities saturate the DM density of the Universe, with $\Omega h^2 \simeq 0.12$, for different values of the CP-even coupling $\lambda_5$ and DM mass $m_\chi$. (Right Panel:) Contours of the constant final asymmetry parameter $r_\infty=\lvert Y_{\Delta\chi} \rvert / Y_S$ (blue dashed lines), with a representative value of $\lambda_5=4 \pi$ and DM mass $m_\chi=14$ TeV. We have set the two coupling parameters determining the DM self-scattering rates to be equal ($\lambda_2 = \lambda_4$) for simplicity. In both the figures, we clearly see the strong interplay of the DM annihilation and self-scattering processes in determining both the DM density and asymmetry.}}
\label{fig:results_with_ann}
\end{figure}

With the introduction of the CP-conserving pair annihilation, the asymmetry parameter $r_\infty$ is seen to take values of the order of $10^{-1}$ (or higher, as we shall see in the following). Thus we can obtain a large asymmetry in the DM sector in this scenario. As discussed earlier, this is due to the fact that after the CP-violating processes decouple, thereby freezing out the difference in the DM and anti-DM yields $Y_{\Delta \chi}$, the CP-conserving pair annihilation processes with larger rates can still remain active. These pair annihilations at later epochs reduce the sum of the DM and anti-DM yields $Y_S$. This leads to the enhancement of the final asymmetry parameter $r_\infty = \lvert Y_{\Delta\chi} \rvert / Y_S$. The $r_\infty$ contours shown in Fig.~\ref{fig:mass_lambda_5} are colour coded, with the blue dashed contours corresponding to the $\lambda_2=\lambda_3=\lambda_4=1$ scenario, while the red dashed ones corresponding to the $\lambda_2=\lambda_3=\lambda_4=10^{-3}$ scenario. 

We now study the variation of the DM density and the asymmetry parameter as a function of the pair-annihilation ($\lambda_3$) and self-scattering couplings ($\lambda_2, \lambda_4$), in the presence of the $\lambda_5$ coupling. As shown in the left panel of Fig.~\ref{fig:results_with_ann}, there are strong correlations between the annihilation and self-scattering in such a scenario. We have shown the contours in this plane where the relic abundance is satisfied, with different sets of values for the CP-conserving coupling $\lambda_5$ and the DM mass $m_\chi$. For each choice of $\lambda_5$, we have chosen $m_\chi$ such that the relic density condition can be satisfied in some region in the coupling plane. The correlations are especially pronounced for larger values of $\lambda_5$, as seen from the shown representative values of $\lambda_5=10$ (pink dot-dashed line) and $\lambda_5=4\pi$ (red solid line), both of which are within the limits of perturbation theory, such that our computations can be reliable. There are small correlations for $\lambda_5=5$ as well (blue dashed line), while no observable correlations are found for $\lambda_5=2$ (violet dotted line). The $\lambda_5=2$ scenario is thus similar to the previously discussed case of no CP-conserving annihilation, as both lead to a rather small final asymmetry. 

In order to understand the physics behind these strong correlations, we pick out one representative value of $\lambda_5=4\pi$, and study  the variation of the asymmetry parameter $r_\infty$ in the right panel of Fig.~\ref{fig:results_with_ann}. We see that $r_\infty$, and hence the asymmetry in the DM sector can be large in this scenario, with the contours upto $r_\infty=0.45$ spanning a large range of the parameter space. This explains the reason behind the strong correlations being observed. For with such a large asymmetry, the total DM and anti-DM yield strongly depends upon the asymmetric yield $Y_{\Delta \chi}$ itself, as can be seen from the analytic approximation in Eq.~\ref{eq:YS_approx_2}, i.e., the asymmetry plays a dominant role in fixing the relic density. This is the reason we observe the strong interplay between the DM self-scatterings and the annihilations in fixing not only the DM asymmetry, but the DM relic density itself. This role of the DM self-scatterings is a novel observation, which should be of considerable interest in the study of  DM cosmology. We would like to emphasize again here that within this scenario of complex singlet scalar DM, we cannot switch off the self-scattering processes, but have only the CP-violating annihilations, or vice versa, and generate an asymmetry in the DM sector. The simultaneous presence of both these types of processes is necessary in order to have a non-zero CP-violation, as dictated by the unitarity sum rules. 

\begin{figure}[htb!]
\centering
\includegraphics[scale=0.52]{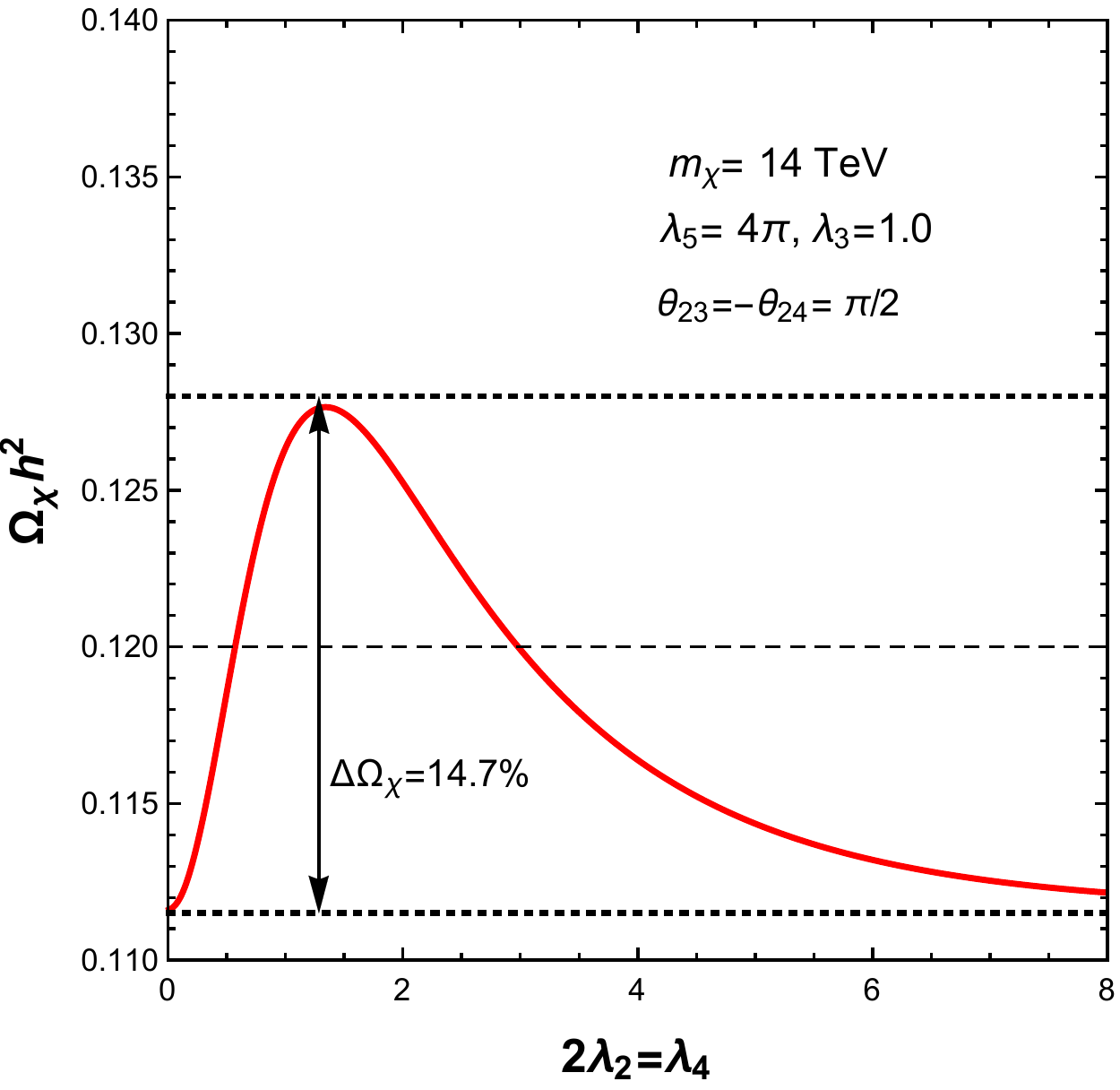}
\includegraphics[scale=0.52]{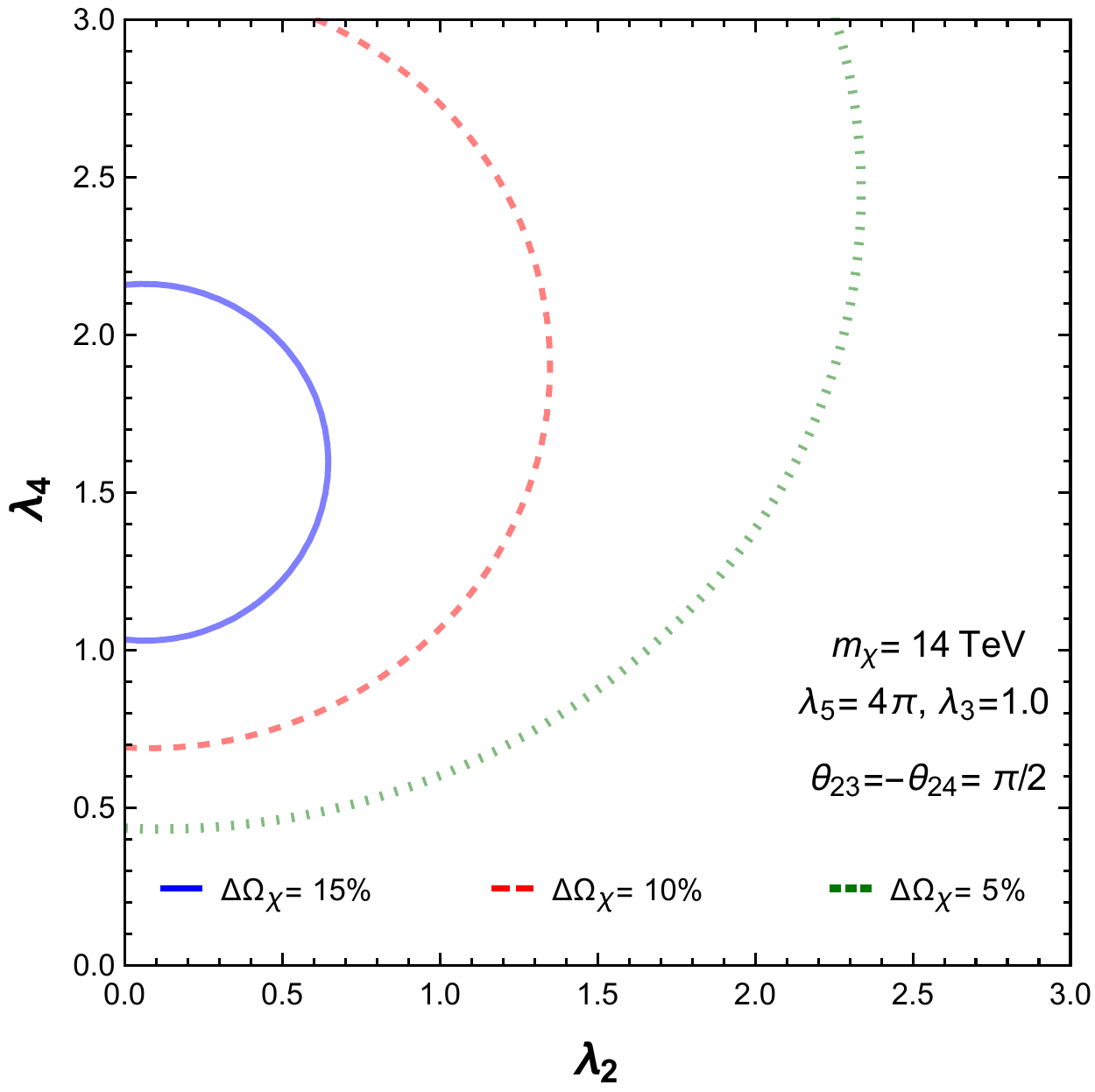}
\caption{\small{\em Role of DM self-scattering in determining the total DM relic abundance $\Omega_\chi h^2$. (Left panel:) $\Omega_\chi h^2$ as a function of the parameters $\lambda_2$ and $\lambda_4$ which determine the self-scattering rate. The maximum deviation in the relic density from the case with no self-scattering, for the choice of parameters in this figure, is $\mathcal{O}(15\%)$, where we have defined the deviation as $\Delta \Omega_\chi (\lambda_2,\lambda_4) =( \Omega_\chi (\lambda_2,\lambda_4) -  \Omega_\chi (\lambda_2=0, \lambda_4=0))/ \Omega_\chi (\lambda_2=0, \lambda_4=0)$. (Right panel:) Contours of fixed values of $\Delta \Omega_\chi$ in the $\lambda_2 - \lambda_4$ plane, with $\Delta \Omega_\chi=15\%, 10\%$ and $5\%$.}}
\label{fig:self_vs_relic}
\end{figure}

In order to further understand the impact of the self-scattering in determining the relic density, we show in Fig.~\ref{fig:self_vs_relic} (left panel) the variation in the total DM density $\Omega_\chi h^2$ as a function of the parameters determining the self-scattering rate, namely $\lambda_2$ and $\lambda_4$. For this figure, we have used a fixed ratio of these two parameters, with $2\lambda_2=\lambda_4$. All other parameters have been kept fixed. We observe that the relic abundance can vary significantly as a function of $\lambda_2$ and $\lambda_4$. For these choice of parameters, the deviation in $\Omega_\chi h^2$ from its value with no self-scattering, defined as, $\Delta \Omega_\chi (\lambda_2,\lambda_4) =( \Omega_\chi (\lambda_2,\lambda_4) -  \Omega_\chi (\lambda_2=0, \lambda_4=0))/ \Omega_\chi (\lambda_2=0, \lambda_4=0)$ takes a maximum value of around $\mathcal{O}(15\%)$. 

Such significant deviations in $\Omega_\chi h^2$ are obtained for a wide range of values for the self-scattering parameters. This is shown in Fig.~\ref{fig:self_vs_relic} (right panel), where we plot contours of fixed values of $\Delta \Omega_\chi$ in the $\lambda_2 - \lambda_4$ plane. The contours are shown for three values of $\Delta \Omega_\chi$, $15\%$ (blue solid line), $10\%$ (red dashed line) and $5\%$ (green dotted line). All other parameters are kept fixed as before. As we can see from this figure, for a fixed value of  $\Delta \Omega_\chi$, the values of $\lambda_2$ and $\lambda_4$ lie on a half-circle, and $\mathcal{O}(15\%)$ deviations are observed for a wide range of values of these parameters. Larger deviations in $\Omega_\chi h^2$ are also possible, in circles of smaller radii. Thus, these figures clearly demonstrate the significant role of the self-scatterings in determining the total relic density itself, by comparing the relic densities in the scenarios with and without DM self-scatterings. 

%%%%%%%%%%%%%%%%%%%%%%%%%%%%%%%%%%%%%%%
\subsection{Summary and discussion}
\label{sec:sec6}
To summarize, we studied the scenario of a complex scalar dark matter, which is a singlet under the SM gauge interactions, and is stabilized by an effective $Z_2$ reflection symmetry. We considered all the renormalizable interactions of the singlet with the SM Higgs doublet, as well as an SM singlet real scalar, including terms that preserve the reflection symmetry, but break the larger global $U(1)_\chi$ symmetry of DM number. Since the interactions of the DM with the Higgs boson are already strongly constrained, we are forced to introduce the minimal scenario with the additional real scalar which is even under the $Z_2$ symmetry $-$ in order to thermalize the DM sector with the SM sector, and thereby obtain a thermal mechanism for producing the DM density.

We find that both CP-violating and CP-conserving scattering reactions involving the DM and the real scalar field ($\phi$) can take place in the thermal bath. The CP-violating scatterings include both DM self-scattering processes as well as DM annihilations to a pair of $\phi$ particles. Since these CP-violating processes involve the $U(1)_\chi$ breaking terms, they can violate DM number as well, and thereby create a possible asymmetry in the DM sector. 

In order to understand the role of the various scattering processes in determining the DM density and composition, we set up the Boltzmann kinetic equations for the (anti-)DM yields, and study them first in terms of the thermally averaged symmetric and asymmetric scattering rates. We obtain approximate analytic solutions to the system of Boltzmann equations, which are found to be in good agreement with the exact numerical results, with a maximum difference of upto $10 \%$. We utilize the unitarity sum rules in relating the CP-violation in the three different scattering processes. These sum rules imply that the CP-violation is non-zero only in the simultaneous presence of both the self-scatterings and the pair-annihilation, and would vanish if either one of them is absent. 

We then go on to compute the relevant CP-violations, and thermally averaged reaction rates in the complex singlet scenario, including relevant Feynman graphs upto one-loop level, with the CP-violation and asymmetric reaction rates ensuing from the interference of the tree level graphs with the one-loop graphs. With these, we compute the relic abundance and asymmetry in the DM sector as a function of the coupling and mass parameters. There are broadly two different regimes observed: one in which the CP-violating pair-annihilation dominantly decides the DM relic abundance, and the other in which the CP-conserving pair-annihilation plays the major role. In the first scenario, the resulting asymmetries are found to be small, so are the correlations between the self-scattering and annihilation couplings in determining the DM density. 

In the second scenario, the asymmetries can be large, with the asymmetry parameter taking values upto $\mathcal{O}(0.5)$. Furthermore, in this scenario, we observe strong correlations between the self-scattering and annihilation processes in obtaining the required DM abundance. This role of DM self-scatterings in determining the DM density is a novel observation, which is the primary result of this chapter. The reason behind this effect is that when the asymmetry in the DM sector becomes large, the asymmetric yield, which is the difference between the particle and antiparticle yields, plays a dominant role in determining the total DM density, as we also show through our analytic solutions. And this asymmetric yield is affected by both the self-scattering couplings as well as the annihilation ones that violate CP. Thus, DM self-scatterings can become important in deciding both the DM density and composition, being strongly correlated with the annihilation. This is further strengthened by the unitarity sum rules mentioned above $-$ we cannot have a non-zero asymmetry unless both these types of processes are simultaneously present. 

Although the cosmological, and the related particle physics aspects of the complex scalar singlet scenario were the focus of this chapter, there are important astrophysical consequences of this scenario which should be explored in future studies. The primary direction would be to look into the indirect detection prospects in the present Universe, which requires a detailed analysis depending upon the mass of the DM $\chi$ and the real singlet scalar $\phi$. Although such an analysis is beyond the scope of this work, a few relevant comments are in order. As we have already seen, the asymmetry generated in this scenario is not maximal, and therefore, we shall have a density of both DM and anti-DM particles existing in the galaxies and clusters. The exact ratio of the densities would be dependent upon both the asymmetry parameter at the generation level, and any possible oscillation that may happen at post-freeze out epochs due to the $U(1)_\chi$ breaking mass terms, if these are present with sufficient strength. 

All the four relevant annihilation processes involving the $\chi$ and $\chi^\dagger$ particles will therefore take place, with $\chi \chi \rightarrow \phi \phi$ (and its conjugate $\chi^\dagger \chi^\dagger \rightarrow \phi \phi$) and $\chi \chi^\dagger \rightarrow \phi \phi$ leading to the production of $\phi$ particles in the final state. Since the $\phi$ particles can have a small mixing with the SM Higgs boson, they will eventually decay to SM particles through this mixing. The exact decay branching ratio to different final states will depend upon the $\phi$ mass, while the energy spectra of the decay products will depend both on the decay mode, as well as on the $\chi$ mass, thus leading to a wide range of possibilities. Given the nature of the Higgs coupling to SM particles, if kinematically allowed, the possible annihilation modes for one pair of (anti-)DM particles are then to four final state particles, drawn in two pairs primarily from the following list  $-\{\gamma \gamma$, $\tau^+ \tau^-$, $b \bar{b}$, $W^+ W^-$, $ZZ$, $t \bar{t}$ and $hh\}$. If $\phi$ is very light, for example, it may decay to $\gamma \gamma$, thereby leading to four-photon final states in (anti-)DM annihilations. 

Such a process, with $\{\chi\chi, \chi^\dagger\chi^\dagger,\chi\chi^{\dagger}\}  \rightarrow \phi \phi \rightarrow 4\gamma$ leads to an interesting feature in the spectra of the gamma ray signal. Each of the produced photons will have an energy in the band $E^{min}_\gamma \leq E_\gamma \leq E^{max}_\gamma$ and therefore the corresponding gamma-ray spectra will be of a box shape of width $\Delta E = E^{max}_\gamma-E^{min}_\gamma = \sqrt{m^2_\chi-m^2_\phi}$, see, for example~\cite{Ibarra:2012dw}. Apart from the DM mass $m_\chi$, the photon flux will depend upon the parameters $r_\infty$, $\langle \sigma v \rangle_3$ and $\langle \sigma v \rangle_A$. Therefore, compared to the standard WIMP scenario, the correspondence between the observed flux and the pair-annihilation rate is expected to be different in this model. In particular, if $\mathcal{F}_{\rm ADM}$ is the photon flux expected in our scenario from $\{\chi\chi, \chi^\dagger\chi^\dagger,\chi\chi^{\dagger}\}  \rightarrow \phi \phi$ processes and $\mathcal{F}_{\rm WIMP}$ is the corresponding flux in case of WIMPs from the same set of $\{\chi\chi, \chi^\dagger\chi^\dagger,\chi\chi^{\dagger}\}  \rightarrow \phi \phi$ processes (here, WIMPs represent the symmetric limit of the ADM in which the final particle and anti-particle number densities are the same), then we obtain the ratio between the two fluxes to be:
\begin{equation}
\frac{\mathcal{F}_{\rm ADM}}{\mathcal{F}_{\rm WIMP}}=\frac{1}{\expval{\sigma v}_A+\expval{\sigma v}_3}  \bigg[(1-r^2_\infty) \expval{\sigma v}_A + (1+r^2_\infty) \expval{\sigma v}_3 + 2 r_\infty \expval{\epsilon\sigma v}_3 \bigg]
\label{flux}
\end{equation} 
We see that while $\mathcal{F}_{\rm ADM} = \mathcal{F}_{\rm WIMP}$ for the completely symmetric case of $r_\infty=0$, $\mathcal{F}_{\rm ADM} \neq \mathcal{F}_{\rm WIMP}$ for $r_\infty \neq 0$, where the difference stems from the different number densities of the $\chi$ and $\chi^\dagger$ particles in the present epoch, while keeping their sum fixed.

Such a spectrum of photons will be discernible above the astrophysical background, for example in the Fermi-LAT satellite looking either into the galactic center or into dwarf-spheroidal galaxies~\cite{Ibarra:2012dw}. However, the relevant constraints will strongly depend upon the DM mass, and for very heavy DM particles, only a future experiment such as the CTA probing a higher photon energy range will be promising~\cite{Ibarra:2015tya}. The other possible final states do not lead to any distinct spectral feature in the gamma ray or charged particle energy distribution, and therefore would have to be searched for in the corresponding diffuse flux. As it is clear from this brief discussion, given the wide possible range of DM and $\phi$ masses, and thus the varied possible final states, there are multiple possible indirect detection probes that would require a detailed independent study.

The direct detection prospect of the present scenario relies also on the mixing of $\phi$ and SM Higgs in the absence of $\lambda_{\chi H}$  coupling. In particular, through the mixing $\phi$ couples to the SM quarks which eventually facilitates the elastic collision between DM particles and target nuclei. To note, there is no qualitative difference between an ADM and a WIMP scenario in the context of direct detection experiments albeit the mass of DM and relevant coupling can be very different from WIMP scenario. Nevertheless, the current bound on the scalar dark matter particle from direct detection probes is very much applicable to our scenario. For a comprehensive study see Refs.\cite{He:2008qm,Barger:2010yn,Schumann:2019eaa} and references therein.   

The complex singlet scalar constitutes one of the simplest scenarios for non-self-conjugate thermal DM. Our study shows that in its full generality, this scenario offers many novel phenomena as far as the DM cosmology is concerned. The role of DM self-scatterings, and their interplay with the other processes in driving the cosmological evolution of the DM properties is a novel effect of considerable interest which should be explored in other scenarios of thermal DM as well.

\newpage
\section{\Large{Revisiting the role of CP-conserving processes in cosmological particle–antiparticle asymmetries}}
\label{chap:chap5}
\subsection{Introduction}
\label{sec:sec52}

In earlier chapters, we have primarily learned about different CP-violating scatterings in deciding possible particle-antiparticle asymmetry in the DM sector, simultaneously generating the present-day density of the same. In our example models, the CP-conserving processes are naturally present, which may or may not be relevant in deciding the particle-antiparticle asymmetry. In this chapter, we focus on the role of CP-conserving processes in deciding the particle-antiparticle asymmetry in the context of the observed baryon-antibaryon asymmetry of the Universe (BAU) as well as a probable asymmetric DM (ADM) sector. 

Two of the most well studied BAU generation mechanisms are baryogenesis in grand unified theories (GUT)~\cite{Yoshimura:1978ex,Weinberg:1979bt, Nanopoulos:1979gx, Yoshimura:1979gy}, and baryogenesis through leptogenesis~\cite{Fukugita:1986hr}. In both scenarios, most of the implementations considered involve the CP-violating out-of-equilibrium decay of a heavy particle in the early Universe. Thus, the primary quantities that determine the net asymmetry generated are the rates of the CP-violating decay and its inverse process. If in addition CP-violating scattering processes are present, they are found to affect both the generated asymmetry, and its subsequent wash-out, depending upon the hierarchy of the decoupling temperatures for the decay and scattering processes; for reviews, see, for example, Refs.~\cite{Kolb:1990vq, Buchmuller:2005eh, Davidson:2008bu}, and references therein. 

The primary source of the CP-violation in a given scenario may also turn out to be scattering processes, with the heavy particle decay processes either sub-dominant or absent~\cite{Bento:2001rc, Nardi:2007jp, Gu:2009yx}. Such scattering mechanisms have been studied in contexts involving the dark matter (DM) particles as well, in achieving baryogenesis through DM annihilations~\cite{Cui:2011ab, Bernal:2012gv, Bernal:2013bga, Kumar:2013uca},  in relating the baryon and dark matter sector asymmetries~\cite{Baldes:2014gca, Baldes:2015lka}, in realizing the ADM scenario through scatterings~\cite{Ghosh:2021qbo}, and through semi-annihilations~\cite{Ghosh:2020lma, DEramo:2010keq}. 

The role of CP-conserving processes have also been studied in the context of leptogenesis in see-saw models of neutrino mass. In leptogenesis from type-I see-saw, CP-conserving scatterings can arise from additional new interactions of the SM gauge singlets~\cite{Gu:2009hn, Sierra:2014sta}. As discussed in Ref.~\cite{Sierra:2014sta}, these additional interactions are CP-conserving by assumption, that is in general they can violate CP if the relevant couplings have complex phases. Furthermore, it is a specific dependence of the reaction rates on the temperature and masses that leads to this competing effect, in spite of one process being a decay and the other a scattering, as analyzed in detail in~\cite{Sierra:2014sta}. In leptogenesis from type-II and type-III see-saw, there are additional gauge interactions, which lead to CP-conserving scatterings, see, for example~\cite{Hambye:2003rt, Hambye:2005tk, Hambye:2012fh}. The rate of these gauge scatterings are usually smaller than the CP-violating decay rates, and therefore as shown in Ref.~\cite{Hambye:2005tk}, the role of the gauge scatterings in determining the asymmetry is small, and the efficiency factor remains close to maximal.

In this chapter, we revisit the role of CP-conserving reactions in determining cosmological particle-antiparticle asymmetries, and point out qualitatively novel scenarios in which their role is significantly exemplified. In particular, to begin with, we may classify the possible scenarios into three broad categories:
\begin{enumerate}
\item[\bf{A.}] both the CP-violating and CP-conserving processes are decays

\item[\bf{B.}] both the CP-violating and CP-conserving processes are scatterings

\item[\bf{C.}] one of the processes is a decay and the other is a scattering
\end{enumerate}
In scenarios A and B, the two processes can naturally have comparable rates, and therefore the CP-conserving reactions can in principle play a significant role in generating the particle-antiparticle asymmetries. In scenario C, which is the case so far considered in the literature, it is expected that the CP-conserving scatterings will have a suppressed reaction rate compared to the CP-violating decays, due to the Boltzmann suppression stemming from the presence of an additional non-relativistic particle in the initial state. In this chapter, we shall discuss examples of both scenarios A and B. 

In each of these scenarios, one can further have two distinct possibilities. For processes of the form $M... \rightarrow D...$, where $M$ is heavier than $D$, we can have a scenario in which
\begin{enumerate}
\item[\bf{I.}] the asymmetry is generated in the mother sector, i.e., in the $M$ sector,

\item[\bf{II.}] the asymmetry is generated in the daughter sector, i.e., in the $D$ sector.
\end{enumerate}
In conventional scenarios of baryogenesis, we usually find examples of type II. In this chapter, we shall show examples of both types, belonging to scenarios A and B above. In particular, we shall point out examples of type I, in which the CP-conserving processes play a dual role. Until the decoupling of the CP-violating scatterings, the CP-conserving processes tend to suppress the asymmetry generation by controlling the out-of-equilibrium number densities of the bath ($M$) particles. However, subsequently, once the net particle antiparticle yield difference has been frozen, the same CP-conserving pair-annihilation reactions modify the ratio of particle anti-particle yields by eliminating the symmetric component of the mother particles~\footnote{Since either the DM particle ($\chi$) or the DM anti-particle ($\chi^\dagger$) may dominate the current DM density, we quantify this ratio by the asymptotic value of the asymmetry parameter $r_\infty=\lvert Y_{\chi} - Y_{\chi^\dagger}\rvert / (Y_{\chi} + Y_{\chi^\dagger})$ in the present epoch, where $Y_i$ are the respective yields; see Sec.~\ref{sec:sec54} for details.}. This leads to a competing effect shown by the same scattering process at different epochs, since, unlike in the scenarios of type II, in scenarios of type I the asymmetry is generated in the sector which initiates both the CP-conserving and violating scatterings in the first place. This dual role is highlighted for the first time in this thesis.

We shall now demonstrate the above scenarios through illustrative examples. In Sec.~\ref{sec:sec53} we discuss a leptogenesis scenario of type A-II, in which both the CP-conserving and violating processes are decays, and the asymmetry is generated in the daughter sector. In Sec.~\ref{sec:sec54} we discuss two scenarios of asymmetric dark matter production from scattering of type B-I, in which both the processes are scatterings, and the asymmetry is generated in the mother sector. We summarize our findings in Sec.~\ref{sec:sec55}.

\subsection{CP-violating and conserving decays and daughter sector asymmetry}
\label{sec:sec53}
We first discuss an illustrative model which belongs to the type A-II in the classification described in the introduction, i.e., both the CP-conserving and violating processes are two-body decays, and the asymmetry is generated in the daughter sector. To this end, consider a leptogenesis model involving two standard model (SM) singlet heavy Majorana neutrinos $N_1$ and $N_2$, with their mass values satisfying $M_{N_1}>M_{N_2}$. For the cosmology of $N_1$, the following decays of $N_1$ to SM charged lepton ($\ell^\pm$), charged scalar boson ($H^\pm$), neutral scalar boson ($h$) and $N_2$ are important:
\begin{align}
N_1 &\rightarrow \ell^{\pm} H^{\mp}~~~~~{\rm (CP-violating)} \nonumber \\
N_1 &\rightarrow N_2 h ~~~~~{\rm (CP-conserving)}. \nonumber
\end{align}

In order to realize these processes obeying the SM gauge symmetries, we need to extend the SM field content in the scalar sector. In addition to the dominantly SM-like scalar doublet $\Phi_1$ which gives mass to the SM fermions, we introduce a second Higgs doublet $\Phi_2$, and an SM singlet scalar field $S$. The  relevant interaction Lagrangian terms before electroweak symmetry breaking include the following:
\begin{equation}
\mathcal{L}_{\rm int} \supset -\frac{g}{2} \overline{N_1} N_2 S - y_i \overline{L} \tilde{\Phi}_1 N_i -  \mu_i |\Phi_i|^2 S -  \frac{\lambda^S_i}{2}  |\Phi_i|^2 S^2 - \lambda (\Phi_1^\dagger \Phi_1) (\Phi_2^\dagger \Phi_2)
\end{equation}
where $\tilde{\Phi}_1=i\sigma_2 \Phi_1^*$. Here, we have written the interaction terms involving the SM singlet Majorana neutrinos using the mass eigenstates $N_1$ and $N_2$. In order to avoid tree-level flavour-changing neutral currents, we have also assumed an additional $Z_2$ symmetry, under which $\Phi_2 \rightarrow -\Phi_2$, that is softly broken by a $m_{12} \Phi_1^\dagger \Phi_2$ term, as in the type-I two Higgs doublet model~\cite{Branco}. This symmetry prohibits the $y^\prime_i \overline{L} \tilde{\Phi}_2 N_i$ terms. After electroweak symmetry breaking, the singlet scalar mixes with the neutral components of the scalar doublets, giving rise to the decay mode $ N_1 \rightarrow N_2 h $, while the mixing of the charged scalars leads to the decay $N_1 \rightarrow \ell^{\pm} H^{\mp}$ through the Yukawa interaction.

Such a model was proposed in Ref.~\cite{Bhattacharya:2011sy} as a low-scale model for leptogenesis that utilizes the quartic scalar couplings~\cite{Kayser:2010fc}. In particular, the lepton number conserving decay mode $N_1 \rightarrow N_2 h$ helps in satisfying the requirements of generating a non-zero CP-violation at the one-loop level, being consistent with the Nanopoulous-Weinberg theorem~\cite{Nanopoulos:1979gx, Bhattacharya:2011sy}. However, the role of the CP-conserving decay mode $N_1 \rightarrow N_2 h$ in the cosmology of $N_1$ --- and therefore in the generated lepton asymmetry --- has not been studied earlier, and is the focus of the present study.

The Boltzmann equations determining the number density of $N_1$, and the lepton asymmetry produced are given by\footnote{Since $M_{N_1}>M_{N_2}$, during the out-of-equilibrium decay of $N_1$ at a temperature $T\simeq M_{N_1}$, the $N_2$ particles are in thermal equilibrium through rapid scatterings in the thermal bath, and hence follow a Maxwell-Boltzmann distribution. Furthermore, for simplicity, we assume that $N_2$ decays after the freeze-out of $N_1$ decay do not wash-out the generated lepton asymmetry, which can be ensured, for example, with $M_{N_2}<M_{H^{\pm}}$.}
\begin{align}
\dfrac{dY_{N_1}}{dx} &\simeq -\dfrac{\expval{\Gamma}_0+\expval{\Gamma}_A}{Hx}\, (Y_{N_1}-Y_{N_1,0}) \nonumber\\
\dfrac{dY_L}{dx} &\simeq -\dfrac{\epsilon\expval{\Gamma}_0}{Hx} \,(Y_{N_1}-Y_{N_1,0}).
\label{eq:Boltz_Decay}
\end{align}
Here, $Y_i = n_i/s$ is the yield of the species $i$, with $n_i$ being its number density and $s$ the entropy density in the radiation bath. We also have $Y_L = Y_{\ell^-}-Y_{\ell^+}$ as the lepton asymmetry produced, and $Y_{N_1,0}$ the equilibrium yield of $N_1$. Here it is assumed that at high temperatures $T>M_{N_1}$, $N_1$ achieves a thermal distribution through rapid scattering processes in the thermal plasma. The thermally averaged symmetric decay width is given by $\expval{\Gamma}_0$, where, $\Gamma_0=\Gamma(N_1 \rightarrow \ell^+ H^-)+\Gamma(N_1 \rightarrow \ell^- H^+)$, and $\Gamma_A$ is the decay width of the CP-conserving mode $N_1 \rightarrow N_2 h$. Finally, the CP-violation parameter is defined as 
\begin{equation}
\epsilon = \frac{|M(N_1 \rightarrow \ell^- H^+)|^2-|M(N_1 \rightarrow \ell^+ H^-)|^2}{|M(N_1 \rightarrow \ell^- H^+)|^2+|M(N_1 \rightarrow \ell^+ H^-)|^2}
\label{eq:eps}
\end{equation}
Here, $|M|^2$ denotes the matrix element squared for the corresponding process. In writing the Boltzmann equation for $Y_L$, we have made the approximation that at the epoch of the generation of the asymmetry, the two-body scattering reactions have decoupled, which is as expected due to their lower rates. We have also dropped terms in the right hand side proportional to $Y_L$, as at the epoch when $Y_L$ is being generated, those are subdominant.

\begin{figure}[htb!]
\centering
\includegraphics[scale=0.55]{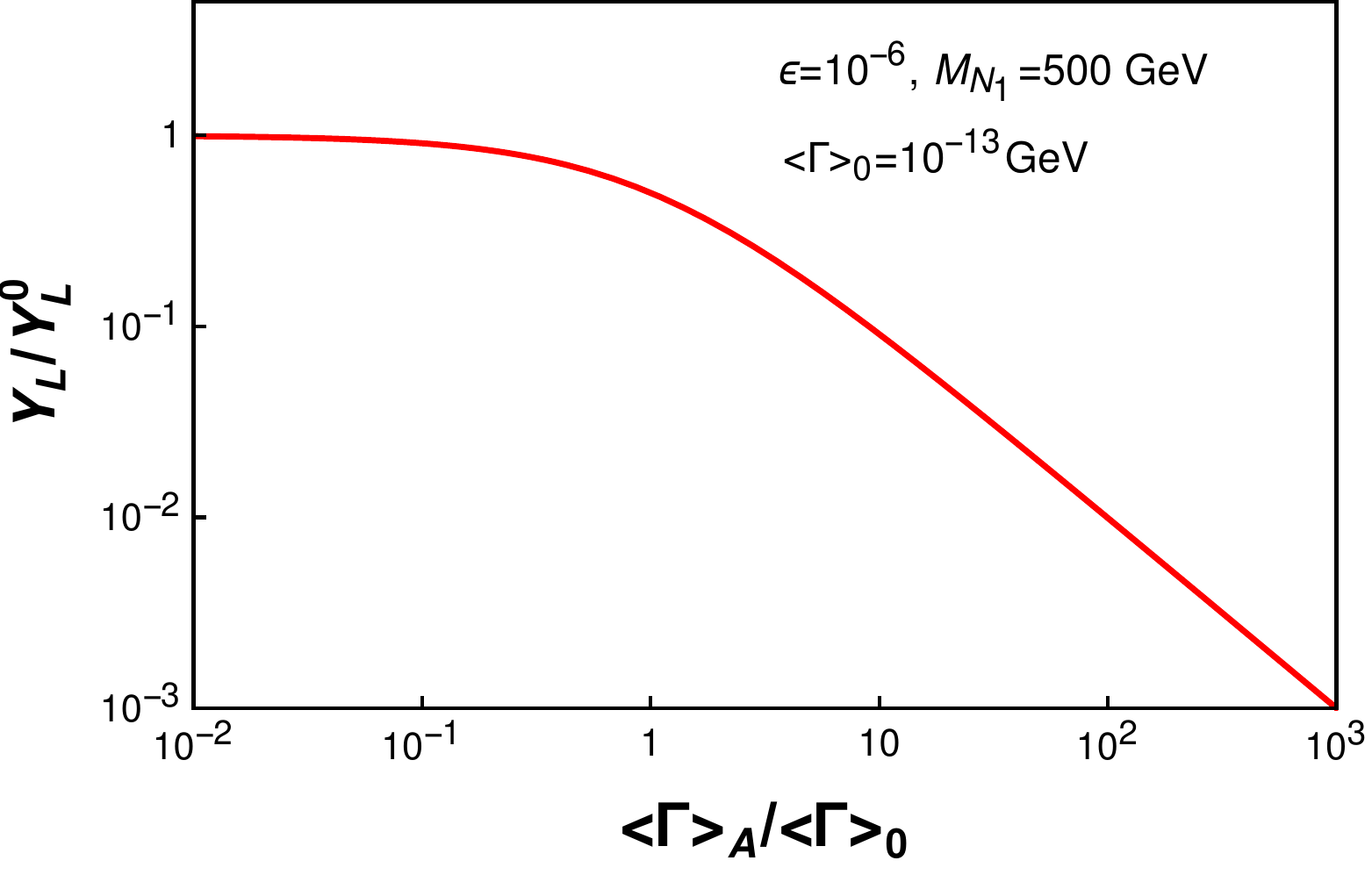}
\caption{\small{\it{The dependence of the lepton asymmetry $Y_L$ (scaled by $Y_{L,0})$, as a function of the ratio of CP-conserving and violating decay widths, $\expval{\Gamma}_A/\expval{\Gamma}_0$. See text for details on the choice of the other parameters.}}}
\label{Fig:lepto}
\end{figure}

We solve Eqs.~\ref{eq:Boltz_Decay} for the parameter choices $M_{N_1}=500$ GeV, $\epsilon=10^{-6}$ and $\Gamma_0=10^{-13}$ GeV, where the parameters are chosen such that the required baryon asymmetry of the Universe may be reproduced. We show the dependence of $Y_L$ (scaled by $Y_{L}^0$, which is the asymmetric yield with $\Gamma_A=0$) as a function of $\expval{\Gamma}_A/\expval{\Gamma}_0$ in Fig.~\ref{Fig:lepto}. As we can clearly see from this figure, the lepton asymmetry $Y_L$ changes by one order of magnitude if $\Gamma_A \sim 10 \Gamma_0$, compared to the scenario where $\Gamma_A$ is negligible. This simply stems from the fact that the longer the CP-conserving process is in equilibrium, the smaller the number density of the mother particle $N_1$, and hence the lepton asymmetry generated from its decays. 

\begin{figure}[htb!]
\centering
\includegraphics[scale=0.55]{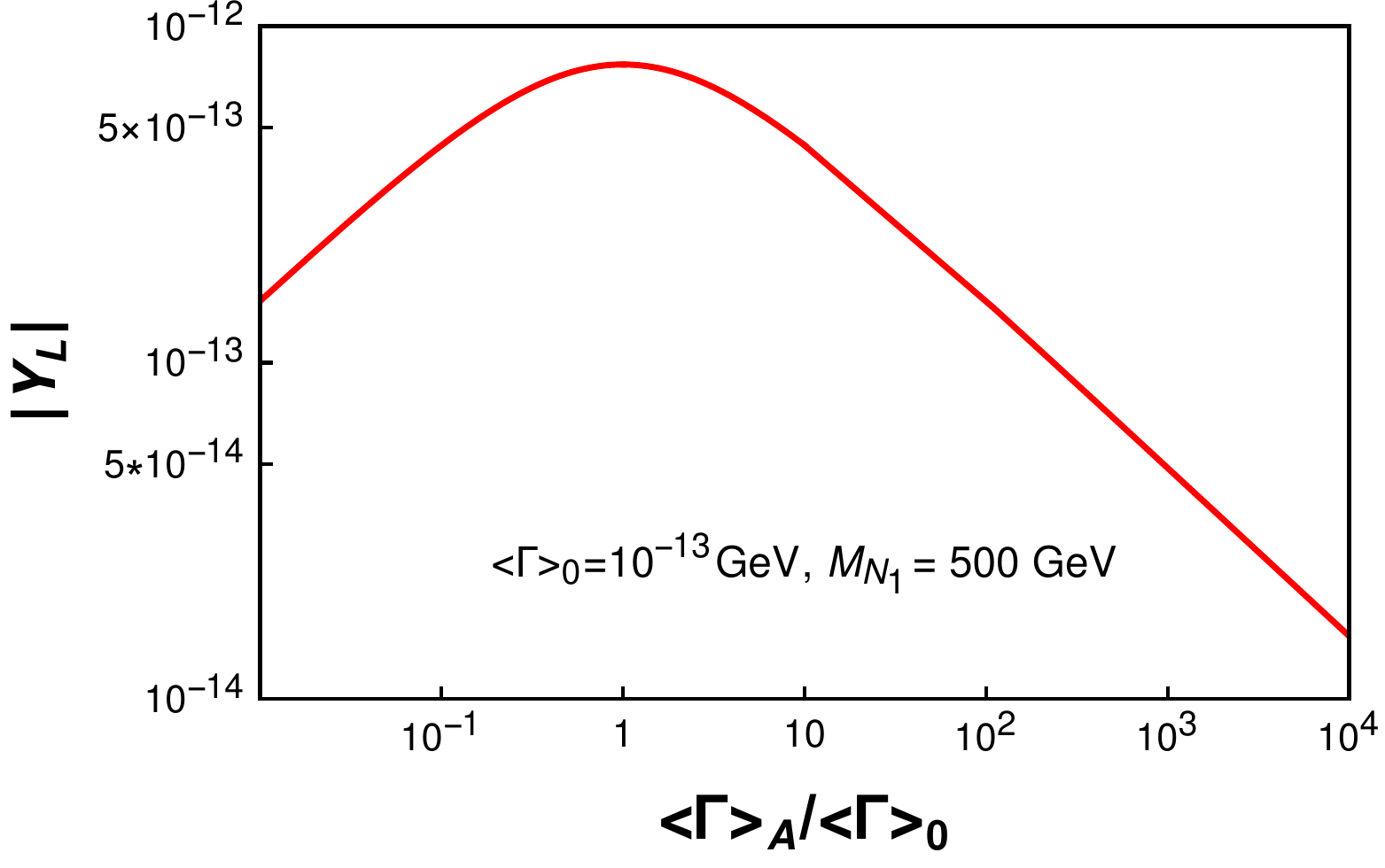}
\caption{\small{\it{The dependence of the lepton asymmetry $|Y_L|$ in the model for low-scale leptogenesis, as a function of the ratio of CP-conserving and violating decay widths, $\expval{\Gamma}_A/\expval{\Gamma}_0$. In this model, the CP-violation parameter $\epsilon$ can be expressed in terms of $\Gamma_A$, for fixed values of the scalar quartic coupling $\lambda$ and $\tan \beta$. See text for details.}}}
\label{Fig:lepto2}
\end{figure}

One can simplify the analysis further in the particular low-scale leptogenesis model discussed above. In this scenario, we can express the CP-violation parameter $\epsilon$ defined in Eq.~\ref{eq:eps}, stemming from the interference of the tree and loop level amplitudes, as follows:
\begin{equation}
\epsilon \simeq -2\lambda\, v\, \sin\beta \,\tilde{g}\, {\rm Im} (L).
\label{eq:lepto_eps}
\end{equation}
Here, $v=246$ GeV is the electroweak symmetry breaking scale and $\tan \beta = v_1/v_2$ is the ratio of the vacuum expectation values of the neutral CP-even components of the two Higgs doublets $\Phi_1$ and $\Phi_2$. We have also defined $\tilde{g}=g\, \sin\alpha$, where $\alpha$ is the mixing angle of the scalar singlet $S$ with the lighter CP-even Higgs boson. For simplicity, here we have assumed that $S$ dominantly mixes with the SM-like lighter Higgs state. Finally, ${\rm Im} (L)$ is the imaginary part of the loop-factor, which is found to be:
\begin{equation}
{\rm Im} (L) \simeq -\frac{M_{N_2}}{8\pi M^2_{N_1}}\frac{1}{1-\xi}\log \left(\frac{1}{\xi} \right),
\end{equation}
where, $\xi = \left(M_{H^\pm}/M_{N_1}    \right)^2$. For example, with $M_{N_1}=500$ GeV, $M_{N_2}=300$ GeV and $M_{H^\pm}=350$ GeV, we obtain ${\rm Im} (L) \simeq -6.7 \times 10^{-5}$. 

Since the width of the CP-conserving decay $N_1 \rightarrow N_2 h$ is $\Gamma_A = \frac{\tilde{g}^2}{8\pi}M_{N_1}$, we can trade the coupling $\tilde{g}$ in Eq.~\ref{eq:lepto_eps} with the square-root of the CP-conserving decay width $\Gamma_A$. In addition, if we fix the scalar quartic coupling $\lambda$ and $\tan \beta$, $\epsilon$ is determined in terms of $\Gamma_A$. The residual dependence of the lepton asymmetry $|Y_L|$ in this model as a function of $\expval{\Gamma}_A/\expval{\Gamma}_0$ is shown in Fig.~\ref{Fig:lepto2}, where $\expval{\Gamma}_0$ is chosen to be $10^{-13}$ GeV, $\sin \beta \sim \mathcal{O}(1)$ and the quartic coupling $\lambda \sim \mathcal{O}(0.1)$. As we can see from this figure, for $\expval{\Gamma}_A<\expval{\Gamma}_0$, the out-of-equilibrium condition for $N_1$ is determined by the CP-violating decay. Therefore, 
$|Y_L|$ increases with increasing $\expval{\Gamma}_A$, simply because the CP-violation $\epsilon$ increases as $\sqrt{\Gamma_A}$, for a fixed scalar quartic. On the other hand, for $\expval{\Gamma}_A > \expval{\Gamma}_0$, $|Y_L|$ decreases rapidly with increasing 
$\expval{\Gamma}_A$, as $N_1$ remains in equilibrium for a longer period, thereby leading to a reduction in its number density. We would like to emphasize that $\Gamma_A >>  \Gamma_0$ is easily achieved in the model described above for natural choices of the parameters.

Thus, CP-conserving processes, when at the same footing as the CP-violating ones (in this case both of them being decays of $N_1$), can play a major role in deciding the lepton asymmetry production in the early Universe. As discussed in the introduction, this is in contrast to the role of such processes observed earlier in the leptogenesis models from different see-saw mechanisms, as in those cases the CP-conserving modes played a subdominant role, mostly because they were scattering reactions, whereas the violating ones were decays, and hence of a higher rate.

%%%%%%%%%%%%%%%%%%%%%%%%%%%%%%%%%%%%%%%%%%%%
\subsection{CP-violating and conserving scatterings and mother sector asymmetry}
\label{sec:sec54}
We shall now discuss two scenarios of asymmetric dark matter production from scattering. These two examples belong to the type B-I explained in the Introduction, in which both the CP-violating and conserving processes are scatterings, and the asymmetry is generated in the mother sector. We shall see that due to the latter feature, there is a dual role played by the CP-conserving processes. The ADM models under discussion have been proposed in Ref.~\cite{Ghosh:2021qbo}, which discusses ADM from scattering, and in Ref.~\cite{Ghosh:2020lma} on ADM from semi-annihilation. 

Both the ADM models share some common features. There is a complex scalar DM $\chi$, which is stabilized by a $Z_N$ symmetry, interacting with itself and a $Z_N$ even real scalar $\phi$. The scalar $\phi$, taken to be lighter than $\chi$,  can mix with or decay to SM states, thereby maintaining kinetic equilibrium of the DM sector with the SM bath. In particular, for a GeV DM particle, the suitable choice for the mass of $\phi$ is around a MeV, as explained in Chapter \ref{chap:chap1}. This is motivated from the constraints on the effective relativistic degrees of freedom at the time of BBN ($T_\gamma \sim 1~ {\rm MeV} $). In both cases, the existence of a CP-conserving scattering $\chi+\chi^\dagger \rightarrow  \phi+\phi$ immediately follows, which is the focus of this discussion. In the ADM model from scattering, the stabilizing symmetry is $Z_2$, while for the ADM from semi-annihilation scenario, it is $Z_3$. Let us briefly discuss each model in the following. 

\subsubsection{Asymmetric DM from scattering}
\label{sec:ADM_1}
In this case, we have the following interaction Lagrangian (detailed in chapter \ref{chap:chap4}) consistent with the $Z_2$ symmetry:
\begin{align}
-\mathcal{L_{\rm int}}  \supset  \mu  \chi^\dagger \chi  \phi+ \left(\frac{\mu_1}{2} \chi^2 \phi+ {\rm h.c.} \right)  + \frac{\lambda_1}{4} \left(\chi^\dagger \chi \right)^2 +   \left(\frac{\lambda_2}{4!}\chi^4 +{\rm h.c.} \right)+\left(\frac{\lambda_3}{4}\chi^2 \phi^2 +{\rm h.c.} \right)\nonumber\\
+\left(\frac{\lambda_4}{3!}\chi^3\chi^{\dagger} +{\rm h.c.} \right)
+ \frac{\lambda_5}{2} \phi^2 \chi^\dagger \chi  +  \frac{\mu_\phi}{3!} \phi^3 +  \frac{\lambda_\phi}{4!} \phi^4 + \frac{\lambda_{\phi H}}{2} \phi^2 |H|^2 + \mu_{\phi H} \phi |H|^2 .\,\,\,
\label{eq:lag_z2}
\end{align}
Here, $H$ is the SM Higgs doublet. The neutral particle $\phi$ can mix with the Higgs boson after electroweak symmetry breaking, thus inducing direct detection signals for DM through the trilinear interactions with couplings $\mu$ and $\mu_1$. To evade the current direct detection bounds, we have set $\mu\simeq 0$ and $\mu_1\simeq 0$, rendering only the contact interactions to be relevant for our discussion.

Several relevant processes for the DM cosmology resulting from these interactions are listed in Table~\ref{Tab:tab1}, along with their properties and the notation used to denote the different thermally averaged symmetric and asymmetric reaction rates. The asymmetric and symmetric thermal averages are defined with and without $\epsilon_f$ for each process (the definition of $\epsilon_f$  is analogous to Eq.~\ref{eq:eps}, which, for scatterings, is an explicit function of the particle momenta $p_i$), where $\expval{\epsilon \sigma v}_{f}$ can be written as:
\begin{equation}
 \expval{\epsilon \sigma v}_{f} = \dfrac{\int \prod^{4}_{i=1} \frac{d^3 p_i}{(2\pi)^3 2 E_{p_i}}  (2 \pi)^4 \delta^{(4)}(p_1+p_2-p_3-p_4) \,\epsilon_f(p_i) |M_0|^2_f f_0(p_1)f_0(p_2)}{\int \dfrac{d^3 p_1}{(2\pi)^3} \dfrac{d^3 p_2}{(2\pi)^3} f_0(p_1)f_0(p_2)} \hspace{0.5cm},
\label{eq:cross}
\end{equation}
with   $|M_0|^2_f = |M|^2_{\chi\chi\rightarrow f}+|M|^2_{\chi^{\dagger}\chi^{\dagger}\rightarrow f^{\dagger}} $ and $f_0(p)$ being the equilibrium distribution function.
\begin{table}[htb!]
\begin{center}
\begin{tabular}{|c|c|c|c|c|}
\hline
Process & CP & DM Number & Rate & Rate \\
              &      & Violation  & (Symmetric) & (Asymmetric) \\
\hline
$\chi+\chi \rightarrow \chi^{\dagger} + \chi^{\dagger}$ & violating & 4 units & $\expval{\sigma v}_1$ &  $\expval{\epsilon \sigma v}_1$\\
\hline
$\chi+\chi \rightarrow \chi + \chi^{\dagger}$ & violating & 2 units & $\expval{\sigma v}_2$  &  $\expval{\epsilon \sigma v}_2$ \\
\hline
$\chi+\chi \rightarrow \phi+\phi$ & violating & 2 units & $\expval{\sigma v}_3$ & $\expval{\epsilon \sigma v}_3$  \\
\hline
$\chi+\chi^\dagger \rightarrow  \phi+\phi$ & conserving & 0 units & $\expval{\sigma v}_A$ & NA \\
\hline
\end{tabular}
\end{center}
\caption{\small{\it{Relevant processes for DM cosmology for the ADM from scattering scenario, their properties, and notation used to denote the corresponding thermally averaged symmetric and asymmetric reaction rates.}}}
\label{Tab:tab1}
\end{table}%

With the above reactions in the thermal plasma, the evolution equations for the symmetric ($Y_S=Y_\chi+Y_{\chi^\dagger}$) and asymmetric yields ($Y_{\Delta \chi}=Y_\chi -Y_{\chi^\dagger}$) in this scenario are given as follows:

\begin{align}
\label{asym}
\dfrac{dY_S}{dx}&=-\dfrac{s}{2Hx}\left[\expval{\sigma v}_A\left(Y^2_S-Y^2_{\Delta \chi}-4Y^2_0\right)+\expval{\sigma v}_3\bigg(\dfrac{Y^2_S+Y^2_{\Delta \chi}-4Y^2_0}{2}\bigg)-\expval{\epsilon\sigma v}_S Y_S Y_{\Delta \chi}\right]\nonumber\\
\dfrac{dY_{\Delta \chi}}{dx}&=-\dfrac{s}{2Hx}\left[\expval{\epsilon\sigma v}_S\left(\dfrac{Y^2_S-4Y^2_0}{2}\right)+\expval{\epsilon\sigma v}_D \dfrac{Y^2_{\Delta \chi}}{2}+\expval{\sigma v}_{all}\,Y_S Y_{\Delta \chi}\right].
\end{align} 
Here, we have defined, $\expval{\epsilon\sigma v}_S =\expval{\epsilon \sigma v}_1+\expval{\epsilon \sigma v}_2 $, $\expval{\epsilon\sigma v}_D=\expval{\epsilon \sigma v}_1-\expval{\epsilon \sigma v}_2$ and $\expval{\sigma v}_{all}=2\expval{\sigma v}_1+\expval{\sigma v}_2+\expval{\sigma v}_3$. In writing these equations, we have used the unitarity sum rule relating the $\expval{\epsilon \sigma v}_i$'s, namely, $\expval{\epsilon \sigma v}_1+\expval{\epsilon \sigma v}_2+\expval{\epsilon \sigma v}_3=0$, and have eliminated $\expval{\epsilon \sigma v}_3$, in terms of the other two asymmetric rates.

\begin{figure}[htb!]
\centering
\includegraphics[scale=0.55]{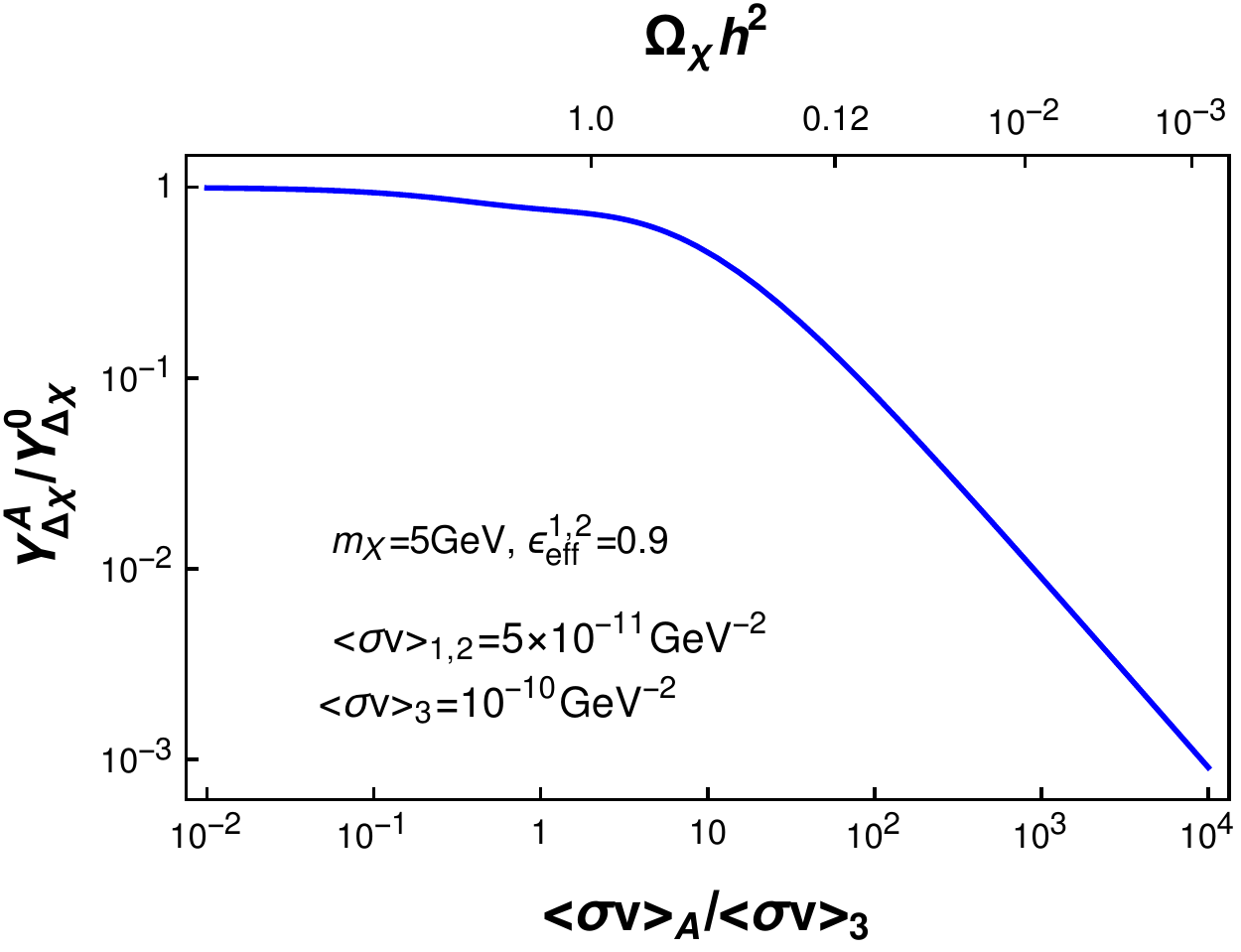}
\includegraphics[scale=0.5]{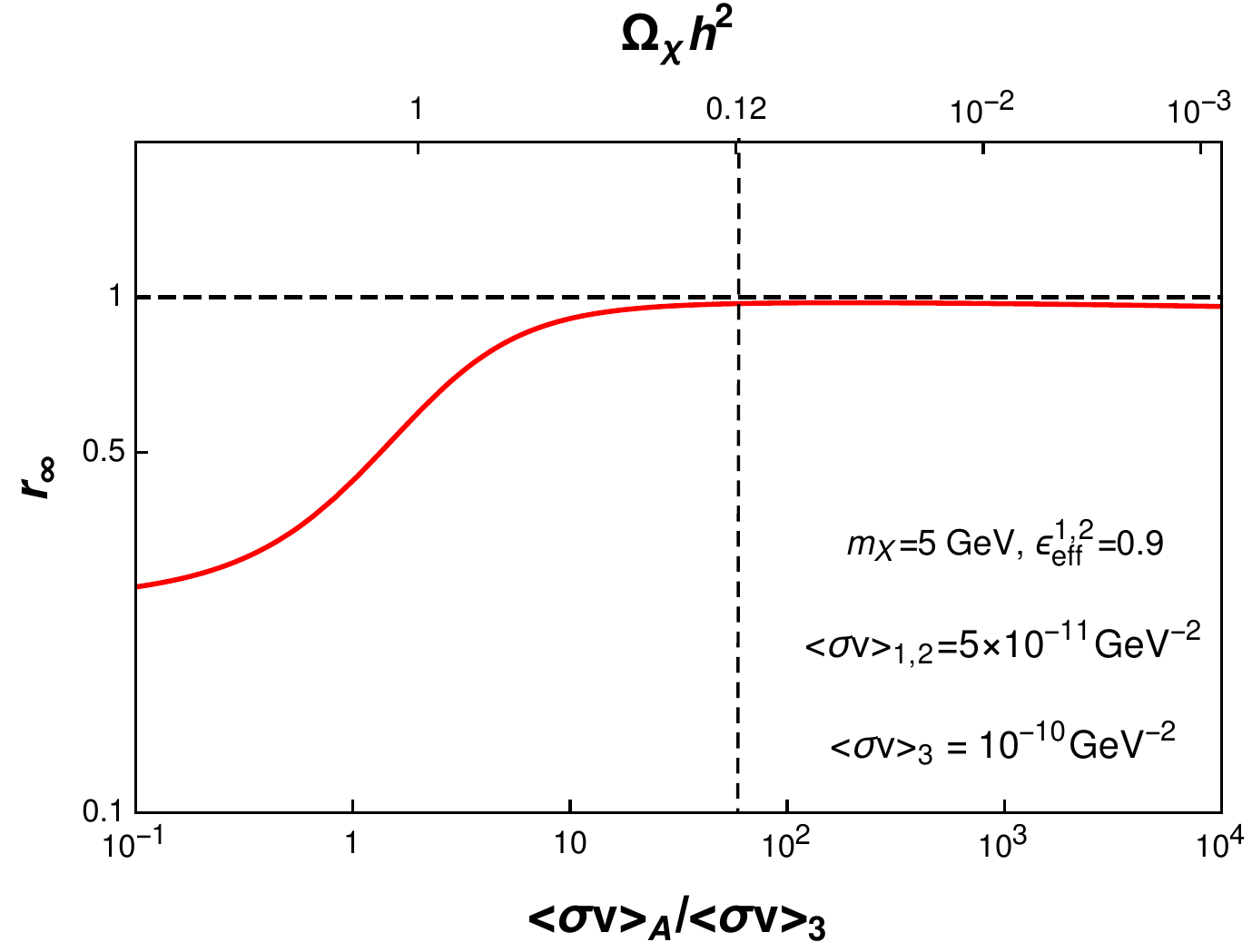}
\caption{\small{\it{Left Panel:  The ratio, $Y_{\Delta \chi}^A/Y_{\Delta \chi}^0$, where $Y_{\Delta \chi}^A$ and $Y_{\Delta \chi}^0$ are the asymptotic values of the yield $Y_{\Delta \chi}$, with and without the CP-conserving annihilation process, respectively, as a function of $\expval{\sigma v}_A/\expval{\sigma v}_3$. Also shown are the corresponding values of the DM relic abundance $\Omega_\chi h^2$. Right Panel: The final particle-antiparticle asymmetry parameter $r_\infty=\lvert Y_{\Delta\chi} \rvert / Y_S$, as a function of $\expval{\sigma v}_A/\expval{\sigma v}_3$. See text for details. Both the figures are for the asymmetric DM model from scattering discussed in Sec.~\ref{sec:ADM_1}.}}}
\label{FigZ2}
\end{figure}

Solving Eqs.~\ref{asym} numerically we can determine the impact of the CP-conserving annihilation process in modifying the asymmetric yield. To this end, we show the ratio $Y_{\Delta \chi}^A/Y_{\Delta \chi}^0$, where $Y_{\Delta \chi}^A$ and $Y_{\Delta \chi}^0$ are the asymptotic values of the yield $Y_{\Delta \chi}$, with and without the CP-conserving annihilation process, respectively. We show the yield ratio as a function of $\expval{\sigma v}_A/\expval{\sigma v}_3$ in Fig.~\ref{FigZ2} (left panel), which is the ratio of the CP-conserving and CP-violating annihilations that control the out-of-equilibrium number densities. Here, we have kept the symmetric and asymmetric reaction rates for all the CP-violating processes, and the DM mass, as fixed. For 
$\expval{\sigma v}_A \gtrsim \expval{\sigma v}_3$, we see that $Y_{\Delta \chi}$ reduces significantly, by a factor of $10$ for $\expval{\sigma v}_A/\expval{\sigma v}_3 \sim 75$. This feature can also be seen by solving Eqs.~\ref{asym} analytically near freeze-out, with the following result, as discussed in detail in Ref.~\cite{Ghosh:2021qbo}:
\begin{equation}
|Y_{\Delta\chi}(x)|= \dfrac{2 H x}{s}\,\dfrac{\expval{\epsilon\sigma v}_S}{\left(2\expval{\sigma v}_A + \expval{\sigma v}_3 \right)\expval{\sigma v}_{all}+\expval{\epsilon \sigma v}^2_S} .
\label{analytic}
\end{equation}
Thus we see that as the CP-conserving reaction rate $\expval{\sigma v}_A $ increases, $Y_{\Delta\chi}(x)$ is reduced, which, in turn, tends to reduce the observed particle-antiparticle asymmetry as well. However, there is a second role of these CP-conserving pair-annihilations, which appears in the epoch after the CP-violating reactions are frozen out. The symmetric component of the DM is then subsequently eliminated by the $\chi+\chi^\dagger \rightarrow  \phi+\phi$ reaction, thus modifying the ratio of particle anti-particle yields at the late epochs.

In order to demonstrate the second effect, it is useful to define the final particle-antiparticle asymmetry parameter $r_\infty$ as follows: 
\begin{equation}
r_\infty =\lvert Y_{\Delta\chi} \rvert / Y_S
\end{equation}
where the asymptotic values of the yields with $x=M_{N_1}/T \rightarrow \infty$ have been used. Clearly, $0 \leq r_\infty \leq 1$, where $r_\infty=0$ corresponds to the completely symmetric limit, in which the asymptotic yields of the DM and anti-DM are the same, while $r_\infty=1$ corresponds to the completely asymmetric limit, in which only either the DM or the anti-DM species survives. In the right panel of Fig.~\ref{FigZ2}, we show the $r_\infty$ parameter as a function of $\expval{\sigma v}_A/\expval{\sigma v}_3$, for fixed values of all the CP-violating rates. Similar to the example for leptogenesis from decays, we have defined an \textit{effective} CP-violation parameter (independent of the particle momenta) for each channel as $\epsilon^f _{\rm eff}=\expval{\epsilon \sigma v}_f/\expval{\sigma v}_f$. We find that as we increase the value of $\expval{\sigma v}_A/\expval{\sigma v}_3$, $r_\infty \rightarrow 1$, and a completely asymmetric DM with the required relic abundance is obtained for $\expval{\sigma v}_A/\expval{\sigma v}_3 \sim 60$. This feature is obtained due to the second role of the CP-conserving process. However, the competition between the two roles is not visible in this figure, which becomes apparent in the second example of ADM production from semi-annihilation discussed in the next subsection.

We note in passing that for our choice of parameters, $\expval{\sigma v}_A/\expval{\sigma v}_3 \sim 60$ is also necessary to satisfy the DM relic abundance requirement of $\Omega_\chi h^2 =0.12$. We have shown the value of $\Omega_\chi h^2$ for different values of $\expval{\sigma v}_A/\expval{\sigma v}_3 $ in Fig.~\ref{FigZ2} (both panels) as well. For $\expval{\sigma v}_A/\expval{\sigma v}_3 > 100$, the DM species is underabundant. This is again due to the fact that the CP-conserving process here plays a dual role of both changing  $Y_{\Delta \chi}$ at the earlier epoch, as well as reducing the symmetric DM component, and the latter eventually reduces the net relic density.

\subsubsection{Asymmetric DM from semi-annihilation}
\label{sec:ADM_2}
In the second example of type B-I, we consider the possibility of producing asymmetric dark matter from semi-annihilation~\cite{Ghosh:2020lma}, in which the dual role of the CP-conserving processes is demonstrated clearly. At this point we remind ourselves with the interaction Lagrangian, consistent with the $Z_3$ symmetry as in Chapter \ref{chap:chap3}. 
\begin{equation}
-\mathcal{L}_{\rm int} \supset \frac{1}{3!} \left(\mu \chi^3 + {\rm h.c.} \right) + \frac{1}{3!} \left(\lambda \chi^3 \phi + {\rm h.c.} \right) + \frac{\lambda_1}{4} \left(\chi^\dagger \chi \right)^2 +   \frac{\lambda_2}{2} \phi^2 \chi^\dagger \chi + \mu_1 \phi  \chi^\dagger \chi +  \frac{\mu_2}{3!} \phi^3 +  \frac{\lambda_3}{4!} \phi^4.
\label{eq:lag_z3}
\end{equation}
In addition, there will be interactions between $\phi$ and the SM Higgs doublet $H$, exactly as in Eq.~\ref{eq:lag_z2}. To realize the semi-annihilation scenario, for the necessary CP-violation through the interference of the tree and one-loop graphs, the two complex couplings $\mu$ and $\lambda$ are required in general, so that one complex phase remains after field redefinitions.  

\begin{table}[h!]
\begin{center}
\begin{tabular}{|c|c|c|c|c|}
\hline
Process & CP & DM Number & Rate & Rate \\
              &      & Violation  & (Symmetric) & (Asymmetric) \\
\hline
$\chi+\chi \rightarrow \chi^{\dagger}+\phi$ & violating & 3 units & $\expval{\sigma v}_1$ &  $\expval{\epsilon \sigma v}_1$\\
\hline
$\chi+\chi \rightarrow \chi^{\dagger}+\phi+\phi$ & violating & 3 units & $\expval{\sigma v}_2$  &  $\expval{\epsilon \sigma v}_2$ \\
\hline
$\chi+\chi \rightarrow \chi^{\dagger}+\chi^{\dagger}+\chi$ & violating & 3 units & $\expval{\sigma v}_3$ & $\expval{\epsilon \sigma v}_3$  \\
\hline
$\chi+\chi^{\dagger} \rightarrow \phi +\phi$ & conserving & 0 units & $\expval{\sigma v}_A$ & NA \\
\hline
\end{tabular}
\end{center}
\caption{\small{\it{Relevant processes for DM cosmology for the ADM from semi-annihilation scenario, their properties, and notation used to denote the corresponding thermally averaged symmetric and asymmetric reaction rates.}}}
\label{Tab:tab2}
\end{table}%

These interactions lead to several relevant processes for the DM cosmology, which are shown in Table~\ref{Tab:tab2}, along with their properties and the notation used to denote the different thermally averaged reaction rates.

The Boltzmann equations for the symmetric ($Y_S=Y_\chi+Y_{\chi^\dagger}$) and asymmetric yields ($Y_{\Delta \chi}=Y_\chi -Y_{\chi^\dagger}$) in this scenario are given by:

\begin{align}
\dfrac{dY_S}{dx} &= -\dfrac{s}{8Hx}\bigg[\expval{\sigma v}_S \left(Y^2_S+Y^2_{\Delta\chi}-2 Y_0 Y_S\right)+\expval{\sigma v}_3\left(Y^2_S\left(\dfrac{Y_S}{2Y_0}-1\right)-Y^2_{\Delta\chi}\left(\dfrac{Y_S}{2Y_0}+1\right)\right) \nonumber \\
 &+ 4\expval{\sigma v}_A\left(Y^2_S-Y^2_{\Delta \chi}-4Y^2_0\right)+\expval{\epsilon\sigma v}_{3}\dfrac{Y_{\Delta\chi}}{2Y_0}\left(Y^2_S-Y^2_{\Delta\chi}+ 4Y^2_0-8 Y_0 Y_S\right)\bigg ] \nonumber\\
\dfrac{dY_{\Delta \chi}}{dx} &= -\dfrac{3 s}{4Hx}\bigg[\expval{\sigma v}_S Y_{\Delta \chi} \left(Y_S+Y_0\right)+\expval{\sigma v}_3 Y_{\Delta \chi}\left(Y_S+\dfrac{Y^2_S-Y^2_{\Delta\chi}}{4Y_0}\right) \nonumber \\
&+ \expval{\epsilon\sigma v}_{3}\dfrac{Y_S}{4Y_0}(Y^2_S-Y^2_{\Delta\chi}-4Y^2_0)\bigg],
\label{boltz:Z3}
\end{align}  
where, we have defined $\expval{\sigma v}_S = \expval{\sigma v}_1+\expval{\sigma v}_2$. Here, the thermally averaged rate $\expval{\sigma v}_3$ has a significant $x-$dependence, given by  $\expval{\sigma v}_3 =\dfrac{1+2x}{\sqrt{\pi x}} e^{-x} (\sigma v_3)_{s}$, where $(\sigma v_3)_{s}$ is an $x-$independent  s-wave piece, and the exponential suppression factor stems from the phase-space cost for producing an extra particle in the final state~\cite{Bhatia:2020itt}. 

\begin{figure}[htb!]
\centering
\includegraphics[scale=0.55]{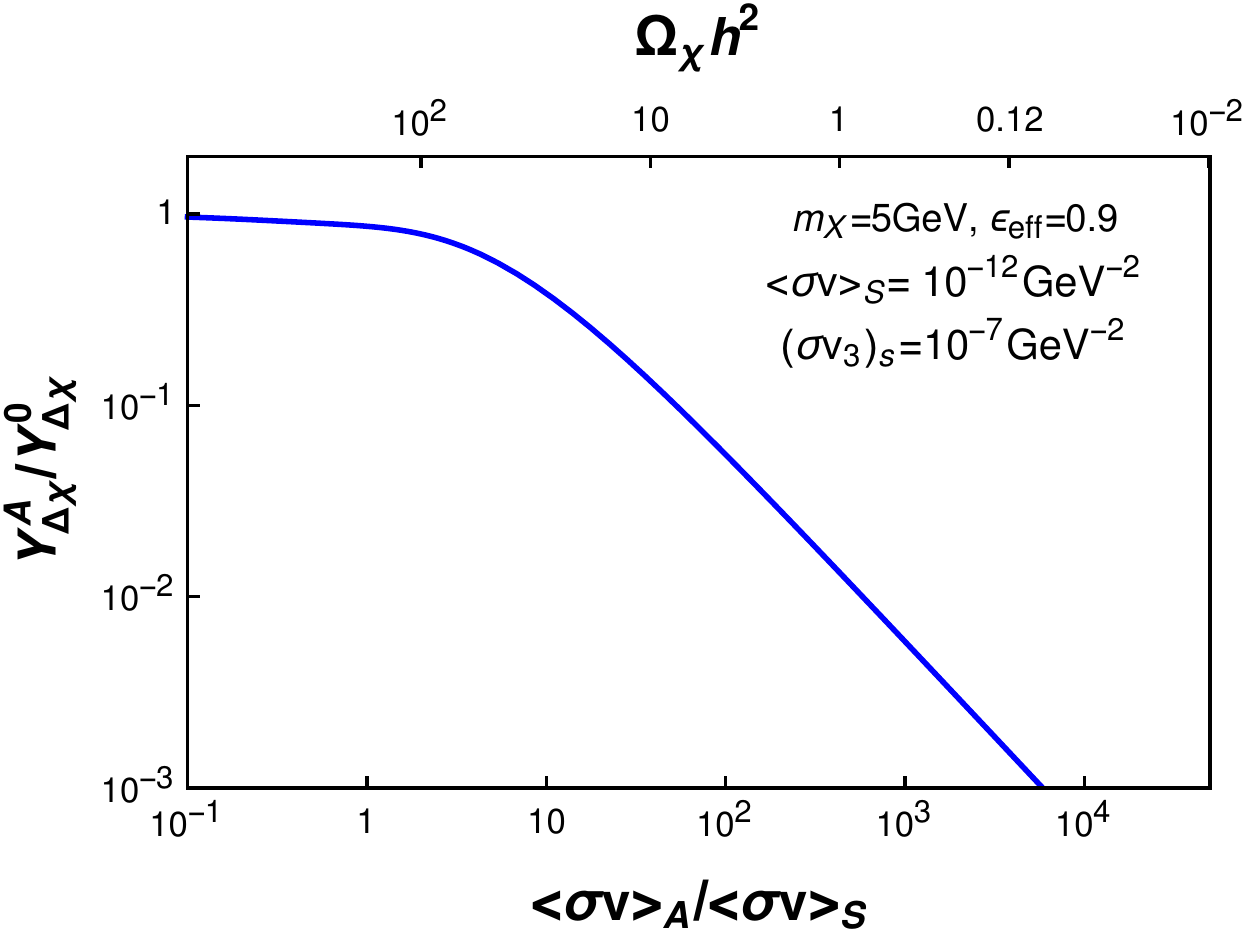}
\includegraphics[scale=0.55]{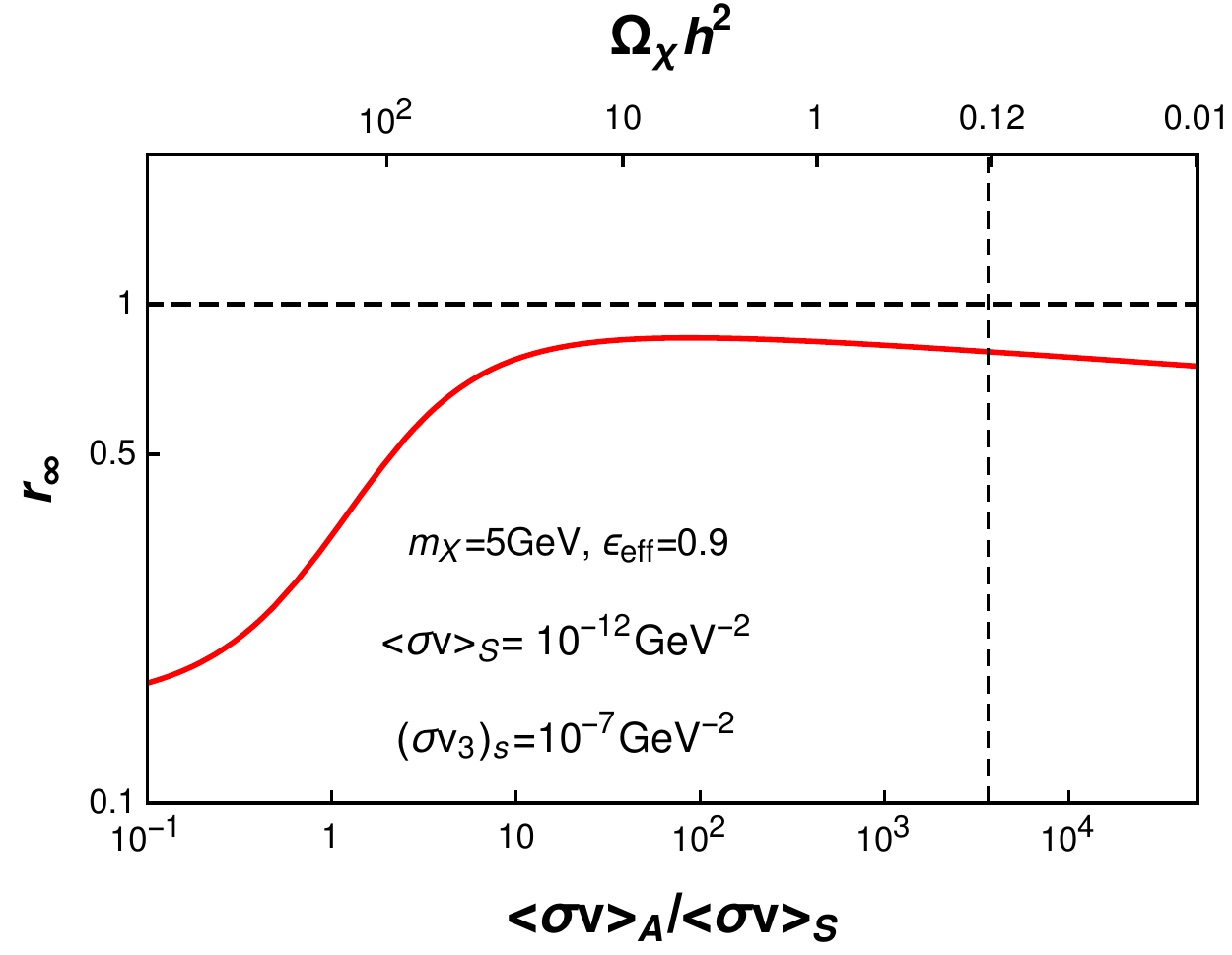}
\caption{\small{\it {Same as Fig.~\ref{FigZ2}, for the asymmetric DM  from semi-annihilation scenario discussed in Sec.~\ref{sec:ADM_2}. See text for details.}}}
\label{FigZ3}
\end{figure}

In passing, we note that the  $\chi+\chi \rightarrow \chi^{\dagger}+\chi^{\dagger}+\chi$ reaction decouples earlier than the semi-annihilation processes, but it is essential for the thermal equilibrium of DM species, $\chi$. In particular, the ratio of the the thermally averaged cross-section of $\chi+\chi \rightarrow \chi^{\dagger}+\chi^{\dagger}+\chi$ to that of the semi-annihilation is given by 
\begin{align}
\frac{\expval{\sigma v_3}}{\expval{\sigma v}_S}\simeq 10^{-3}~~\dfrac{1+2x}{\sqrt{\pi x}} e^{-x}\left(\frac{\mu}{m_\chi}\right)^2 \left(\frac{\lambda_1}{\lambda}\right)^2
\end{align}
where, $\lambda_1$ is related to the pair annihilation and $\lambda$ is related to the semi-annihilation. For $\mu < m_\chi$, the semi-annihilation dominates in the non-relativistic regime due to exponential suppression in the former process. 

Now, by solving Eqs.~\ref{boltz:Z3} numerically, we find the ratio of the asymmetric yields, as before, with and without the CP-conserving annihilation process, and show the results in the left panel of Fig.~\ref{FigZ3}, as a function of $\expval{\sigma v}_A /\expval{\sigma v}_S $. In the right panel of the same figure, we also show the final asymmetry parameter $r_\infty$ as defined in the previous subsection. 

Compared to the scenario in the previous subsection, the $r_\infty$ curve shows more features with the change in $\expval{\sigma v}_A /\expval{\sigma v}_S $. Initially, as we increase $\expval{\sigma v}_A$, $r_\infty$ approaches the completely asymmetric limit of $1$. However, on further increase of the ratio $\expval{\sigma v}_A /\expval{\sigma v}_S \gtrsim 100$, the value of $r_\infty$ is reduced instead. As before, there is a dual role played by the CP-conserving process -- that of reducing the value of $Y_{\Delta \chi}$ to begin with, and thus trying to reduce the asymmetry on the one hand, and at the same time of removing the symmetric component of the DM particles, and thus generating a competing effect to increase the final asymmetry parameter $r_\infty$ signifying the modification of the relative particle anti-particle yields. The initial increase of $r_\infty$ is observed as the second effect dominates for moderately large values of $\expval{\sigma v}_A /\expval{\sigma v}_S$. However, for very large values of $\expval{\sigma v}_A /\expval{\sigma v}_S$, the first effect dominates, reducing the $r_\infty$ to be lower than $1$. Thus the competition between the two effects at different epochs due to the same CP-conserving process is brought out clearly in this example of asymmetric dark matter production from semi-annihilation.

\subsection{Summary}
\label{sec:sec55}
To summarize, in this chapter, we have revisited the role of CP-conserving processes in generating particle-antiparticle asymmetries in the early Universe -- either in the lepton sector for baryogenesis through leptogenesis, or in the dark matter sector. We have focussed on scenarios in which either both the CP-violating and conserving processes are decays, or both of them are scatterings, thus naturally being of comparable rates. For this reason, in the examples considered by us, the effect of the CP-conserving reactions is found to be highly significant. This is in contrast to the scenarios for leptogenesis in different see-saw models of neutrino mass, in which the role of CP-conserving reactions were explored earlier, and found to be mostly sub-dominant. In those scenarios, the primary source of CP-violation was a decay process, while the CP-conserving reaction was a scattering, and thus the latter process is generally Boltzmann suppressed. 

Within each scenario above, the asymmetry may be produced either in the mother sector, or in the daughter sector, leading to distinct effects of the CP-conserving processes in each case. As an example for a scenario in which both the CP-violating and conserving processes are decays and the asymmetry is generated in the daughter sector, we discussed a low-scale model for leptogenesis. It is shown that as the rate for the CP-conserving decays is increased compared to that of the CP-violating one, the lepton asymmetry yield proportionately decreases, and can vary by orders of magnitude for natural values of the model parameters. This is simply because, in such cases the CP-conserving decays remain longer in equilibrium, reducing the number density of the non-relativistic mother particles, and thereby reducing the generated lepton asymmetry. 

We then discussed two examples of asymmetric dark matter production in which both the CP-violating and conserving reactions are scatterings, and the asymmetry is generated in the mother sector. The latter feature leads to an interesting dual role played by the CP-conserving reactions. Initially, when the CP-violating reactions are active, the larger the rate of the CP-conserving scatterings, the smaller is the asymmetric yield of the particle-antiparticle system. However, subsequent to the freezing out of the CP-violating reactions, the CP-conserving pair annihilations remove the symmetric component of the DM, thereby enhancing the final asymmetry parameter, signifying the modification of the relative particle anti-particle yields. This dual role played by the same CP-conserving reaction leads to a novel competing effect observed in the ADM models. Thus CP-conserving processes can play qualitatively distinct roles in generating cosmological particle-antiparticle asymmetries, and can modify the asymmetric yields by orders of magnitude.

%%%%%%%%%%%%%%%%%%%%%%%%%%%%%%%%%%%%%%%

\newpage
\section{\Large{Summary of the thesis}}
In this thesis, we have studied a general class of particle DM production in the early universe. In particular, we have demonstrated several scenarios of producing symmetric, completely asymmetric and partially asymmetric densities of DM from the decoupling of thermal processes. Thermal production of DM final density is independent of initial conditions, thereby the paradigm is the most studied one in the literature. In this context, we have broadly discussed about the genesis of a non-zero chemical potential of DM and its implication in deciding the observed relic. It is clear that simple pair annihilation of DM is insufficient to generate particle-antiparticle asymmetry in the early universe because it violates neither the DM number nor the $CP$-symmetry, which are essential requirements for generating an asymmetry. 

In most propositions, the DM asymmetry is generated simultaneously with the visible sector asymmetry, considering the both sectors share an equal and opposite charge related to some global $U(1)$ symmetry. Consequently, the final number densities of DM and the visible sector become comparable, which in turn constrain the DM mass to be $\mathcal{O}(1-10)GeV$. In this type of scenarios, the non-zero chemical potential in each of the sectors is generated either from the decay or scattering of some heavy particle which couples with both the sectors. Thus, at low energies one can integrate out the heavy particle to generate $O_{SM} O_{DM}$ type operators which preserve the combined charge of the both sector. In the absence of subsequent charge violating process the individual asymmetries remain preserved for the rest of the cosmic history. We have described a different class of models in which the asymmetry in the DM sector is generated independent of the visible sector asymmetry. In particular, DM asymmetry is generated via DM scatterings only, like semi-annihilation, self-scatterings and annihilation. These processes are in chemical equilibrium in the early universe and eventually decouple so that the generated asymmetry does not wash out. The initial abundance of DM particles is thermal as the thermalization is ensured by sufficient interaction with the SM bath. In contrast with the former cases, the mass of DM is rather relaxed and can vary from GeV to TeV, depending on the final asymmetry contained in the relic density. 

As an example scenario, we first discussed the role of the semi-annihilation of DM in producing an asymmetry between particle-antiparticle, at the same time reducing the total particle and antiparticle number. When the CP-violating in semi-annihilation process is a maximal one, the DM sector becomes completely asymmetric at the decoupling of the process. Consequently, a subsequent pair annihilation process becomes irrelevant, unlike in most  baryogenesis and leptogenesis scenarios, where the symmetric component of the relic is removed by the same. The importance of semi-annihilation-type scenarios open the possibility of indirect signal of a completely asymmetric dark sector, as a semi-annihilation produces a neutral particle which decays or mixes with the SM particles. Depending on the mass of the neutral particle and mixing strength with the SM sector, it can produce cosmic rays and gamma rays. This should be contrasted with the general lore about ADM that it is a nightmare situation for indirect detection of DM as it does not undergo pair annihilation in the present universe. Nevertheless, if there is an partial asymmetry then both the pair annihilation and the semi-annihilation contribute to the probable indirect detection in general, making the distinction between a WIMP DM and an ADM a bit obscure.   

The next example is based on $\mathcal{Z}_2$ symmetric DM model in which we get a plethora of DM scatterings. There are several CP-violating processes which include self-scatterings and annihilation of DM. In particular, the CP-violating self-scattering and annihilation coming from the same initial states are related by $S$-matrix unitarity. In addition, there are CP-conserving self-scatterings and annihilation are also present. The DM asymmetry is generated due to a novel interplay between CP-violating self-scatterings and annihilation. Unlike the semi-annihilation scenario, the pair annihilation is essential to remove symmetric part. In presence of sufficient pair annihilation the relic density is decided predominantly by the asymmetry generated by the self-scatterings. Therefore, the relic density is controlled by the self-scatterings though these do not change the total particle-antiparticle number. 

In above scenario, we have found a dual role of the CP-conserving process. In our model, the rate of CP-conserving process is naturally of the same order that of CP-violating ones.  The CP-conserving process initially suppresses the asymmetry generation by controlling the out-of-equilibrium number densities of the DM particles, but subsequently modify the ratio of particle antiparticle yields at the present epoch by eliminating the symmetric component of the DM sector through pair-annihilation, leading to a competing effect stemming from the same process at different epochs. This competing effect is a ramification of the fact that the DM asymmetry is generated from DM scatterings. Nevertheless, we observe the first role of the CP-conserving process in baryogenesis and leptogenesis scenarios too. We have demonstrated a low scale leptogenesis in which the CP-conserving decay naturally can compete with the usual CP-violating one. The generated lepton asymmetry can vary several order of magnitude depending upon the relative size of the CP-conserving and violating reaction rates.

\newpage
\appendix
\section{\Large{Illustration of Boltzmann H-theorem}}
\label{app:A}

For the illustration of Boltzmann H-theorem, we take two processes, namely $\chi\chi \rightarrow \psi \psi$, $\chi\chi \rightarrow \phi \phi$ and their corresponding backward processes in consideration. Now, we need Boltzmann equations for all three particles as the $H$-function involves all possible collision terms as the following.
\begin{align}
\frac{d\tilde{H}}{dt} =\sum_\alpha  (1+\log f_\alpha)\, \frac{df_\alpha}{dt}=\sum_\alpha (1+\log f_\alpha)\, C[f_\alpha].
\label{eq:htheorem}
\end{align}
As stated in the Chapter \ref{chap:chap2}, $\alpha$ runs over particle species and all possible momenta and quantum numbers, then the collision terms for $\chi$, $\psi$ and $\phi$ are given by
\begin{align}
\int\frac{d^3 p_1}{(2\pi)^3}C[f_\chi(p_1)](1+f_\chi(p_1)) = -\int d\Pi\, (1+f_\chi(p_1))\bigg[f_{\chi}(p_1)f_\chi(p_2)\bigg(|M|^2_{\chi\chi\rightarrow \psi\psi}+|M|^2_{\chi\chi\rightarrow \phi\phi}\bigg)\hspace{2.0cm}\nonumber\\
-f_{\psi}(p_3)f_\psi(p_4) |M|^2_{\psi\psi\rightarrow \chi\chi}
-f_{\phi}(p_3)f_\phi(p_4) |M|^2_{\phi\phi\rightarrow \chi\chi}\bigg],\hspace{2.0cm}
\end{align}
\vspace{-0.8cm}
\begin{align}
\int\frac{d^3 p_1}{(2\pi)^3}C[f_\psi(p_1)](1+f_\psi(p_1)) = -\int d\Pi (1+f_\psi(p_3))\bigg[f_{\psi}(p_3)f_\psi(p_4) |M|^2_{\psi\psi\rightarrow \chi\chi}\nonumber\\
-f_{\chi}(p_1)f_\chi(p_2)|M|^2_{\chi\chi\rightarrow \psi\psi}\bigg],
\end{align}
\vspace{-0.8cm}
\begin{align}
\int\frac{d^3 p_1}{(2\pi)^3}C[f_\phi(p_1)](1+f_\phi(p_1)) = -\int d\Pi (1+f_\phi(p_3))\bigg[f_{\phi}(p_3)f_\phi(p_4) |M|^2_{\phi\phi\rightarrow \chi\chi}\nonumber\\
-f_{\chi}(p_1)f_\chi(p_2)|M|^2_{\chi\chi\rightarrow \phi\phi}\bigg],
\end{align} 
where $\displaystyle{ d\Pi=(2\pi)^4 \delta^4(p_1+p_2-p_3-p_4)\prod^4_{i=1}\frac{d^3p_i}{(2\pi)^3 2E_i}}$ and we have suppressed time symbol for brevity. Now, we put the above expressions in Eq.\ref{eq:htheorem} and exploit the energy-momentum conservation law wherever is necessary.
\begin{align}
\frac{d\tilde{H}}{dt} = -\int \frac{d\Pi}{2} \bigg[\log\left(\frac{f_\chi(p_1)f_\chi(p_2)}{f_\psi(p_3)f_\psi(p_4)}\right)\bigg(f_\chi(p_1)f_\chi(p_2) |M|^2_{\chi\chi\rightarrow \psi\psi}- f_\psi(p_3)f_\psi(p_4) |M|^2_{\psi\psi\rightarrow \chi\chi}\bigg)\nonumber\\
+ \log\left(\frac{f_\chi(p_1)f_\chi(p_2)}{f_\phi(p_3)f_\phi(p_4)}\right)\bigg(f_\chi(p_1)f_\chi(p_2) |M|^2_{\chi\chi\rightarrow \phi\phi}- f_\phi(p_3)f_\phi(p_4) |M|^2_{\phi\phi\rightarrow \chi\chi}\bigg)\bigg].
\end{align}
Now, we can relate various scattering amplitudes using Eq.\ref{eq:defasy} and the $CPT$ theorem namely, $|M|^2_{\chi\chi\rightarrow f}=\frac{1+\epsilon_f}{2}|M|^2_f$,  $|M|_{f\rightarrow \chi \chi}=|M|^2_{\chi^\dagger\chi^\dagger\rightarrow f^\dagger}=\frac{1-\epsilon_f}{2}|M|^2_f$ and the above expression becomes after defining $F_i(p_1,p_2)=f_i(p_1)f_i(p_2)$,
\begin{align}
\frac{d\tilde{H}}{dt} = -\int \frac{d\Pi}{4} \bigg[\log\left(\frac{F_\chi(p_1,p_2)}{F_\psi(p_3,p_4)}\right)|M|^2_\psi\bigg(F_\chi(p_1,p_2)(1-\epsilon_\psi)-F_\psi(p_3,p_4)(1+\epsilon_\psi)\bigg)\nonumber\\
+ \log\left(\frac{F_\chi(p_1,p_2)}{F_\phi(p_3,p_4)}\right)|M|^2_\phi\bigg(F_\chi(p_1,p_2)(1-\epsilon_\phi)-F_\phi(p_3,p_4)(1+\epsilon_\phi)\bigg)\bigg].
\end{align}   
For a simplified demonstration, we assume $f_\psi(p)=f_\phi(p)$, and the above expression becomes using unitarity relation as in Eq.\ref{eq:epr} as,
\begin{align}
\frac{d\tilde{H}}{dt} = -\int \frac{d\Pi}{2} \bigg[\log\left(\frac{F_\chi(p_1,p_2)}{F_\psi(p_3,p_4)}\right)\bigg(F_\chi(p_1,p_2)-F_\psi(p_3,p_4) \bigg) |M|^2_T\bigg] \leq 0,
\end{align}
where, $|M|^2_T~ (=|M|^2_\psi+|M|^2_\phi)$ is a positive quantity.  Now, the integrand becomes positive quantity for $F_\chi(p_1,p_2)> F_\psi(p_3,p_4)$, $F_\chi(p_1,p_2)< F_\psi(p_3,p_4)$ and identically zero for $F_\chi(p_1,p_2)= F_\psi(p_3,p_4)$ . Hence, the Boltzmann $H$-theorem is proved for classical particles. Similar proof can be done including the quantum effect by appropriately defining the $H$-function.

\newpage
\section{\Large{CP-violation in the $\chi+\phi \rightarrow \chi^{\dagger}+\phi$ channel}}
\label{App.A}
\begin{figure}[h!]
\includegraphics[scale=0.30]{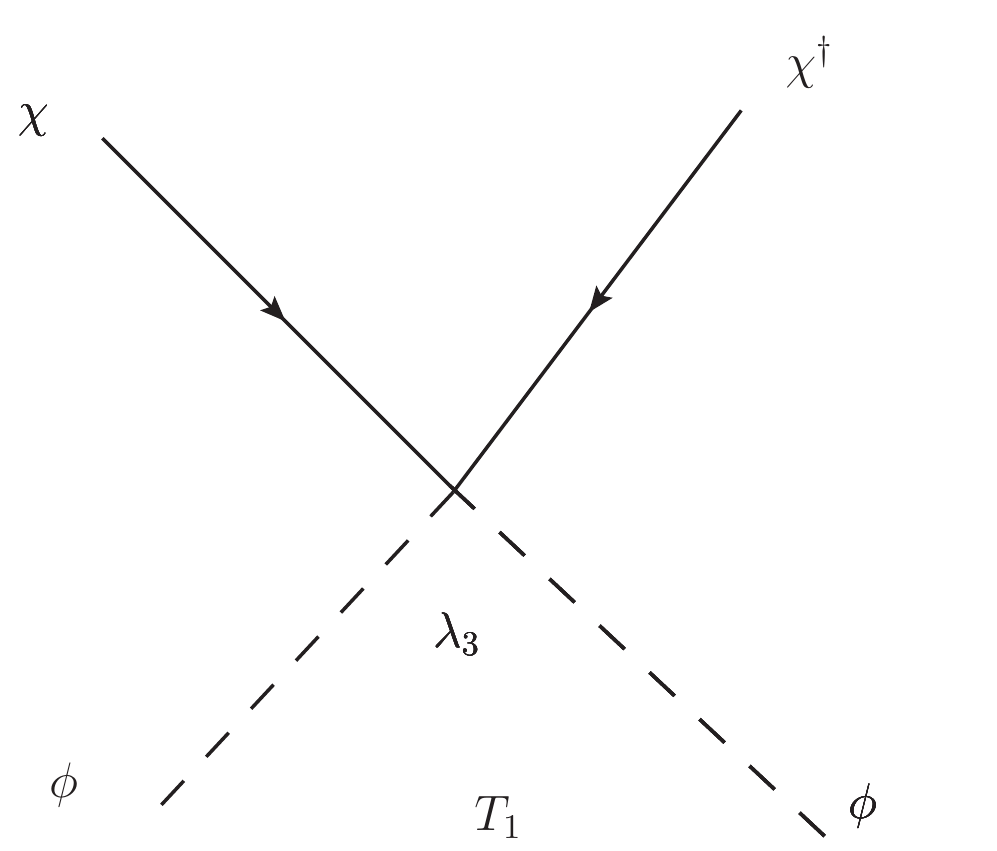}
\includegraphics[scale=0.30]{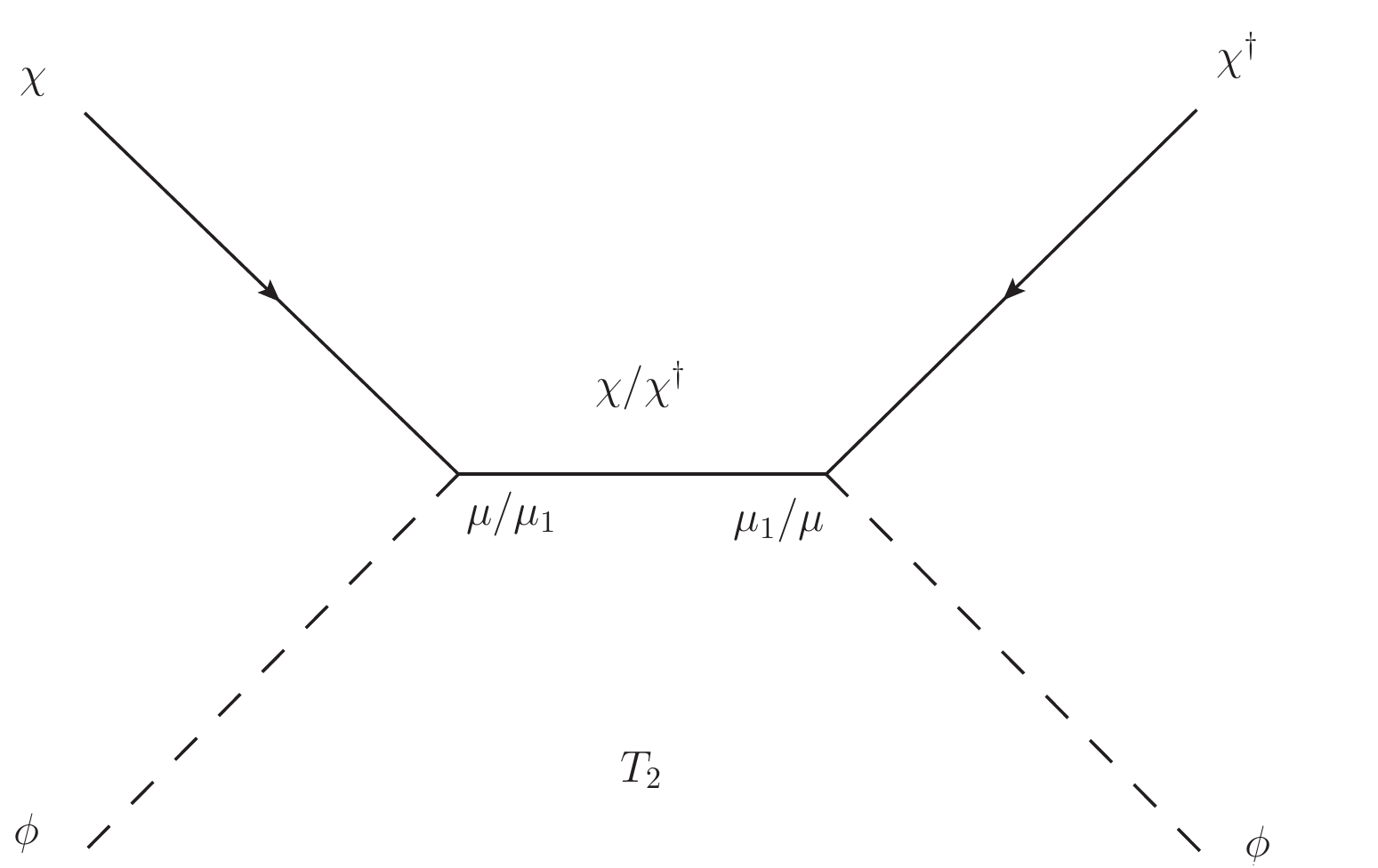}
\includegraphics[scale=0.45]{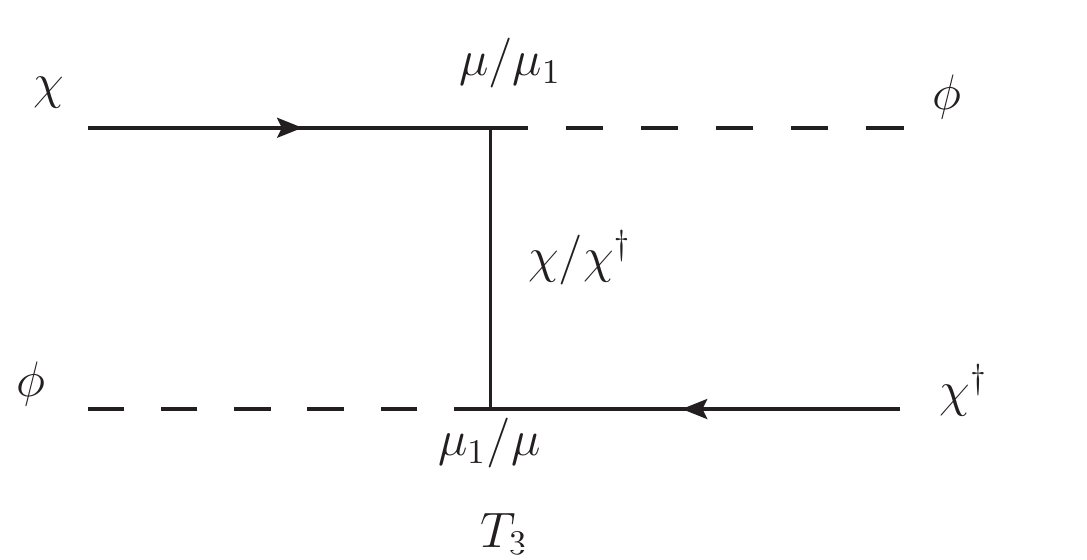}
\includegraphics[scale=0.4]{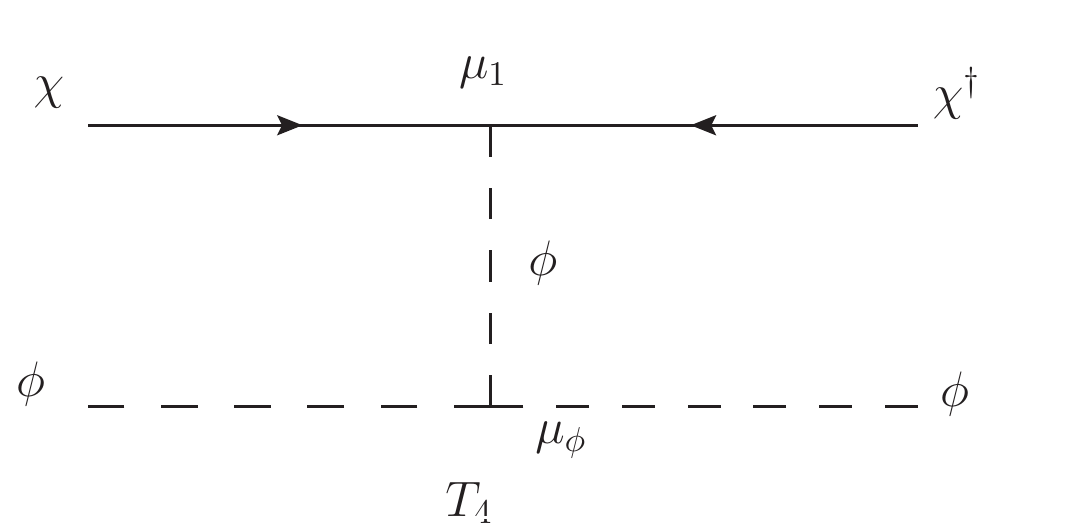}
\includegraphics[scale=0.39]{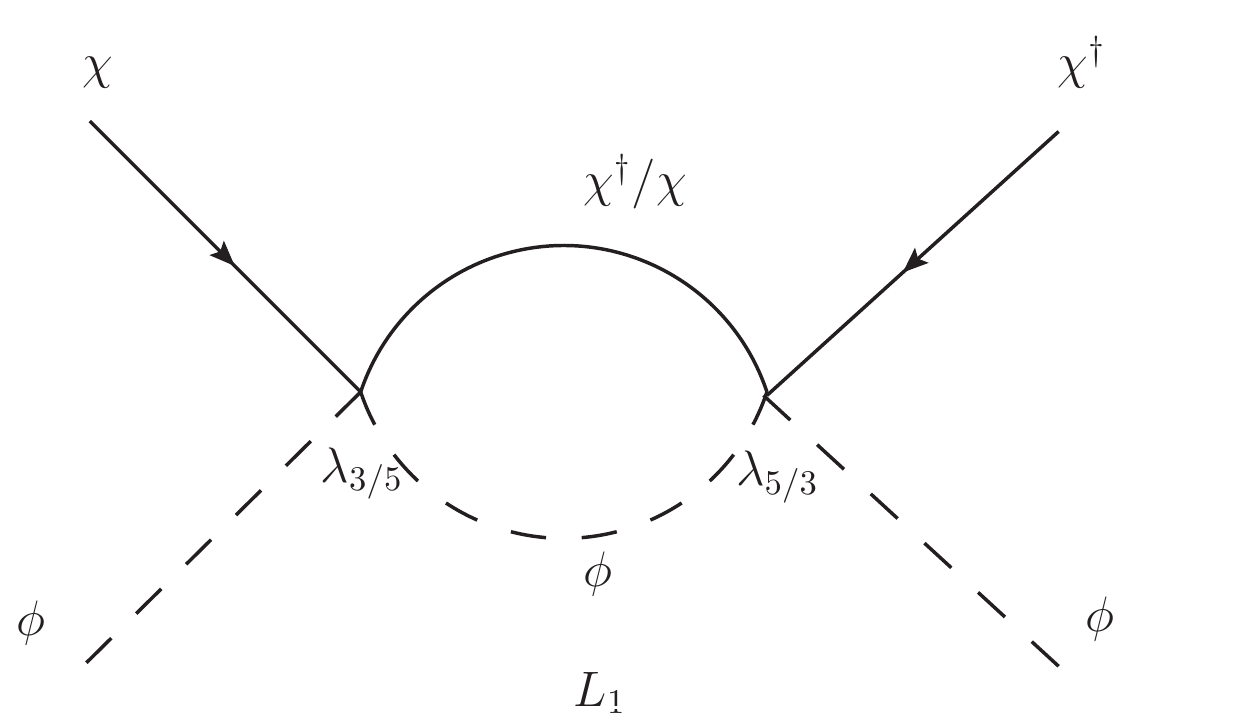}
\includegraphics[scale=0.39]{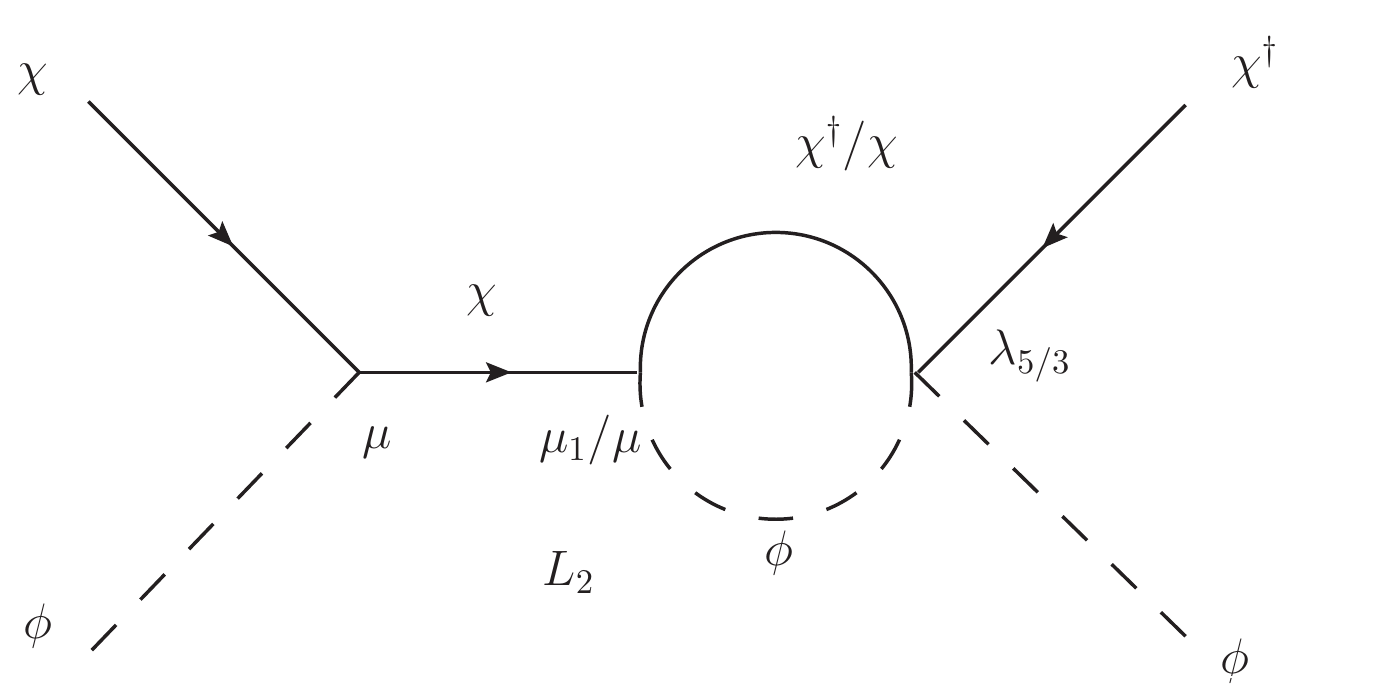}
\includegraphics[scale=0.35]{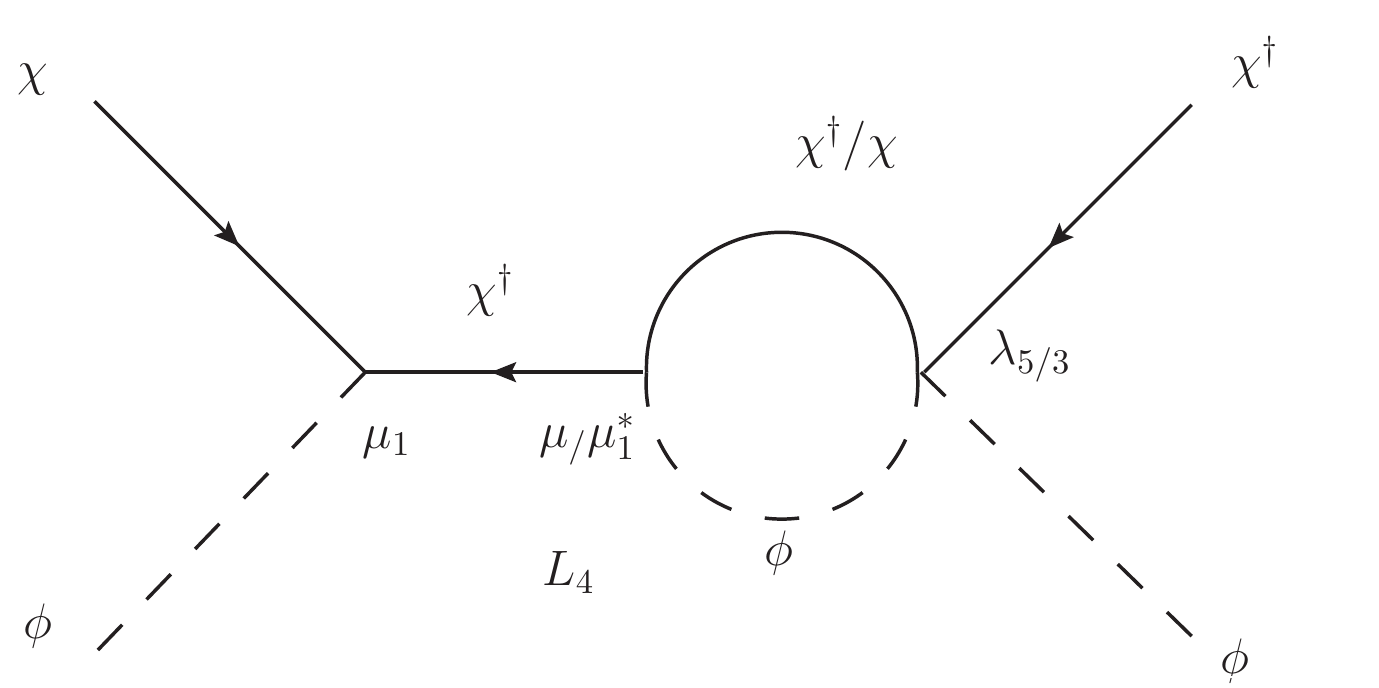}
\includegraphics[scale=0.35]{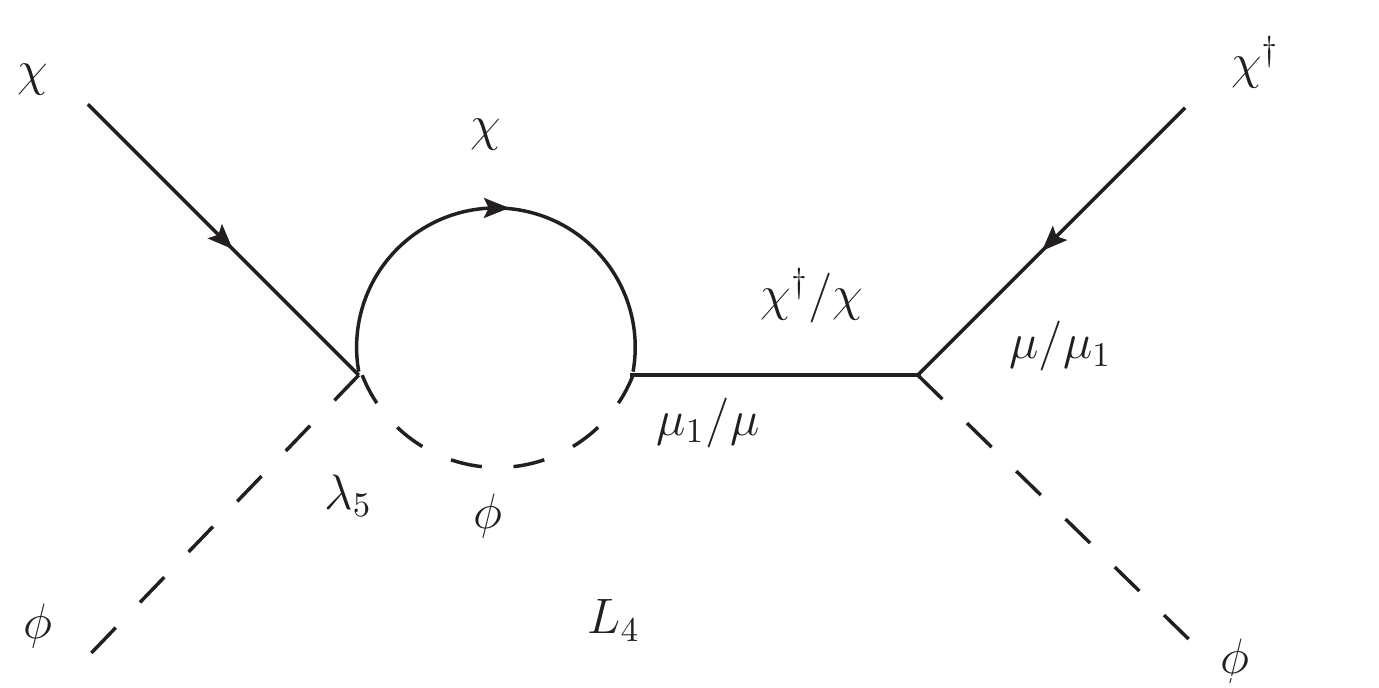}
\includegraphics[scale=0.35]{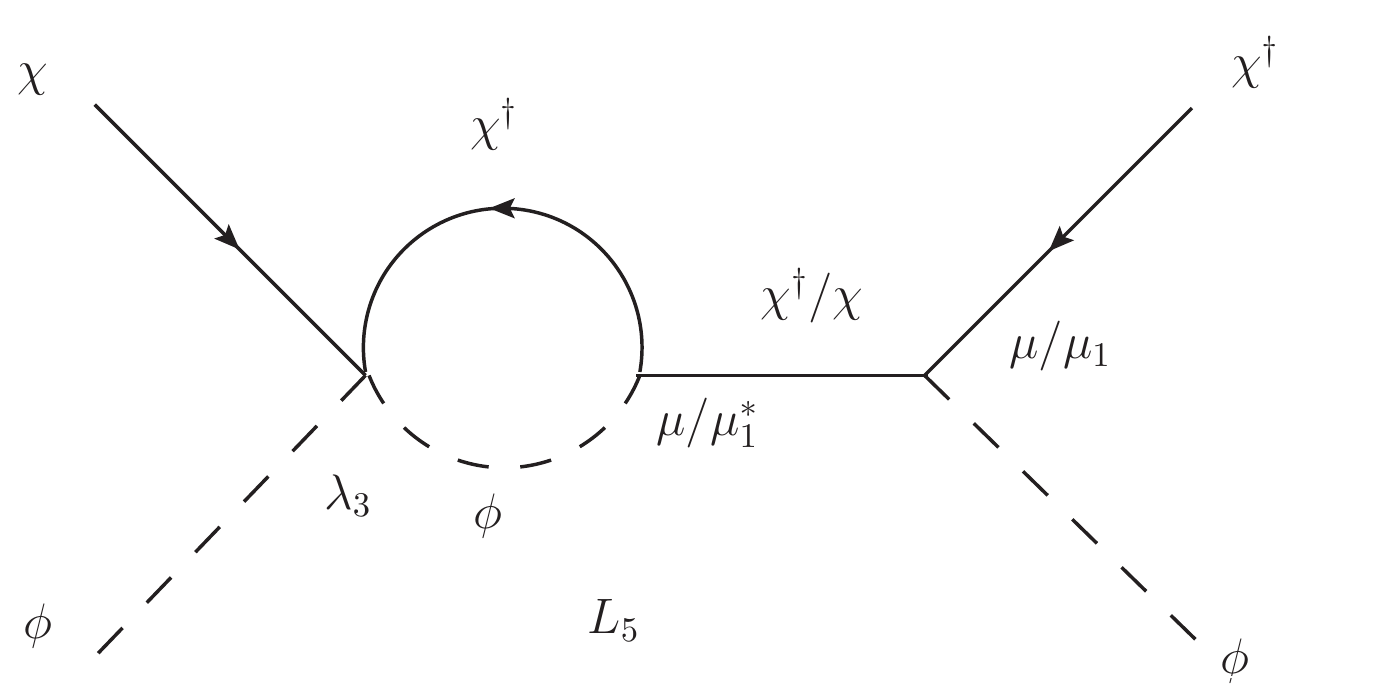}
\includegraphics[scale=0.36]{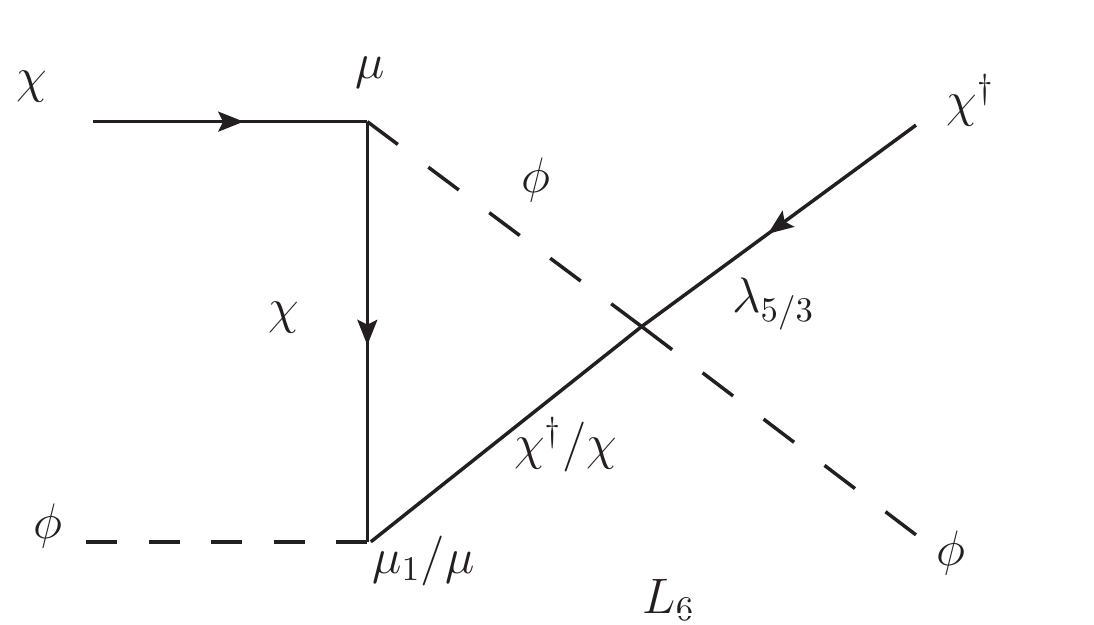}
\includegraphics[scale=0.36]{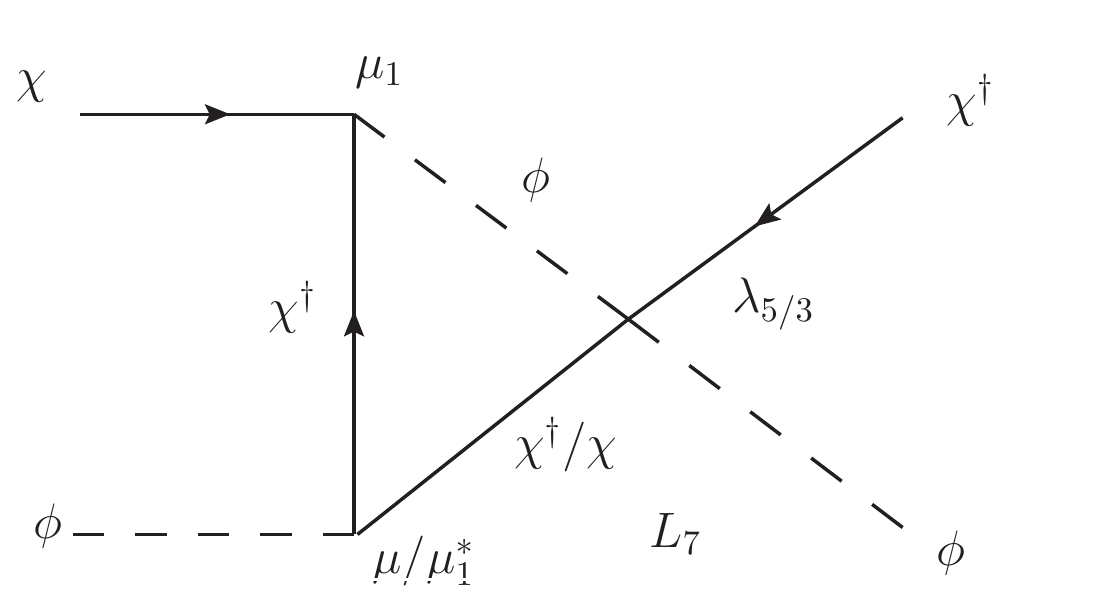}
\includegraphics[scale=0.36]{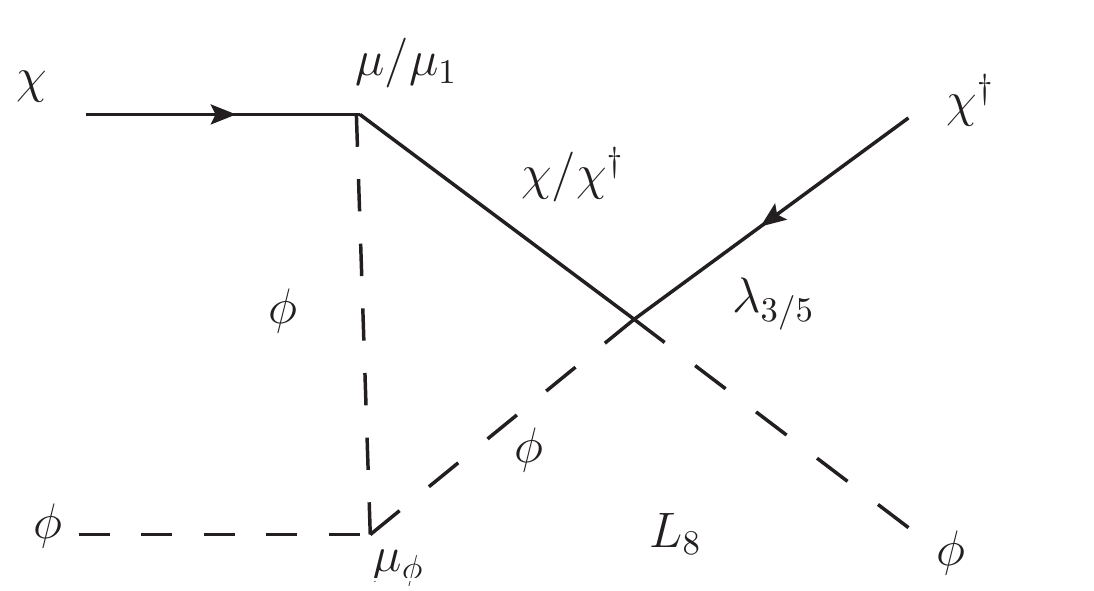}
\includegraphics[scale=0.36]{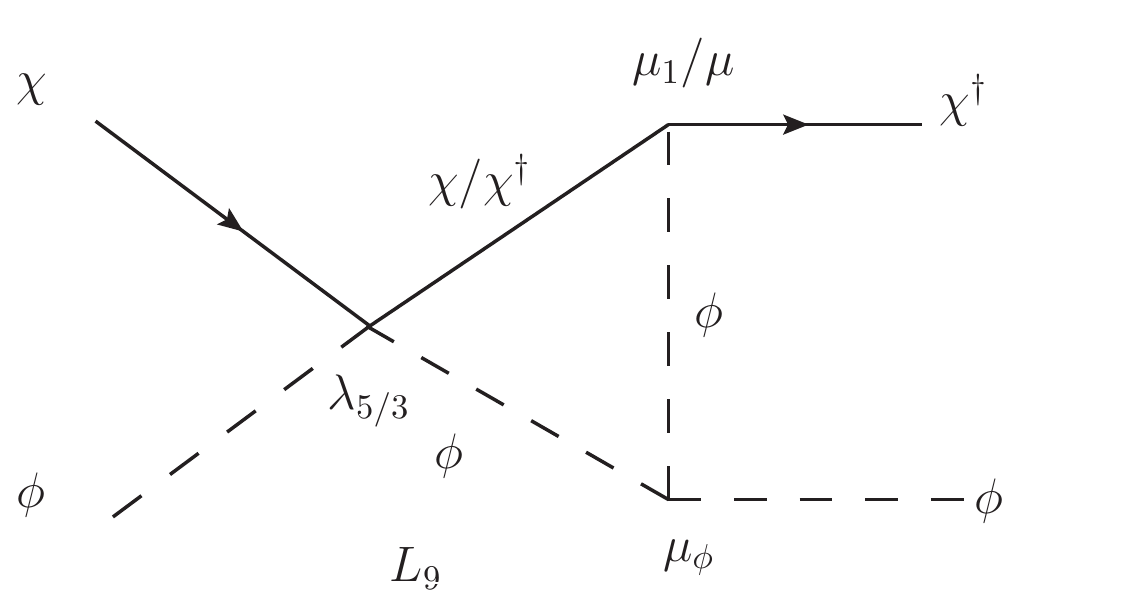}
\includegraphics[scale=0.36]{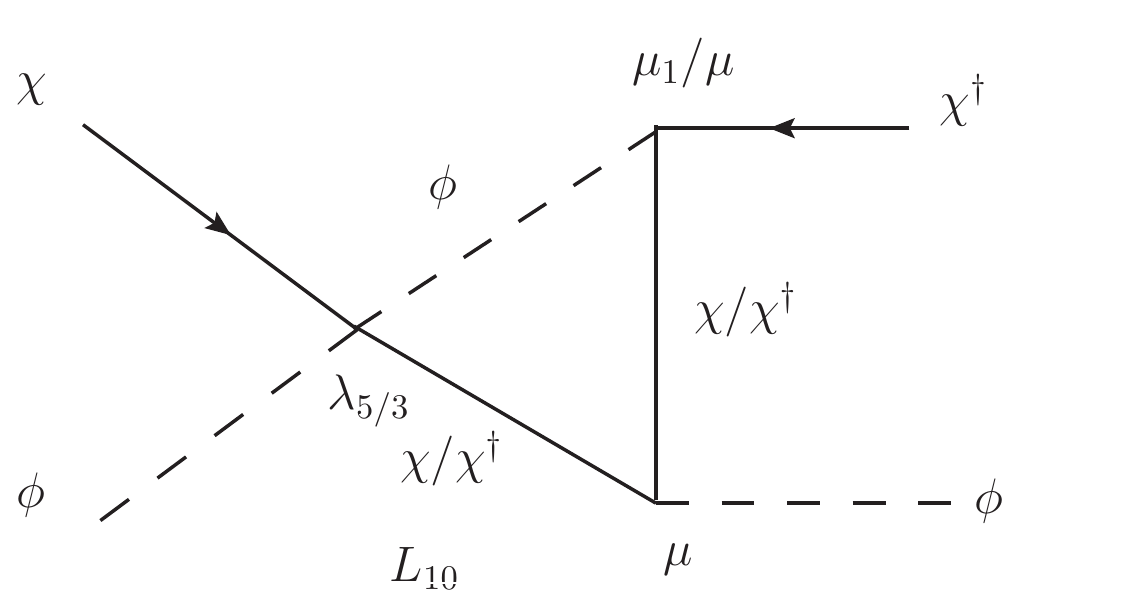}
\includegraphics[scale=0.36]{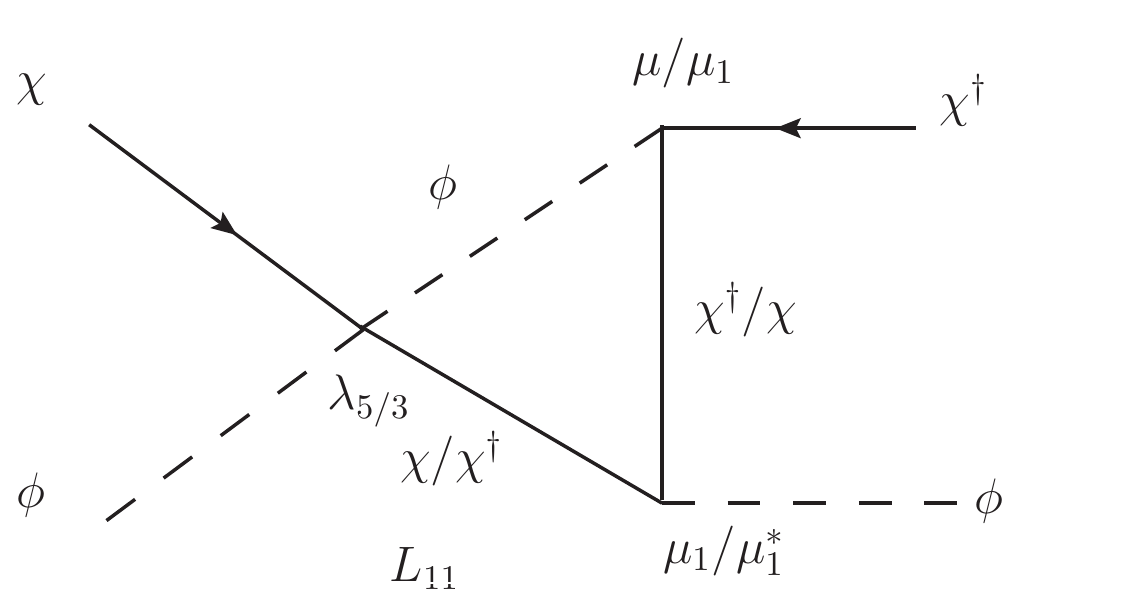}
\caption{\small{\em Relevant tree level and one-loop Feynman diagrams for the CP-violating DM-anti DM conversion process, $\chi+\phi \rightarrow \chi^{\dagger} + \phi$. Apart from $T_1$ and $T_4$, each diagram represents two distinct topologies, depending upon whether the intermediate state  is a particle or an antiparticle state. The particle-flow arrows and the relevant couplings are chosen accordingly.}}
\label{fig:diag10}
\end{figure}

In this appendix, we describe the computational details of the CP-violation in the process $\chi+\phi \rightarrow \chi^{\dagger}+\phi$, the results of which were quoted in Secs.~\ref{sec:sec2} and ~\ref{sec:sec4}. The relevant Feynman diagrams for the process $\chi+\phi \rightarrow \chi^{\dagger}+\phi$ are shown in Fig.~\ref{fig:diag10}.

The CP-violation, resulting  from the interference of the tree-level diagrams with the one-loop ones, can be divided into two categories. The first category is the interference of the contact interaction-induced tree diagram $T_1$ with the one-loop ones, which gives
\vspace{-0.2cm}
\begin{align}
\int dPS_2\left(|M|^2_{\chi\phi\rightarrow \chi^\dagger\phi}-|M|^2_{\chi^{\dagger}\phi \rightarrow \chi\phi}\right)_{T_1-{\rm Loop}}= 4\, \rm Im(\lambda^*_3\mu_1)\int dPS_2\,\bigg[\mu\lambda_5\, Im \big(L_2+L_3+2\,L_4 \nonumber\\
+L_6+L_7+L_{10}+L_{11}\big)+\mu_\phi \lambda_5\, \rm Im \left(L_8+L_{9}\right) \bigg].
\end{align}

Using the Cutkosky rules \cite{Peskin:1995ev}, we find that $\rm Im L_2=Im L_3=Im L_4$, $\rm Im L_6=Im L_7=Im L_{10}=Im L_{11}$ and  $\rm Im L_8=Im L_9$. Thus, the above expression can be simplified as 
\begin{align}
\int dPS_2\left(|M|^2_{\chi\phi\rightarrow \chi^\dagger\phi}-|M|^2_{\chi^{\dagger}\phi \rightarrow \chi\phi}\right)_{T_1-{\rm Loop}}= 4\, \rm Im(\lambda^*_3\mu_1)\int dPS_2\,\bigg[4\,\mu\lambda_5\, Im \left( L_2+L_6\right) \nonumber \\
+2\,\mu_\phi \lambda_5\, \rm Im \left(L_8\right) \bigg]  
\label{eq:eq.A3} 
\end{align}
Similarly, the CP-violation coming from the interference of the other three tree-level diagrams $T_2,T_3$ and $T_4$ with the loop-amplitudes is given by
\begin{align}
\sum^4_{i=2}\int dPS_2\left(|M|^2_{\chi\phi\rightarrow \chi^\dagger\phi}-|M|^2_{\chi^{\dagger}\phi \rightarrow \chi\phi}\right)_{T_i-{\rm Loop}}= - 4\, \rm Im(\lambda^*_3\mu_1)\int dPS_2\,\bigg[4 \mu \lambda _5 \rm Im\, L_1\,\bigg(\frac{1}{s-m_\chi^2} \nonumber\\
+\frac{1}{t-m_\chi^2}\bigg)+2 \mu_\phi \lambda_5\, \rm Im\, L_1\frac{1}{t-m^2_\phi}+ \mathcal{O}(|\hat\mu/m_\chi|^3)\bigg].
\label{eq:eq.A4} 
\end{align}
From the explicit calculations of the integrals in Eq.\ref{eq:eq.A3} and Eq.\ref{eq:eq.A4}  we find the following relation 
\begin{align}
\int dPS_2\,\bigg[4\,\mu\lambda_5\,\rm Im \left( L_2+L_6\right)+
2\,\mu_\phi \lambda_5\, \rm Im \left(L_8\right) \bigg]=\int dPS_2\,\bigg[4 \mu \lambda _5 \rm Im\, L_1\,\left(\frac{1}{s-m_\chi^2} +\frac{1}{t-m_\chi^2}\right )\nonumber\\
+2 \mu_\phi \lambda_5\, \rm Im\, L_1\frac{1}{t-m^2_\phi}\bigg],
\end{align}
where, $s$,$t$ are the Mandelstam variables. This shows that the leading terms cancel identically between the contribution from $T_1$ and the combined contributions from $T_2,T_3$ and $T_4$, whereas the surviving sub-leading terms are found to be 
\begin{align}
\sum^4_{i=1}\int dPS_2\left(|M|^2_{\chi\phi\rightarrow \chi^\dagger\phi}-|M|^2_{\chi^{\dagger}\phi \rightarrow \chi\phi}\right)_{T_i-{\rm Loop}} \propto{\rm Im(\lambda^*_3\mu_1)} \times {\rm ~Terms ~of}~\mathcal{O}({\hat\mu^3/m_\chi^4}),
\end{align}
where, $\hat{\mu}$ is either $\mu$, $\mu_1$ or $\mu_\phi$. 

\newpage
\section{\Large{Invisible Decay of Higgs : $h \rightarrow \chi^{\dagger}+\chi$ channel}}
\label{App.B}
The neutral scalar introduced in the context of ADM in Chapter \ref{chap:chap3} and Chapter \ref{chap:chap4} mixes with SM Higgs due to $(1/2)~\mu_{\phi H}\phi\,|H|^2$ term after EWSB. Consequently, the mass matrix is constructed from the following part of potential, i.e. given by
\begin{align}
\mathcal{L}_M = \frac{1}{2}\left(m^2_{\tilde{h}} \tilde{h}^2+m^2_\phi\phi^2+\mu_{\phi H}\phi\tilde{h}\right).
\end{align}  
After proper diagonalization, the unphysical fields ($\tilde{h},\phi$) are written in terms of physical basis ($h,\tilde{\phi}$) as the following.
\begin{align}
\tilde{h}&= \cos\beta ~ h - \sin\beta ~\tilde{\phi},\nonumber\\
\phi &= \sin \beta ~ h + \cos\beta ~\tilde{\phi} ,
\end{align}  
where, $\beta$ is the mixing angle between $\phi$ and $\tilde{h}$ fields, which is given by
\begin{align}
\tan 2\beta = \frac{\mu_{\phi H}~ v_H}{m^2_{\tilde{h}}-m^2_\phi}.
\end{align}
This mixing introduces interaction between $\chi$ and $h$ via $\lambda \chi^3 \phi$, $\mu_1 \chi^\dagger \chi \phi$ and $\lambda_2 \phi^2 \chi^\dagger \chi$ terms in Eq.\ref{eq:lag1}. For sufficiently small $\beta$ the third term in the above line is suppressed. For sufficiently light DM scenarios, Higgs might decay to three body ($h \rightarrow 3 \chi (\chi^\dagger)$) and two final states ($h\rightarrow \chi^\dagger \chi$) of $\chi$ particles, which will be constrained from the upper bound on the invisible decay width of Higgs given by LHC. The two-body decay would be dominant for $m_\chi = 5 ~\rm GeV$, $|\lambda|=1.3\times 10^{-5}$ and $\mu_1=9 ~\rm MeV$, as given in the scenario shown in the right panel of Fig.\ref{fig:epsilon_model}. The decay width of $h\rightarrow \chi^\dagger \chi$ is given by,
\begin{align}
\Gamma_{h\rightarrow \chi^\dagger\chi} = \frac{\mu^2_1}{16 \pi m_h} \sqrt{1-\frac{4 m^2_\chi}{m^2_h}}~\sin^2 \beta = (12.8 ~{\rm eV})\sin^2 \beta  ~~; ~~~ (m_h = 125 ~\rm GeV) . 
\end{align} 
Now, to put an upper bound on $|\sin \beta|$, the constraint on the invisible decay width of Higgs ($\Gamma_{inv} \lesssim 0.95 ~\rm MeV $ \cite{CMS:2018yfx}) is usually exploited. For this case, we might have looked at $h \rightarrow \tilde{\phi}\tilde{\phi}$ channel. However, for very light mass of $\phi$, the stringent bound on $|\sin \beta|$ comes from the decay of B-mesons, namely $|\sin \beta| \lesssim 10^{-2}$ for $m_{\tilde{\phi}} \leq 5 ~\rm GeV$ \cite{OConnell:2006rsp,Falkowski:2015iwa,CLEO:1989aus}. To remind, for the freeze-out of $\chi$ we demanded earlier that $m_\chi > m_{\tilde{\phi}}$. It is apparent that $\Gamma_{h\rightarrow \chi^\dagger\chi}$ is ridiculously smaller than the $\Gamma_{inv}$. Hence, $m_\chi = 5 ~\rm GeV$ is very much allowed from the experimental data.

%\bibliographystyle{JHEP}
%\bibliography{Thesis}

\end{document}